\newcommand*{\ATLASLATEXPATH}{latex/}
\pdfoutput=1
\documentclass[UKenglish,texlive=2011,cernpreprint,txfonts]{\ATLASLATEXPATH atlasdoc}

\usepackage{\ATLASLATEXPATH atlaspackage}
\usepackage{lscape}
\usepackage{\ATLASLATEXPATH atlasbiblatex}

\usepackage{\ATLASLATEXPATH atlasphysics}

\graphicspath{{logos/}{figures/}}

\usepackage{mydocument-defs}

\AtlasTitle{Measurement of the angular coefficients in $\Zboson$-boson events using electron and muon pairs from data taken at~$\sqrt{s}=8$~TeV with the ATLAS detector }

\author{The ATLAS Collaboration}

\PreprintIdNumber{CERN-EP-2016-087}

 \AtlasJournal{JHEP}

\AtlasAbstract{%
The angular distributions of Drell--Yan charged lepton pairs in the vicinity of the \Zboson-boson mass peak probe the underlying QCD dynamics of $\Zboson$-boson production. This paper presents a measurement of the complete set of angular coefficients $A_{0-7}$ describing these distributions in the $Z$-boson Collins--Soper frame. The data analysed correspond to 20.3~\ifb\ of $pp$~collisions at~$\sqrt{s} = 8$~TeV, collected by the~ATLAS detector at the CERN~LHC. The measurements are compared to the most precise fixed-order calculations currently available~($\mathcal{O}(\alpha^{2}_{\text{s}})$) and with theoretical predictions embedded in Monte Carlo generators. The measurements are precise enough to probe QCD corrections beyond the formal accuracy of these calculations and to provide discrimination between different parton-shower models. A significant deviation from the $\mathcal{O}(\alpha^{2}_{\text{s}})$ predictions is observed for $A_{0}-A_{2}$. Evidence is found for non-zero~$A_{5,6,7}$, consistent with expectations. 
}

\hypersetup{pdftitle={ATLAS draft},pdfauthor={The ATLAS Collaboration}}

\begin{document}

\maketitle



\clearpage
\section{Introduction}
\label{sec:intro}

The angular distributions of charged lepton pairs produced in hadron--hadron collisions via the Drell--Yan neutral current process provide a portal to precise measurements of the production dynamics through spin correlation effects between the initial-state partons and the final-state leptons mediated by a spin-1 intermediate state, predominantly the~\Zboson~boson. In the \Zboson-boson rest frame, a plane spanned by the directions of the incoming protons can be defined, e.g. using the Collins--Soper (CS) reference frame~\cite{Collins:1977prd}. The lepton polar and azimuthal angular variables, denoted by~$\cos\theta$ and~$\phi$ in the following formalism, are defined in this reference frame. The spin correlations are described by a set of nine helicity density matrix elements, which can be calculated within the context of the parton model using perturbative quantum chromodynamics~(QCD). The theoretical formalism is elaborated in~Refs.~\cite{Mirkes92, arXiv9406381, Mirkes94PhysRevD, Mirkes94prep}.

The full five-dimensional differential cross-section describing the kinematics of the two Born-level leptons from the \Zboson-boson decay can be  decomposed as a sum of nine harmonic polynomials, which depend on~$\cos\theta$ and~$\phi$, multiplied by corresponding helicity cross-sections that depend on the \Zboson-boson transverse momentum ($\ptz$), rapidity ($\yz$), and invariant mass ($\mz$). It is a standard convention to factorise out the unpolarised cross-section, denoted in the literature by~$\sigma^{U+L}$, and to present the five-dimensional differential cross-section as an expansion into nine harmonic polynomials $\poli(\cos\theta,\phi)$ and dimensionless angular coefficients $A_{0-7}(\ptz, \yz, \mz)$, which represent ratios of helicity cross-sections with respect to the unpolarised one, $\sigma^{U+L}$, as explained in detail in~Appendix~\ref{appendix:theory}:

\begin{align}
 \label{Eq:master2}
  \frac{ \mathrm{d}\sigma}{\mathrm{d}\ptz~\mathrm{d}\yz~\mathrm{d}\mz~\mathrm{d}\cos\theta~\mathrm{d}\phi} =& 
     \frac{3}{ 16 \pi} \frac{\mathrm{d}\sigma ^{ U+ L }}{\mathrm{d}\ptz~\mathrm{d}\yz~\mathrm{d}\mz}  \nonumber \\
            \bigg\{&(1 + \cos^2\theta) + \frac{1}{2}\ A_0 (1 - 3 \cos^2\theta) + A_1 \ \sin 2 \theta\ \cos\phi \\
   +& \frac{1}{2}\ A_2 \ \sin^2\theta\ \cos 2 \phi +  A_3 \ \sin \theta\ \cos \phi +     A_4 \ \cos\theta  \nonumber \\
    +&     A_5 \ \sin^2\theta \ \sin 2 \phi +     A_6 \ \sin 2 \theta\ \sin \phi +     A_7 \ \sin \theta \ \sin \phi \bigg\}\nonumber .
\end{align}

The dependence of the differential cross-section on $\cos\theta$ and $\phi$ is thus completely manifest analytically. In contrast, the dependence on $\ptz$,~$\yz$, and~$\mz$ is entirely contained in the $\Ai$~coefficients and $\sigma^{U+L}$. Therefore, all hadronic dynamics from the production mechanism are described implicitly within the structure of the $\Ai$~coefficients, and are factorised from the decay kinematics in the \Zboson-boson rest frame. This allows the measurement precision to be essentially insensitive to all uncertainties in QCD, quantum electrodynamics (QED), and electroweak (EW) effects related to \Zboson-boson production and decay. In particular, EW~corrections that couple the initial-state quarks to the final-state leptons have a negligible impact (below~0.05\%) at the \Zboson-boson pole. This has been shown for the LEP~precision measurements~\cite{EWLEP1,EWLEP2}, when calculating the interference between initial-state and final-state QED~radiation. 

When integrating over $\cos\theta$ or $\phi$, the information about the~$A_1$ and $A_6$~coefficients is lost, so both angles must be explicitly used to extract the full set of eight coefficients. Integrating Eq.~(\ref{Eq:master2}) over $\cos\theta$ yields:

\begin{equation}
\frac{\mathrm{d}\sigma}{\mathrm{d}\ptz~\mathrm{d}\yz~\mathrm{d}\mz~\mathrm{d} \phi}   = \frac{1}{2 \pi} \frac{\mathrm{d}\sigma^{U+L}}{\mathrm{d}\ptz~\mathrm{d}\yz~\mathrm{d}\mz}\left\{ 1 + \frac{1}{4} A_2 \cos 2\phi + \frac{3 \pi}{16} A_3 \cos \phi + \frac{1}{2} A_5 \sin 2 \phi + \frac{3 \pi}{16} A_7 \sin \phi \right\},
\label{Eq:master_phi}
\end{equation}

while integrating over $\phi$ yields:

\begin{equation}
\frac{\mathrm{d}\sigma}{\mathrm{d}\ptz~\mathrm{d}\yz~\mathrm{d}\mz~\mathrm{d}\cos\theta}   = \frac{3}{8} \frac{d\sigma^{U+L}}{\mathrm{d}\ptz~\mathrm{d}\yz~\mathrm{d}\mz}\left\{  (1 + \cos^2\theta) + \frac{1}{2}A_0 (1-3 \cos^2\theta) + A_4 \cos\theta \right\}.
\label{Eq:master_theta}
\end{equation}

At leading order (LO) in QCD, only the annihilation diagram $q\bar{q}\rightarrow Z$ is present and only $A_4$ is non-zero. At next-to-leading order (NLO) in~QCD~($\mathcal{O}(\alpha_{\text{s}})$), $A_{0-3}$~also become non-zero. The Lam--Tung relation~\cite{Lam:1978pu,Lam:1978zr,Lam:1980uc}, which predicts that $A_0-A_2=0$ due to the spin-1 of the gluon in the $qg\rightarrow Zq$ and $q\bar q\rightarrow Zg$ diagrams, is expected to hold up to~$\mathcal{O}(\alpha_{\text{s}})$, but can be violated at higher orders. The coefficients~$A_{5,6,7}$ are expected to become non-zero, while remaining small, only at next-to-next-to-leading order (NNLO) in~QCD~($\mathcal{O}(\alpha^{2}_{\text{s}})$), because they arise from gluon loops that are included in the calculations~\cite{Hagiwara1,Hagiwara2}. The coefficients $A_3$ and $A_4$ depend on the product of vector and axial couplings to quarks and leptons, and are sensitive to the Weinberg angle~$\sin^2\theta_{\text{W}}$. The explicit formulae for these dependences can be found in~Appendix~\ref{appendix:theory}. 

The full set of coefficients has been calculated for the first time at~$\mathcal{O}(\alpha^{2}_{s})$ in~Refs.~\cite{Mirkes92, arXiv9406381, Mirkes94PhysRevD, Mirkes94prep}. More recent discussions of these angular coefficients may be found in Ref.~\cite{arXiv:1407.2940}, where the predictions in the NNLOPS~scheme of the \POWHEG~\cite{arXiv:0409146,Frixione:2007vw,Powheg1,Powheg2} event generator are shown for \Zboson-boson production, and in Ref.~\cite{arXiv:1103.5445}, where the coefficients are explored in the context of \Wboson-boson production, for which the same formalism holds.

The CDF Collaboration at the Tevatron published~\cite{PhysRevLett.106.241801} a measurement of some of the angular coefficients of lepton pairs produced near the \Zboson-boson mass pole, using  2.1~\ifb\ of proton--anti-proton collision data at a centre-of-mass energy $\sqrt{s}=1.96$ TeV. Since the measurement was performed only in projections of $\cos\theta$ and $\phi$, the coefficients $A_{1}$ and $A_{6}$ were inaccessible. They further assumed $A_{5}$ and $A_{7}$  to be zero since the sensitivity to these coefficients was beyond the precision of the measurements; the coefficients $A_{0,2,3,4}$ were measured as a function of $\ptz$. These measurements were later used by CDF~\cite{arXiv1307.0770} to infer an indirect measurement of $\sin^2\theta_{\text{W}}$, or equivalently, the \Wboson-boson mass in the on-shell scheme, from the average $A_4$~coefficient. These first measurements of the angular coefficients demonstrated the potential of this not-yet-fully explored experimental avenue for investigating hard~QCD and EW~physics.    

Measurements of the \Wboson-boson angular coefficients at the~LHC were published by both ATLAS~\cite{WpolATLAS} and~CMS~\cite{WpolCMS}. More recently, a measurement of the \Zboson-boson angular coefficients with $Z\rightarrow\mu\mu$ decays was published by CMS~\cite{Khachatryan:2015paa}, where the first five coefficients were measured with 19.7~\ifb of proton--proton ($pp$) collision data at $\sqrt{s}$ = 8~TeV. The measurement was performed in two $\yz$ bins, $0 < |\yz| < 1$ and $1 < |\yz| < 2.1$, each with eight bins in $\ptz$ up to 300 GeV. The violation of the Lam--Tung relation was observed, as predicted by QCD calculations beyond~NLO.

This paper presents an inclusive measurement of the full set of eight $\Ai$ coefficients using charged lepton pairs (electrons or muons), denoted hereafter by~$\ell$. The measurement is performed in the \Zboson-boson invariant mass window of 80--100~GeV, as a function of $\ptz$, and also in three bins of~$\yz$. These results are based on 20.3~\ifb\ of $pp$ collision data collected at  $\sqrt{s} = 8$~TeV by the ATLAS experiment~\cite{atlas-detector} at the LHC. With the measurement techniques developed for this analysis, the complete set of coefficients is extracted with fine granularity over 23~bins of~$\ptz$ up to~600~GeV. The measurements, performed in the~CS reference frame~\cite{Collins:1977prd}, are first presented as a function of  $\ptz$, integrating over $\yz$. Further measurements divided into three bins of~$\yz$ are also presented: $0 < |\yz| < 1$, $1 < |\yz| < 2$, and $2 < |\yz| < 3.5$. The \Zee\ and \Zmm\ channels where both leptons fall within the pseudorapidity range~$|\eta|<2.4$ (hereafter referred to as the central--central or $ee_{\text{CC}}$ and $\mu\mu_{\text{CC}}$ channels) are used for the $\yz$-integrated measurement and the first two $\yz$~bins. The \Zee\ channel where one of the electrons instead falls in the region $|\eta|>2.5$ (referred to hereafter as the central--forward or $ee_{\text{CF}}$ channel) is used to extend the measurement to the high-$\yz$ region encompassed by the third $\yz$ bin. In this case, however, because of the fewer events available for the measurement itself and to evaluate the backgrounds (see Section~\ref{sec:DataAnalysis}), the measurement is only performed for $\ptz$ up to~100~GeV using projections of~$\cos\theta$ and~$\phi$, making $A_{1}$ and $A_{6}$ inaccessible in the $2 < |\yz| < 3.5$ bin. 

The high granularity and precision of the specific measurements presented in this paper provide a stringent test of the most precise perturbative QCD~predictions for \Zboson-boson production in $pp$ collisions and of Monte Carlo (MC) event generators used to simulate \Zboson-boson production. This paper is organised as follows. Section~\ref{sec:theory} summarises the theoretical formalism used to extract the angular coefficients and presents the fixed-order QCD predictions for their variations as a function of~$\ptz$. Section~\ref{sec:dataMC} describes briefly the ATLAS detector and the data and MC~samples used in the analysis, while Section~\ref{sec:DataAnalysis} presents the data analysis and background estimates for each of the three channels considered. Section~\ref{sec:Methodology} describes the fit methodology used to extract the angular coefficients in the full phase space as a function of~$\ptz$ and Section~\ref{sec:Uncertainties} gives an overview of the statistical and systematic uncertainties of the measurements. Sections~\ref{sec:results} and~\ref{sec:comparisons} present the results and compare them to various predictions from theoretical calculations and MC~event generators, and Section~\ref{sec:conclusions} summarises and concludes the paper.

\section{Theoretical predictions}
\label{sec:theory}


The differential cross-section in~Eq.~(\ref{Eq:master2}) is written for pure \Zboson~bosons, although it also holds for the contribution from~$\gamma^*$ and its interference with the~$Z$~boson. The tight invariant mass window of~80--100~GeV is chosen to minimise the $\gamma^*$~contribution, although the predicted $\Ai$~coefficients presented in this paper are effective coefficients, containing this small contribution from~$\gamma^*$. This contribution is not accounted for explicitly in the detailed formalism described in~Appendix~\ref{appendix:theory}, which is presented for simplicity for pure $Z$-boson production. Throughout this paper, the leptons from $Z$-boson decays are defined at the Born level, i.e.~before final-state QED radiation, when discussing theoretical calculations or predictions at the event-generator level.

The $\ptz$ and $\yz$ dependence of the coefficients varies strongly with the choice of spin quantisation axis in the \Zboson-boson rest frame ($z$-axis). In the CS reference frame adopted for this paper, the $z$-axis is defined in the \Zboson-boson rest frame as the external bisector of the angle between the momenta of the two protons, as depicted in~Fig.~\ref{CSframe}. The positive direction of the $z$-axis is defined by the direction of positive longitudinal \Zboson-boson momentum in the laboratory frame. To complete the coordinate system, the $y$-axis is defined as the normal vector to the plane spanned by the two incoming proton momenta and the $x$-axis is chosen to define a right-handed Cartesian coordinate system with the other two axes. Polar and azimuthal angles are calculated with respect to the negatively charged lepton and are labelled~$\theta_{\text{CS}}$ and~$\phics$, respectively. In the case where~$\ptz = 0$, the direction of the $y$-axis and the definition of~$\phics$ are arbitrary. Historically, there has been an ambiguity in the definition of the sign of the $\phics$~angle in the CS~frame: this paper adopts the recent convention followed by Refs.~\cite{arXiv:1407.2940,Khachatryan:2015paa}, whereby the coefficients~$A_1$ and~$A_3$ are positive.

\begin{figure}[htp]
  \begin{center}                               
{
    \includegraphics[width=10cm,angle=0]{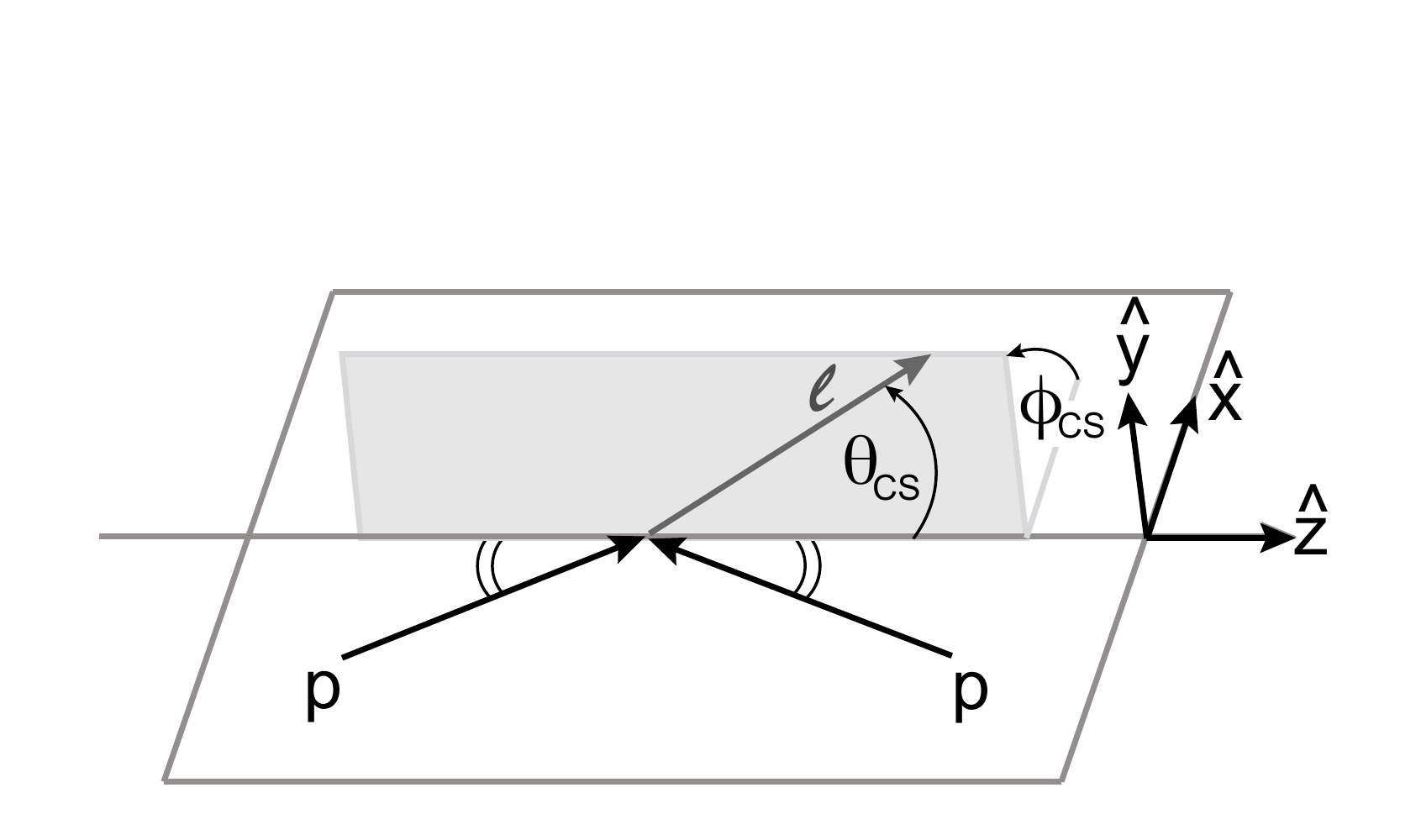}
}
\end{center}
\caption{Sketch of the Collins-Soper reference frame, in which the angles ~$\theta_{\text{CS}}$ and~$\phics$ are defined with respect to the negatively charged lepton~$\ell$~(see text). The notations $\hat x, \hat y$ and $\hat z$ denote the unit vectors along the corresponding axes in this reference frame.  
\label{CSframe} }
\end{figure}

The coefficients are not explicitly used as input to the theoretical calculations nor in the MC~event generators. They can, however, be extracted from the shapes of the angular distributions with the method proposed in~Ref.~\cite{arXiv9406381}, owing to the orthogonality of the $\poli$~polynomials. The weighted average of the angular distributions with respect to any specific polynomial isolates an average reference value or moment of its corresponding coefficient. The moment of a polynomial $P(\cos\theta,\phi)$ over a specific range of~$\ptz$, $\yz$, and~$\mz$ is defined to be:

\begin{equation}
\langle P(\cos\theta,\phi)\rangle = \frac{\int  P(\cos\theta,\phi) \mathrm{d}\sigma(\cos\theta,\phi) \mathrm{d}\cos\theta \mathrm{d}\phi}{\int \mathrm{d}\sigma(\cos\theta,\phi) \mathrm{d}\cos\theta \mathrm{d}\phi} .
\end{equation}

The moment of each harmonic polynomial can thus be expressed as (see~Eq.~(\ref{Eq:master2})):

\begin{equation}
\begin{split}
\langle\frac{1}{2}(1-3 \cos^2\theta)\rangle & = \frac{3}{20} (A_0 - \frac{2}{3} ); \ \ \ 
\langle\sin2\theta \cos\phi\rangle  = \frac{1}{5} A_1; \ \ \
\langle\sin^2\theta \cos2\phi\rangle  = \frac{1}{10} A_2; \\
\langle\sin\theta \cos\phi\rangle & = \frac{1}{4} A_3; \ \ \
\langle\cos\theta\rangle  = \frac{1}{4} A_4; \ \ \
\langle\sin^2\theta \sin 2 \phi\rangle = \frac{1}{5} A_5; \\
\langle\sin 2\theta \sin \phi\rangle & = \frac{1}{5} A_6; \ \ \
\langle\sin \theta \sin \phi\rangle  = \frac{1}{4} A_7 .
\end{split}
\label{Eq:moments}
\end{equation}

One thus obtains a representation of the effective angular coefficients for $Z/\gamma^*$~production. These effective angular coefficients display in certain cases a dependence on~$\yz$, which arises mostly from the fact that the interacting quark direction is unknown on an event-by-event basis. As the method of~Ref.~\cite{arXiv9406381} relies on integration over the full phase space of the angular distributions, it cannot be applied directly to data, but is used to compute all the theoretical predictions shown in this paper.

The inclusive fixed-order perturbative QCD predictions for \Zboson-boson production at NLO and NNLO were obtained with \DYNNLO~v1.3~\cite{Catani:2009sm}. These inclusive calculations are formally accurate to~$\mathcal{O}(\alpha^2_{\text{s}})$. The Z-boson is produced, however, at non-zero transverse momentum only at~$\mathcal{O}(\alpha_{\text{s}})$, and therefore the calculation of the coefficients as a function of~$\ptz$ is only~NLO. Even though the fixed-order calculations do not provide reliable absolute predictions for the $\ptz$~spectrum at low values, they can be used for $\ptz~>$~2.5~GeV for the angular coefficients. The results were cross-checked with NNLO predictions from FEWZ~v3.1.b2~\cite{Gavin:2010az, Gavin:2012sy, Li:2012wn} and agreement between the two programs was found within uncertainties. The renormalisation and factorisation scales in the calculations were set to $E_{\text{T}}^Z=\sqrt{(\mz)^{2}+(\ptz)^{2}}$~\cite{Aad:2014xaa} on an event-by-event basis. The calculations were done using the CT10~NLO or~NNLO~parton distribution functions~(PDFs)~\cite{Gao:2013xoa}, depending on the order of the prediction.

The NLO EW corrections affect mostly the leading-order QCD cross-section normalisation in the \Zboson-pole region and have some impact on the $\ptz$ distribution, but they do not affect the angular correlations at the \Zboson-boson vertex. The \DYNNLO~calculation was done at leading order in~EW, using the $G_{\mu}$ scheme~\cite{Gmureference}. This choice determines the value of $A_4$ at low~$\ptz$, and for the purpose of the comparisons presented in this paper, both $A_3$ and $A_4$ obtained from \DYNNLO are rescaled to the values predicted when using the measured value of~$\sin^2\theta_{\text{W}}^{\rm{eff}} = 0.23113$~\cite{PDG14}.  

The theoretical predictions are shown in Fig.~\ref{Fig::TheoryEvgen:AngCoeff} and tabulated in~Table~\ref{Tab:PredSummaryMGV} for three illustrative $\ptz$~bins. The binning in $\ptz$ is chosen based on the experimental resolution at low~$\ptz$ and on the number of events at high~$\ptz$ and has the following boundaries (in~GeV) used consistently throughout the measurement:

\begin{equation}
\begin{array}{l l l l l l l l l l l}
\ptz^{,\rm{boundary}} [\rm{GeV}]
    & = & \{0,  & 2.5,  & 5.0,  & 8.0,   & 11.4,  & 14.9, & 18.5, & 22.0,\\
    &   & 25.5, & 29.0, & 32.6, & 36.4,  & 40.4,  & 44.9, & 50.2, & 56.4,\\
    &   & 63.9, & 73.4, & 85.4, & 105.0, & 132.0, & 173.0,& 253.0,& 600.0\} .\\
\end{array}
\label{Tab:MC:pTbins}
\end{equation}

The predictions show the following general features. The $A_0$ and $A_2$ coefficients increase as a function of~$\ptz$ and the deviations from lowest-order expectations are quite large, even at modest values of~$\ptz~=$~20--50~GeV. The~$A_1$ and $A_3$~coefficients are relatively small even at large $\ptz$, with a maximum value of~$0.08$. In the limit where $\ptz =  0$, all coefficients except~$A_4$ are expected to vanish at~NLO. The NNLO~corrections are typically small for all coefficients except~$A_2$, for which the largest correction has a value of~$-0.08$, in agreement with the original theoretical studies~\cite{Mirkes92}. The theoretical predictions for $A_{5,6,7}$ are not shown because these coefficients are expected to be very small at all values of~$\ptz$: they are zero at~NLO and the NNLO~contribution is large enough to be observable, namely of the order of~$0.005$ for values of~$\ptz$ in the range~20--200~GeV.

The statistical uncertainties of the calculations, as well as the factorisation and renormalisation scale and PDF uncertainties, were all considered as sources of theoretical uncertainties. The statistical uncertainties of the NLO and NNLO predictions in absolute units are typically~$0.0003$ and~$0.003$, respectively. The larger statistical uncertainties of the NNLO predictions are due to the longer computational time required than for the NLO~predictions. The scale uncertainties were estimated by varying the renormalisation and factorisation scales simultaneously up and down by a factor of two. As stated in Ref.~\cite{Mirkes92}, the theoretical uncertainties due to the choice of these scales are very small for the angular coefficients because they are ratios of cross-sections. The resulting variations of the coefficients at NNLO were found in most cases to be comparable to the statistical uncertainty. The PDF uncertainties were estimated using the CT10 NNLO~eigenvector variations, as obtained from~FEWZ and normalised to~$68\%$~confidence level. They were found to be small compared to the NNLO statistical uncertainty, namely of the order of~$0.001$ for~$A_{0-3}$ and~$0.002$ for~$A_4$.

\begin{figure}[p]
  \begin{center}                               
{
    \includegraphics[width=7.5cm,angle=0]{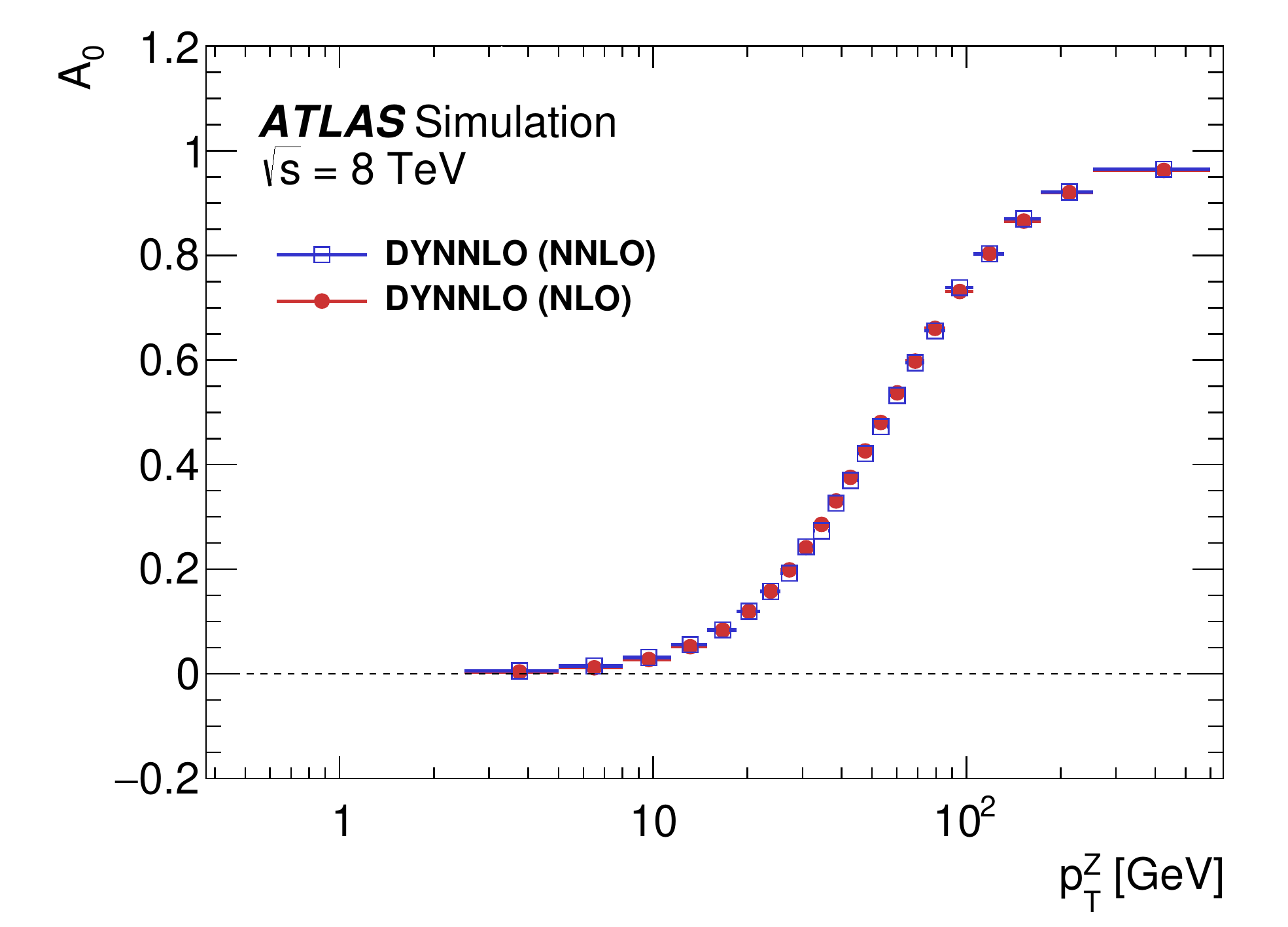}
    \includegraphics[width=7.5cm,angle=0]{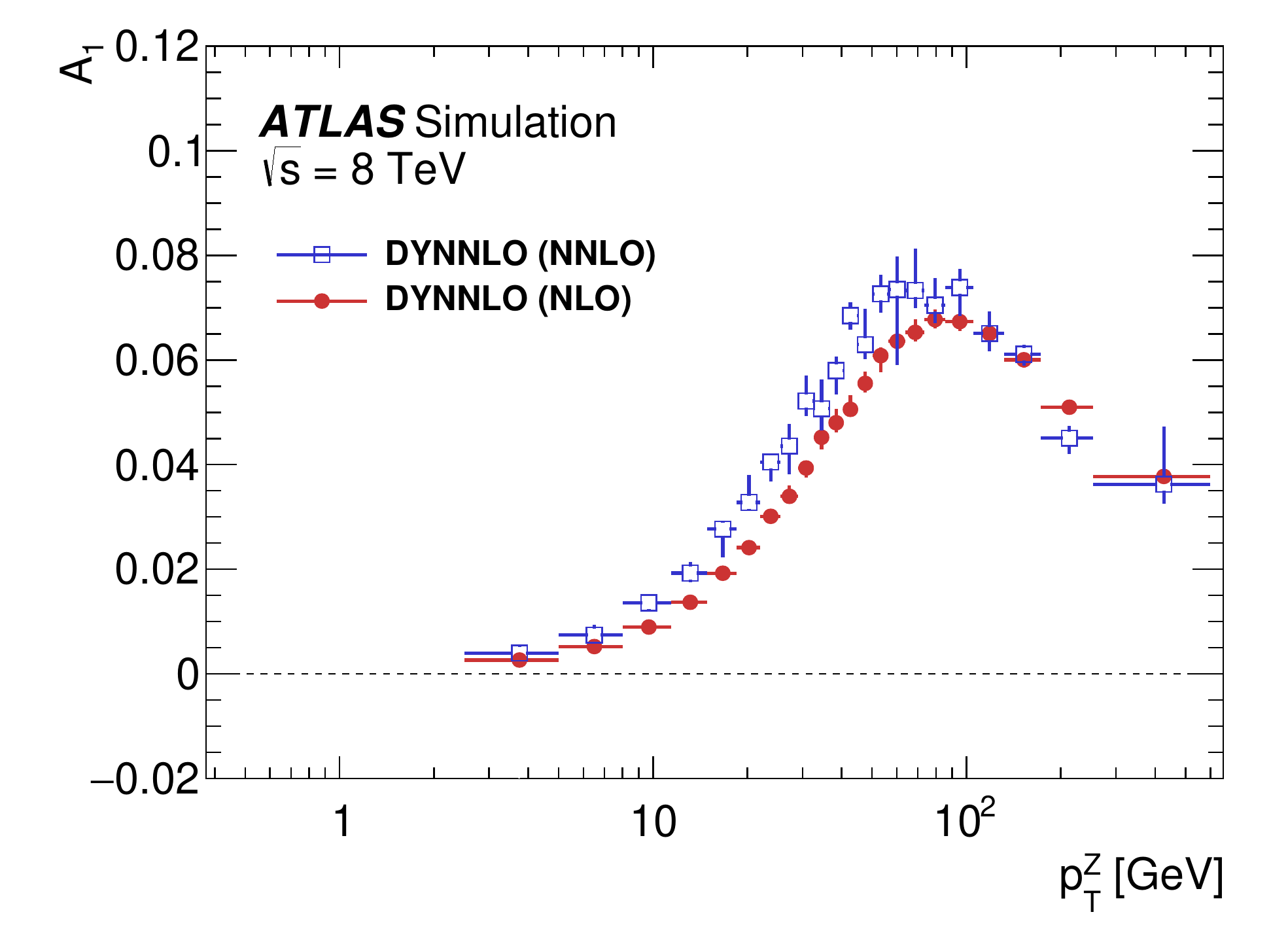}
    \includegraphics[width=7.5cm,angle=0]{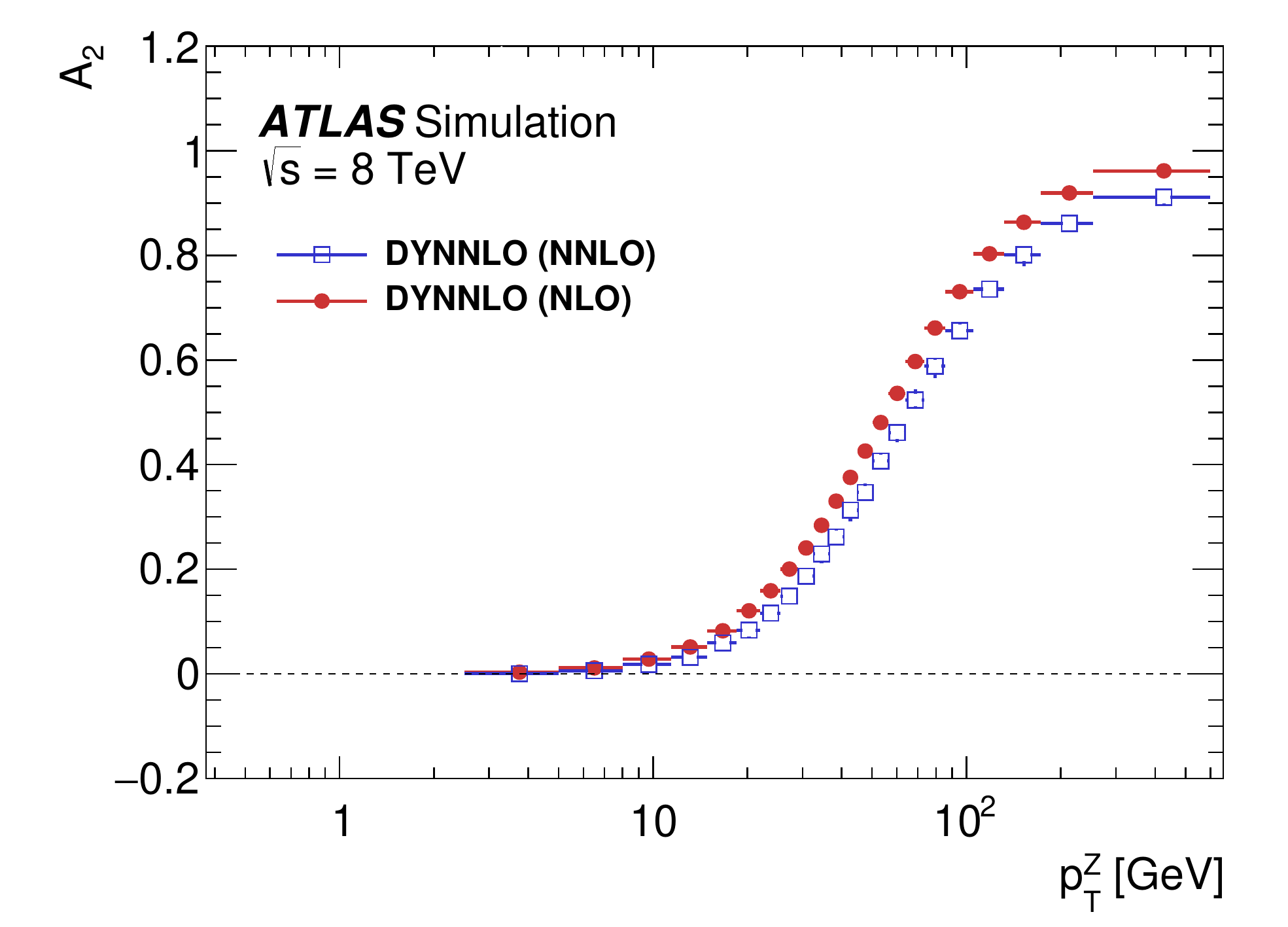}
    \includegraphics[width=7.5cm,angle=0]{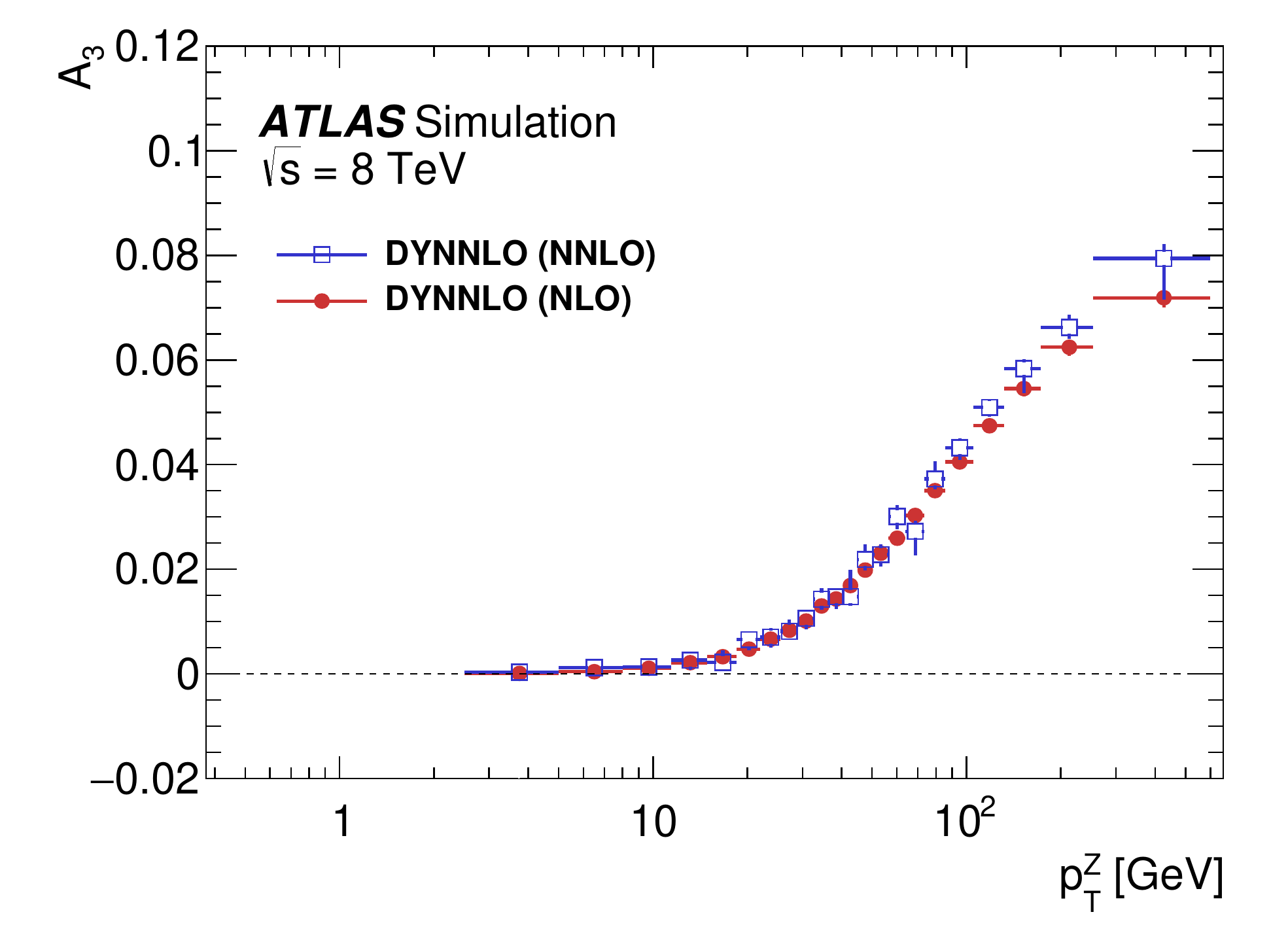}
    \includegraphics[width=7.5cm,angle=0]{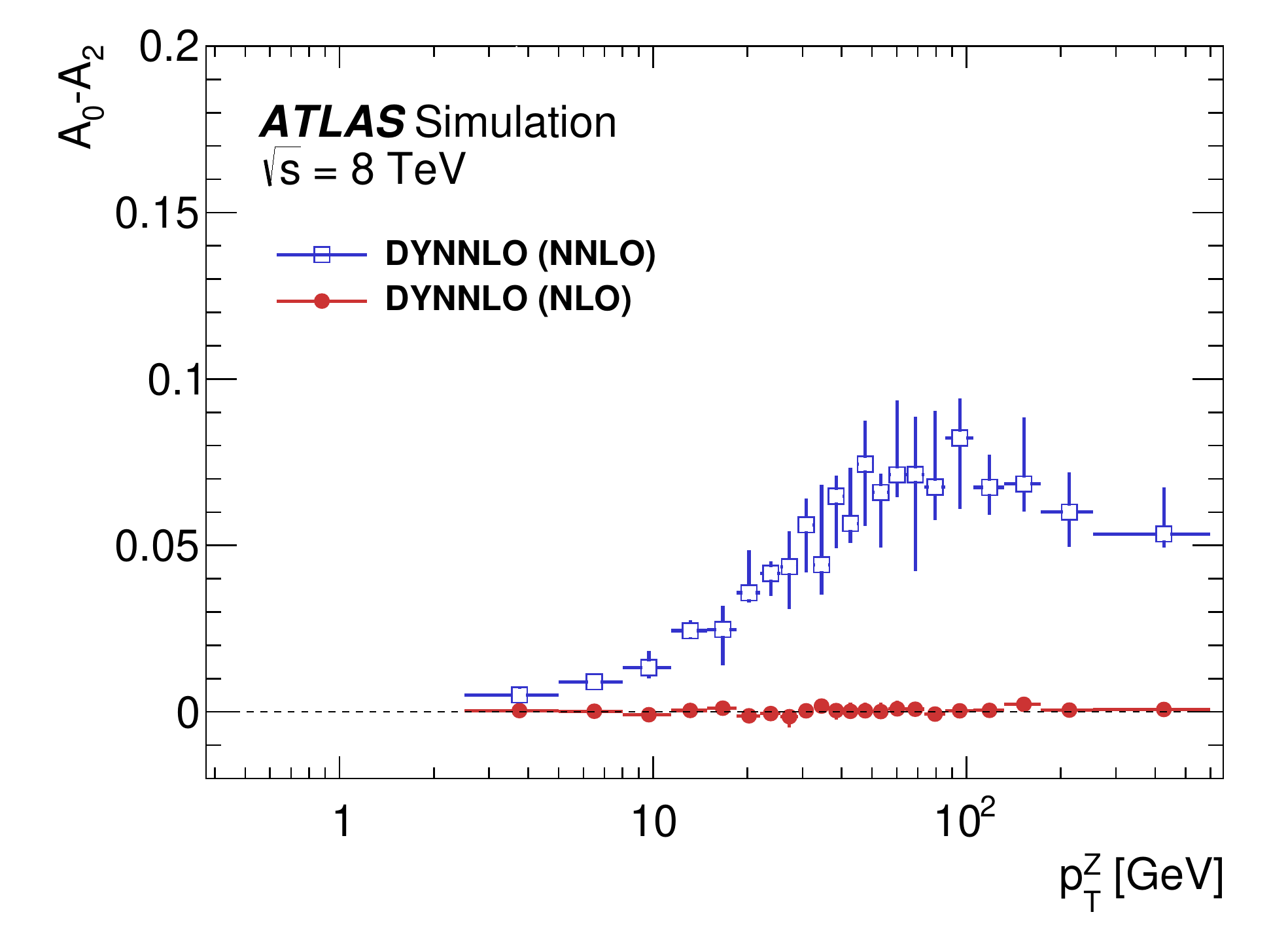}
    \includegraphics[width=7.5cm,angle=0]{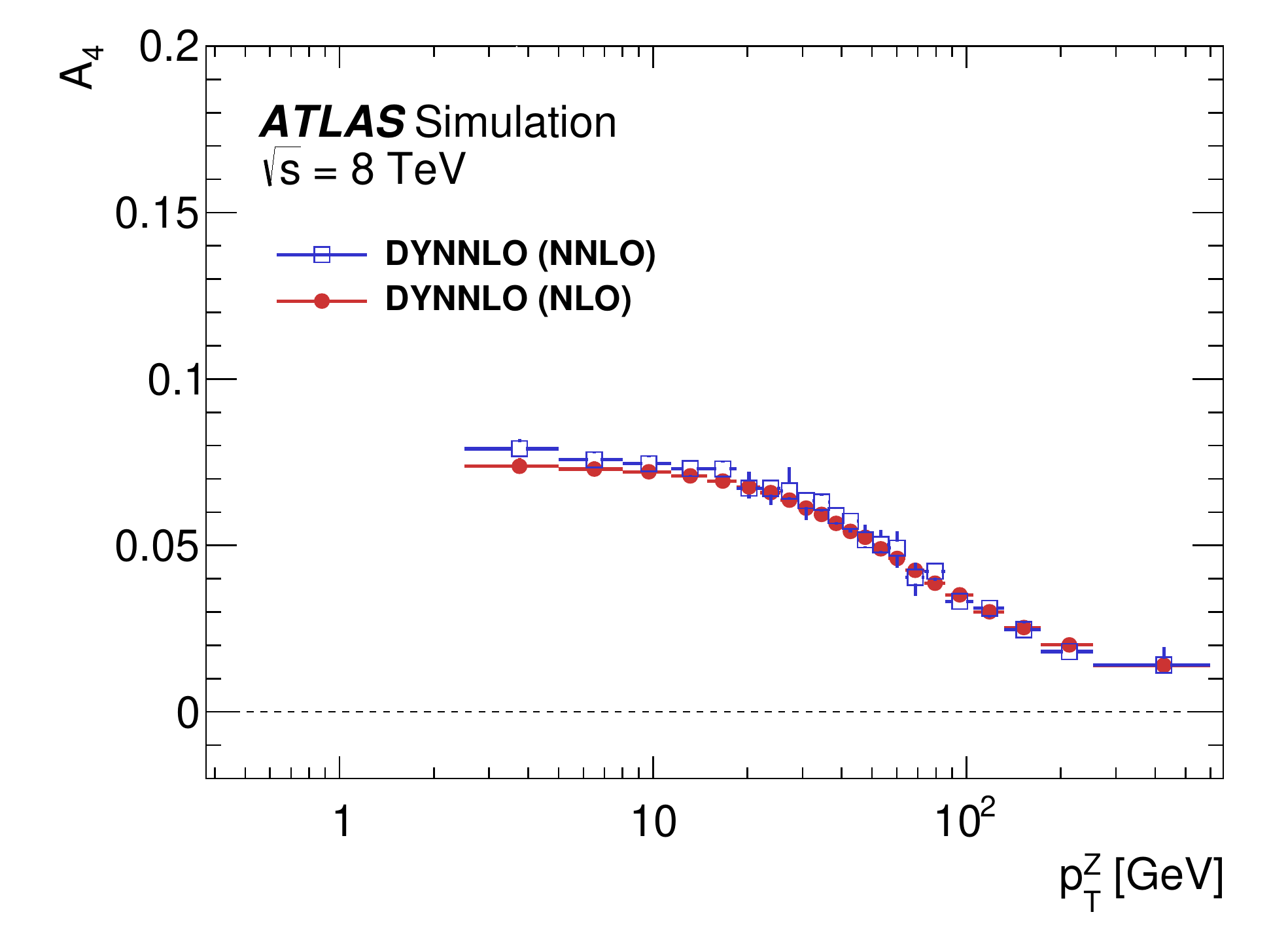}

}
\end{center}
\caption{The angular coefficients $A_{0-4}$  and the difference $A_0-A_2$, shown as a function of~\ptz, as predicted
from \DYNNLO calculations at~NLO and~NNLO in~QCD. The NLO~predictions for $A_0-A_2$ are compatible with zero, as expected from the Lam--Tung relation~\cite{Lam:1978pu,Lam:1978zr,Lam:1980uc}. The error bars show the total uncertainty 
of the predictions, including contributions from statistical uncertainties, QCD~scale variations and~PDFs. The statistical uncertainties
of the NNLO~predictions are dominant and an order of magnitude larger than those of the NLO~predictions. 
\label{Fig::TheoryEvgen:AngCoeff} }
\end{figure}

\begin{table}
\caption{Summary of predictions from DYNNLO at NLO and NNLO for $A_{0}$, $A_{2}$, $A_{0}-A_{2}$, $A_{1}$, $A_{3}$, $A_{4}$, $A_{5}$, $A_{6}$, and $A_{7}$ at low (5--8~GeV), mid (22--25.5~GeV), and high (132--173~GeV) \ptz\ for the $\yz$-integrated configuration. The uncertainty represents the sum of statistical and systematic uncertainties.\label{Tab:PredSummaryMGV}}
\begin{center}
\small
    \setlength\extrarowheight{3pt}
\begin{tabular}{l||c|c||c|c||c|c}
\hline
\hline
& \multicolumn{2}{c||}{$\ptz=5-8$ GeV} & \multicolumn{2}{c||}{$\ptz=22-25.5$~GeV} & \multicolumn{2}{c}{$\ptz=132-173$ GeV}\\ 
& NLO & NNLO & NLO & NNLO & NLO & NNLO \\ \hline
$A_{0}$ &   $0.0115^{+0.0006}_{-0.0003}$ & $0.0150^{+0.0006}_{-0.0008}$ & $0.1583^{+0.0008}_{-0.0009}$ & $0.1577^{+0.0041}_{-0.0018}$ & $0.8655^{+0.0008}_{-0.0006}$ & $0.8697^{+0.0017}_{-0.0023}$ \\
$A_{2}$ &   $0.0113^{+0.0004}_{-0.0004}$ & $0.0060^{+0.0010}_{-0.0017}$ & $0.1588^{+0.0014}_{-0.0009}$ & $0.1161^{+0.0092}_{-0.0028}$ & $0.8632^{+0.0013}_{-0.0009}$ & $0.8012^{+0.0073}_{-0.0215}$ \\
$A_{0}-A_{2}$ & $0.0002^{+0.0007}_{-0.0005}$ & $0.0090^{+0.0014}_{-0.0013}$ & $-0.0005^{+0.0016}_{-0.0012}$ & $ 0.0416^{+0.0036}_{-0.0067}$ & $0.0023^{+0.0015}_{-0.0011}$ & $0.0685^{+0.0200}_{-0.0082}$ \\[3pt] \hline
$A_{1}$ & $0.0052^{+0.0004}_{-0.0003}$ & $0.0074^{+0.0020}_{-0.0008}$ & $0.0301^{+0.0013}_{-0.0013}$ & $0.0405^{+0.0014}_{-0.0038}$ & $0.0600^{+0.0013}_{-0.0015}$ & $0.0611^{+0.0018}_{-0.0023}$ \\
$A_{3}$ & $0.0004^{+0.0002}_{-0.0001}$ & $0.0012^{+0.0003}_{-0.0006}$ & $0.0066^{+0.0003}_{-0.0005}$ & $0.0070^{+0.0017}_{-0.0020}$ & $0.0545^{+0.0003}_{-0.0016}$ & $0.0584^{+0.0018}_{-0.0047}$ \\
$A_{4}$ & $0.0729^{+0.0023}_{-0.0006}$ & $0.0757^{+0.0021}_{-0.0025}$ & $0.0659^{+0.0019}_{-0.0003}$ & $0.0672^{+0.0018}_{-0.0050}$ & $0.0253^{+0.0007}_{-0.0002}$ & $0.0247^{+0.0024}_{-0.0018}$\\[3pt] \hline
$A_{5}$ & $0.0001^{+0.0002}_{-0.0002}$ & $0.0001^{+0.0007}_{-0.0007}$ & $< 0.0001$ & $0.0011^{+0.0013}_{-0.0030}$ & $-0.0004^{+0.0005}_{-0.0005}$ & $ 0.0044^{+0.0042}_{-0.0026}$ \\
$A_{6}$ & $-0.0002^{+0.0002}_{-0.0003}$ & $ 0.0013^{+0.0006}_{-0.0005}$ & $0.0004^{+0.0006}_{-0.0004}$ & $0.0017^{+0.0043}_{-0.0015}$ & $0.0003^{+0.0003}_{-0.0006}$ & $0.0028^{+0.0017}_{-0.0018}$ \\ 
$A_{7}$ & $< 0.0001$ & $0.0014^{+0.0007}_{-0.0004}$ & $0.0002^{+0.0003}_{-0.0007}$ & $0.0024^{+0.0013}_{-0.0013}$ & $0.0003^{+0.0004}_{-0.0007}$ & $0.0048^{+0.0027}_{-0.0012}$ \\[3pt] \hline
\hline
\end{tabular}
\end{center}
\end{table}

\section{The ATLAS experiment and its data and Monte Carlo samples}
\label{sec:dataMC}

\newcommand{\AtlasCoordFootnote}{%
ATLAS uses a right-handed coordinate system with its origin at the nominal interaction point (IP) in the centre of the detector and the $z$-axis along the beam pipe. The $x$-axis points from the IP to the centre of the LHC ring, and the $y$-axis points upwards. Cylindrical coordinates $(r,\phi)$ are used in the transverse plane, $\phi$ being the azimuthal angle around the $z$-axis. The pseudorapidity is defined in terms of the polar angle $\theta$ as $\eta = -\ln \tan(\theta/2)$. Angular distance is measured in units of $\Delta R \equiv \sqrt{(\Delta\eta)^{2} + (\Delta\phi)^{2}}$.}

\subsection{ATLAS detector}

The ATLAS experiment~\cite{atlas-detector} at the LHC is a multi-purpose particle detector
with a forward-backward symmetric cylindrical geometry and a near $4\pi$ coverage in solid angle.\footnote{\AtlasCoordFootnote}
It consists of an inner tracking detector, electromagnetic (EM) and hadronic calorimeters, and a muon spectrometer.
The inner tracker provides precision tracking of charged particles in the pseudorapidity range $|\eta| < 2.5$.
This region is matched to a high-granularity EM sampling calorimeter covering the pseudorapidity range $|\eta| < 3.2$ 
and a coarser granularity calorimeter up to $|\eta| = 4.9$.
A hadronic calorimeter system covers the entire pseudorapidity range up to  $|\eta| = 4.9$.
The muon spectrometer provides triggering and tracking capabilities in the range $|\eta| < 2.4$ and $|\eta| < 2.7$, respectively.
A first-level trigger is implemented in hardware, followed by two software-based trigger levels that
together reduce the accepted event rate to \SI{400}{\hertz} on average. For this paper, a central lepton is one found in the region $|\eta| < 2.4$ (excluding, for electrons, the electromagnetic calorimeter barrel/end-cap transition region~$1.37 < |\eta| < 1.52$), while a forward electron is one found in the region $2.5 < |\eta| < 4.9$ (excluding the transition region~$3.16 < |\eta| < 3.35$ between the electromagnetic end-cap and forward calorimeters).

\subsection{Data and Monte Carlo samples}

The data were collected by the ATLAS detector in 2012 at a centre-of-mass energy of $\sqrt{s} = 8$~TeV, and correspond to an integrated luminosity of~20.3~fb$^{-1}$. The mean number of additional $pp$ interactions per bunch crossing (pile-up events) in the data set is approximately~20. 

The simulation samples used in the analysis are shown in Table~\ref{Tab:Zall_MC_samples}. The four event generators used to produce the $Z/\gamma^*\rightarrow \ell\ell$ signal events are listed in Table~\ref{Tab:Zall_MC_samples}. The baseline \POWHEGBOX~(v1/r2129) sample~\cite{arXiv:0409146,Frixione:2007vw,Powheg1,Powheg2}, which uses the {\sc CT10 NLO} set of~PDFs~\cite{Lai:2010vv}, is interfaced to \PYTHIA~8~(v.8.170)~\cite{arXiv:0710.3820} with the AU2~set of tuned parameters~\cite{ATL-PHYS-PUB-2012-003} to simulate the parton shower, hadronisation and underlying event, and to \PHOTOS~(v2.154)~\cite{Golonka:2005pn} to simulate QED~final-state radiation~(FSR) in the $\Zboson$-boson decay. The alternative signal samples are from \POWHEGBOX interfaced to \HERWIG~(v.6.520.2)~\cite{Corcella:2000bw} for the parton shower and hadronisation, \JIMMY~(v4.31)~\cite{Butterworth:1996zw} for the underlying event, and \PHOTOS for~FSR. The \SHERPA~(v.1.4.1)~\cite{Gleisberg:2008ta,Hoeche:2009rj,Gleisberg:2008fv,Schumann:2007mg} generator is also used, and has its own implementation of the parton shower, hadronisation, underlying event and FSR, and uses the {\sc CT10 NLO} PDF~set. These alternative samples are used to test the dependence of the analysis on different matrix-element calculations and parton-shower models, as discussed in~Section~\ref{sec:Uncertainties}. The \POWHEG~(v2.1)~+~\MINLO event generator~\cite{arXiv:1206.3572} was used for the $Z$+jet process at~NLO to normalise certain reference coefficients for the $ee_{CF}$~analysis, as described in~Section~\ref{sec:Methodology}. The number of events available in the baseline \POWHEGBOX+~\PYTHIA8 signal sample corresponds to approximately~4~(25)~times that in the data below (above) $\ptz=105$~GeV.

Backgrounds from EW (diboson and $\gamma\gamma\to \ell\ell$ production) and top-quark (production of top-quark pairs and of single top quarks) processes are evaluated from the MC samples listed in Table~\ref{Tab:Zall_MC_samples}. The $\Wboson + \mathrm{jets}$ contribution to the background is instead included in the data-driven multijet background estimate, as described in~Section~\ref{sec:DataAnalysis}; \Wboson-boson samples listed in Table~\ref{Tab:Zall_MC_samples} are thus only used for studies of the background composition. 

All of the samples are processed with the {\textsc{Geant}4}-based simulation~\cite{Agostinelli:2002hh} of the ATLAS detector~\cite{SOFT-2010-01}. The effects of additional $pp$ collisions in the same or nearby bunch crossings are simulated by the addition of so-called minimum-bias events generated with~\PYTHIA~8. 

\begin{table} 
\caption{ MC samples used to estimate the signal and backgrounds in the analysis. }
\vspace{2mm}
\begin{center} 
\small
\begin{tabular}{ l|l|l|l} 
\hline \hline 
 Signature                        & Generator   &PDF & Refs.\\  \hline \hline 
$Z/\gamma^* \rightarrow \ell \ell$ & \POWHEGBOX +~\PYTHIA8     &  {\sc CT10 NLO}  & \cite{arXiv:0409146,Frixione:2007vw,Powheg1,Powheg2,arXiv:0710.3820,Lai:2010vv}\\
$Z/\gamma^* \rightarrow \ell \ell$ & \POWHEGBOX +~\JIMMY/\HERWIG   &  {\sc CT10 NLO}  & \cite{Corcella:2000bw}\\
$Z/\gamma^* \rightarrow \ell \ell$ & \SHERPA             &  {\sc CT10 NLO}  & \cite{Gleisberg:2008ta,Hoeche:2009rj,Gleisberg:2008fv,Schumann:2007mg}\\ 
$Z/\gamma^* \rightarrow \ell \ell$ + jet & \POWHEG +~\MINLO      &  {\sc CT10 NLO}  & \cite{arXiv:1206.3572}\\
$W\rightarrow \ell\nu$            & \POWHEGBOX +~\PYTHIA8    &  {\sc CT10 NLO}  & \\
$W\rightarrow \ell\nu$            & \SHERPA             &  {\sc CT10 NLO}  & \\
$t\bar{t}$ pair                   & \MCATNLO +~\JIMMY/\HERWIG   & {\sc CT10 NLO}  & \cite{Frixione:2002ik,Butterworth:1996zw}\\
Single top quark:                 &                            &              & \\
 \: $t$ channel                   & \ACERMC  +~\PYTHIA6    & {\sc CTEQ6L1}         & \cite{Kersevan:2004yg,Pumplin:2002vw}\\
 \: $s$ and $Wt$ channels         & \MCATNLO+~\JIMMY/\HERWIG   & {\sc CT10 NLO}          &\\
Dibosons                           & \SHERPA             & {\sc CT10 NLO}          & \\
Dibosons                          & \HERWIG           & {\sc CTEQ6L1}         & \\
$\gamma\gamma\to \ell\ell$        & \PYTHIA8            & {\sc MRST2004QED NLO}   & \cite{Martin:2004dh}\\
\hline \hline 
\end{tabular}
\end{center} 
\label{Tab:Zall_MC_samples}
\end{table}

\section{Data analysis}
\label{sec:DataAnalysis}

\subsection{Event selection}

As mentioned in~Sections~\ref{sec:intro} and~\ref{sec:dataMC}, the data are split into three orthogonal channels, namely the $ee_{\text{CC}}$ channel with two central electrons, the $\mu\mu_{\text{CC}}$ channel with two central muons, and the $ee_{\text{CF}}$ channel with one central electron and one forward electron. Selected events are required to be in a data-taking period in which the beams were stable and the detector was functioning well, and to contain a reconstructed primary vertex with at least three tracks with~$\pT > 0.4$~GeV. 

Candidate $ee_{\text{CC}}$~events are obtained using a dielectron trigger requiring two electron candidates with $\pT~>~12$~GeV, combined with  high-\pT\ single-electron triggers. Electron candidates are required to have $\pT~>~25$~GeV and are reconstructed from clusters of energy in the electromagnetic calorimeter matched to inner detector tracks. The electron candidates must satisfy a set of ``medium'' selection criteria~\cite{Electrons2011, ATLAS-CONF-2014-032}, which have been optimised for the level of pile-up present in the 2012 data. Events are required to contain exactly two electron candidates of opposite charge satisfying the above criteria.

Candidate $\mu\mu_{\text{CC}}$~events are retained for analysis using a dimuon trigger requiring two muon candidates with~$\pT > 18$~GeV and 8 GeV, respectively, combined with single high-$\pT$ muon triggers. Muon candidates are required to have $\pT > 25$~GeV and are identified as tracks in the inner detector which are matched and combined with track segments in the muon spectrometer~\cite{Muons20112012}. Track-quality and longitudinal and transverse impact-parameter requirements are imposed for muon identification to suppress backgrounds, and to ensure that the muon candidates originate from a common primary $pp$ interaction vertex. Events are required to contain exactly two muon candidates of opposite charge satisfying the above criteria.

Candidate $ee_{\text{CF}}$~events are obtained using a single-electron trigger, requiring an isolated central electron candidate with~$\pT > 24$~GeV, combined with a looser high-\pT\ single-electron trigger. The central electron candidate is required to have $\pT > 25$~GeV. Because the expected background from multijet events is larger in this channel than in the $ee_{\text{CC}}$~channel, the central electron candidate is required to satisfy a set of ``tight'' selection criteria~\cite{Electrons2011}, which are optimised for the level of pile-up observed in the 2012 data. The forward electron candidate is required to have $\pT > 20$~GeV and to satisfy a set of ``medium'' selection criteria, based only on the shower shapes in the electromagnetic calorimeter~\cite{Electrons2011} since this region is outside the acceptance of the inner tracker. Events are required to contain exactly two electron candidates satisfying the above criteria.

Since this analysis is focused on the $Z$-boson pole region, the lepton pair is required to have an invariant mass ($\mll$) within a narrow window around the $Z$-boson mass, $80 < \mll < 100$~GeV. Events are selected for $\yz$-integrated measurements without any requirements on the rapidity of the lepton pair~($\yll$). For the $\yz$-binned measurements, events are selected in three bins of rapidity: $|\yll| < 1.0$, $1.0 < |\yll| < 2.0$, and $2.0 < |\yll| < 3.5$. Events are also required to have a dilepton transverse momentum ($\ptll$) less than the value of~600~(100)~GeV used for the highest bin in the $ee_{\text{CC}}$ and $\mu\mu_{\text{CC}}$ ($ee_{\text{CF}}$) channels. The variables $\mll$, $\yll$, and $\ptll$, which are defined using reconstructed lepton pairs, are to be distinguished from the variables $\mz$, $\yz$, and $\ptz$, which are defined using lepton pairs at the Born level, as described in Section~\ref{sec:theory}.

The simulated events are required to satisfy the same selection criteria, after applying small corrections to account for the differences between data and simulation in terms of reconstruction, identification and trigger efficiencies and of energy scale and resolution for electrons and muons~\cite{Electrons2011,ATLAS-CONF-2014-032,ElectronCalibration,Muons20112012}. All simulated events are reweighted to match the distributions observed in data for the level of pile-up and for the primary vertex longitudinal position.

Figure~\ref{Fig:yZacceptance} illustrates the different ranges in $\ptz$ and $\yz$ expected to be covered by the three channels along with their acceptance times selection efficiencies, which is defined as the ratio of the number of selected events to the number in the full phase space. The difference in shape between the $ee_{\text{CC}}$ and $\mu\mu_{\text{CC}}$ channels arises from the lower reconstruction and identification efficiency for central electrons at high values of $|\eta|$ and from the lower trigger and reconstruction efficiency for muons at low values of $|\eta|$. The central--central and central--forward channels overlap in the region $1.5 < |\yz| < 2.5$.

\begin{figure}
  \begin{center}                               
{
   \includegraphics[width=0.45\textwidth,angle=0]{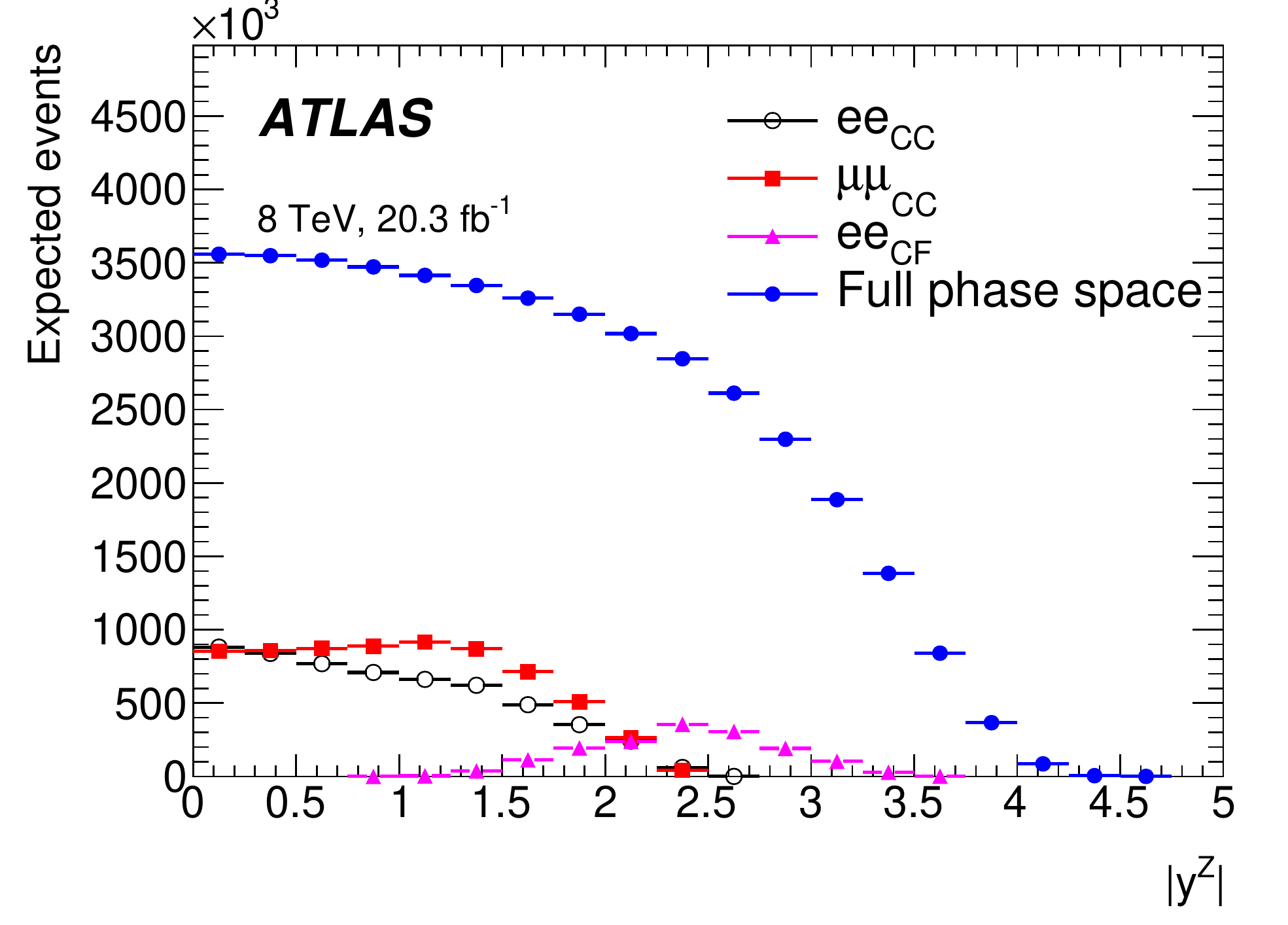}
   \includegraphics[width=0.45\textwidth,angle=0]{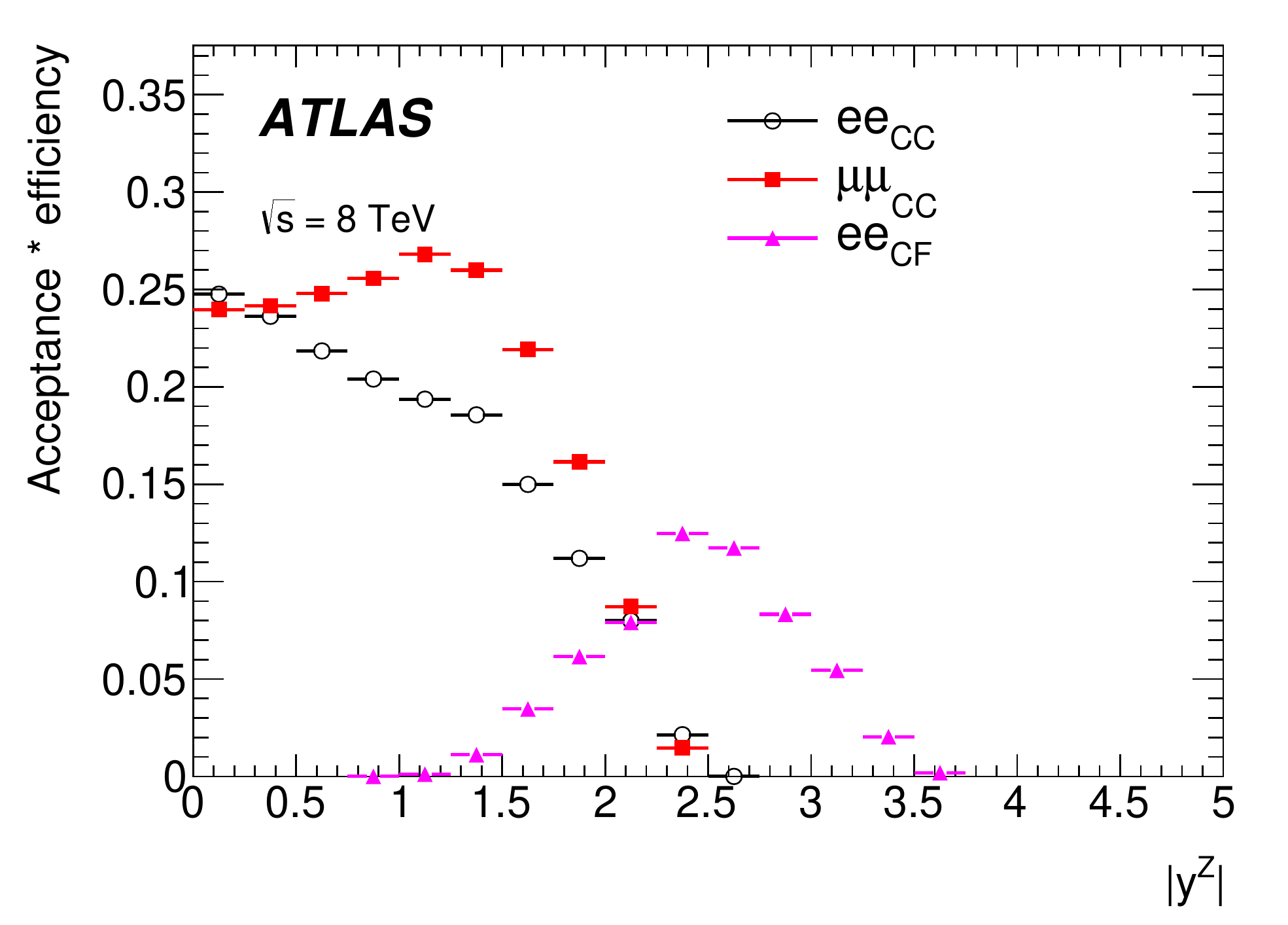}
   \includegraphics[width=0.45\textwidth,angle=0]{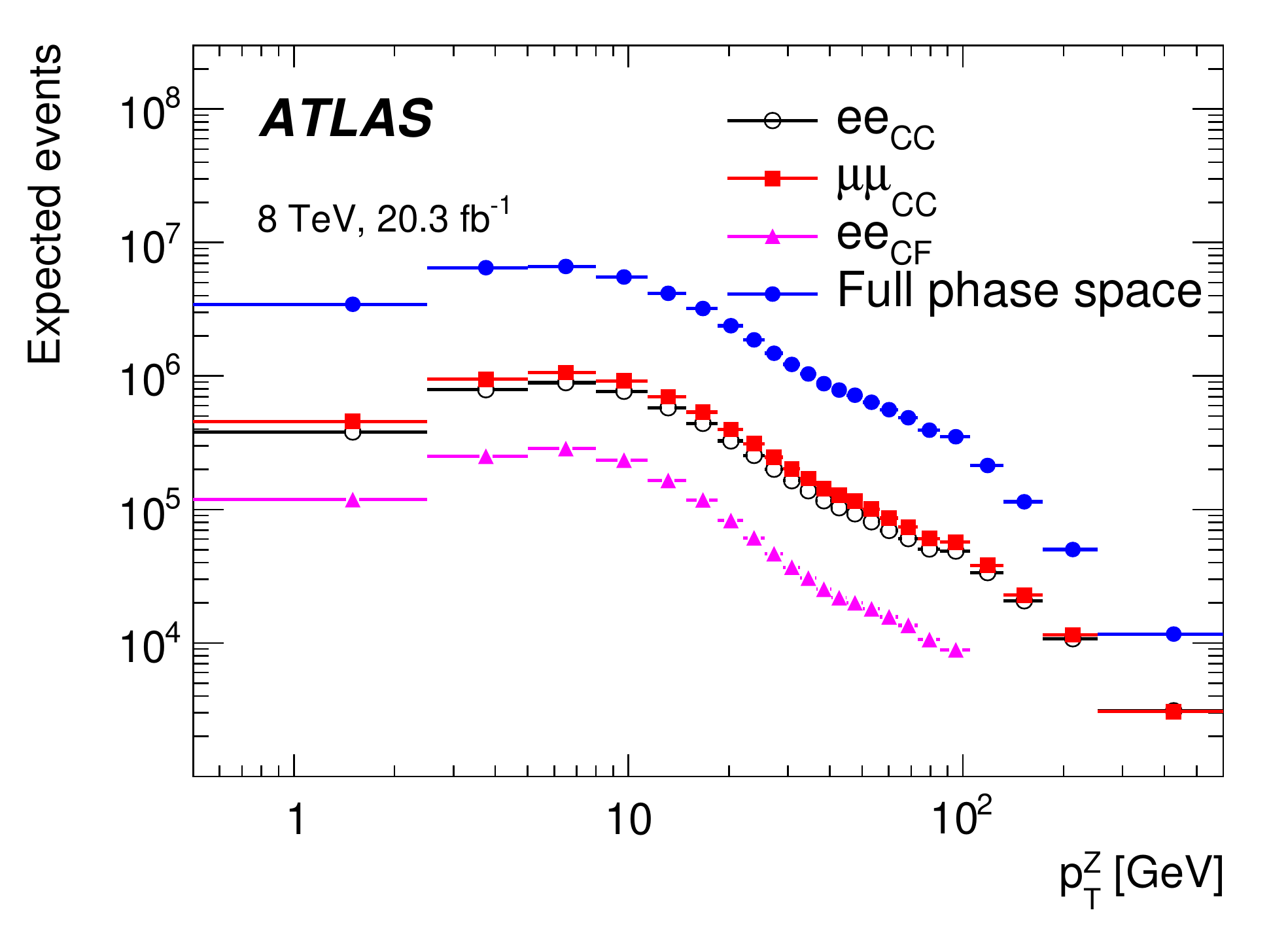}
   \includegraphics[width=0.45\textwidth,angle=0]{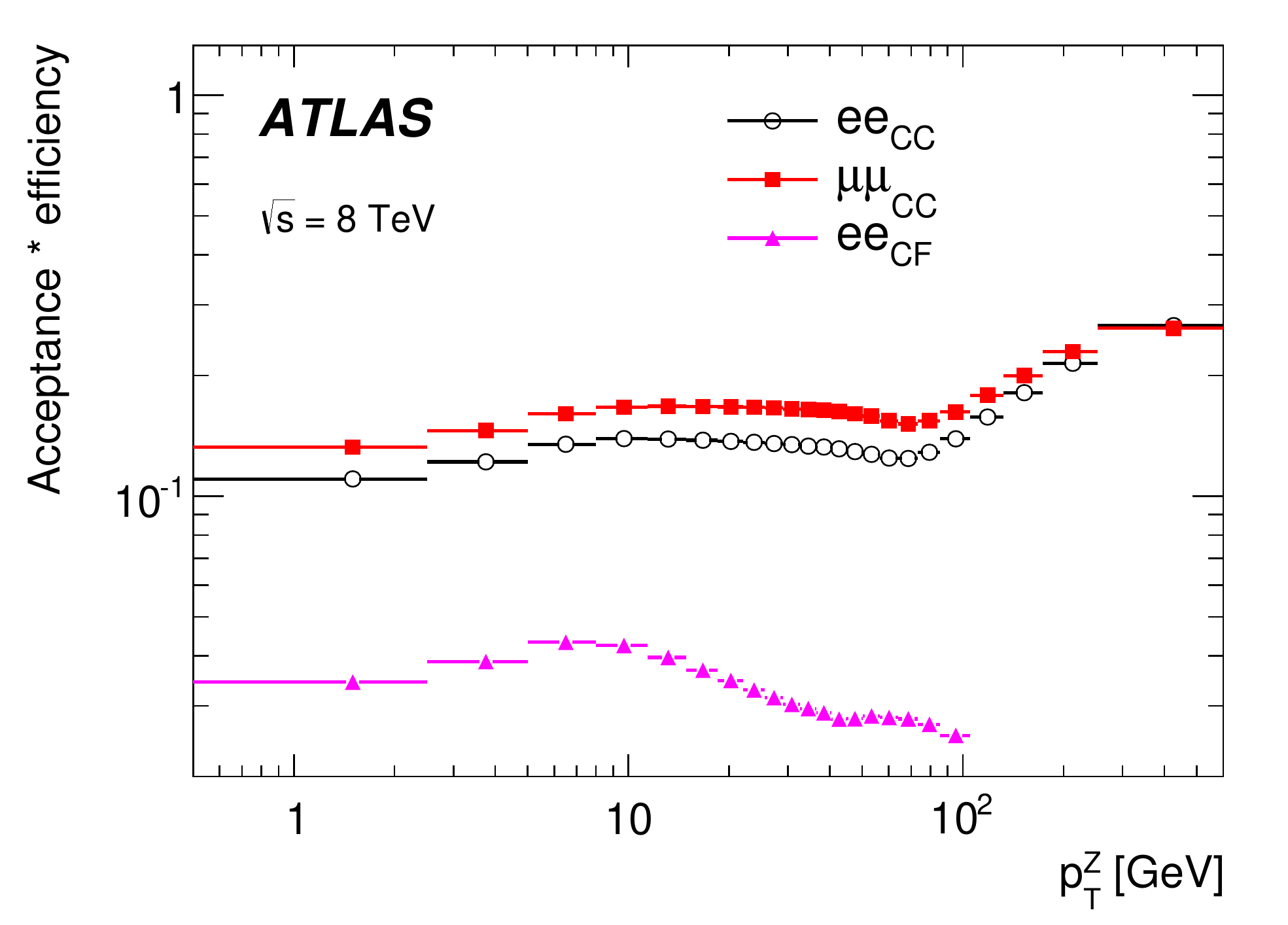}
}
\end{center}
\caption{  Comparison of the expected yields (left) and acceptance times efficiency of selected events (right) as a function of $\yz$ (top) and $\ptz$ (bottom), for the $ee_{\text{CC}}$, $\mu\mu_{\text{CC}}$, and $ee_{\text{CF}}$ events. Also shown are the expected yields at the event generator level over the full phase space considered for the measurement, which corresponds to all events with a dilepton mass in the chosen window, $80 < \mz < 100$~GeV.}
\label{Fig:yZacceptance}
\end{figure}

\subsection{Backgrounds}
\label{sec:backgrounds}

In the $Z$-boson pole region, the backgrounds from other processes are small, below the half-percent level for the $ee_{\text{CC}}$ and $\mu\mu_{\text{CC}}$ channels and at the level of~2\% for the $ee_{\text{CF}}$ channel. The backgrounds from prompt isolated lepton pairs are estimated using simulated samples, as described in~Section~\ref{sec:dataMC}, and consist predominantly of lepton pairs from top-quark processes and from diboson production with a smaller contribution from~$Z \to \tau\tau$~decays. The other background source arises from events in which at least one of the lepton candidates is not a prompt isolated lepton but rather a lepton from heavy-flavour hadron decay (beauty or charm) or a fake lepton in the case of electron candidates (these may arise from charged hadrons or from photon conversions within a hadronic jet). This background consists of events containing two such leptons (multijets) or one such lepton ($\Wboson +\mathrm{jets}$ or top-quark pairs) and is estimated from data using the lepton isolation as a discriminating variable, a procedure described for example in~Ref.~\cite{Electrons2011} for electrons. For the central--central channels, the background determination is carried out in the full two-dimensional space of~($\costhetacs,\phics$) and in each bin of~$\ptll$. In the case of the central--forward channel, the multijet background, which is by far the dominant one, is estimated separately for each projection in~$\costhetacs$ and~$\phics$ because of the limited amount of data. This is the main reason why the angular coefficients in the central--forward channel are extracted only in projections, as described in~Section~\ref{sec:intro}.

\begin{figure}[p]
  \begin{center}                               
{
 \includegraphics[width=7.5cm,angle=0]{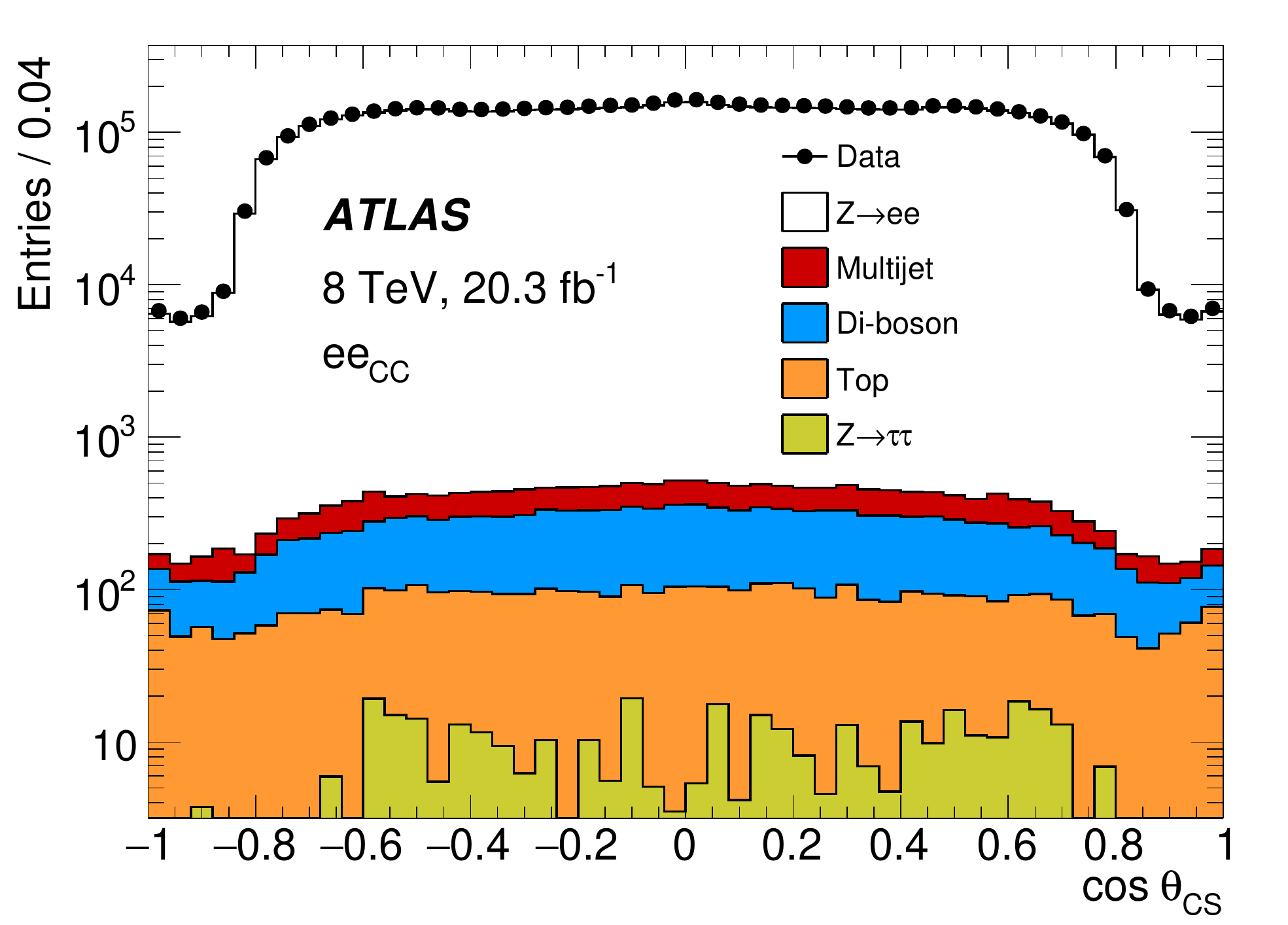}
 \includegraphics[width=7.5cm,angle=0]{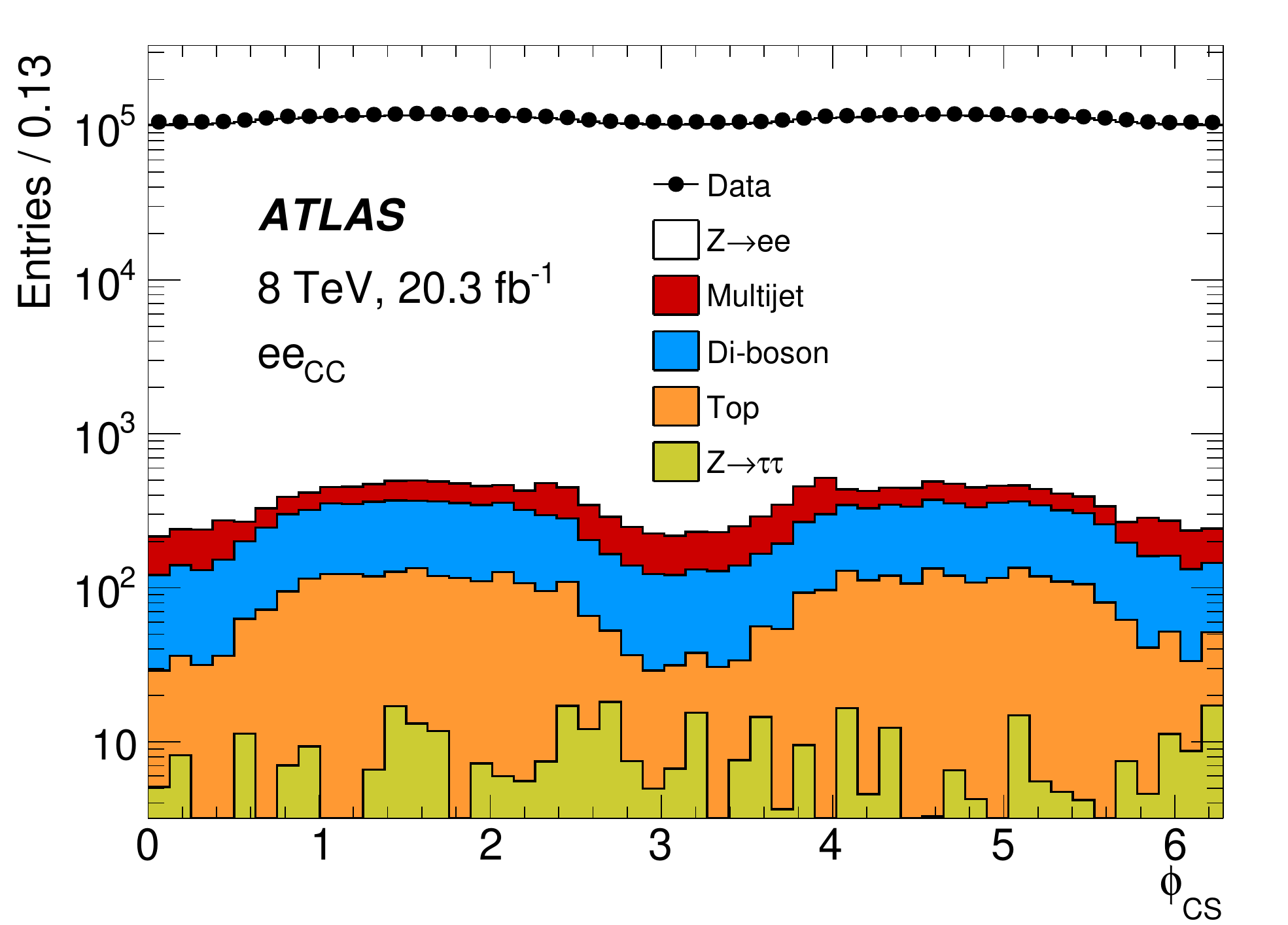}
 \includegraphics[width=7.5cm,angle=0]{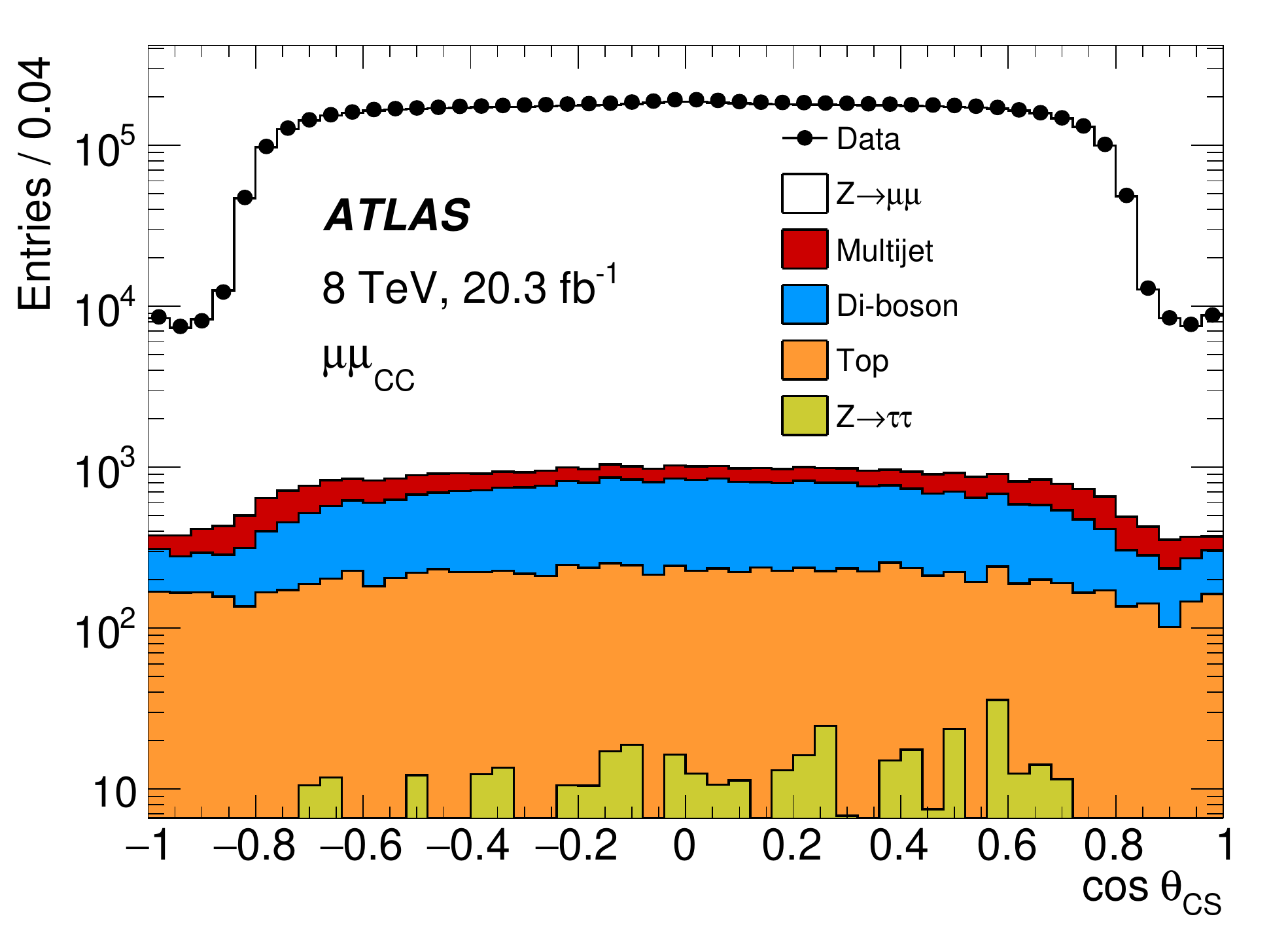}
 \includegraphics[width=7.5cm,angle=0]{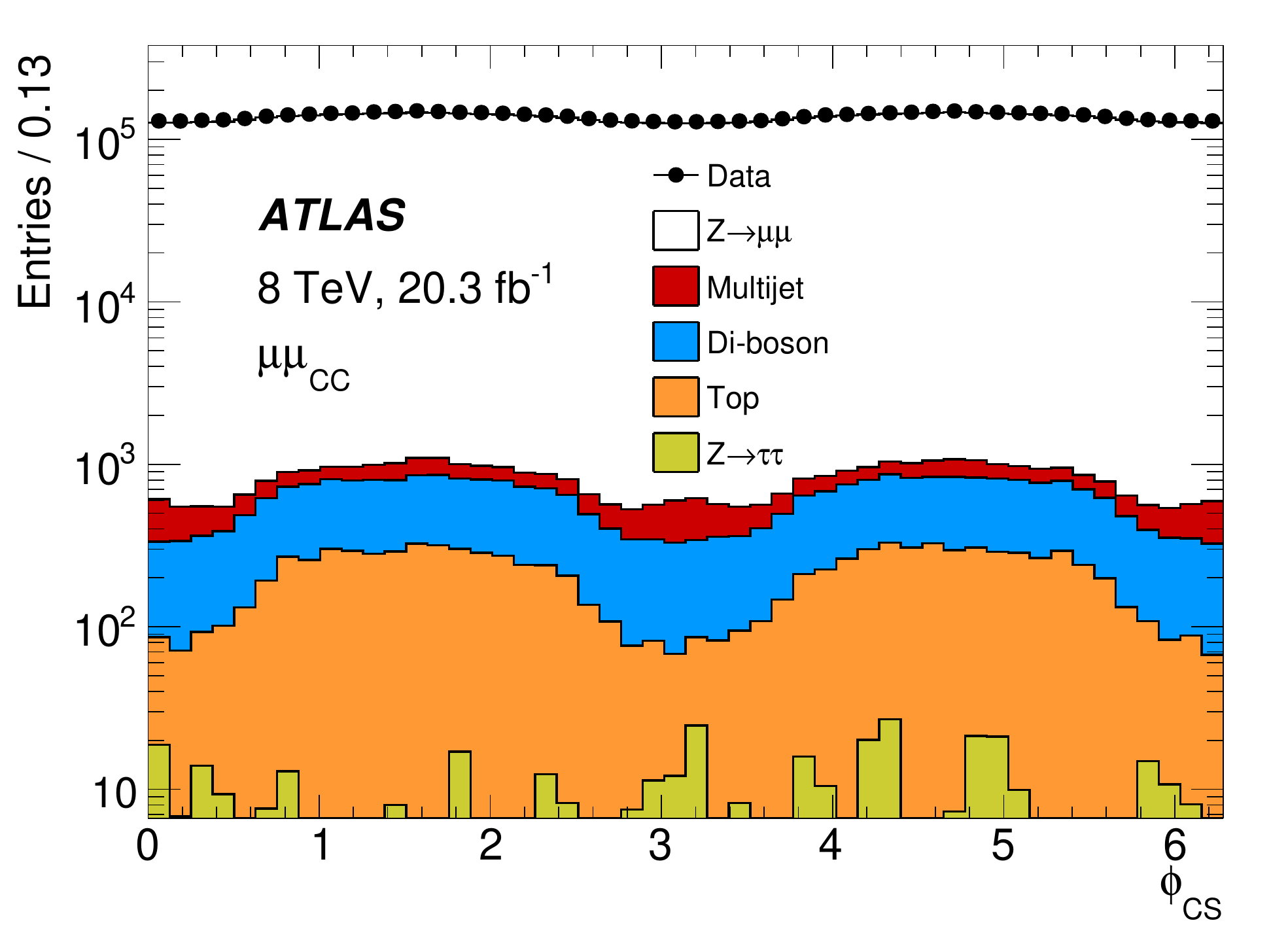}
 \includegraphics[width=7.5cm,angle=0]{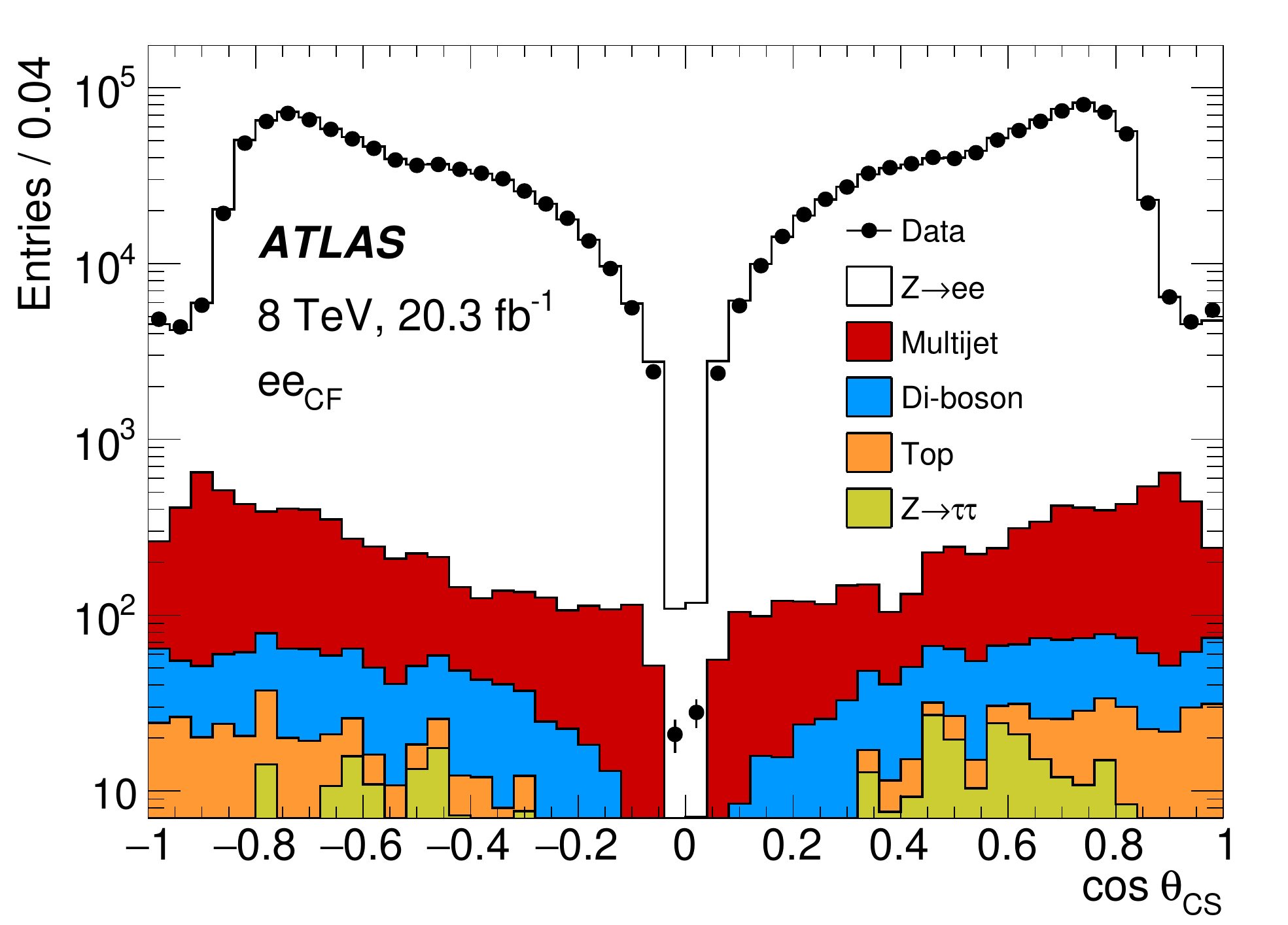}
 \includegraphics[width=7.5cm,angle=0]{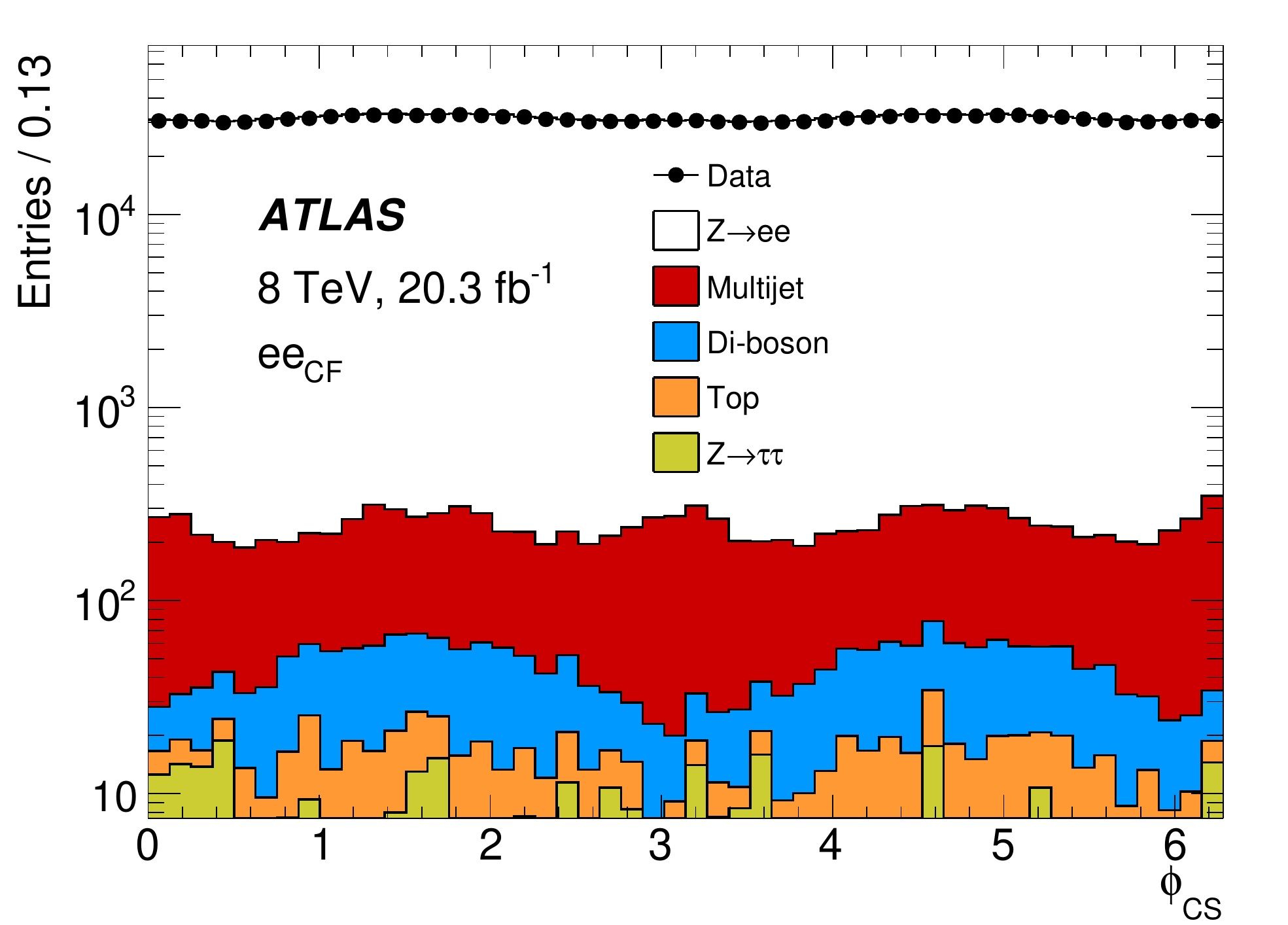}
}
  \end{center}
\caption{ 
The $\costhetacs$ (left) and  $\phics$  (right) angular distributions, averaged over all $Z$-boson \pt, for the $ee_{\text{CC}}$ (top), $\mu\mu_{\text{CC}}$ (middle) and $ee_{\text{CF}}$ (bottom) channels. The distributions are shown separately for the different background sources contributing to each channel. The multijet background is determined from data, as explained in the text.
\label{Fig:ControlPlot_bgd} }
\end{figure}

\begin{figure}[p]
  \begin{center}                               
{
 \includegraphics[width=9.5cm,angle=0]{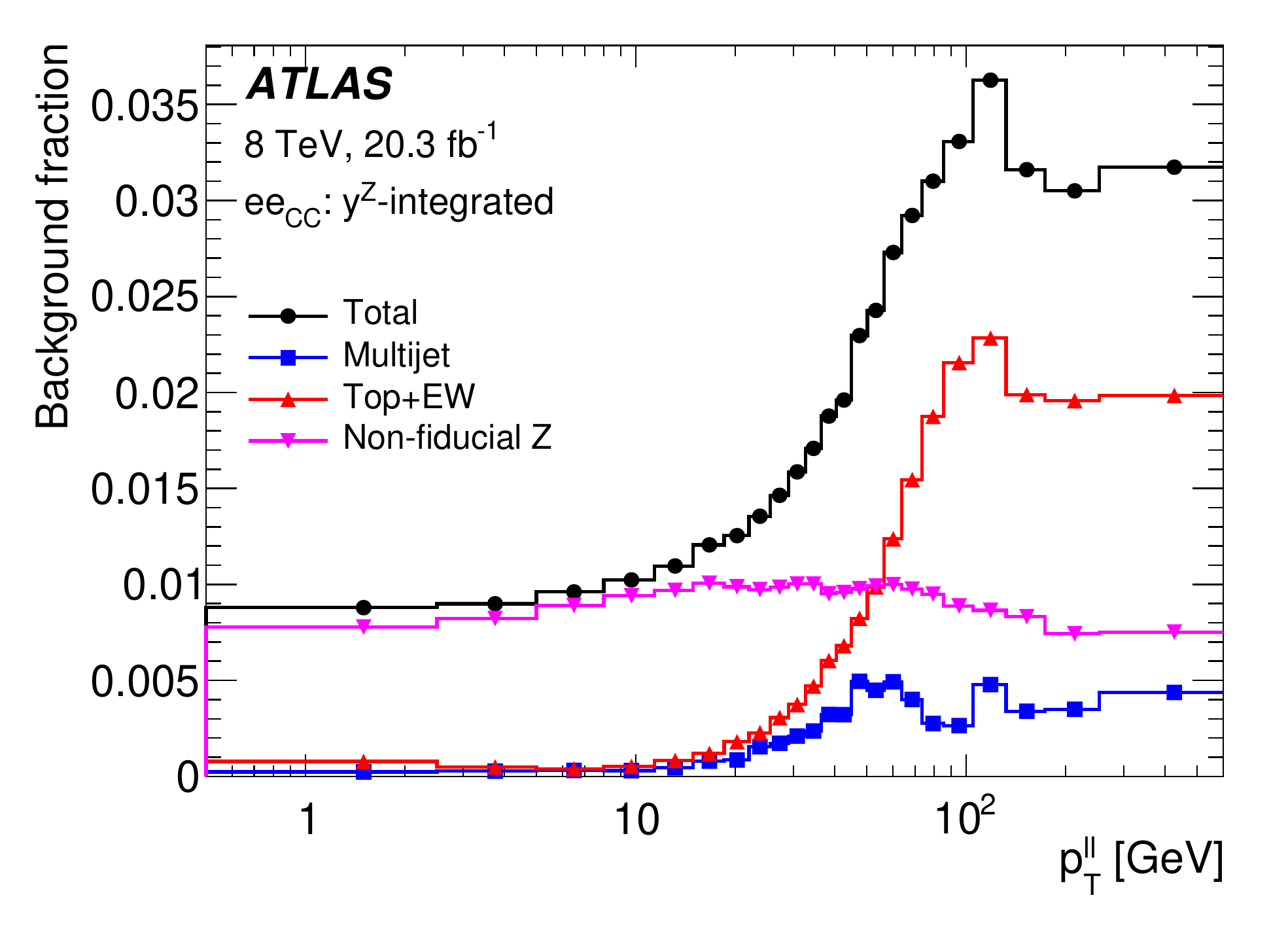}
 \includegraphics[width=9.5cm,angle=0]{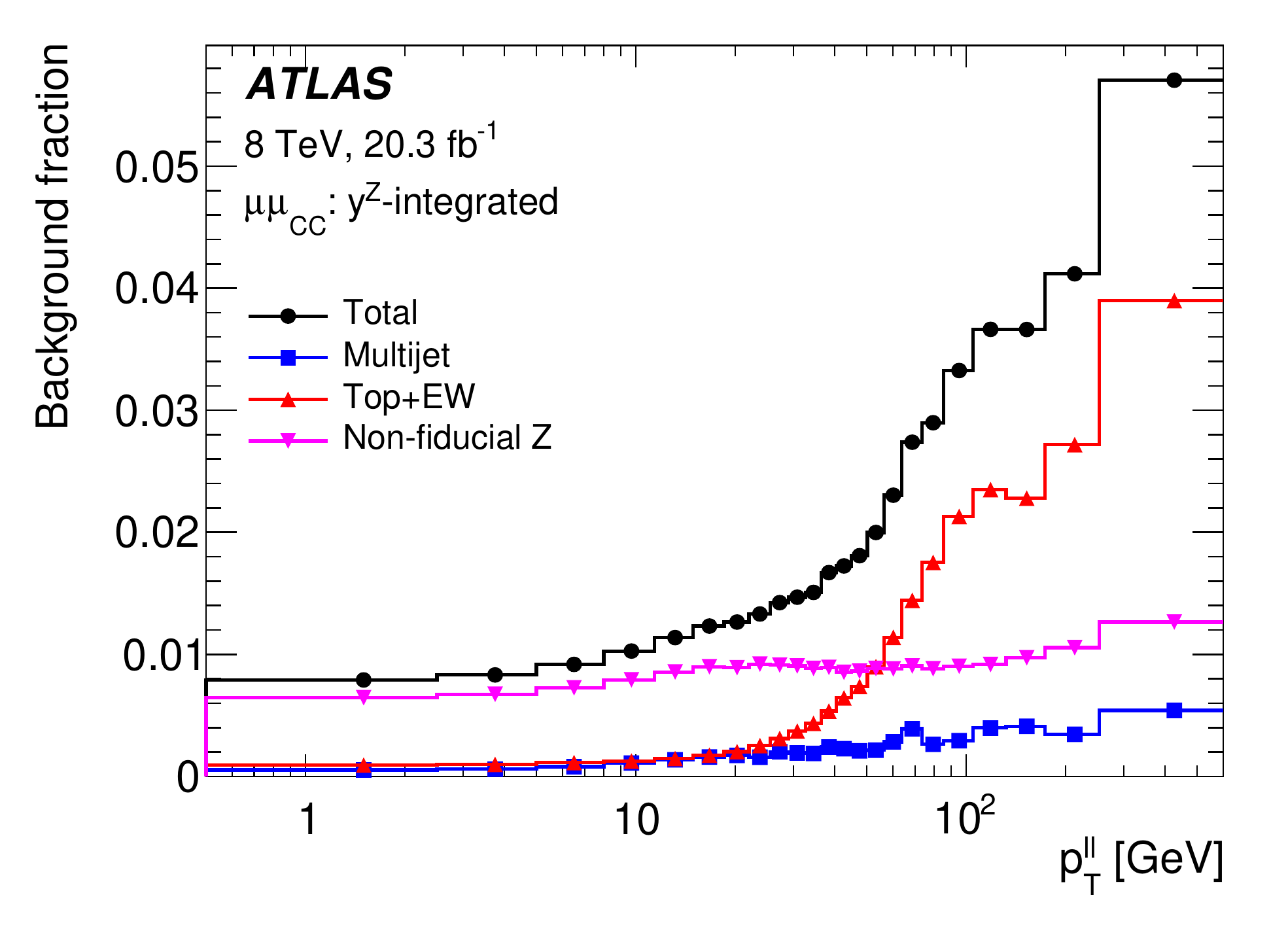}  
 \includegraphics[width=9.5cm,angle=0]{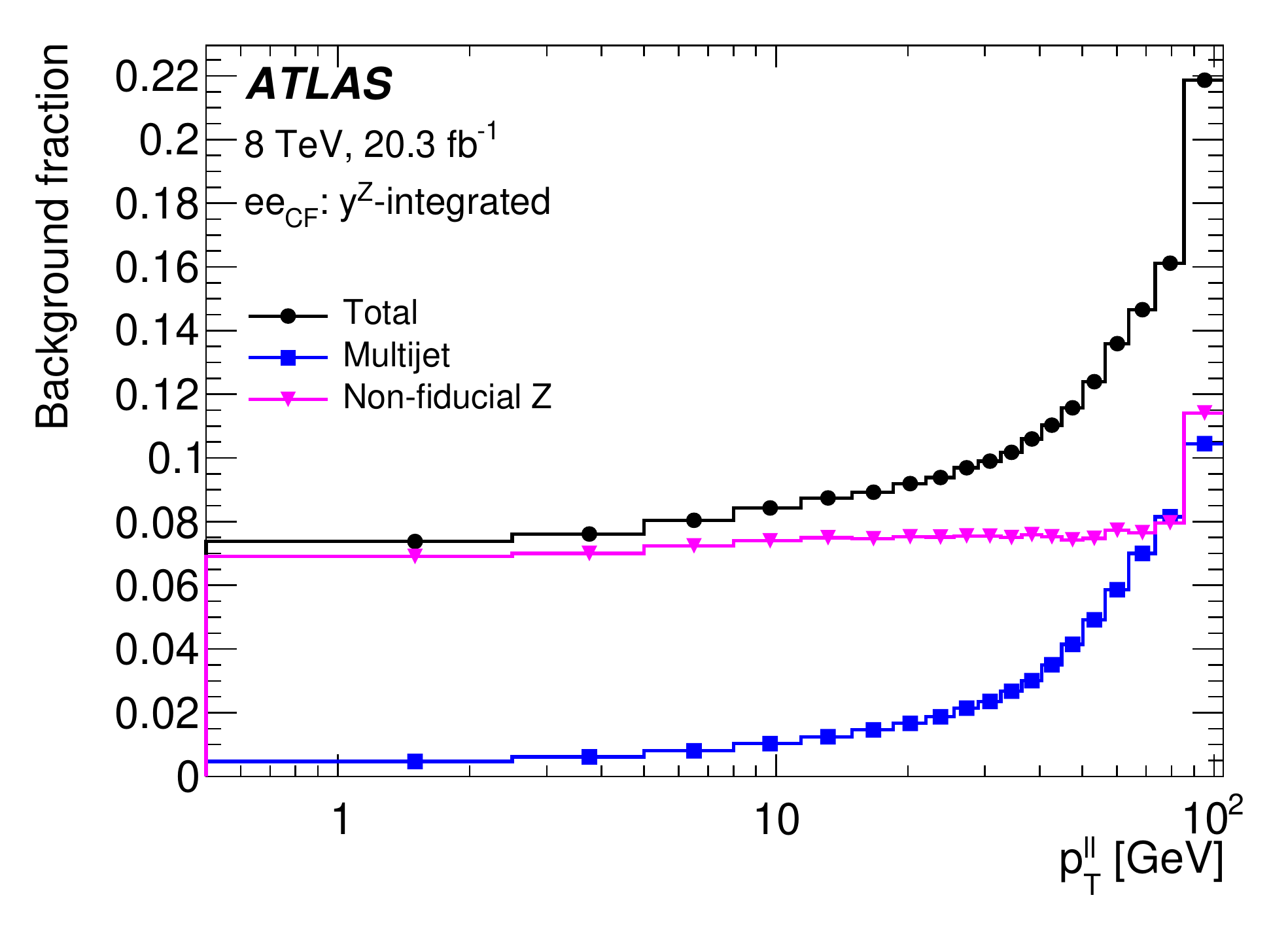}
}
  \end{center}
\vspace{-3mm}
\caption{ 
Fractional background contributions as a function of $\ptll$, for the $ee_{\text{CC}}$ (top), $\mu\mu_{\text{CC}}$ (middle) and $ee_{\text{CF}}$ (bottom) channels. The distributions are shown separately for the relevant background contributions to each channel together with the summed total background fraction. The label ``Non-fiducial~$Z$'' refers to signal events which are generated outside the phase space used to extract the angular coefficients (see text).  
\label{Fig:Fraction_bgd} }
\end{figure}

Figure~\ref{Fig:ControlPlot_bgd} shows the angular distributions, $\costhetacs$ and  $\phics$, for the three channels for the data, the $Z$-boson signal MC~sample, and the main sources of background discussed above. The total background in the central--central events is below~0.5\% and its uncertainty is dominated by the large uncertainty in the multijet background of approximately~50\%. The uncertainty in the top+electroweak background is taken conservatively to be~20\%. In the case of the central--forward electron pairs, the top+electroweak background is so small compared to the much larger multijet background that it is neglected for simplicity in the fit procedure described in~Section~\ref{sec:Methodology}. 
Table~\ref{Tab:cutflow} summarises the observed yields of events in data for each channel, integrated over all values of~\ptll, together with the expected background yields with their total uncertainties from multijet events and from top+electroweak sources. More details of the treatment of the background uncertainties are discussed in~Section~\ref{sec:Uncertainties}.

\begin{table}
\caption{ For each of the three channels, yield of events observed in data and expected background yields (multijets, top+electroweak, and total) corresponding to the 2012 data set and an integrated luminosity of~20.3~fb$^{-1}$. The uncertainties quoted include both the statistical and systematic components (see text).}
\setlength\extrarowheight{4pt}
\label{Tab:cutflow} 
\vspace{2mm}
\begin{tabular}{|l|c|c|c|c|}
\hline
  Channel & Observed & \multicolumn{3}{c|}{Expected background}   \\ \cline{3-5}
        &   & Multijets (from data)    &  Top+electroweak (from MC)  &  Total       \\    \hline
    $ee_{\text{CC}}$  & $5.5 \times 10^{6}$     &   6000 $\pm$ 3000       &   13000 $\pm$ 3000      &  19000 $\pm$ 4000   \\    
    $\mu\mu_{\text{CC}}$ & $7.0 \times 10^{6}$ &  9000 $\pm$ 4000   &  19000 $\pm$ 4000     &    28000 $\pm$ 6000             \\
    $ee_{\text{CF}}$  &  $1.5 \times 10^{6}$  &   28000 $\pm$ 14000     &  1000 $\pm$ 200     &   29000 $\pm$ 14000       \\        \hline
\end{tabular}
\end{table}

There are also signal events that are considered as background to the measurement because they are present in the data only due to the finite resolution of the measurements, which leads to migrations in mass and rapidity. These are denoted ``Non-fiducial~$Z$'' events and can be divided into four categories: the dominant fraction consists of events that have~$\mz$ at the generator level outside the chosen $\mll$~mass window but pass event selection, while another contribution arises from events that do not belong to the $\yz$~bin considered for the measurement at generator level. The latter contribution is sizeable only in the $ee_{\text{CF}}$~channel. Other negligible sources of this type of background arise from events for which the central electron has the wrong assigned charge in the $ee_{\text{CF}}$~channel or both central electrons have the wrong assigned charge in the $ee_{\text{CC}}$~channel, or for which~$\ptz$ at the generator level is larger than~600~GeV. These backgrounds are all included as a small component of the signal MC~sample in~Fig.~\ref{Fig:ControlPlot_bgd}. Their contributions amount to one percent or less for the~$ee_{\text{CC}}$ and $\mu\mu_{\text{CC}}$~channels, increasing to almost~8\% for the $ee_{\text{CF}}$~channel because of the much larger migrations in energy measurements in the case of forward electrons. For the $2<|\yz|<3.5$ bin in the $ee_{\text{CF}}$ channel, the $\yz$ migration contributes~2\% to the non-fiducial~$Z$ background. The fractional contribution of all backgrounds to the total sample is shown explicitly for each channel as a function of~$\ptll$ in~Fig.~\ref{Fig:Fraction_bgd} together with the respective contributions of the multijet and top+electroweak backgrounds. The sum of all these backgrounds is also shown and templates of their angular distributions are used in the fit to extract the angular coefficients, as described in~Section~\ref{sec:Methodology}.

\subsection{Angular distributions}

The measurement of the angular coefficients is performed in fine bins of~\ptz\ and for a fixed dilepton mass window on the same sample as that used to extract from data the small corrections applied to the lepton efficiencies and calibration. The analysis is thus largely insensitive to the shape of the distribution of~\ptz, and also to any residual differences in the modelling of the shape of the dilepton mass distribution. It is, however, important to verify qualitatively the level of agreement between data and MC~simulation for the~$\costhetacs$ and~$\phics$ angular distributions before extracting the results of the measurement. This is shown for the three channels separately in~Fig.~\ref{Fig:ControlPlot_signal}, together with the ratio of the observed data to the sum of predicted events. The data and MC~distributions are not normalised to each other, resulting in normalisation differences at the level of a few percent. The measurement of the angular coefficients is, however, independent of the normalisation between data and simulation in each bin of~\ptz. The differences in shape in the angular distributions reflect the mismodelling of the angular coefficients in the simulation~(see~Section~\ref{sec:results}).

\begin{figure}[p]
  \begin{center}                               
{
 \includegraphics[width=7.5cm,angle=0]{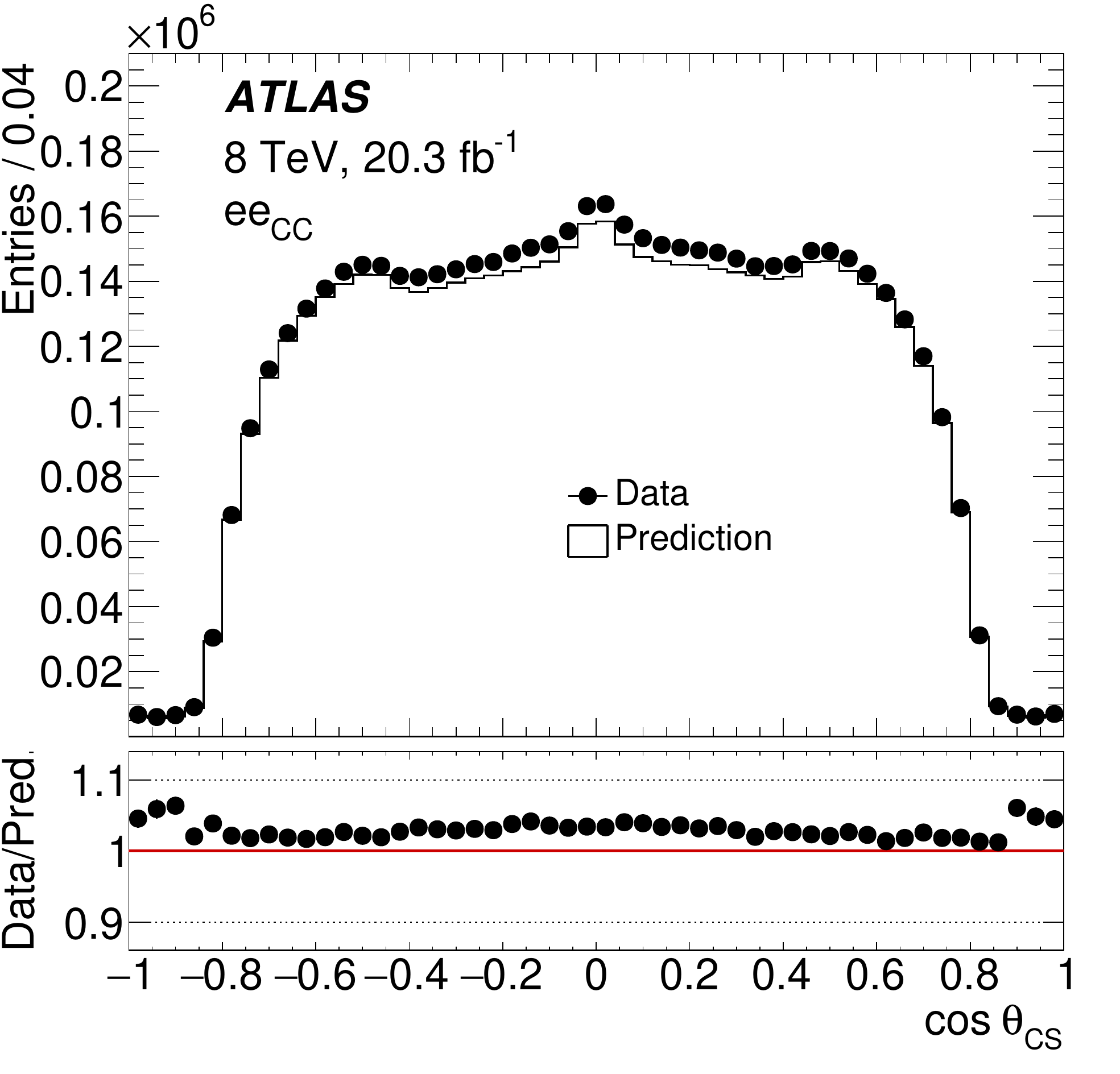}
 \includegraphics[width=7.5cm,angle=0]{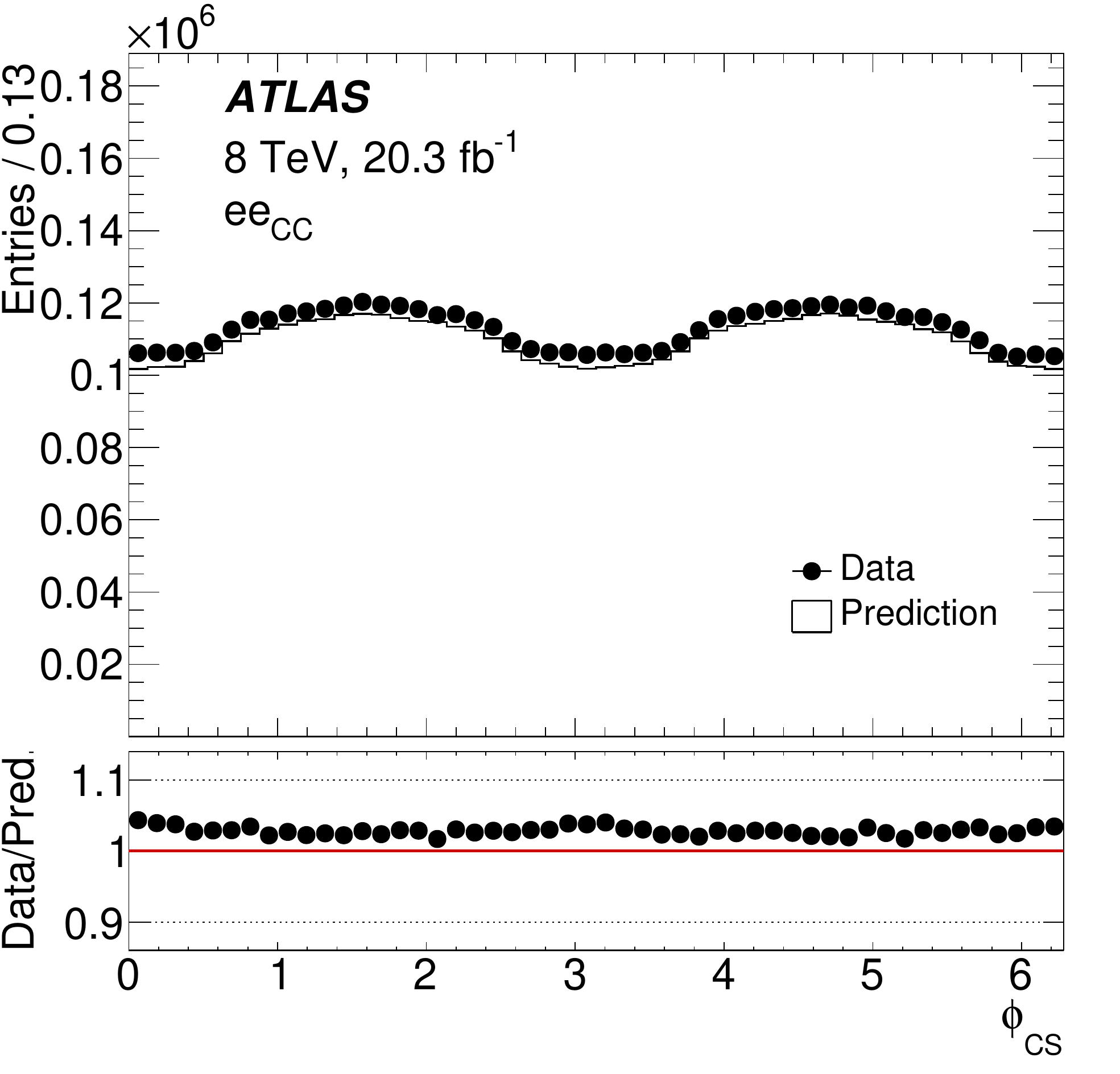}
 \includegraphics[width=7.5cm,angle=0]{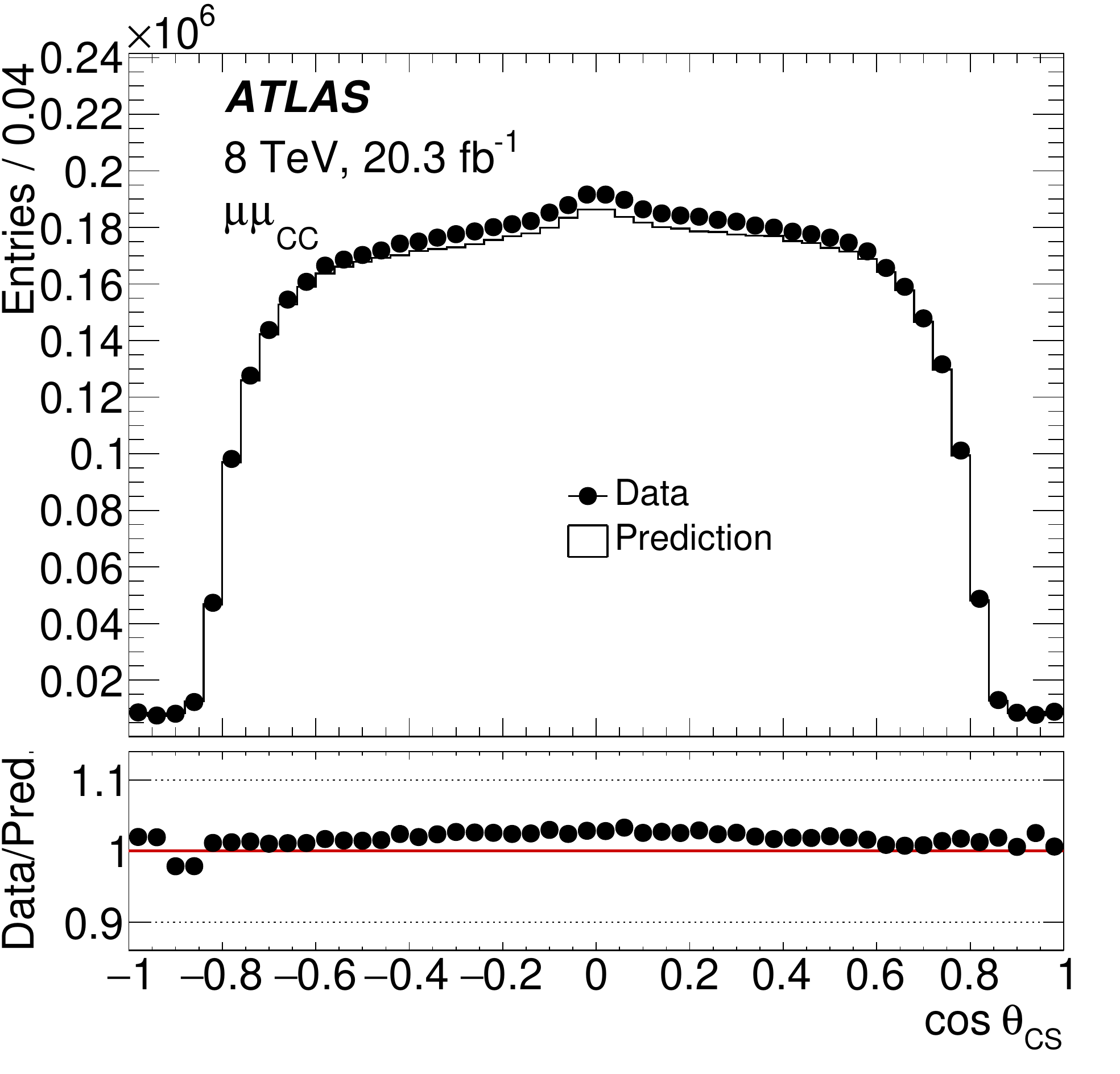}
 \includegraphics[width=7.5cm,angle=0]{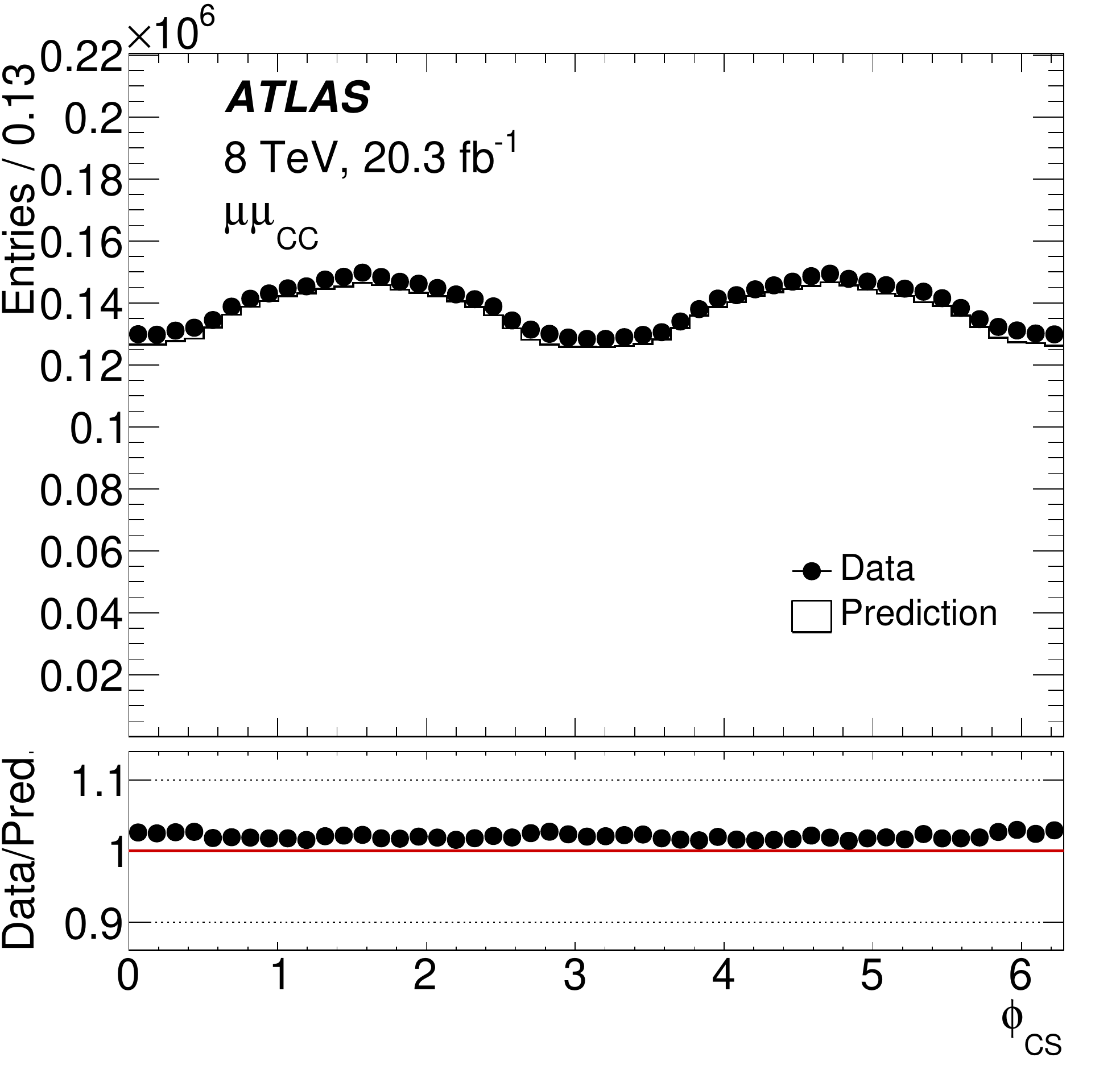}
 \includegraphics[width=7.5cm,angle=0]{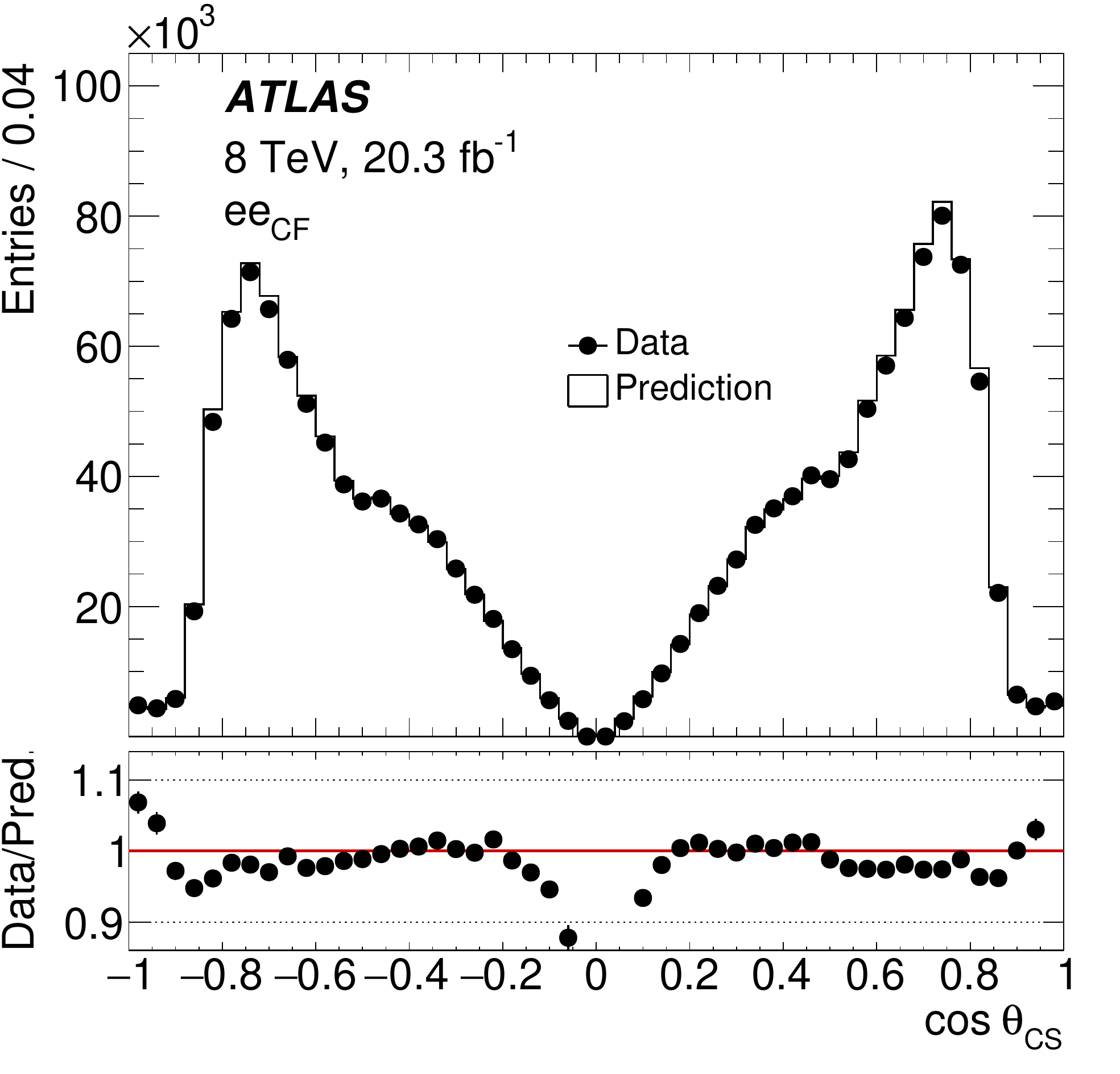}
 \includegraphics[width=7.5cm,angle=0]{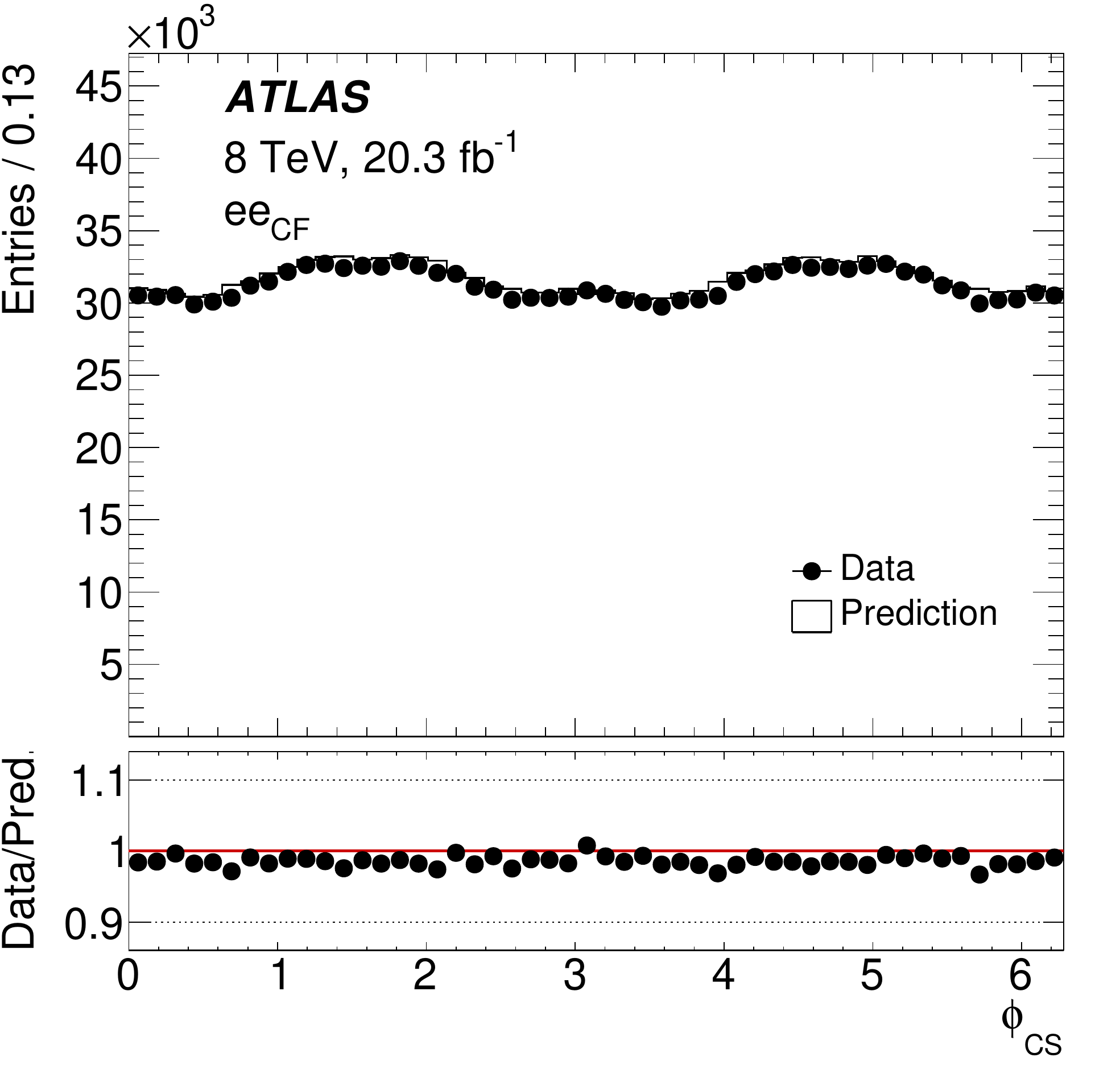}
}
  \end{center}
\caption{ 
The $\costhetacs$ (left) and  $\phics$  (right) angular distributions, averaged over all $\ptll$, for the $ee_{\text{CC}}$ (top), $\mu\mu_{\text{CC}}$  (middle) and $ee_{\text{CF}}$ (bottom) channels. In the panels showing the ratios of the data to the summed signal+background predictions, the uncertainty bars on the points are only statistical.
\label{Fig:ControlPlot_signal} }
\end{figure}

\section{Coefficient measurement methodology}
\label{sec:Methodology}

The coefficients are extracted from the data by fitting templates of the $\poli$ polynomial terms, defined in~Eq.~(\ref{Eq:master2}), to the reconstructed angular distributions. Each template is normalised by free parameters for its corresponding coefficient $\Ai$, as well as an additional common parameter representing the unpolarised cross-section. All of these parameters are defined independently in each bin of $\ptz$. The polynomial $P_8 = 1+\cos^2\theta_{\text{CS}}$ in Eq.~(\ref{Eq:master2}) is only normalised by the parameter for the unpolarised cross-section.

In the absence of selections for the final-state leptons, the angular distributions in the gauge-boson rest frame are determined by the gauge-boson polarisation. In the presence of selection criteria for the leptons, the distributions are sculpted by kinematic effects, and can no longer be described by the sum of the nine $\poli$~polynomials as in Eq.~(\ref{Eq:master2}). Templates of the $\poli$ terms are constructed in a way to account for this, which requires fully simulated signal~MC to model the acceptance, efficiency, and migration of events. This process is described in Section~\ref{sec:templates}. Section~\ref{sec:likelihood} then describes the likelihood that is built out of the templates and maximised to obtain the measured coefficients. The methodology for obtaining uncertainties in the measured parameters is also covered there. The procedure for combining multiple channels is covered in Section~\ref{sec:combination}, along with alternative coefficient parameterisations used in various tests of measurement results from different channels.

\subsection{Templates}
\label{sec:templates}

To build the templates of the $\poli$~polynomials, the reference coefficients $\AiRef$ for the signal MC~sample are first calculated with the moments method, as described in~Section~\ref{sec:theory} and~Eq.~(\ref{Eq:moments}). These are obtained in each of the 23~$\ptz$ bins in Eq.~(\ref{Tab:MC:pTbins}), and also in each of the three $\yz$ bins for the $\yz$-binned measurements. The information about the angular coefficients in the simulation is then available through the corresponding functional form of Eq.~(\ref{Eq:master2}). Next, the MC~event weights are divided by the value of this function on an event-by-event basis. When the MC~events are weighted in this way, the angular distributions in the full phase space at the event generator level are flat. Effectively, all information about the \Zboson-boson polarisation is removed from the MC~sample, so that further weighting the events by any of the $\poli$~terms yields the shape of the polynomial itself, and if selection requirements are applied, this yields the shape of the selection efficiency. The selection requirements, corrections, and event weights mentioned in Section~\ref{sec:DataAnalysis} are then applied. Nine separate template histograms for each $\ptz$ and $\yz$~bin~$j$ at generator level are finally obtained after weighting by each of the $\poli$ terms. The templates $t_{ij}$ are thus three-dimensional distributions in the measured~\costhetacs,~\phics, and~\ptll~variables, and are constructed for each~$\ptz$ and $\yz$~bin. 
Eight bins in~$\costhetacs$ and~$\phics$ are used, while the binning for the reconstructed $\ptll$ is the same as for the 23~bins defined in Eq.~(\ref{Tab:MC:pTbins}). By construction, the sum of all signal templates normalised by their reference coefficients and unpolarised cross-sections agrees exactly with the three-dimensional reconstructed distribution expected for signal MC~events. Examples of templates projected onto each of the dimensions~$\costhetacs$ and~$\phics$ for the $\yz$-integrated $ee_{\text{CC}}$ channel in three illustrative $\ptz$~ranges, along with their corresponding polynomial shapes, are shown in~Fig.~\ref{Fig:templates}. The polynomials $P_1$ and $P_6$ are not shown as they integrate to zero in the full phase space in either projection (see Section~\ref{sec:likelihood}). The effect of the acceptance on the polynomial shape depends on~$\ptz$ because of the event selection, as can be seen from the difference between the template polynomial shapes in each corresponding $\ptz$~bin. This is particularly visible in the $P_{8}$ polynomial, which is uniform in $\phics$, and therefore reflects exactly the acceptance shape in the templated polynomials. In~Appendix~\ref{sec:add_templates}, two-dimensional versions of~Fig.~\ref{Fig:templates} are given for all nine polynomials in~Figs.~\ref{Fig:templates2D_Ai_1}~--~\ref{Fig:templates2D_Ai_3}. These two-dimensional views are required for~$P_1$ and~$P_6$, as discussed above.

\begin{figure}
  \begin{center}
{
    \includegraphics[width=7.5cm,angle=0]{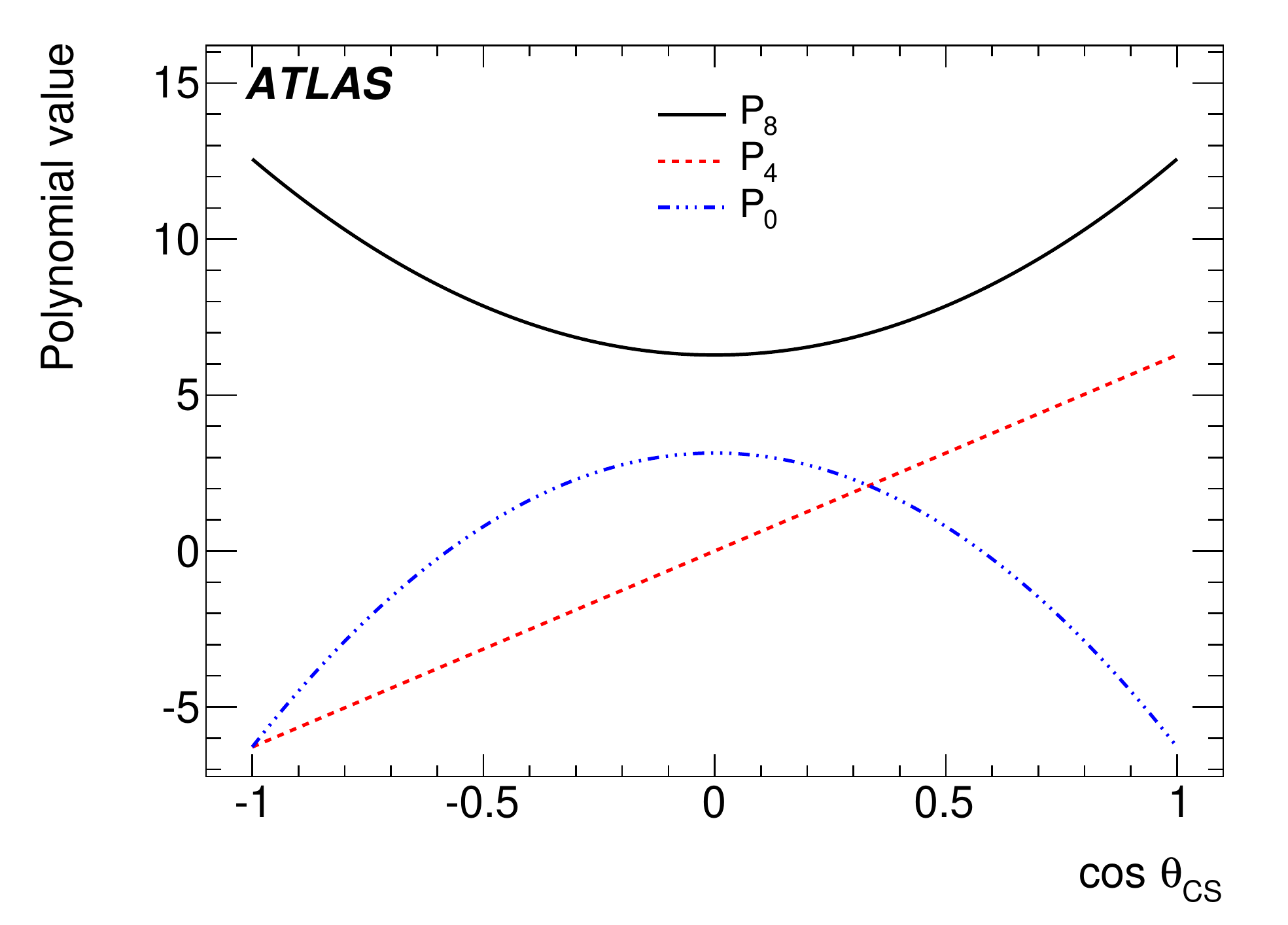}
    \includegraphics[width=7.5cm,angle=0]{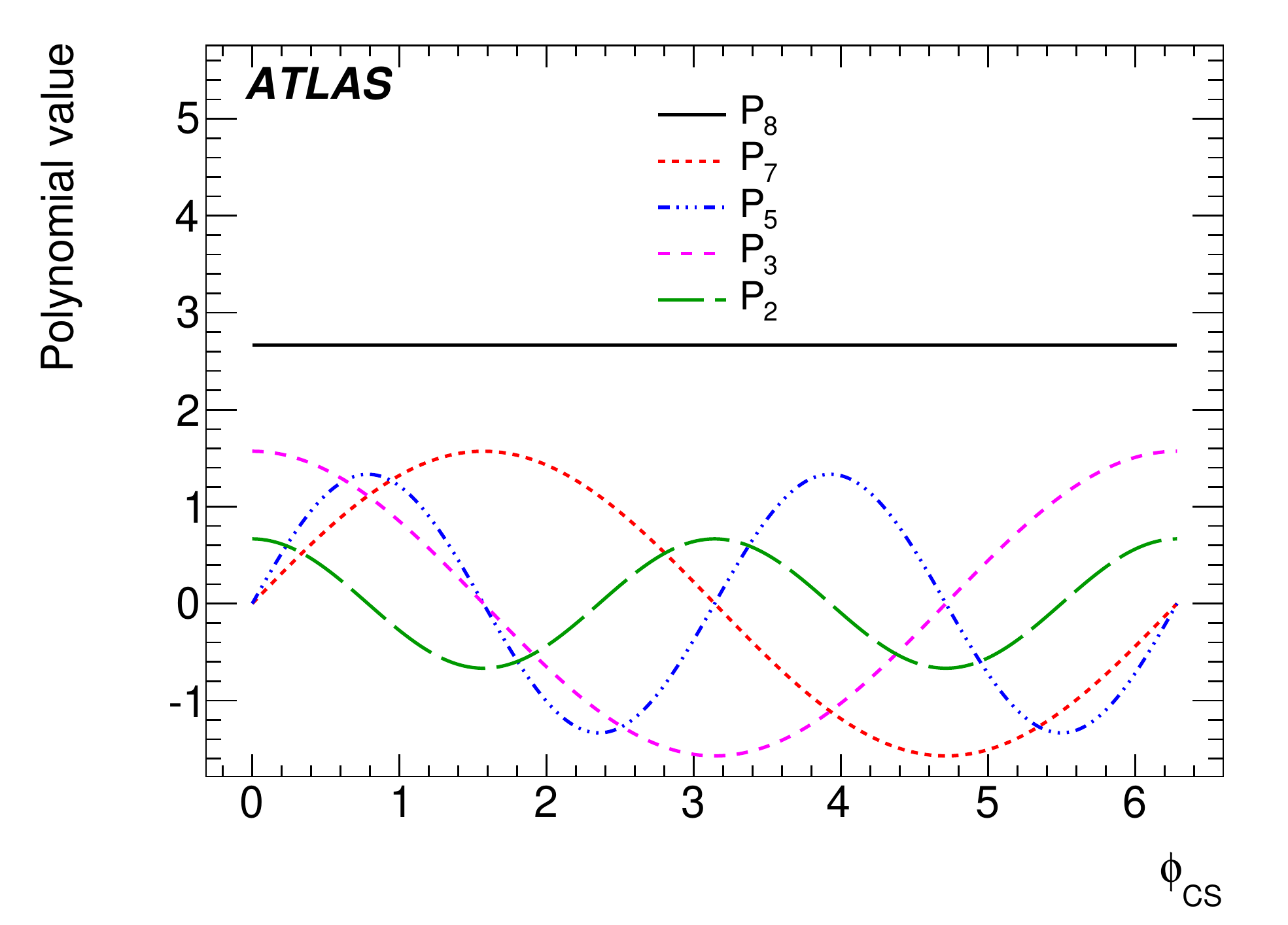}
    \includegraphics[width=7.5cm,angle=0]{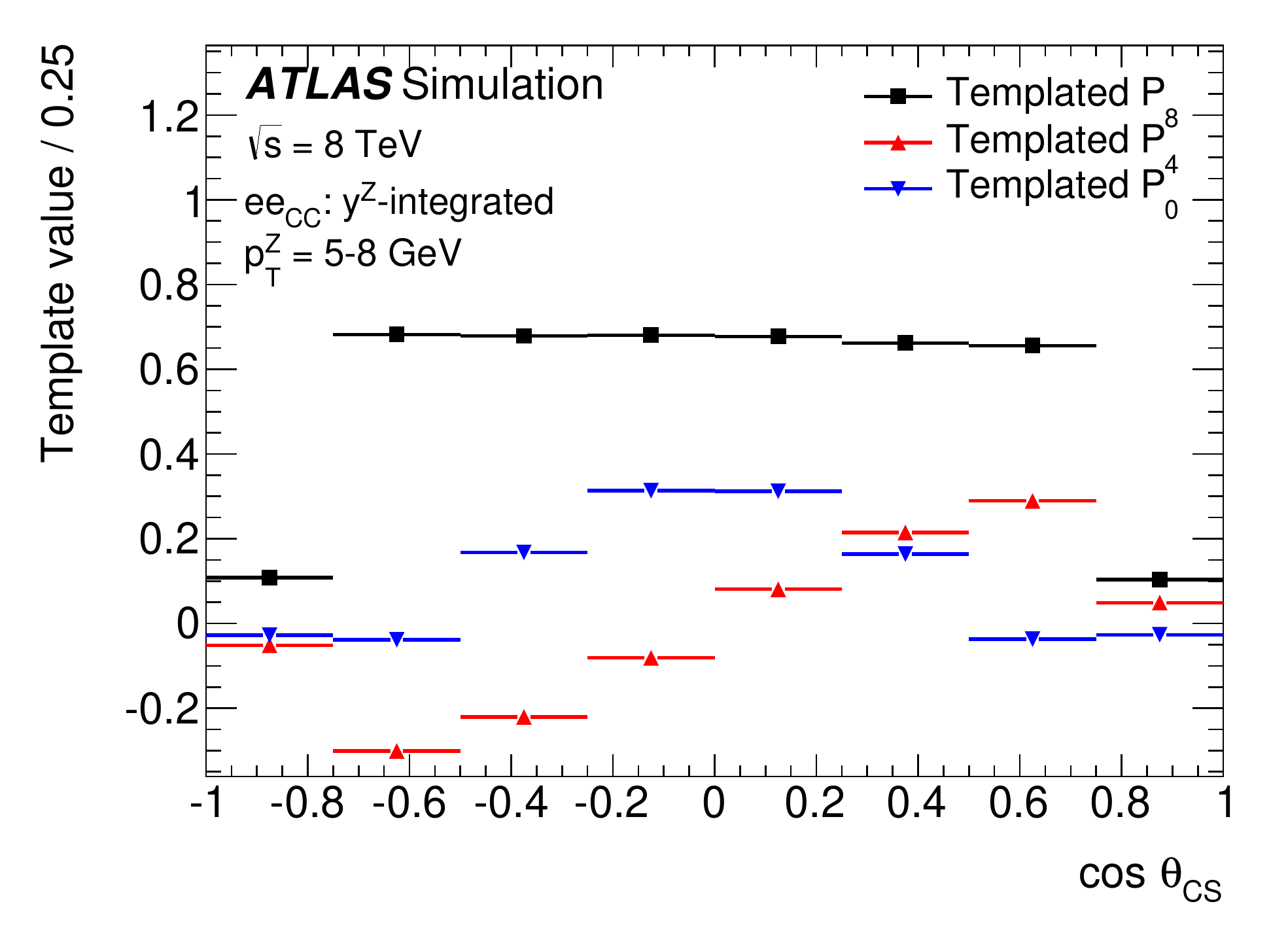}
    \includegraphics[width=7.5cm,angle=0]{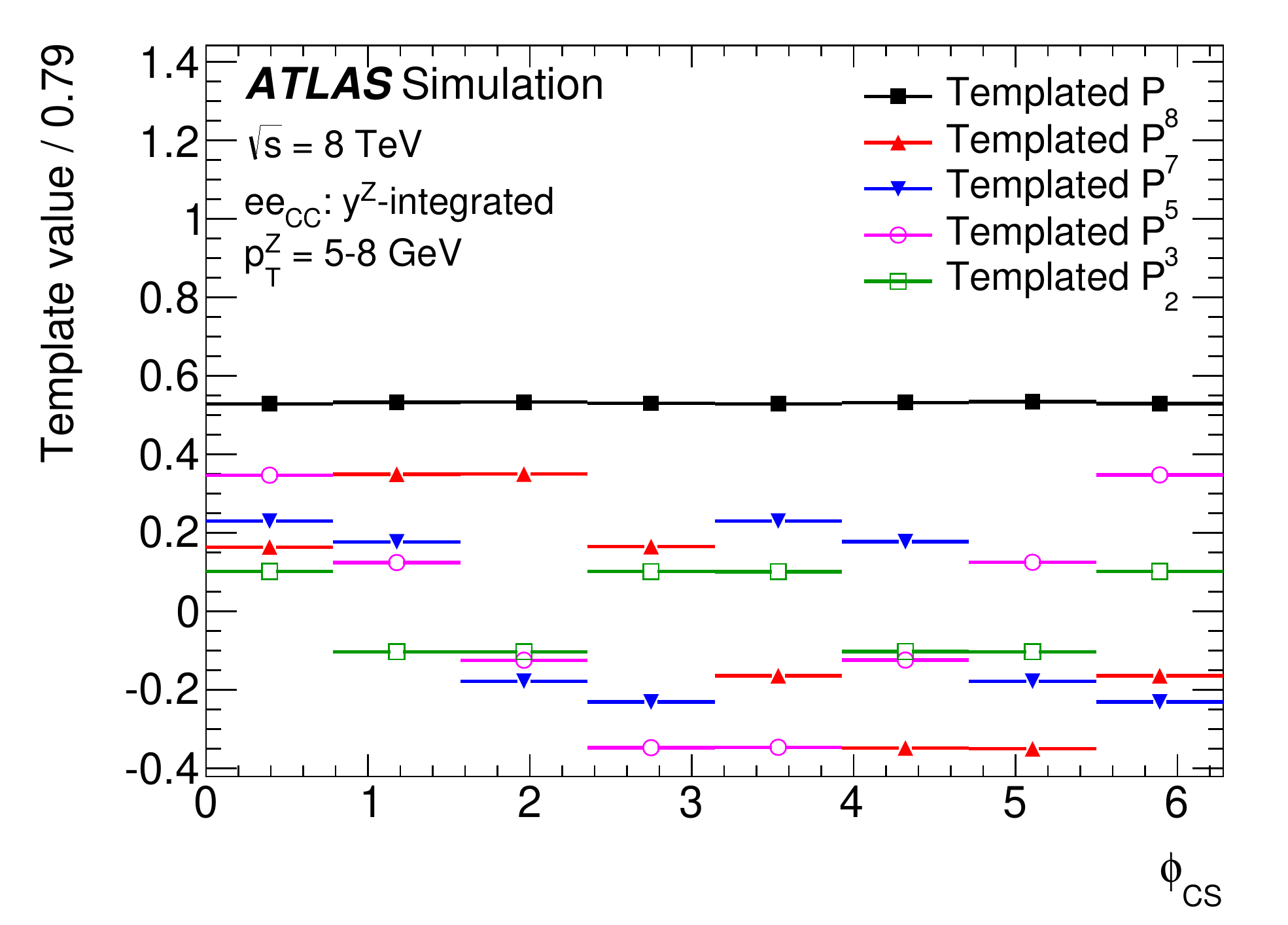}
    \includegraphics[width=7.5cm,angle=0]{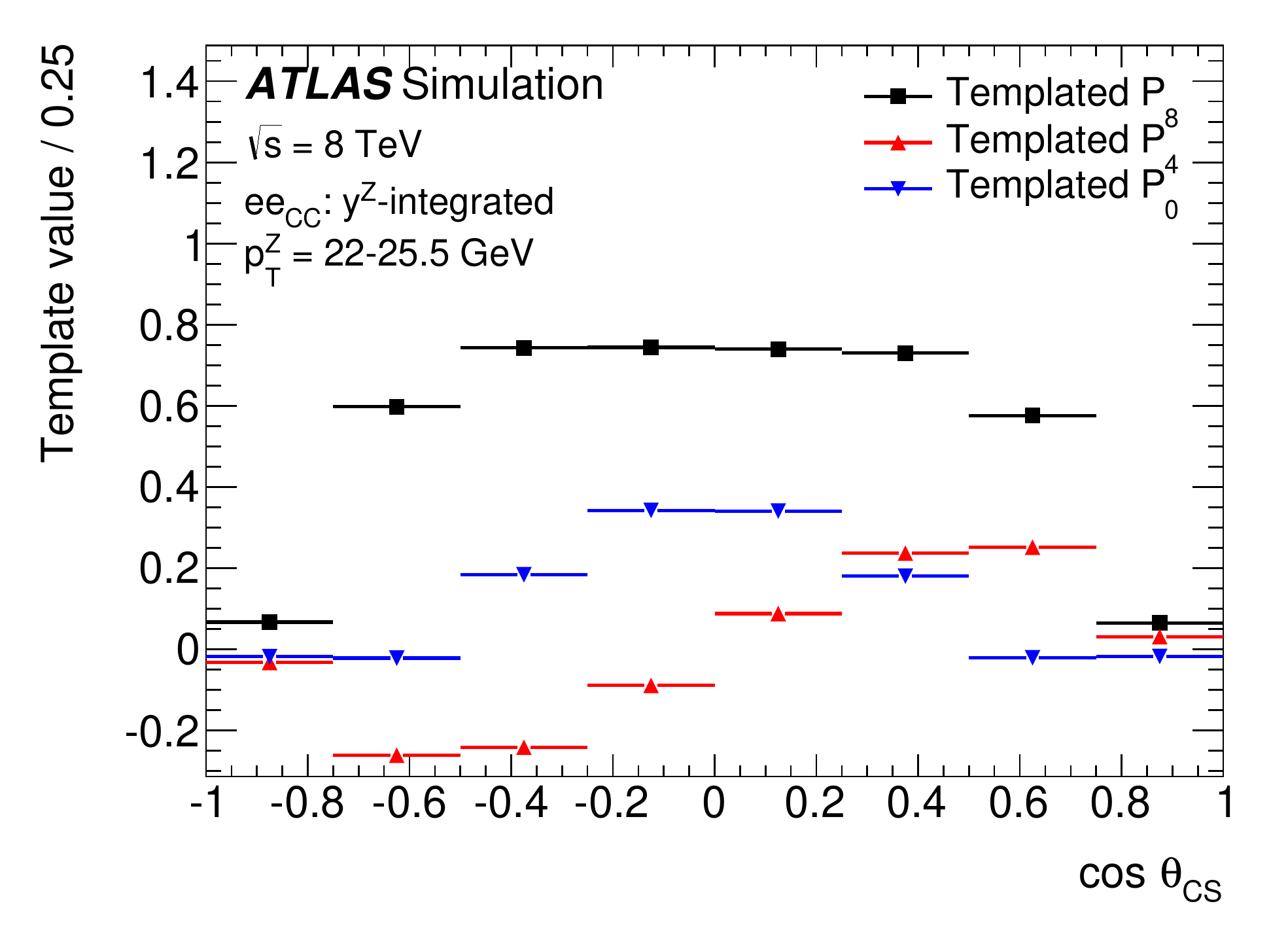}
    \includegraphics[width=7.5cm,angle=0]{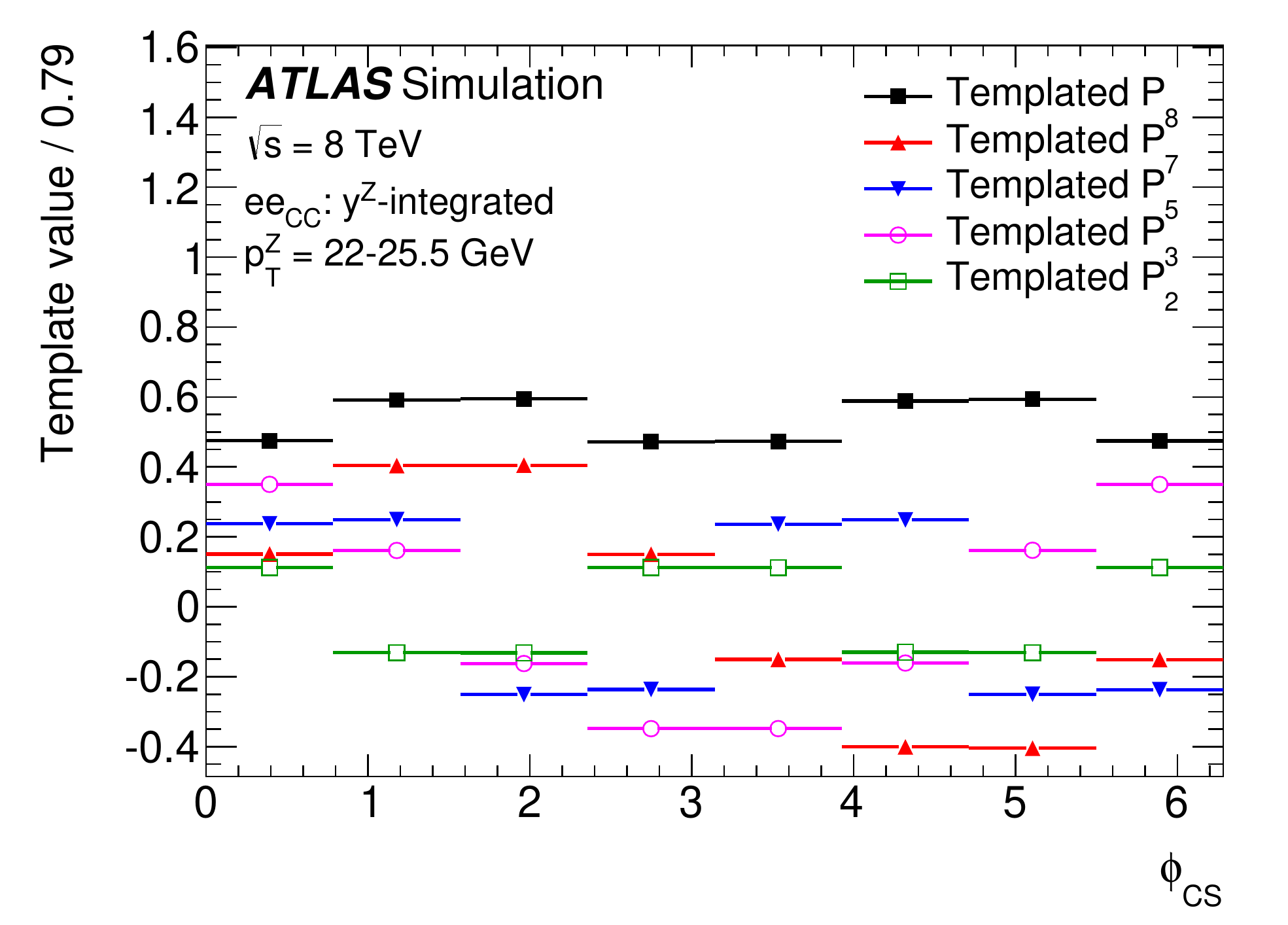}
    \includegraphics[width=7.5cm,angle=0]{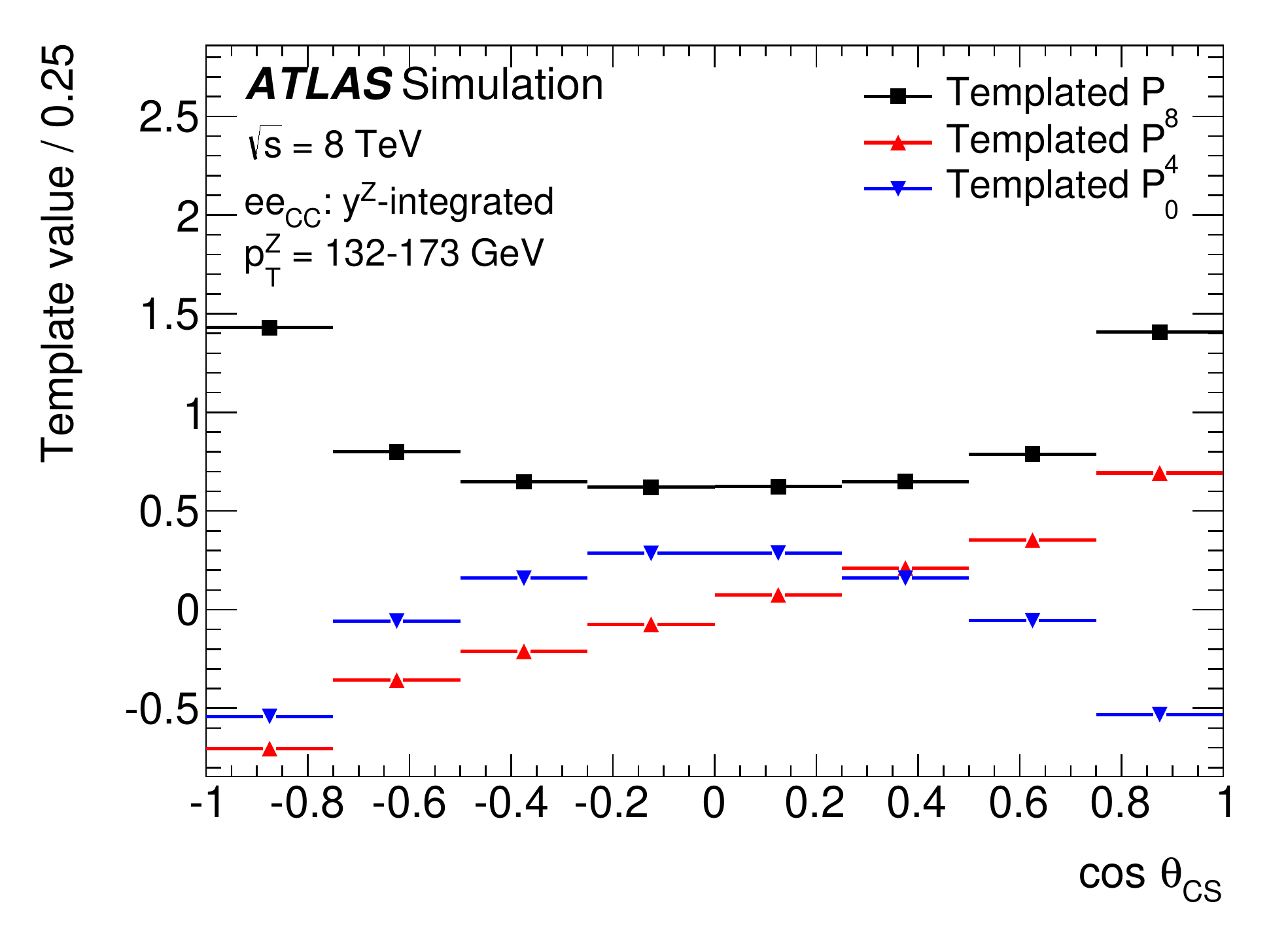}
    \includegraphics[width=7.5cm,angle=0]{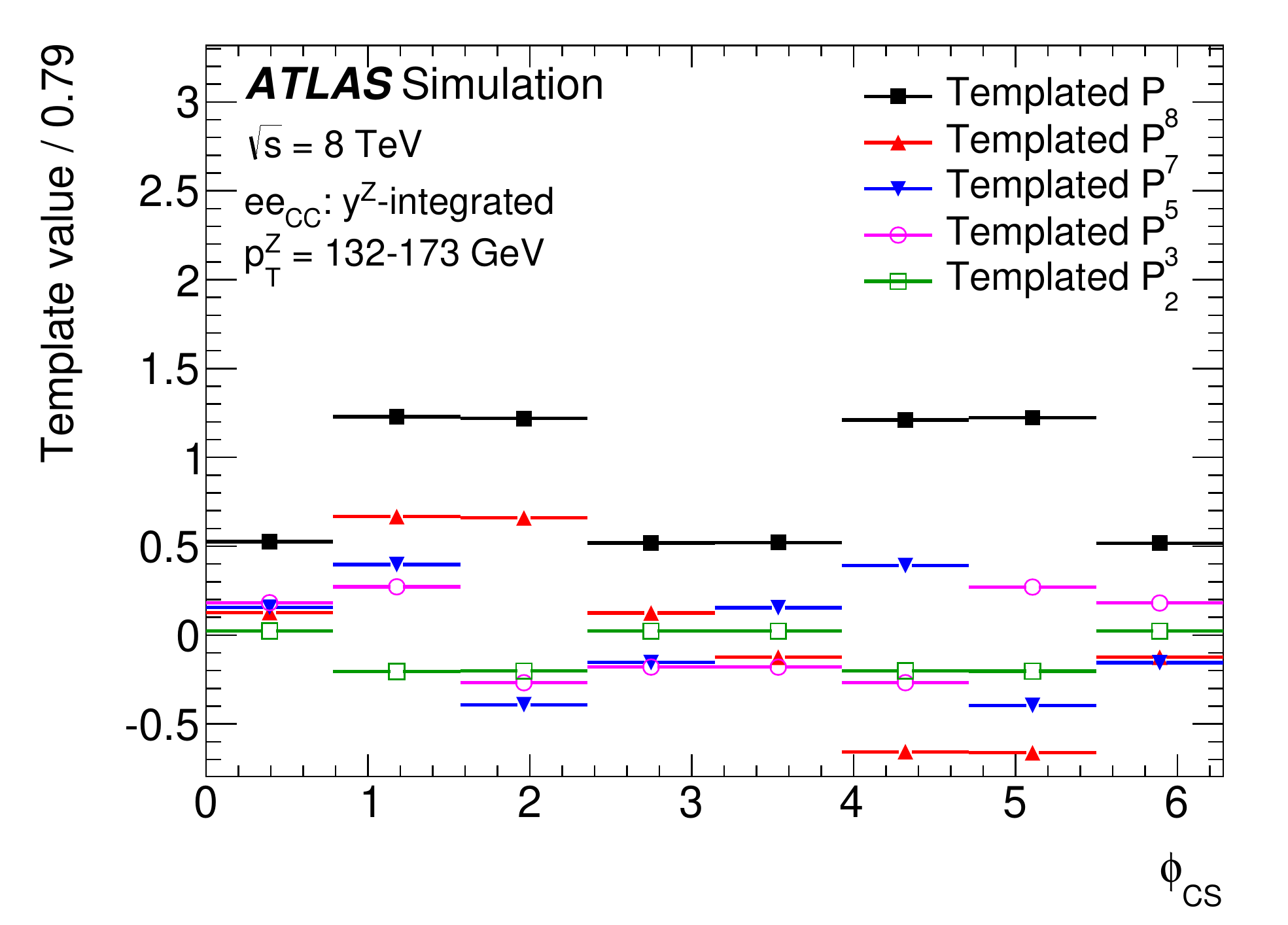}

}
\end{center}
\vspace{-8mm}
\caption{Shapes of polynomials $P_{0,4,8}$ as a function of $\costhetacs$ (top left) and $P_{2,3,5,7,8}$ as a function of~$\phics$ (top right). Shown below are the templated polynomials for the $\yz$-integrated $ee_{\text{CC}}$~events at low~(5--8~GeV), medium~(22--25.5~GeV), and high~(132--173~GeV) values of~$\ptz$ projected onto each of the dimensions~$\costhetacs$ and~$\phics$. The $\ptll$~dimension that normally enters through migrations is also integrated over. The differences between the polynomials and the templates reflect the acceptance shape after event selection.
\label{Fig:templates}}
\end{figure}

Templates $T_{B}$ are also built for each of the multijet, top+electroweak, and non-fiducial \Zboson-boson backgrounds discussed in Section~\ref{sec:backgrounds}. These are normalised by their respective cross-sections times luminosity, or data-driven estimates in the case of the multijet background. The templates for the projection measurements in the $ee_{\text{CF}}$ channel are integrated over either the $\costhetacs$ or $\phics$ axis at the end of the process.

Templates corresponding to variations of the systematic uncertainties in the detector response as well as in the theoretical modelling are built in the same way, after varying the relevant source of systematic uncertainty by~$\pm 1$~standard deviation~($\sigma$). If such a variation changes the $\AiRef$~coefficients in the MC~prediction, for example in the case of~PDF or parton shower uncertainties, the varied $\AiRef$~coefficients are used as such in the weighting procedure. In this way, the theoretical uncertainties on the predictions are not directly propagated to the uncertainties on the measured $\Ai$~coefficients. However, they may affect indirectly the measurements through their impact on the acceptance, selection efficiency, and migration modelling.

\subsection{Likelihood}
\label{sec:likelihood}

A likelihood is built from the nominal templates and the varied templates reflecting the systematic uncertainties. A set of nuisance parameters (NPs) $\theta = \left\{\beta,\gamma\right\}$ is used to interpolate between them. These are constrained by auxiliary probability density functions and come in two categories: $\beta$ and $\gamma$. The first category $\beta$ are the NPs representing experimental and theoretical uncertainties. Each $\beta^m$ in the set $\beta=\left\{\beta^1,...,\beta^M\right\}$ are constrained by unit Gaussian probability density functions~$G(0|\beta^m, 1)$ and linearly interpolate between the nominal and varied templates. These are defined to have a nominal value of zero, with $\beta^m=\pm 1$ corresponding to~$\pm 1\sigma$ for the systematic uncertainty under consideration. The total number of $\beta^{m}$ is $M=171$ for the $ee_{\text{CC}}+\mu\mu_{\text{CC}}$ channel and $M=105$ for the $ee_{\text{CF}}$ channel. The second category $\gamma$ are~NPs that handle systematic uncertainties from the limited size of the MC~samples. For each bin $n$ in the reconstructed $\costhetacs$, $\phics$, and $\ptll$ distribution, $\gamma^n$ in the set $\gamma=\left\{\gamma^1,...,\gamma^{N_{\mathrm{bins}}}\right\}$, where $N_{\mathrm{bins}}=8\times 8\times 23$ is the total number of bins in the reconstructed distribution, has a nominal value of one and normalises the expected events in bin $n$ of the templates. They are constrained by Poisson probability density functions~$P(N_{\mathrm{eff}}^n|\gamma^n N_{\mathrm{eff}}^{n})$, where $N_{\mathrm{eff}}^n$ is the effective number of MC~events in bin $n$. The meaning of ``effective'' here refers to corrections applied for non-uniform event weights. When all signal and background templates are summed over with their respective normalisations, the expected events $N_{\mathrm{exp}}^{n}$ in each bin $n$ can be written as:

\begin{equation}
N_{\mathrm{exp}}^{n}(A,\sigma,\theta) = \left\{\displaystyle\sum_{j=1}^{23}\sigma_{j}\times L\times\left [t_{8j}^{n}(\beta) + \displaystyle\sum_{i=0}^{7}\Aij\times t_{ij}^{n}(\beta)\right ] + \displaystyle\sum_{B}^{\mathrm{bkgs}}T_{B}^{n}(\beta) \right\}\times\gamma^n \ ,
\label{eq:Nexp}
\end{equation}

where:

\begin{itemize}
\item $\Aij$: Coefficient parameter for $\ptz$ bin $j$
\item $A$: Set of all $\Aij$
\item $\sigma_j$: Signal cross-section parameter
\item $\sigma$: Set of all $\sigma_j$
\item $\theta$: Set of all NPs
\item $\beta$: Set of all Gaussian-constrained NPs
\item $\gamma^{n}$: Poisson-constrained NP
\item $t_{ij}$: $\poli$ template
\item $T_{B}$: Background templates
\item $L$: Integrated luminosity constant.
\end{itemize}

The summation over the index $j$ takes into account the contribution of all $\ptz$ bins at generator level in each reconstructed $\ptll$ bin. This is necessary to account for migrations in $\ptll$. The likelihood is the product of Poisson probabilities across all $N_{\mathrm{bins}}$ bins and of auxiliary constraints for each nuisance parameter~$\beta_m$:

\begin{equation}
\mathcal{L}(A,\sigma,\theta|N_{\mathrm{obs}}) = \displaystyle\prod_{n}^{N_{\mathrm{bins}}}\left\{P(N_{\mathrm{obs}}^{n}|N_{\mathrm{exp}}^{n}(A,\sigma,\theta))P(N_{\mathrm{eff}}^{n}|\gamma^n N_{\mathrm{eff}}^{n})\right\}\times\displaystyle\prod_{m}^{M}G(0|\beta^m,1).
\label{eq:likelihood}
\end{equation}

Unlike in the $ee_{\text{CC}}$ and $\mu\mu_{\text{CC}}$ channels that use both angular variables simultaneously, the $ee_{\text{CF}}$ measurements are performed in projections (see Eq.~(\ref{Eq:master_phi}) and Eq.~(\ref{Eq:master_theta})), and therefore the $A_1$ and $A_6$ coefficients are not measured in this channel. The $\poli$~polynomials that normally integrate to zero when projecting onto one angular variable in full phase space may, however, not integrate to zero if their shape is distorted by the event selection. The residual shape is not sufficient to properly constrain their corresponding $\Ai$, and therefore an external constraint is applied to them. For the $\Ai$ that are largely independent of $\yz$ ($A_{0}$ and $A_{2}$), the constraints are taken from the independent $\yz$-integrated measurements in the combined $ee_{\text{CC}}+\mu\mu_{\text{CC}}$ channel. For the $\yz$-dependent coefficients $A_{1}$, $A_{3}$, and $A_{4}$, which are inaccessible to the $ee_{\text{CC}}+\mu\mu_{\text{CC}}$ channels in the $\yz$ bin in which $ee_{\text{CF}}$ is used, predictions from \POWHEG+~\MINLO~\cite{arXiv:1206.3572} are used.

The migration of events between $\ptll$ bins leads to anti-correlations between $\Ai$ in neighbouring bins which enhance the effects of statistical fluctuations. To mitigate this effect and aid in resolving underlying structure in the $\Ai$ spectra, the $\Ai$ spectra are regularised by multiplying the unregularised likelihood by a Gaussian penalty term, which is a function of the significance of higher-order derivatives of the $\Ai$ with respect to $\ptz$. The covariance terms between the $\Aij$~coefficients are taken into account and computed first with the unregularised likelihood. This has parallels with, for example, regularised Bayesian unfolding, where additional information is added through the prior probability of unfolded parameter values~\cite{Blobel,2012arXiv1201.4612C}. As is the case there, the choice of penalty term (or in the Bayesian case, the choice of added information) must be one that leads to a sound result with minimal bias. See Appendix~\ref{sec:regularisation} for more details.

The uncertainties in the parameters are obtained through a likelihood scan. For each parameter of interest $\Aij$, a likelihood ratio is constructed as

\begin{equation}
\Lambda(\Aij) = \frac{\mathcal{L}(\Aij,\hat{A}( \Aij ),\hat{\theta}( \Aij ))}{\mathcal{L}(\hat{A},\hat{\theta})}.
\label{eq:likelihood_ratio}
\end{equation}

In the denominator, the likelihood is maximised unconditionally across all parameters of interest and NPs. In the numerator, the likelihood is maximised for a specific value of a single $\Aij$. The maximum likelihood estimators for the other parameters of interest $\hat{A}$ and NPs $\hat{\theta}$ are in general a function of $\Aij$, hence the explicit dependence is shown in the numerator. The \MINUIT\ package is used to perform numerical minimisation~\cite{James:1975dr} of~$-2\log\Lambda(\Aij)$, and a two-sided test statistic is built from the likelihood ratio:

\begin{equation}
\qA = -2\log\Lambda(\Aij).
\label{eq:test_stat}
\end{equation}

This is asymptotically distributed as a $\chi^{2}$ with one degree of freedom~\cite{Cowen2011}. In this case, the $\pm 1\sigma$ confidence interval of~$\Aij$ is defined by the condition $q_{A^{\pm}_{ij}} = 1$, where $A_{ij}^{\pm}\equiv \hat{A}_{ij} \pm \sigma^{\pm}$.

\subsection{Combinations and alternative parameterisations}
\label{sec:combination}

When applicable, multiple channels are combined through a simple likelihood multiplication. Each likelihood can be decomposed into three types of terms: those that contain the observed data in each channel, denoted $\mathcal{L}_{i}(A,\sigma,\theta)$, the auxiliary terms that constrain the nuisance parameters $\theta$, denoted $\mathcal{A}_{i}(\theta_{i})$, and the auxiliary term that imposes the regularisation, $\mathcal{A}_{\mathrm{reg}}(A)$. There are a total of $M^{\rm{cb}}$ NPs, corresponding to the total number of unique NPs, including the total number of bins, across all combined channels. With this notation the combined likelihood can be written as:

\begin{equation}
\mathcal{L}_{\mathrm{cb}}(A,\sigma,\theta) = \left\{\displaystyle\prod_{i}^{\mathrm{channels}}\mathcal{L}_{i}(A,\sigma,\theta)\right\}\left\{\displaystyle\prod_{i}^{M^{\rm{cb}}}\mathcal{A}_{i}(\theta_{i})\right\}\mathcal{A}_{\mathrm{reg}}(A).
\label{formula:cblikelihood}
\end{equation}

There are several instances in which a combination of two channels is performed. Within these combinations, the compatibility of the channels is assessed. The measurements in the first two $\yz$ bins and the $\yz$-integrated configuration are obtained from a combination of the $ee_{\text{CC}}$ and $\mu\mu_{\text{CC}}$ channels. The $\yz$-integrated $\mu\mu_{\text{CC}}$ and $ee_{\text{CF}}$ channels are also combined in order to assess the compatibility of the high $\yz$ region probed by the $ee_{\text{CF}}$ channel and the lower rapidity region probed by the central--central channels.

The compatibility of channels is assessed through a reparameterisation of the likelihood into parameters that represent the difference between the coefficients in two different channels. For coefficients~$A_{ij}^a$ and~$A_{ij}^b$ in respective channels~$a$ and~$b$, difference parameters $\Delta A_{ij}\equiv A_{ij}^{a}-A_{ij}^{b}$ are defined that effectively represent the difference between the measured coefficients in the two channels. Substitutions are made in the form of $A_{ij}^{a}\rightarrow \Delta A_{ij}+A_{ij}^b$. These new parameters are measured with the same methodology as described in Section~\ref{sec:likelihood}. Similar reparameterisations are also done to measure the difference between the $A_{0}$ and $A_{2}$ coefficients. These reparameterisations have the advantage that the correlations between the new parameters are automatically taken into account.

\section{Measurement uncertainties}
\label{sec:Uncertainties}

Several sources of statistical and systematic uncertainty play a role in the precision of the measurements presented in this paper. 
In particular, some of the systematic uncertainties impact the template-building procedure described in Section~\ref{sec:templates}. For this reason, templates are rebuilt after each variation accounting for a systematic uncertainty, and the difference in shape between the varied and nominal templates is used to evaluate the resulting uncertainty. 

A description of the expected statistical uncertainties (both in data in Section~\ref{sec:data-stat} and in simulation in Section~\ref{sec:mc_stat-sys}) and systematic uncertainties (experimental in Section~\ref{sec:exp-sys}, theoretical in Section~\ref{sec:th-sys}, and those related to the methodology in Section~\ref{sec:method_sys}) associated with the measurement of the $\Ai$~coefficients is given in this section. These uncertainties are summarised in Section~\ref{sec:expectedresults} in three illustrative \ptz\ bins for the $ee_{\text{CC}}$, $\mu\mu_{\text{CC}}$ (and their combination), and $ee_{\text{CF}}$ channels. The evolution of the uncertainty breakdown as a function of \ptz\ is illustrated there as well.

\subsection{Uncertainties from data sample size}
\label{sec:data-stat}

 Although the harmonic polynomials are completely orthogonal in the full phase space, resolution and acceptance effects lead to some non-zero correlation between them. Furthermore, the angular distributions in a bin of reconstructed~$\ptll$ have contributions spanning several generator-level $\ptz$ bins. This leads to correlations between the measured coefficients which increase their statistical uncertainties. The amount of available data is the largest source of uncertainty, although the resolution and binning in the angular variables also play a role. A discussion of the categorisation of this uncertainty may be found in~Appendix~\ref{appendix:stat-cat}.

\subsection{Uncertainties from Monte Carlo sample size}
\label{sec:mc_stat-sys}

Statistical uncertainties from the simulated MC~samples are treated as uncorrelated between each bin of the three-dimensional ($\ptll$,~$\costhetacs$,~$\phics$)  distribution. Although the events used to build each template are the same, they receive a different weight from the different polynomials, and are therefore only partially correlated. It was verified that assuming that the templates are fully correlated yields slightly more conservative uncertainties, but central values identical to those obtained using the fully correct treatment. For simplicity, this assumption is used for this uncertainty.

\subsection{Experimental systematic uncertainties}
\label{sec:exp-sys}

Experimental systematic uncertainties affect the migration and efficiency model of the detector simulation, impacting the variables used to construct the templates and the event weights applied to the simulation.  

{\bf Lepton-related systematic uncertainties:} Scale factors correcting the lepton reconstruction, identification, and trigger efficiencies to those observed in data~\cite{ATLAS-CONF-2014-032,Electrons2011,Muons20112012} are applied to the simulation as event weights. The statistical uncertainties of the scale factors tend to be naturally uncorrelated in the kinematic bins in which they are measured, while the systematic uncertainties tend to be correlated across these bins. Lepton calibration (electron energy scale and resolution as well as muon momentum scale and resolution)~\cite{ElectronCalibration,Muons20112012} and their associated uncertainties are derived from data-driven methods and applied to the simulation. Whereas the charge misidentification rate for muons is negligible, the probability for the electron charge to be misidentified can be significant for central electrons at high~$|\eta|$,  due to bremsstrahlung in the inner detector and the subsequent conversion of the photon. This uncertainty is a potential issue in particular for the $ee_{\text{CF}}$ channel, where the measured charge of the central electron sets the charge of the forward electron (where no charge determination is possible). Measurements of the per-electron charge misidentification rate using same-charge electron pairs have been done in data and compared to that in simulation; the systematic uncertainty coming from this correction has a negligible impact on the measurement. 

{\bf Background-related systematic uncertainties:} Uncertainties in the multijet background estimate come from two sources. The first source is the statistical uncertainty in the normalisation of the background in each bin of reconstructed~$\ptll$.  The second is the systematic uncertainty of the overall background normalisation, which is evaluated using alternative criteria to define the multijet background templates. These uncertainties are applied to all three channels and treated as uncorrelated amongst them. In addition, a 20\% systematic uncertainty uncorrelated across $\ptll$~bins but correlated across the $ee_{\text{CC}}$ and $\mu\mu_{\text{CC}}$ channels is applied to the estimation of the top+electroweak background. 

{\bf Other experimental systematic uncertainties:} Several other potential sources of experimental systematic uncertainty are considered, such as event pileup or possible detector misalignments which might affect the muon momentum measurement, and are found to contribute negligibly to the overall measurement uncertainties. The uncertainty in the integrated luminosity is $\pm$ 2.8\%. It is derived following the same methodology as that detailed in~Ref.~\cite{LumiPaper2011}. It only enters (negligibly) in the scaling of the background contributions evaluated from the Monte Carlo samples.

\subsection{Theoretical systematic uncertainties}
\label{sec:th-sys}

Theoretical systematic uncertainties due to QCD renormalisation/factorisation scale, PDFs, parton-shower modelling, generator modelling, and QED/EW corrections are considered. They are evaluated using either event weights, for example through PDF~reweighting, or templates built from alternative Monte Carlo samples. The templates built after each variation accounting for a systematic uncertainty have their own set of reference coefficients so that each variation starts from isotropic angular distributions. This procedure is done so that uncertainties in the simulation predictions for the coefficients propagate minimally to the uncertainties in the measurement; rather, uncertainties in the measurement are due to the theoretical uncertainty of the migration and acceptance modelling. A small fraction of the acceptance can, however, be attributed to the behaviour of coefficients outside the accessible $\yz$~range. In this specific case, the theoretical predictions used for the coefficients can have a small influence on the uncertainties in the measured coefficients.

{\bf QCD scale: } These systematic uncertainties only affect the predictions over the small region of phase space where no measurements are available. They are evaluated by varying the factorisation and renormalisation scale of the predicted coefficients in the region $|\yz|>3.5$~(see~Fig.~\ref{Fig:yZacceptance}). The changes induced in the templates due to the variation in acceptance are used to assess the impact of this uncertainty, which is found to be negligible.

{\bf PDF:} These systematic uncertainties are computed with the 52 {\sc CT10} eigenvectors representing 26~independent sources. The {\sc CT10} uncertainties are provided at 90\%~CL, and are therefore rescaled by a factor of 1.64 to bring them to 68\%~CL variations. Events are also reweighted using the {\sc NNPDF2.3}~\cite{Ball:2012cx} and {\sc MSTW}~\cite{Martin:2009iq} PDFs and are treated as independent systematics. These two-point variations are symmetrised in the procedure. 

{\bf Parton showers:} The \POWHEG+~\HERWIG samples are used to compute an alternative set of templates. The shape difference between these and the templates obtained from the baseline \POWHEG+~\PYTHIA8 samples are used to evaluate the underlying event and parton shower uncertainty.  

{\bf Event generator:} These systematic uncertainties are evaluated through the reweighting of the rapidity distribution of the nominal \POWHEG+~\PYTHIA8~MC sample to that from~\SHERPA, which corresponds approximately to a 5\%~slope per unit of~$|\yz|$. An alternative set of signal templates is built from this variation, using the new set of reference coefficients averaged over rapidity after the reweighting. 

{\bf QED/EW corrections:} The impact of the QED FSR corrections on the measurements is accounted for by the uncertainties in the lepton efficiencies and scales. The contribution of the EW~corrections to the calculation of the \Zboson-boson decay angular distributions in~Eq.~(\ref{Eq:master2}) is estimated to be negligible around the \Zboson-pole mass, based on the extensive and detailed work done at LEP in this area~\cite{EWLEP1,EWLEP2,Was3}, as discussed in~Section~\ref{sec:intro}. 

The PDF uncertainties were found to be the only non-negligible source of theoretical systematic uncertainty in the measured $\Ai$~coefficients, and are in particular the dominant source of uncertainty in the measurement of~$A_0$ at low~$\ptz$.

\subsection{Systematic uncertainties related to the methodology}
\label{sec:method_sys}

Systematic uncertainties related to the template building, fitting, and regularisation methodology are considered. These could manifest through sensitivity to the $\ptz$~shape in the simulation, the particular shape of the $\Ai$~coefficient being fitted, or possible biases caused by the regularisation scheme.

{\bf $\ptll$ shape:} The sensitivity to the shape of the $\ptll$ spectrum is tested in two different ways. First, the shape of the $\ptll$ spectrum in simulation is reweighted with a polynomial function so that it approximately reproduces the reconstructed spectrum in data. The impact of this procedure is expected to be small, since the signal is normalised to the data in fine bins of $\ptz$. Second, the $\ptll$ shape within each $\ptll$ bin is reweighted to that of the data. Since the binning is fine enough that the $\ptll$ shape does not vary too rapidly within one bin, the impact of this is also small.  

{\bf $\Ai$ shape:} Closure tests are performed by fitting to pseudo-data corresponding to various sets of reference $\AiRef$~coefficients to ensure that the fitted $\Ai$~coefficients can reproduce the reference. The $\AiRef$~coefficients are obtained from~\POWHEG+~\PYTHIA8, from~\SHERPA1.4, or are all set to~zero.

{\bf Regularisation:} The potential bias induced by the regularisation is evaluated with pseudo-experiments. A sixth-order polynomial is fit to the \POWHEGBOX+~\PYTHIA8 set of $\AiRef$~coefficients to obtain a continuous reference spectrum $y_{ij}$. Pseudo-data are randomised around the expected distribution obtained from this fit using a Poisson distribution for each bin. The difference between $y_{ij}$ and the expectation value $E[\Aij]$ of the distribution of fitted and regularised $\Aij$ is an estimate of the potential bias in the regularised $\Aij$. The envelope of $|E[\Aij] - y_{ij}|$ is symmetrised and taken to be the bias uncertainty. (See Appendix~\ref{sec:regularisation} for more details.)

The effect of $\ptz$ reweighting and closure of $\Ai$ spectra were found to be negligible. The only non-negligible source of uncertainty in the methodology was found to be the regularisation bias, which can have a size approaching the statistical uncertainty of $A_{0}$ and $A_{2}$.

\subsection{Summary of uncertainties}
\label{sec:expectedresults}

Tables~\ref{Tab:SystSummary_incl_bin2_reg_A0}--\ref{Tab:SystSummary_ybin3_bin2_reg_A0} show the uncertainties in each measured coefficient in three representative $\ptz$ bins, along with the impact of each category of systematics. The theoretical uncertainties are dominated by the PDF uncertainties, which in a few cases are larger than the statistical uncertainties. The experimental uncertainties are dominated by the lepton uncertainties and are the leading source of systematic uncertainty for low values of $\ptz$ for the $A_{2}$~coefficient. The large uncertainties assigned to the multijet background estimates and their shape have a negligible impact on this measurement.

The dominant uncertainty in the measurements of the $\Ai$ coefficients is in most cases the statistical uncertainty, even in the most populated bins at low $\ptz$, which contain hundreds of thousands of events. The exception is the $A_{0}$ coefficient where PDF and electron efficiency uncertainties dominate for $\ptz$~values below~80~GeV. The next largest uncertainty is due to the signal~MC statistical uncertainty. This is reflected in~Fig.~\ref{Fig:incl_CC_A0_uncertainties}, which shows the uncertainty evolution versus $\ptz$ for $A_{0}-A_{2}$, including a breakdown of the systematic uncertainties for both the unregularised and regularised measurements. The evolution versus $\ptz$ of the total, statistical, and systematic uncertainties is shown for all other coefficients in Fig.~\ref{Fig:incl_uncertainties} for the regularised measurement.

\begin{table}
\caption{Summary of regularised uncertainties expected for $A_{0}$, $A_{2}$, and $A_{0}-A_{2}$ at low (5--8~GeV), mid (22--22.5~GeV), and high (132--173~GeV) \ptz\ for the $\yz$-integrated configuration. The total systematic uncertainty is shown with the breakdown into its underlying components. Entries marked with ``-'' indicate that the uncertainty is below 0.0001.
\label{Tab:SystSummary_incl_bin2_reg_A0} }
\begin{center}
\begin{tabular}{l||c|c|c||c|c|c||c|c|c}
\hline
\hline
\multicolumn{10}{c}{$p_{T}^{Z} =$ 5--8 GeV} \\
\hline
\hline
Coefficient & \multicolumn{3}{c||}{$A_{0}$} & \multicolumn{3}{c||}{$A_{2}$} & \multicolumn{3}{c}{$A_{0}-A_{2}$} \\
\hline
 Channel        & $ee$       & $\mu\mu$ & $ee$+$\mu\mu$    & $ee$    & $\mu\mu$ & $ee$+$\mu\mu$         & $ee$       & $\mu\mu$ & $ee$+$\mu\mu$   \\
\hline
Total & 0.0114 & 0.0123 & 0.0083 & 0.0061 & 0.0045 & 0.0036 & 0.0102 & 0.0107 & 0.0076\\
\hline
Data Stat. & 0.0034 & 0.0029 & 0.0022 & 0.0039 & 0.0034 & 0.0025 & 0.0050 & 0.0043 & 0.0033\\
\hline
Syst. & 0.0109 & 0.0120 & 0.0081 & 0.0047 & 0.0029 & 0.0026 & 0.0089 & 0.0098 & 0.0068\\
\: MC Stat. & 0.0017 & 0.0015 & 0.0011 & 0.0020 & 0.0018 & 0.0013 & - & 0.0023 & 0.0017\\
\: Lepton & 0.0065 & 0.0006 & 0.0014 & 0.0036 & 0.0021 & 0.0017 & 0.0072 & 0.0021 & 0.0022\\
\: Bkg. & 0.0004 & 0.0003 & 0.0002 & 0.0002 & 0.0001 & 0.0001 & - & - & 0.0006\\
\: Theory & 0.0054 & 0.0100 & 0.0042 & 0.0006 & 0.0005 & 0.0005 & 0.0046 & - & 0.0041\\
\: Method. & 0.0016 & 0.0016 & 0.0016 & 0.0013 & 0.0013 & 0.0013 & 0.0001 & 0.0001 & 0.0001\\
\hline
\hline
\multicolumn{10}{c}{$p_{T}^{Z} =$ 22--25.5 GeV} \\
\hline
\hline
Coefficient & \multicolumn{3}{c||}{$A_{0}$} & \multicolumn{3}{c||}{$A_{2}$} & \multicolumn{3}{c}{$A_{0}-A_{2}$} \\
\hline
 Channel        & $ee$       & $\mu\mu$ & $ee$+$\mu\mu$    & $ee$    & $\mu\mu$ & $ee$+$\mu\mu$         & $ee$       & $\mu\mu$ & $ee$+$\mu\mu$   \\
\hline
Total & 0.0101 & 0.0120 & 0.0080 & 0.0067 & 0.0050 & 0.0041 & 0.0102 & 0.0111 & 0.0077\\
\hline
Data Stat. & 0.0049 & 0.0043 & 0.0033 & 0.0047 & 0.0043 & 0.0031 & 0.0064 & 0.0060 & 0.0045\\
\hline
Syst. & 0.0089 & 0.0112 & 0.0073 & 0.0047 & 0.0027 & 0.0026 & 0.0079 & 0.0094 & 0.0063\\
\: MC Stat. & 0.0023 & 0.0021 & 0.0015 & 0.0022 & 0.0020 & 0.0015 & 0.0039 & 0.0035 & 0.0025\\
\: Lepton & 0.0050 & 0.0005 & 0.0013 & 0.0037 & 0.0003 & 0.0013 & 0.0064 & 0.0009 & 0.0019\\
\: Bkg. & 0.0005 & 0.0003 & 0.0002 & 0.0003 & 0.0002 & 0.0002 & - & - & 0.0006\\
\: Theory & 0.0047 & 0.0092 & 0.0038 & 0.0003 & 0.0003 & 0.0002 & 0.0043 & 0.0097 & 0.0039\\
\: Method. & 0.0021 & 0.0021 & 0.0021 & 0.0016 & 0.0016 & 0.0016 & 0.0003 & 0.0003 & 0.0003\\
\hline
\hline
\multicolumn{10}{c}{$p_{T}^{Z} =$ 132--173 GeV} \\
\hline
\hline
Coefficient & \multicolumn{3}{c||}{$A_{0}$} & \multicolumn{3}{c||}{$A_{2}$} & \multicolumn{3}{c}{$A_{0}-A_{2}$} \\
\hline
 Channel        & $ee$       & $\mu\mu$ & $ee$+$\mu\mu$    & $ee$    & $\mu\mu$ & $ee$+$\mu\mu$         & $ee$       & $\mu\mu$ & $ee$+$\mu\mu$   \\
\hline
Total & 0.0143 & 0.0143 & 0.0110 & 0.0400 & 0.0380 & 0.0294 & 0.0326 & 0.0367 & 0.0227\\
\hline
Data Stat. & 0.0113 & 0.0104 & 0.0077 & 0.0324 & 0.0289 & 0.0214 & 0.0295 & 0.0304 & 0.0196\\
\hline
Syst. & 0.0087 & 0.0092 & 0.0079 & 0.0229 & 0.0239 & 0.0202 & 0.0139 & 0.0206 & 0.0116\\
\: MC Stat. & 0.0029 & 0.0060 & 0.0032 & 0.0085 & 0.0167 & 0.0092 & 0.0091 & 0.0181 & 0.0100\\
\: Lepton & 0.0031 & 0.0006 & 0.0012 & 0.0095 & 0.0026 & 0.0040 & 0.0076 & - & 0.0043\\
\: Bkg. & 0.0006 & 0.0009 & 0.0006 & 0.0020 & 0.0033 & 0.0020 & - & - & 0.0009\\
\: Theory & 0.0008 & 0.0015 & 0.0010 & 0.0009 & 0.0021 & 0.0016 & 0.0024 & 0.0047 & 0.0026\\
\: Method. & 0.0067 & 0.0067 & 0.0067 & 0.0172 & 0.0172 & 0.0172 & 0.0067 & 0.0067 & 0.0067\\
\hline
\hline
\end{tabular}
\end{center}
\end{table}

\begin{table}
\caption{Summary of regularised uncertainties expected for $A_{1}$, $A_{3}$ and $A_{4}$ at low (5--8~GeV), mid (22--25.5~GeV), and high (132--173~GeV) \ptz\ for the $\yz$-integrated configuration. The total systematic uncertainty is shown with the breakdown into its underlying components. Entries marked with ``-'' indicate that the uncertainty is below 0.0001.
\label{Tab:SystSummary_incl_bin2_reg_A1} }
\begin{center}
\begin{tabular}{l||c|c|c||c|c|c||c|c|c}
\hline
\hline
\multicolumn{10}{c}{$p_{T}^{Z} =$ 5--8 GeV} \\
\hline
\hline
Coefficient & \multicolumn{3}{c||}{$A_{1}$} & \multicolumn{3}{c||}{$A_{3}$} & \multicolumn{3}{c}{$A_{4}$} \\
\hline
 Channel        & $ee$       & $\mu\mu$ & $ee$+$\mu\mu$    & $ee$    & $\mu\mu$ & $ee$+$\mu\mu$         & $ee$       & $\mu\mu$ & $ee$+$\mu\mu$   \\
\hline
Total & 0.0032 & 0.0027 & 0.0023 & 0.0018 & 0.0016 & 0.0012 & 0.0034 & 0.0030 & 0.0024\\
\hline
Data Stat. & 0.0024 & 0.0021 & 0.0016 & 0.0016 & 0.0015 & 0.0011 & 0.0023 & 0.0020 & 0.0015\\
\hline
Syst. & 0.0021 & 0.0018 & 0.0017 & 0.0008 & 0.0007 & 0.0006 & 0.0025 & 0.0022 & 0.0019\\
\: MC Stat. & 0.0012 & 0.0010 & 0.0008 & 0.0008 & 0.0007 & 0.0006 & 0.0011 & 0.0010 & 0.0008\\
\: Lepton & 0.0015 & 0.0018 & 0.0012 & 0.0002 & 0.0002 & 0.0001 & 0.0002 & - & 0.0001\\
\: Bkg. & 0.0002 & 0.0001 & 0.0001 & - & 0.0001 & - & 0.0001 & - & 0.0001\\
\: Theory & 0.0003 & 0.0003 & 0.0003 & 0.0001 & 0.0001 & 0.0001 & 0.0020 & 0.0018 & 0.0017\\
\: Method. & 0.0006 & 0.0006 & 0.0006 & 0.0001 & 0.0001 & 0.0001 & 0.0002 & 0.0002 & 0.0002\\
\hline
\hline
\multicolumn{10}{c}{$p_{T}^{Z} =$ 22--25.5 GeV} \\
\hline
\hline
Coefficient & \multicolumn{3}{c||}{$A_{1}$} & \multicolumn{3}{c||}{$A_{3}$} & \multicolumn{3}{c}{$A_{4}$} \\
\hline
 Channel        & $ee$       & $\mu\mu$ & $ee$+$\mu\mu$    & $ee$    & $\mu\mu$ & $ee$+$\mu\mu$         & $ee$       & $\mu\mu$ & $ee$+$\mu\mu$   \\
\hline
Total & 0.0042 & 0.0038 & 0.0027 & 0.0023 & 0.0021 & 0.0016 & 0.0039 & 0.0035 & 0.0026\\
\hline
Data Stat. & 0.0033 & 0.0029 & 0.0022 & 0.0021 & 0.0019 & 0.0014 & 0.0032 & 0.0028 & 0.0021\\
\hline
Syst. & 0.0026 & 0.0025 & 0.0016 & 0.0010 & 0.0009 & 0.0007 & 0.0022 & 0.0020 & 0.0015\\
\: MC Stat. & 0.0016 & 0.0014 & 0.0011 & 0.0010 & 0.0009 & 0.0007 & 0.0016 & 0.0014 & 0.0010\\
\: Lepton & 0.0011 & 0.0009 & 0.0007 & 0.0003 & - & 0.0001 & 0.0003 & 0.0001 & 0.0002\\
\: Bkg. & - & - & - & 0.0001 & - & 0.0001 & 0.0001 & 0.0001 & 0.0001\\
\: Theory & 0.0010 & 0.0014 & 0.0005 & 0.0003 & 0.0002 & 0.0002 & 0.0013 & 0.0012 & 0.0012\\
\: Method. & 0.0007 & 0.0007 & 0.0007 & 0.0001 & 0.0001 & 0.0001 & 0.0003 & 0.0003 & 0.0003\\
\hline
\hline
\multicolumn{10}{c}{$p_{T}^{Z} =$ 132--173 GeV} \\
\hline
\hline
Coefficient & \multicolumn{3}{c||}{$A_{1}$} & \multicolumn{3}{c||}{$A_{3}$} & \multicolumn{3}{c}{$A_{4}$} \\
\hline
 Channel        & $ee$       & $\mu\mu$ & $ee$+$\mu\mu$    & $ee$    & $\mu\mu$ & $ee$+$\mu\mu$         & $ee$       & $\mu\mu$ & $ee$+$\mu\mu$   \\
\hline
Total & 0.0127 & 0.0129 & 0.0092 & 0.0113 & 0.0118 & 0.0081 & 0.0074 & 0.0079 & 0.0054\\
\hline
Data Stat. & 0.0113 & 0.0106 & 0.0078 & 0.0108 & 0.0102 & 0.0074 & 0.0071 & 0.0068 & 0.0049\\
\hline
Syst. & 0.0054 & 0.0070 & 0.0049 & 0.0033 & 0.0059 & 0.0034 & 0.0022 & 0.0040 & 0.0022\\
\: MC Stat. & 0.0035 & 0.0060 & 0.0034 & 0.0032 & 0.0059 & 0.0033 & 0.0021 & 0.0037 & 0.0022\\
\: Lepton & 0.0025 & - & 0.0007 & 0.0011 & 0.0015 & 0.0007 & 0.0004 & 0.0004 & 0.0002\\
\: Bkg. & 0.0006 & - & - & 0.0006 & 0.0011 & 0.0003 & 0.0002 & 0.0003 & 0.0001\\
\: Theory & 0.0010 & 0.0010 & 0.0007 & - & - & 0.0005 & 0.0004 & 0.0004 & 0.0004\\
\: Method. & 0.0031 & 0.0031 & 0.0031 & 0.0004 & 0.0004 & 0.0004 & 0.0005 & 0.0005 & 0.0005\\
\hline
\hline
\end{tabular}
\end{center}
\end{table}

\begin{table}
\caption{Summary of regularised uncertainties expected for $A_{5}$, $A_{6}$ and $A_{7}$ at low (5--8~GeV), mid (22--25.5~GeV), and high (132--173~GeV) \ptz\ for the $\yz$-integrated configuration. The total systematic uncertainty is shown with the breakdown into its underlying components. Entries marked with ``-'' indicate that the uncertainty is below 0.0001.
\label{Tab:SystSummary_incl_bin2_reg_A2} }
\begin{center}
\begin{tabular}{l||c|c|c||c|c|c||c|c|c}
\hline
\hline
\multicolumn{10}{c}{$p_{T}^{Z} =$ 5--8 GeV} \\
\hline
\hline
Coefficient & \multicolumn{3}{c||}{$A_{5}$} & \multicolumn{3}{c||}{$A_{6}$} & \multicolumn{3}{c}{$A_{7}$} \\
\hline
 Channel        & $ee$       & $\mu\mu$ & $ee$+$\mu\mu$    & $ee$    & $\mu\mu$ & $ee$+$\mu\mu$         & $ee$       & $\mu\mu$ & $ee$+$\mu\mu$   \\
\hline
Total & 0.0021 & 0.0019 & 0.0015 & 0.0022 & 0.0019 & 0.0015 & 0.0015 & 0.0014 & 0.0010\\
\hline
Data Stat. & 0.0018 & 0.0017 & 0.0013 & 0.0019 & 0.0017 & 0.0013 & 0.0014 & 0.0012 & 0.0009\\
\hline
Syst. & 0.0010 & 0.0009 & 0.0007 & 0.0011 & 0.0009 & 0.0007 & 0.0007 & 0.0006 & 0.0005\\
\: MC Stat. & 0.0009 & 0.0008 & 0.0006 & 0.0010 & 0.0009 & 0.0007 & 0.0007 & 0.0006 & 0.0005\\
\: Lepton & 0.0003 & - & 0.0001 & 0.0001 & 0.0001 & 0.0001 & 0.0001 & - & 0.0001\\
\: Bkg. & 0.0001 & - & - & 0.0001 & - & - & - & - & -\\
\: Theory & 0.0002 & 0.0001 & - & 0.0004 & 0.0004 & 0.0004 & 0.0002 & 0.0002 & 0.0002\\
\: Method. & 0.0003 & 0.0003 & 0.0003 & 0.0002 & 0.0002 & 0.0002 & 0.0001 & 0.0001 & 0.0001\\
\hline
\hline
\multicolumn{10}{c}{$p_{T}^{Z} =$ 22--25.5 GeV} \\
\hline
\hline
Coefficient & \multicolumn{3}{c||}{$A_{5}$} & \multicolumn{3}{c||}{$A_{6}$} & \multicolumn{3}{c}{$A_{7}$} \\
\hline
 Channel        & $ee$       & $\mu\mu$ & $ee$+$\mu\mu$    & $ee$    & $\mu\mu$ & $ee$+$\mu\mu$         & $ee$       & $\mu\mu$ & $ee$+$\mu\mu$   \\
\hline
Total & 0.0026 & 0.0023 & 0.0018 & 0.0028 & 0.0025 & 0.0019 & 0.0020 & 0.0018 & 0.0014\\
\hline
Data Stat. & 0.0023 & 0.0021 & 0.0015 & 0.0025 & 0.0022 & 0.0017 & 0.0018 & 0.0016 & 0.0012\\
\hline
Syst. & 0.0012 & 0.0011 & 0.0009 & 0.0013 & 0.0012 & 0.0009 & 0.0009 & 0.0008 & 0.0006\\
\: MC Stat. & 0.0011 & 0.0010 & 0.0008 & 0.0012 & 0.0011 & 0.0008 & 0.0009 & 0.0008 & 0.0006\\
\: Lepton & - & - & 0.0001 & 0.0002 & 0.0001 & 0.0001 & 0.0001 & 0.0001 & 0.0001\\
\: Bkg. & - & - & - & 0.0001 & 0.0001 & 0.0001 & 0.0001 & 0.0001 & 0.0001\\
\: Theory & 0.0001 & 0.0001 & - & 0.0002 & 0.0001 & 0.0001 & 0.0002 & 0.0001 & 0.0001\\
\: Method. & 0.0004 & 0.0004 & 0.0004 & 0.0002 & 0.0002 & 0.0002 & 0.0001 & 0.0001 & 0.0001\\
\hline
\hline
\multicolumn{10}{c}{$p_{T}^{Z} =$ 132--173 GeV} \\
\hline
\hline
Coefficient & \multicolumn{3}{c||}{$A_{5}$} & \multicolumn{3}{c||}{$A_{6}$} & \multicolumn{3}{c}{$A_{7}$} \\
\hline
 Channel        & $ee$       & $\mu\mu$ & $ee$+$\mu\mu$    & $ee$    & $\mu\mu$ & $ee$+$\mu\mu$         & $ee$       & $\mu\mu$ & $ee$+$\mu\mu$   \\
\hline
Total & 0.0092 & 0.0097 & 0.0069 & 0.0076 & 0.0081 & 0.0056 & 0.0066 & 0.0071 & 0.0048\\
\hline
Data Stat. & 0.0087 & 0.0083 & 0.0060 & 0.0072 & 0.0070 & 0.0050 & 0.0063 & 0.0061 & 0.0044\\
\hline
Syst. & 0.0034 & 0.0052 & 0.0034 & 0.0023 & 0.0041 & 0.0024 & 0.0018 & 0.0035 & 0.0020\\
\: MC Stat. & 0.0024 & 0.0046 & 0.0027 & 0.0020 & 0.0039 & 0.0022 & 0.0018 & 0.0035 & 0.0019\\
\: Lepton & 0.0013 & - & 0.0008 & 0.0006 & 0.0003 & 0.0004 & 0.0005 & 0.0004 & 0.0003\\
\: Bkg. & 0.0004 & 0.0007 & - & 0.0004 & 0.0003 & 0.0002 & 0.0003 & 0.0002 & 0.0002\\
\: Theory & 0.0002 & 0.0008 & 0.0005 & 0.0004 & 0.0003 & 0.0002 & 0.0003 & 0.0003 & 0.0002\\
\: Method. & 0.0022 & 0.0022 & 0.0022 & 0.0008 & 0.0008 & 0.0008 & 0.0002 & 0.0002 & 0.0002\\
\hline
\hline
\end{tabular}
\end{center}
\end{table}

\begin{table}
\caption{Summary of regularised uncertainties expected for the coefficients at low (5--8~GeV), mid (22--25.5~GeV), and high (73.4--85.4~GeV) \ptz\ for the $2<|\yz|<3.5$ configuration. The total systematic uncertainty is shown with the breakdown into its underlying components. Entries marked with ``-'' indicate that the uncertainty is below 0.0001.
\label{Tab:SystSummary_ybin3_bin2_reg_A0} }
\begin{center}
\begin{tabular}{l||c|c|c|c|c|c}
\hline
\hline
\multicolumn{7}{c}{$p_{T}^{Z} =$ 5--8 GeV} \\
\hline
\hline
Coefficient & $A_{0}$ & $A_{2}$ & $A_{3}$ & $A_{4}$ & $A_{5}$ & $A_{7}$ \\
\hline
Total & 0.0377 & 0.0657 & 0.0190 & 0.0097 & 0.0161 & 0.0064\\
\hline
Data Stat. & 0.0169 & 0.0569 & 0.0183 & 0.0090 & 0.0152 & 0.0059\\
\hline
Syst. & 0.0337 & 0.0328 & 0.0054 & 0.0036 & 0.0053 & 0.0026\\
\: Lepton & 0.0282 & 0.0263 & 0.0014 & 0.0015 & 0.0021 & 0.0005\\
\: MC Stat. & 0.0059 & 0.0150 & 0.0047 & 0.0032 & 0.0049 & 0.0026\\
\: Bkg. & 0.0047 & 0.0202 & 0.0010 & 0.0006 & 0.0007 & 0.0002\\
\: Theory & 0.0121 & 0.0032 & 0.0008 & 0.0012 & - & 0.0002\\
\: Method. & 0.0008 & 0.0002 & 0.0006 & 0.0002 & 0.0007 & 0.0001\\
\hline
\hline
\multicolumn{7}{c}{$p_{T}^{Z} =$ 22--25.5 GeV} \\
\hline
\hline
Coefficient & $A_{0}$ & $A_{2}$ & $A_{3}$ & $A_{4}$ & $A_{5}$ & $A_{7}$ \\
\hline
Total & 0.0395 & 0.0724 & 0.0225 & 0.0148 & 0.0193 & 0.0107\\
\hline
Data Stat. & 0.0257 & 0.0660 & 0.0216 & 0.0140 & 0.0183 & 0.0101\\
\hline
Syst. & 0.0300 & 0.0299 & 0.0063 & 0.0048 & 0.0061 & 0.0035\\
\: Lepton & 0.0264 & 0.0203 & 0.0024 & 0.0015 & 0.0014 & 0.0006\\
\: MC Stat. & 0.0083 & 0.0165 & 0.0057 & 0.0045 & 0.0057 & 0.0035\\
\: Bkg. & 0.0062 & 0.0190 & 0.0015 & 0.0006 & 0.0007 & 0.0004\\
\: Theory & 0.0089 & 0.0023 & 0.0002 & 0.0013 & 0.0003 & 0.0004\\
\: Method. & 0.0007 & 0.0011 & 0.0005 & 0.0003 & 0.0002 & 0.0002\\
\hline
\hline
\multicolumn{7}{c}{$p_{T}^{Z} =$ 73.4--85.4 GeV} \\
\hline
\hline
Coefficient & $A_{0}$ & $A_{2}$ & $A_{3}$ & $A_{4}$ & $A_{5}$ & $A_{7}$ \\
\hline
Total & 0.0425 & 0.1242 & 0.0383 & 0.0211 & 0.0355 & 0.0233\\
\hline
Data Stat. & 0.0296 & 0.0991 & 0.0345 & 0.0192 & 0.0323 & 0.0214\\
\hline
Syst. & 0.0304 & 0.0747 & 0.0167 & 0.0089 & 0.0146 & 0.0094\\
\: Lepton & 0.0149 & 0.0399 & 0.0034 & 0.0026 & 0.0015 & 0.0017\\
\: MC Stat. & 0.0125 & 0.0417 & 0.0145 & 0.0083 & 0.0136 & 0.0090\\
\: Bkg. & 0.0301 & 0.0343 & 0.0053 & 0.0018 & 0.0038 & 0.0008\\
\: Theory & 0.0033 & 0.0069 & 0.0008 & 0.0006 & 0.0009 & 0.0009\\
\: Method. & 0.0014 & 0.0041 & 0.0012 & 0.0008 & 0.0012 & 0.0004\\
\hline
\hline
\end{tabular}
\end{center}
\end{table}

\begin{figure}
\begin{center}
{

\includegraphics[width=7.5cm,angle=0]{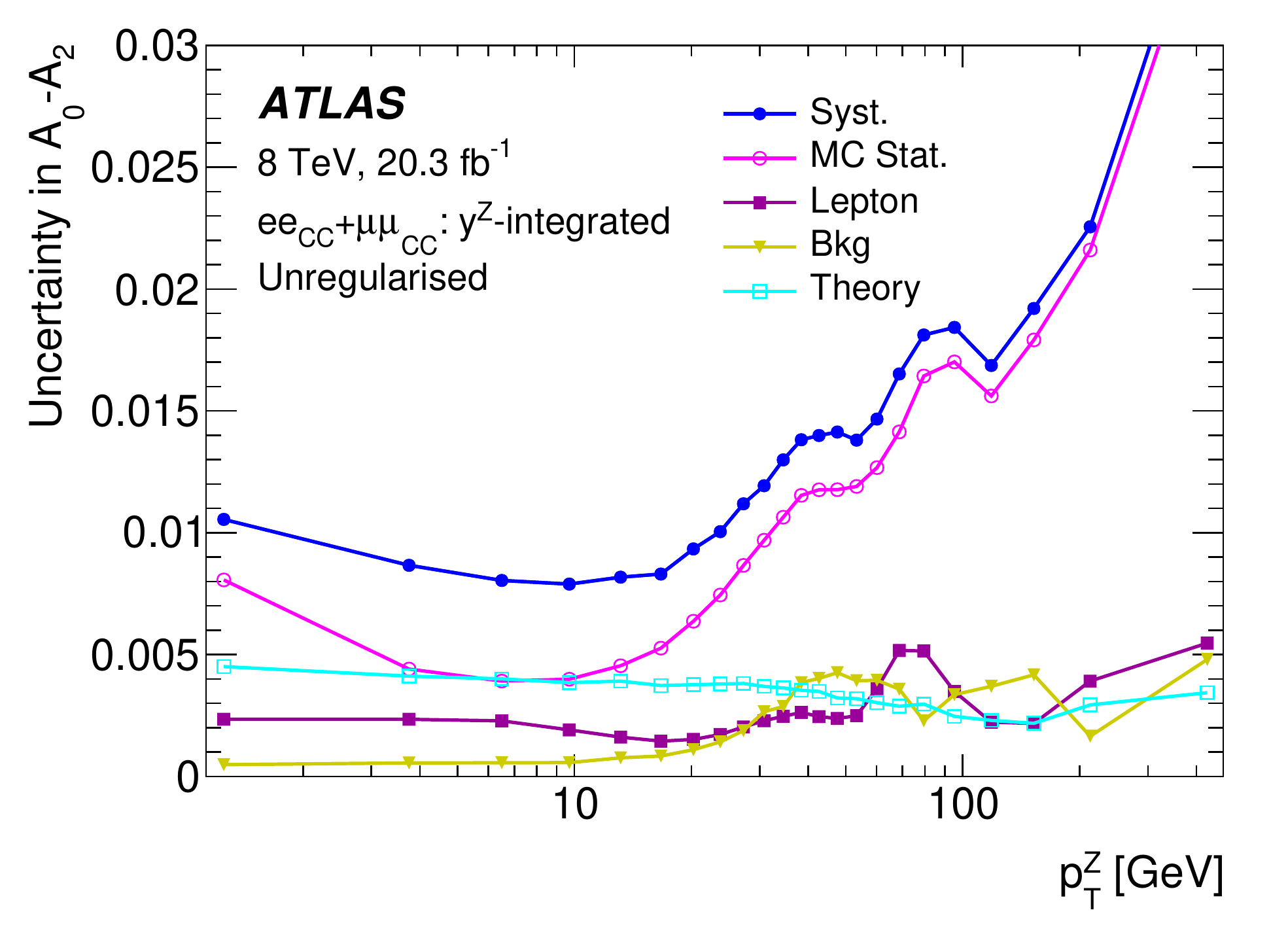}
\includegraphics[width=7.5cm,angle=0]{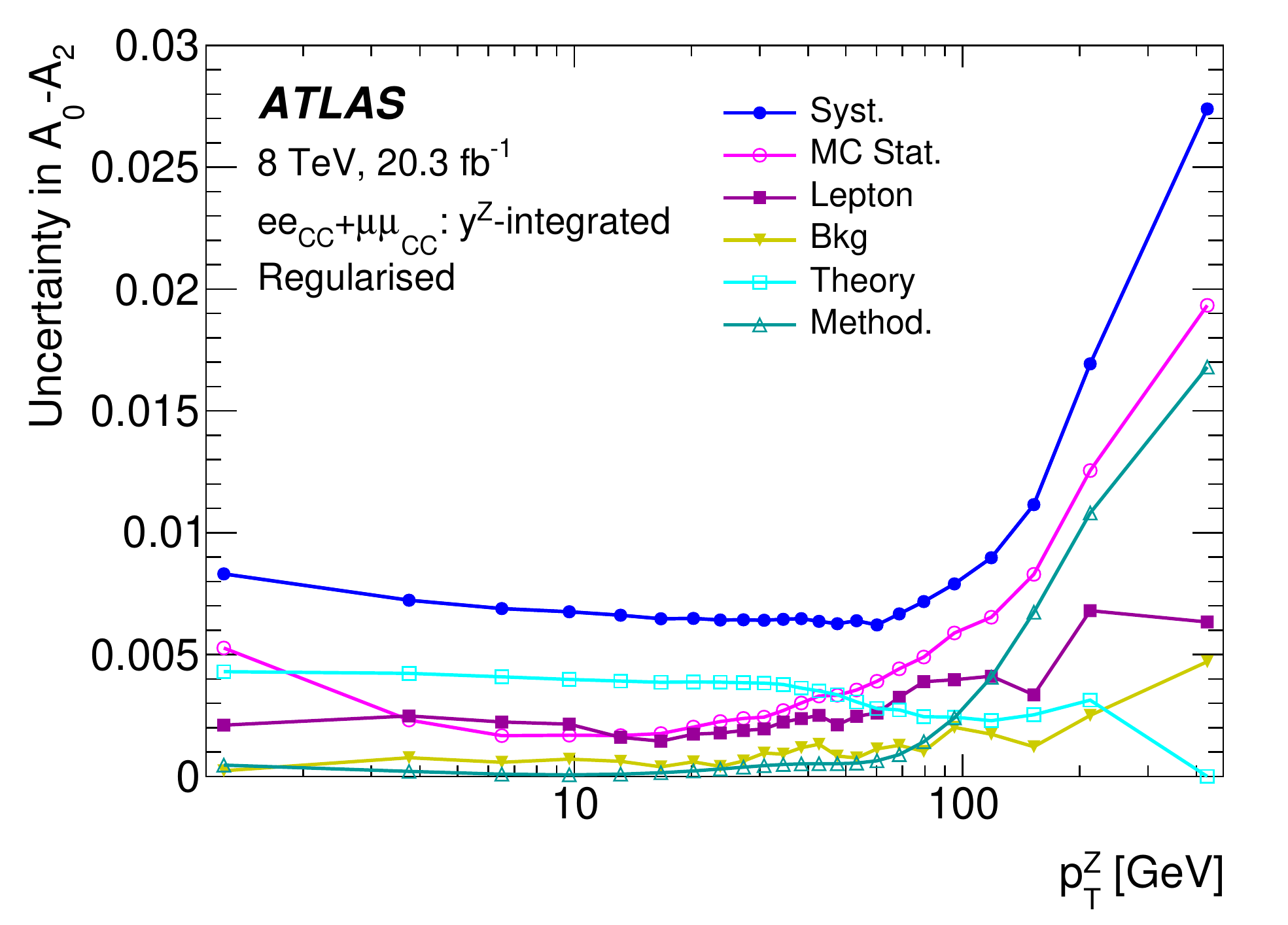}
\includegraphics[width=7.5cm,angle=0]{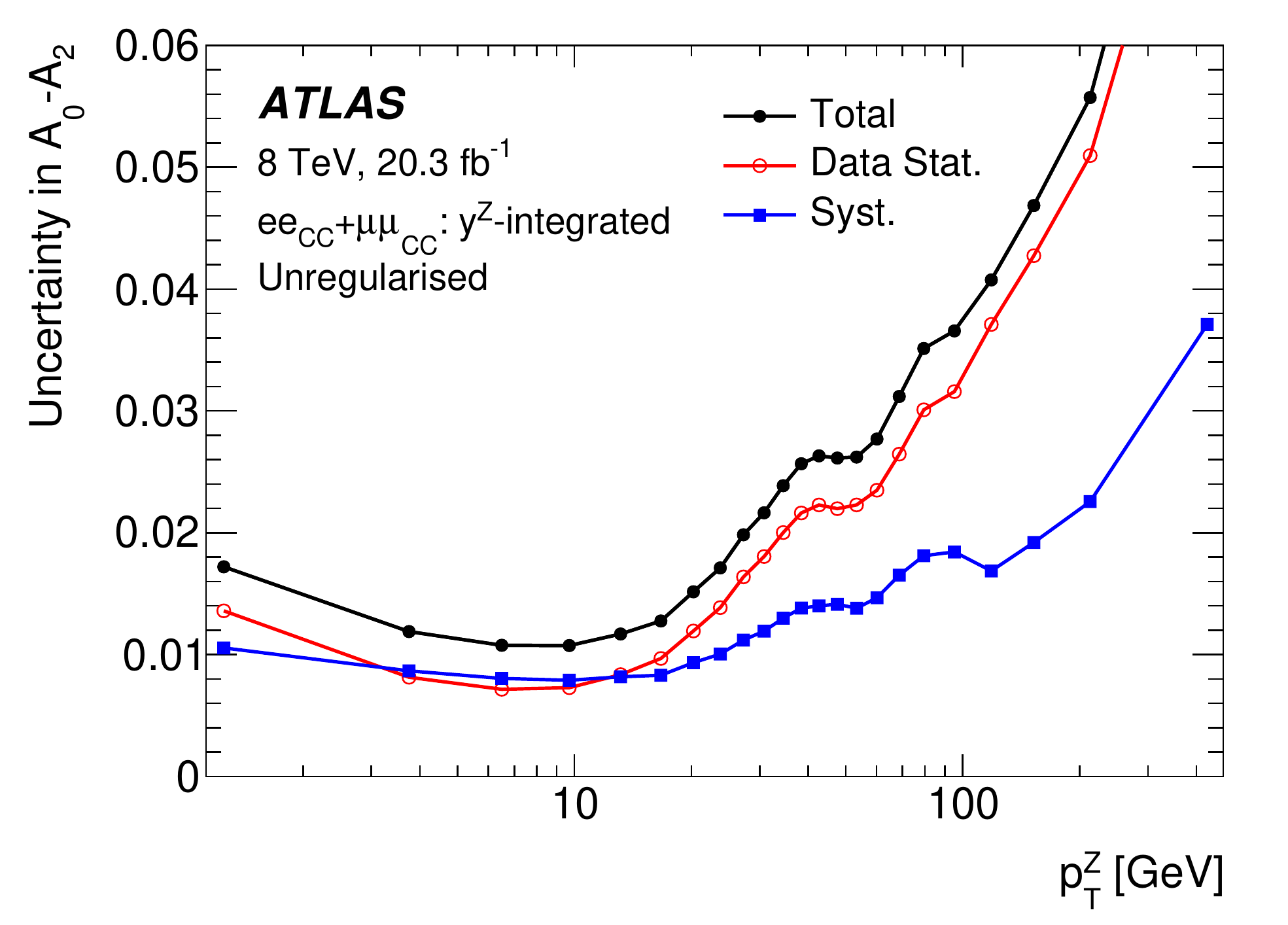}
\includegraphics[width=7.5cm,angle=0]{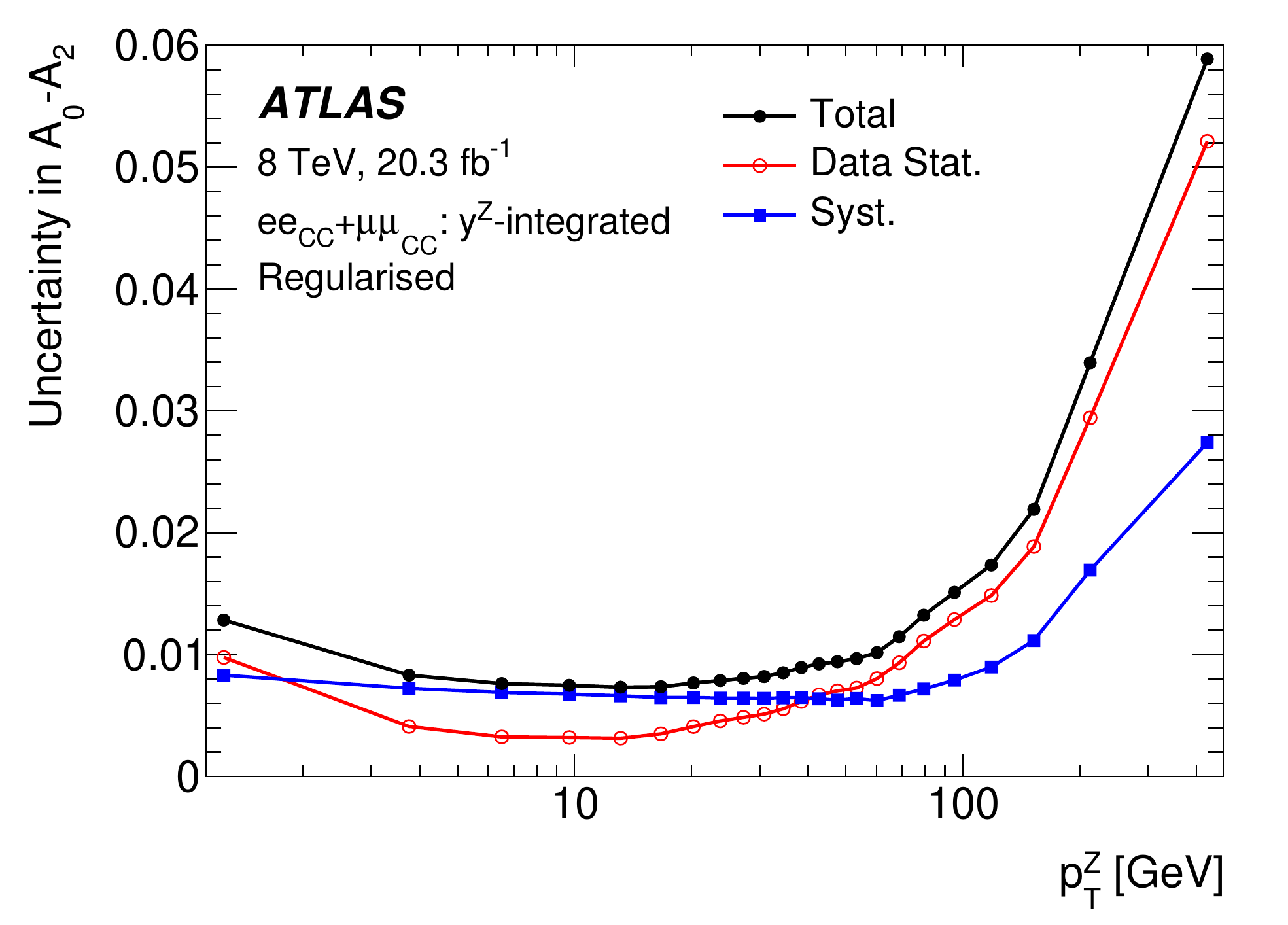}
}
\end{center}
\caption{Uncertainty breakdown for $A_{0}-A_{2}$ as a function of $\ptz$ in the $\yz$-integrated $ee_{\text{CC}}+\mu\mu_{\text{CC}}$ measurement: the systematic uncertainty (top) and the total uncertainty (bottom). The left column shows the unregularised version, while the right column shows the regularised one.
\label{Fig:incl_CC_A0_uncertainties}}
\end{figure}

\begin{figure}
\begin{center}
{
\includegraphics[width=7.5cm,angle=0]{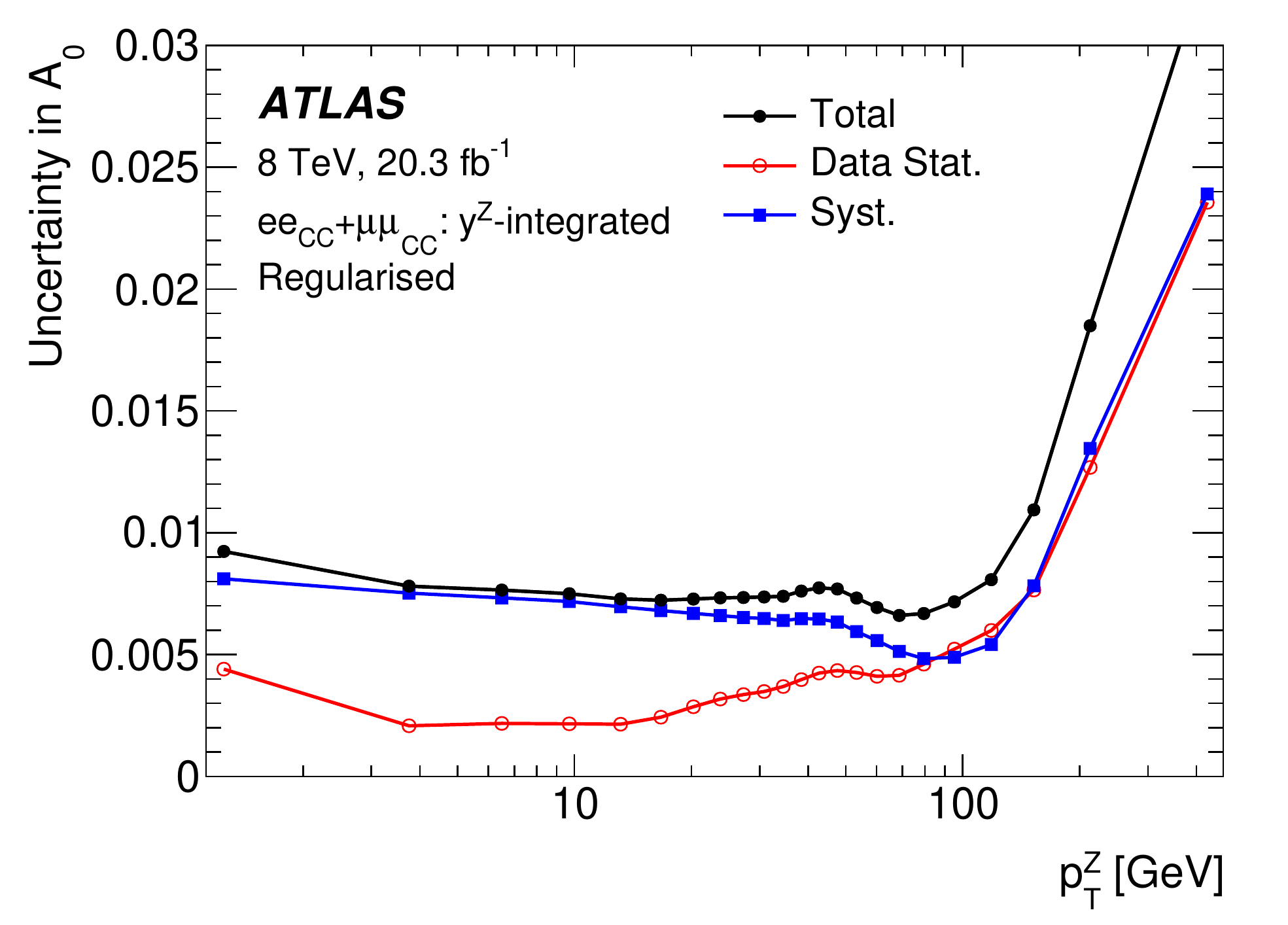}
\includegraphics[width=7.5cm,angle=0]{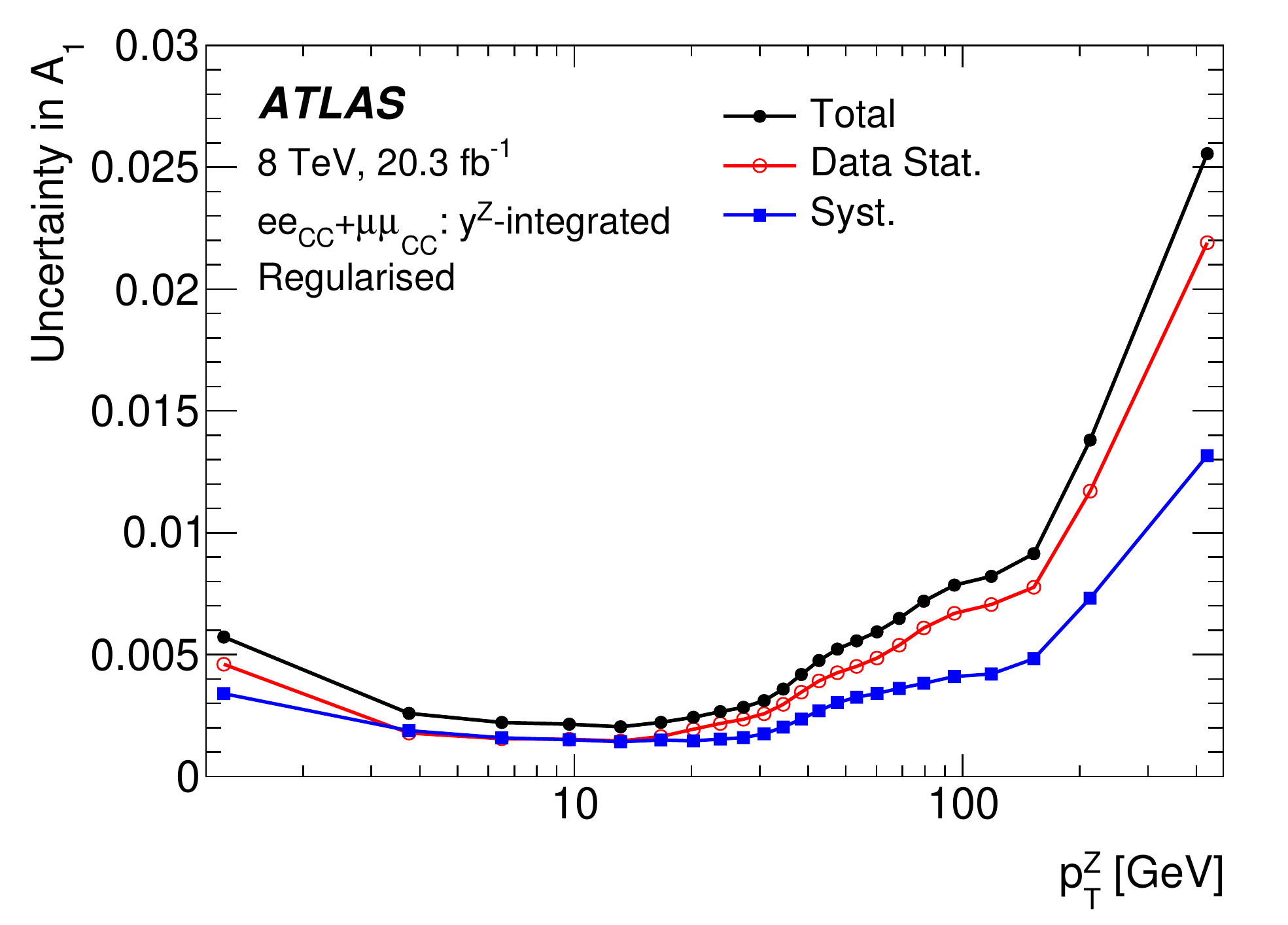}
\includegraphics[width=7.5cm,angle=0]{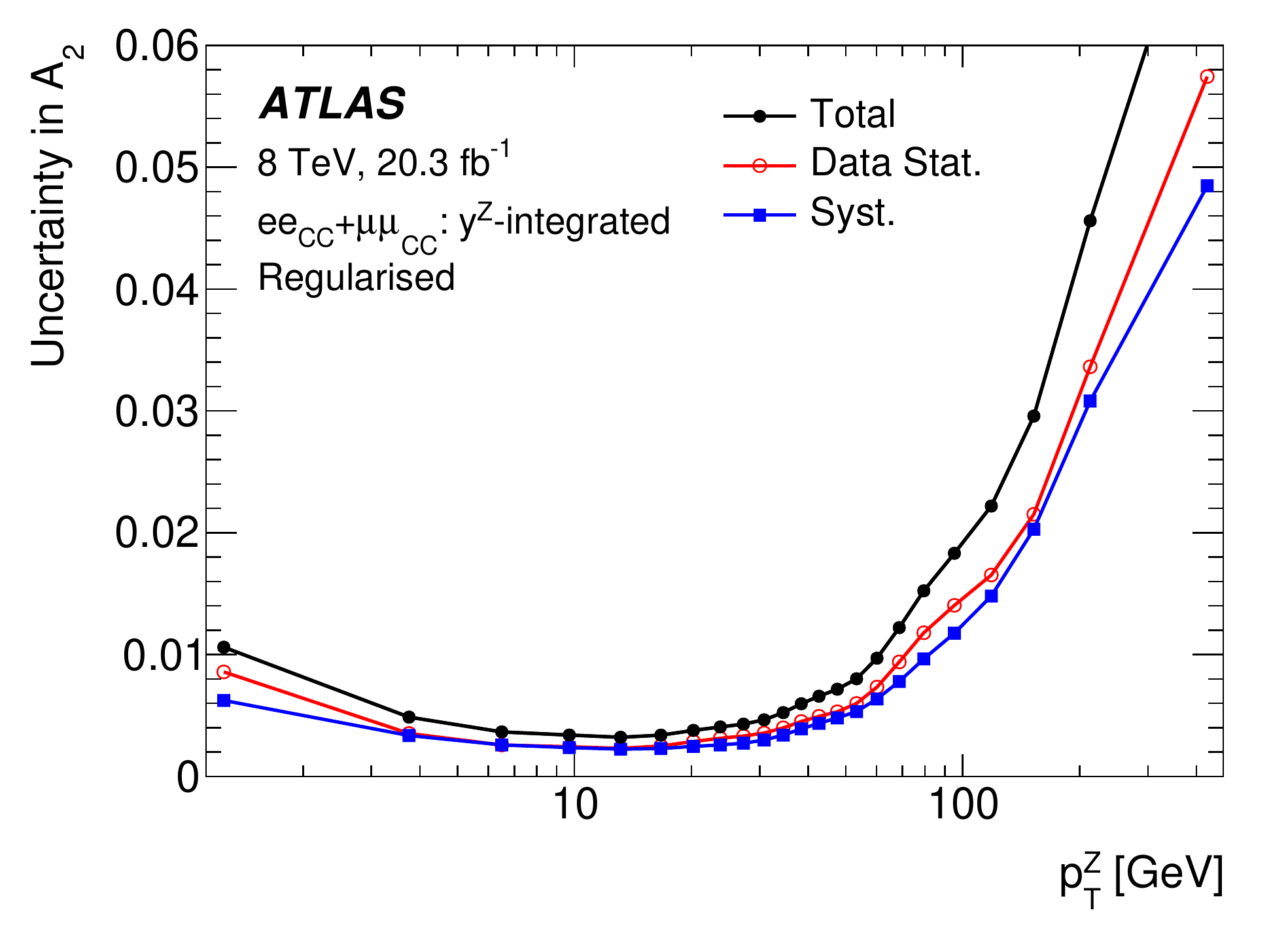}
\includegraphics[width=7.5cm,angle=0]{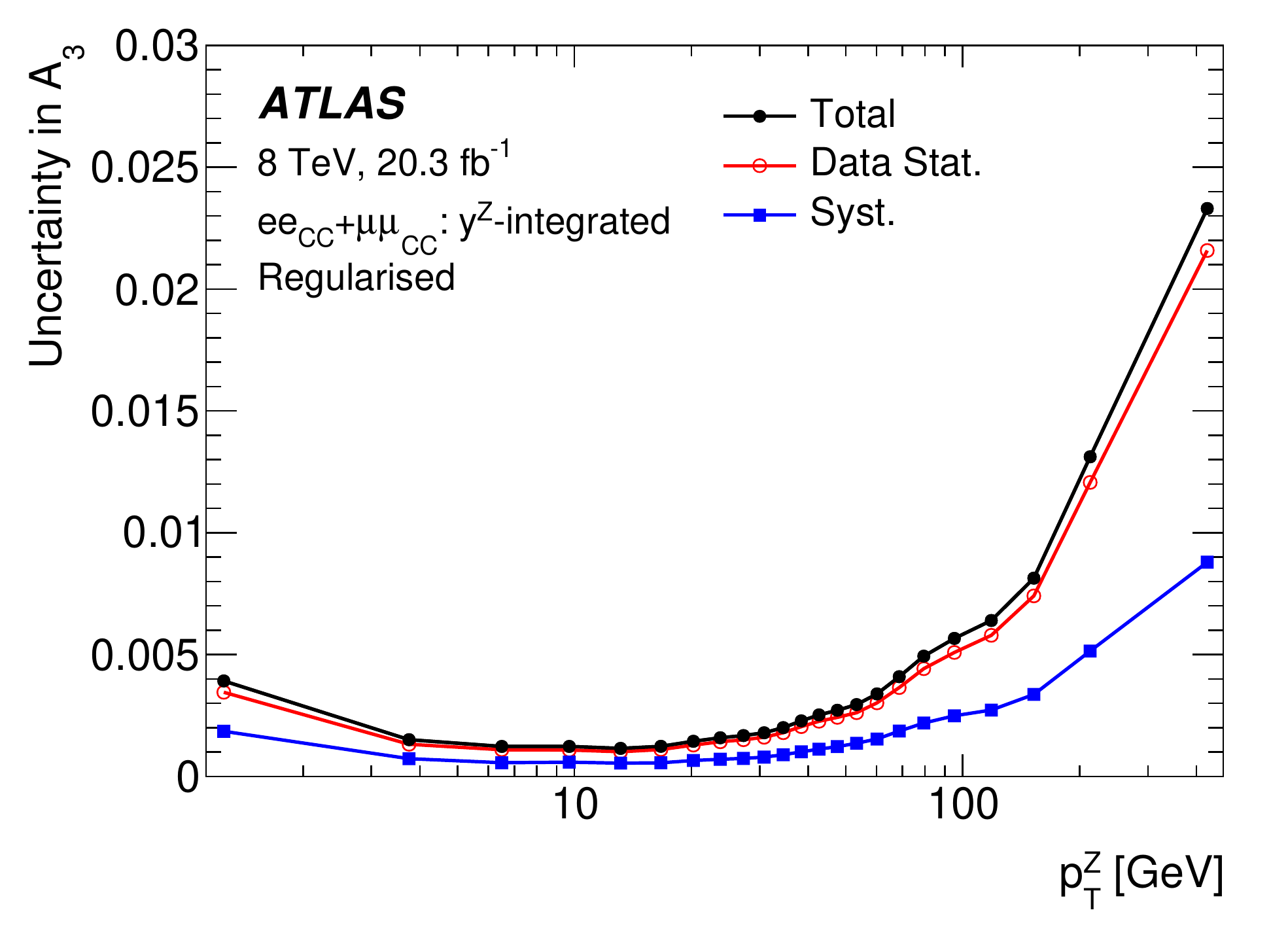}
\includegraphics[width=7.5cm,angle=0]{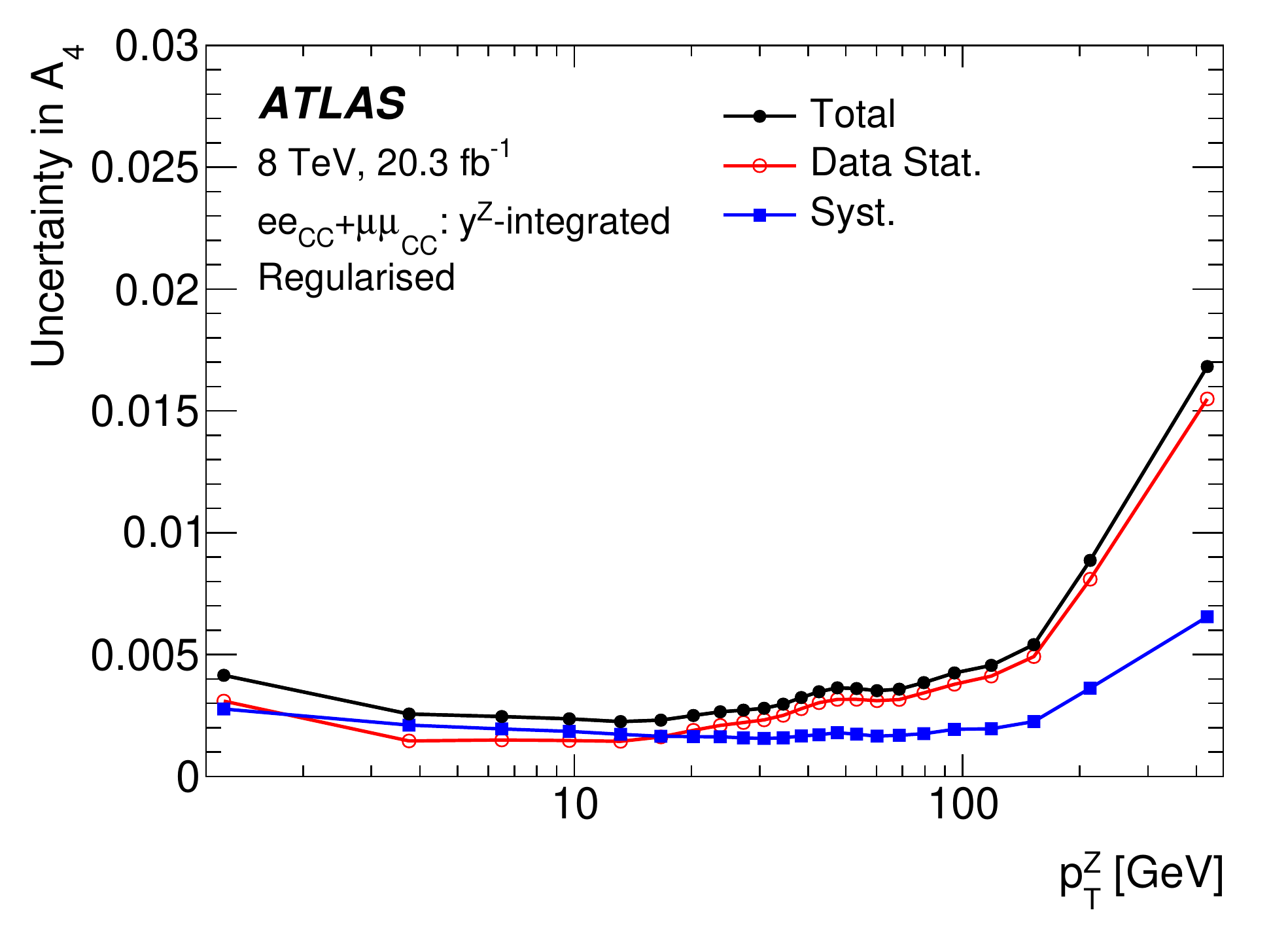}
\includegraphics[width=7.5cm,angle=0]{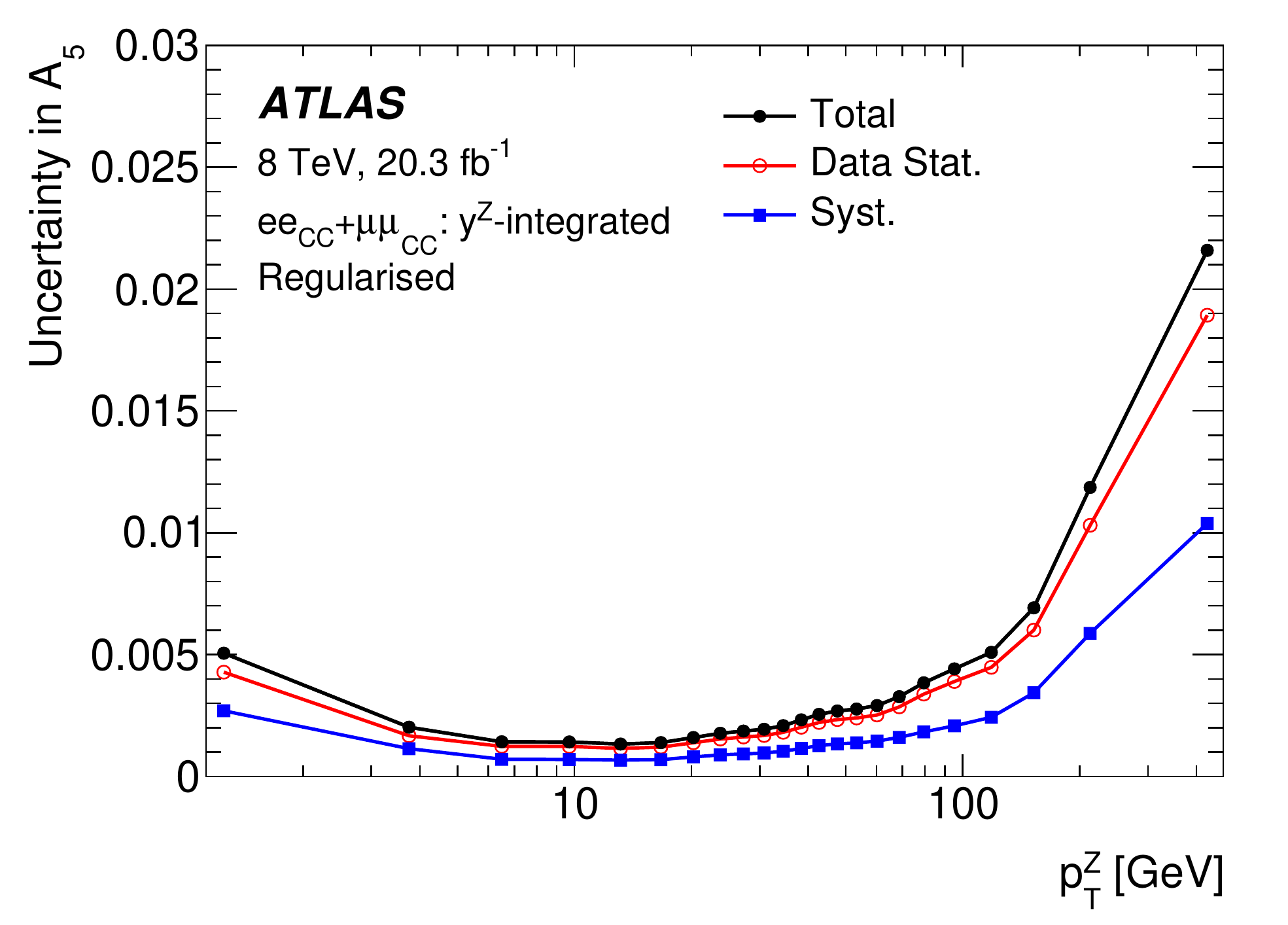}
\includegraphics[width=7.5cm,angle=0]{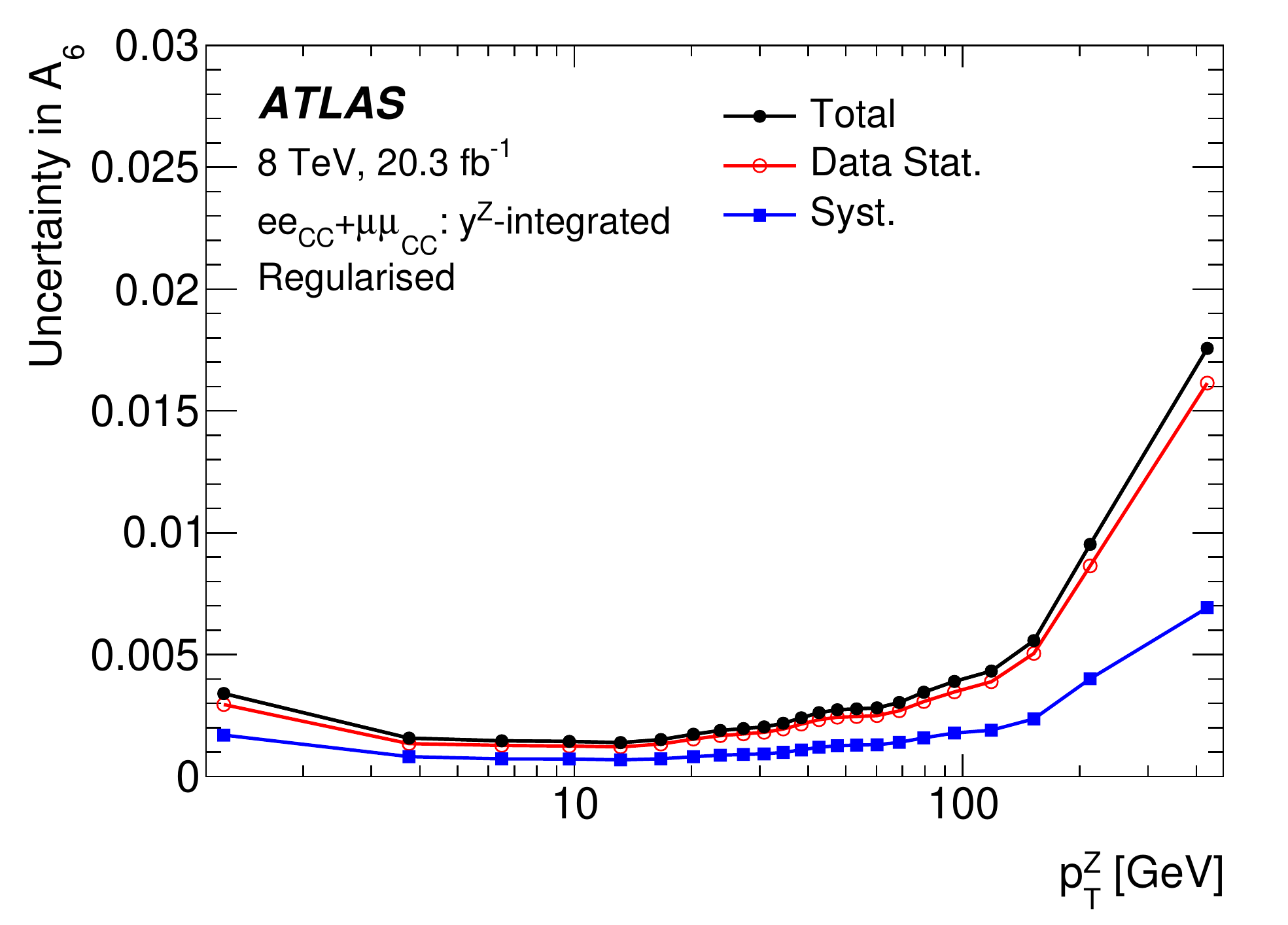}
\includegraphics[width=7.5cm,angle=0]{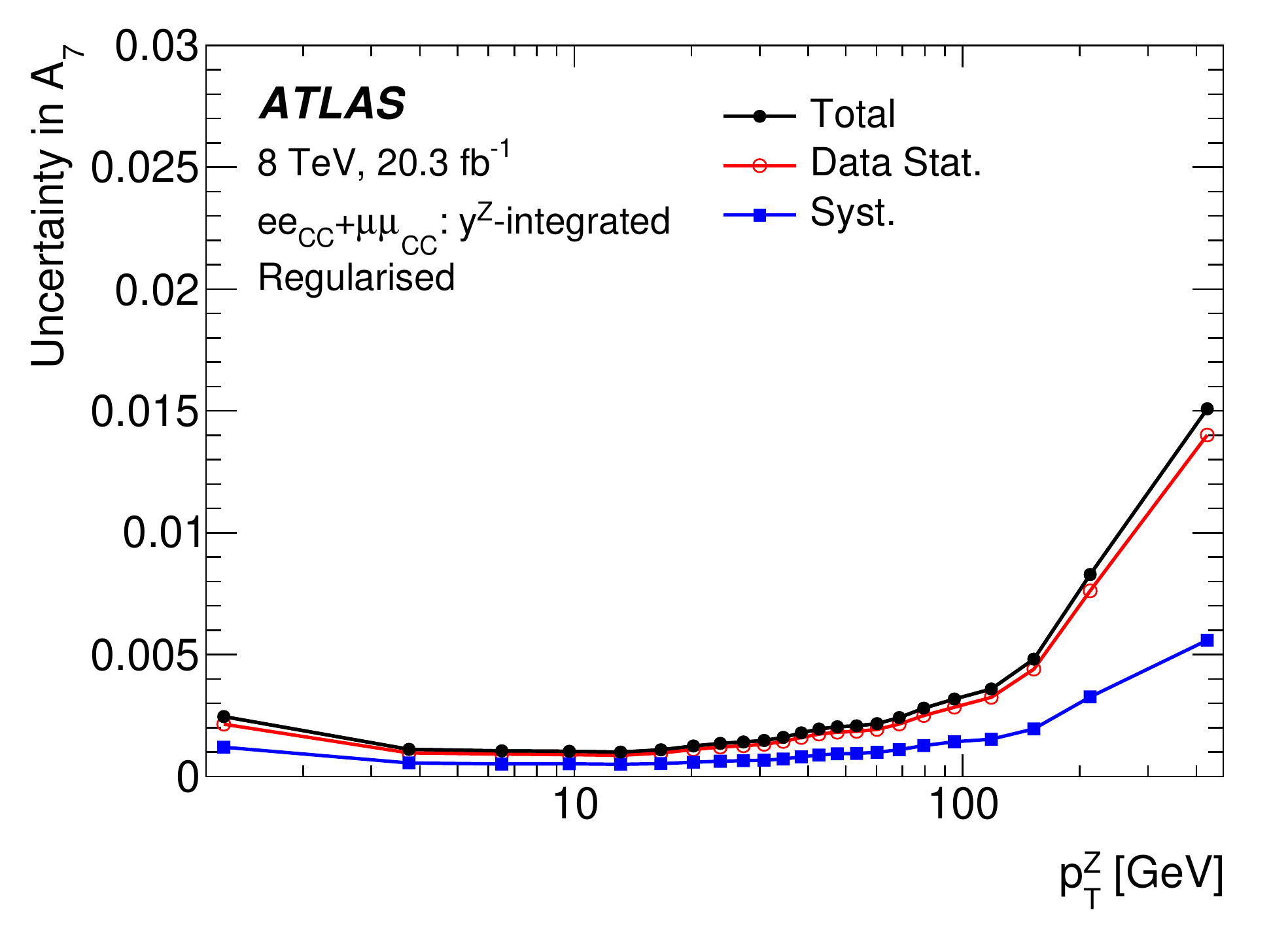}
}
\end{center}
\caption{The total uncertainty as a function of $\ptz$ along with a breakdown into statistical and systematic components for all coefficients in the regularised $\yz$-integrated $ee_{\text{CC}}+\mu\mu_{\text{CC}}$ measurement.
\label{Fig:incl_uncertainties}}
\end{figure}

\section{Results}
\label{sec:results}

This section presents the full set of experimental results. The compatibility between channels is assessed in~Section~\ref{sec:compatibility}. The measured $\Ai$~coefficients are then shown in Section~\ref{sec:baseline_results}. A test is also performed to check for non-zero values of the $A_{5,6,7}$~coefficients. Several cross-checks are presented in~Section~\ref{sec:cross-checks}, including a test of the validity of the nine-term decomposition, probing for the presence of higher-order $\poli$ polynomial terms.

\subsection{Compatibility between channels}
\label{sec:compatibility}

Given that a complex fitting procedure based on reconstructed observables is used, the compatibility between different channels is assessed with a strict quantitative test. The likelihood is parameterised in terms of $\Delta A_{ij}\equiv A_{ij}^{\mu\mu}-A_{ij}^{ee}$ for coefficient index $i$ and $\ptz$ bin $j$, as described in Section~\ref{sec:combination}. The compatibility of the $\Delta A_{ij}$ with zero can be quantified via a $\chi^2$ test taking into account all systematic uncertainty correlations. 
The $\chi^2$ values are first computed for each coefficient $i$ and across all $\ptz$ bins $j$, then for all coefficients and $\ptz$ bins simultaneously. This test is done in the $\yz$-integrated case for the differences between the measurements extracted from the $\mu\mu_{\text{CC}}$ and $ee_{\text{CC}}$~events and from the $\mu\mu_{\text{CC}}$ and $ee_{\text{CF}}$~events, as well as in the first two $\yz$ bins for the $\mu\mu_{\text{CC}}$ and $ee_{\text{CC}}$~events. The $\chi^2$ values are tabulated in Table~\ref{tab:deltaAi_chi2} and indicate almost all the differences are compatible with zero.

The $\Delta A_{ij}$ spectra are shown in Fig.~\ref{Fig:DiffeeCC-mumuCC} for the $\yz$-integrated $ee_{\text{CC}}$ and $\mu\mu_{\text{CC}}$ channels. The regularised and unregularised spectra are overlayed. Visually, it appears that these results are compatible with zero. In some cases, the unregularised $\Delta A_{ij}$ show alternating fluctuations above and below zero due to anti-correlations between neighbouring $\ptz$ bins. These are smoothed out in the regularised results, which come at the expense of larger bin-to-bin correlations.

\begin{table}
\caption{Tabulation of the compatibility of the measured $\Delta A_{i}$ with zero reported as $\chi^2$ per degree of freedom ($N_{\rm{DoF}}$), where $\Delta A_{i}$ represents the difference between the $A_i$~coefficient extracted from the~$\mu\mu_{\text{CC}}$ and $ee_{\text{CC}}$~events~(left) and from the~$\mu\mu_{\text{CC}}$ and $ee_{\text{CF}}$~events~(right). For the $ee_{\text{CC}}$ versus $\mu\mu_{\text{CC}}$ tests, the number of degrees of freedom is~23 for the tests of the individual coefficients and~184 for the tests of all coefficients simultaneously. Likewise, for the $ee_{\text{CF}}$ versus $\mu\mu_{\text{CC}}$ tests, there are 19~degrees of freedom for the tests of the individual coefficients, 38 for the simultaneous test in the $\costhetacs$ projection, and 76 for the simultaneous test in the $\phics$ projection. The comparisons are not performed for the $A_1$ and $A_6$ coefficients between the $\mu\mu_{\text{CC}}$ and $ee_{\text{CF}}$ channels (see Section~\ref{sec:likelihood}).
\label{tab:deltaAi_chi2} }
\begin{center}
\scalebox{0.9}{
\begin{tabular}{c|c|c|c|c|c}
\hline
 & \multicolumn{3}{c|}{$\chi^2 / N_{\rm{DoF}}$ for $\mu\mu_{\text{CC}}$ versus $ee_{\text{CC}}$} & \multicolumn{2}{c}{$\chi^2 / N_{\rm{DoF}}$ for $\mu\mu_{\text{CC}}$ versus $ee_{\text{CF}}$} \\
\hline
\raisebox{-0.4ex}{$A_{i}$} & \raisebox{-0.4ex}{$\yz$-integrated} & \raisebox{-0.4ex}{$0<|\yz|<1$} & \raisebox{-0.4ex}{$1<|\yz|<2$} & \raisebox{-0.4ex}{$\yz$-integrated ($\costhetacs$-proj.)} & \raisebox{-0.4ex}{$\yz$-integrated ($\phics$-proj.)} \\*[0.2cm]
\hline

0     & 15.4 / 23  & 25.0 / 23 & 9.8 / 23  & 18.9 / 19 &   - \\
1     & 32.9 / 23  & 24.9 / 23 & 28.2 / 23 &    -      &   - \\
2     & 17.0 / 23  & 22.7 / 23 & 19.4 / 23 &    -      &  35.0 / 19 \\
3     & 15.8 / 23  & 20.9 / 23 & 19.5 / 23 &    -      &  16.9 / 19 \\
4     & 27.2 / 23  & 31.1 / 23 & 23.4 / 23 & 15.1 / 19 &   - \\
5     & 20.0 / 23  & 23.1 / 23 & 18.4 / 23 &    -      &  17.9 / 19 \\
6     & 21.9 / 23  & 17.7 / 23 & 27.6 / 23 &    -      &   - \\
7     & 18.3 / 23  & 22.9 / 23 & 18.1 / 23 &    -      &  27.4 / 19 \\
\hline
All & 173.1 / 184 & 190 / 184 & 166.1 / 184 & 33.8 / 38 & 94.5 / 76  \\
\end{tabular}
}
\end{center}
\end{table}

\begin{figure}
  \begin{center}
{
    \includegraphics[width=7.3cm,angle=0]{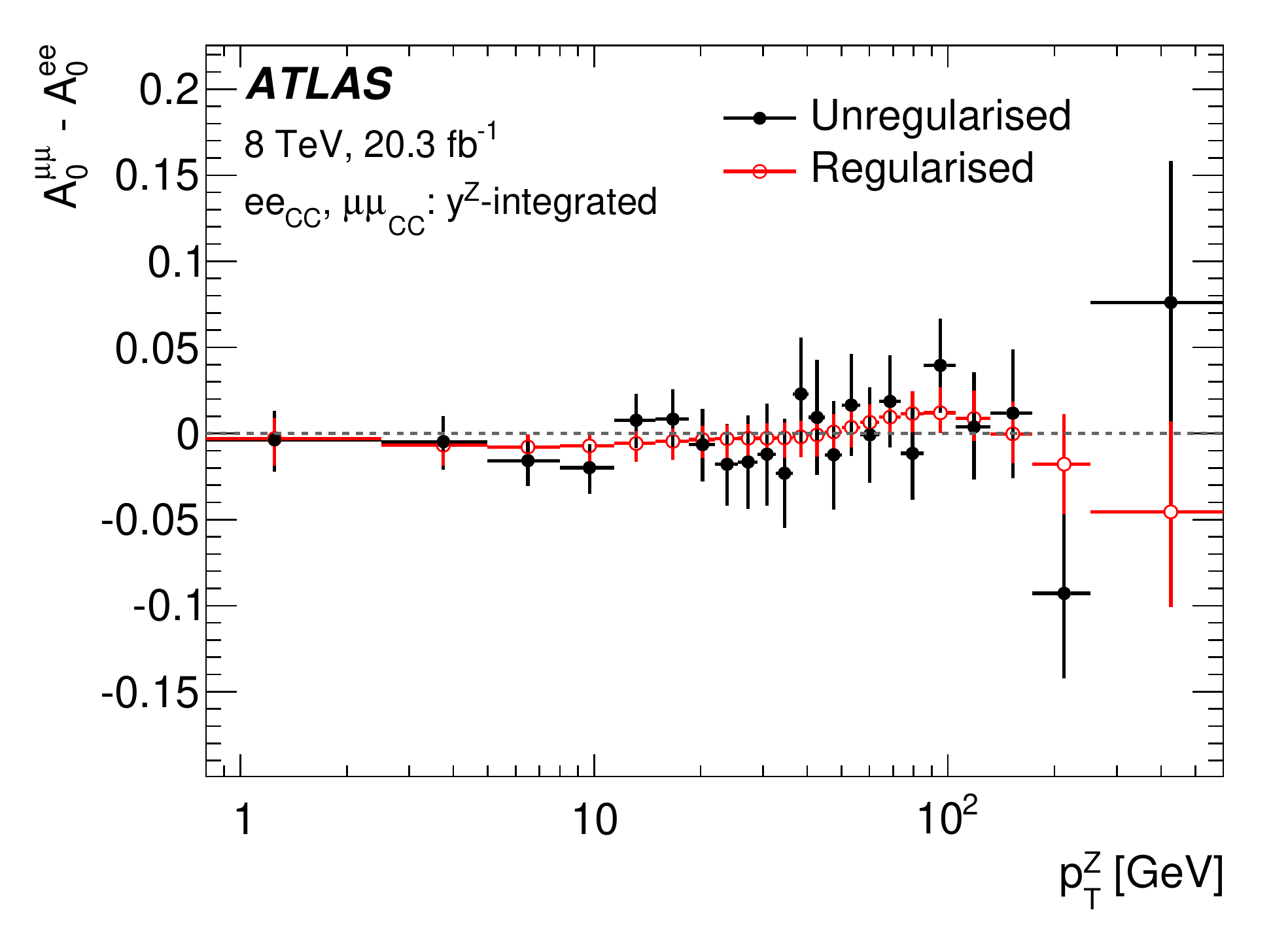}
    \includegraphics[width=7.3cm,angle=0]{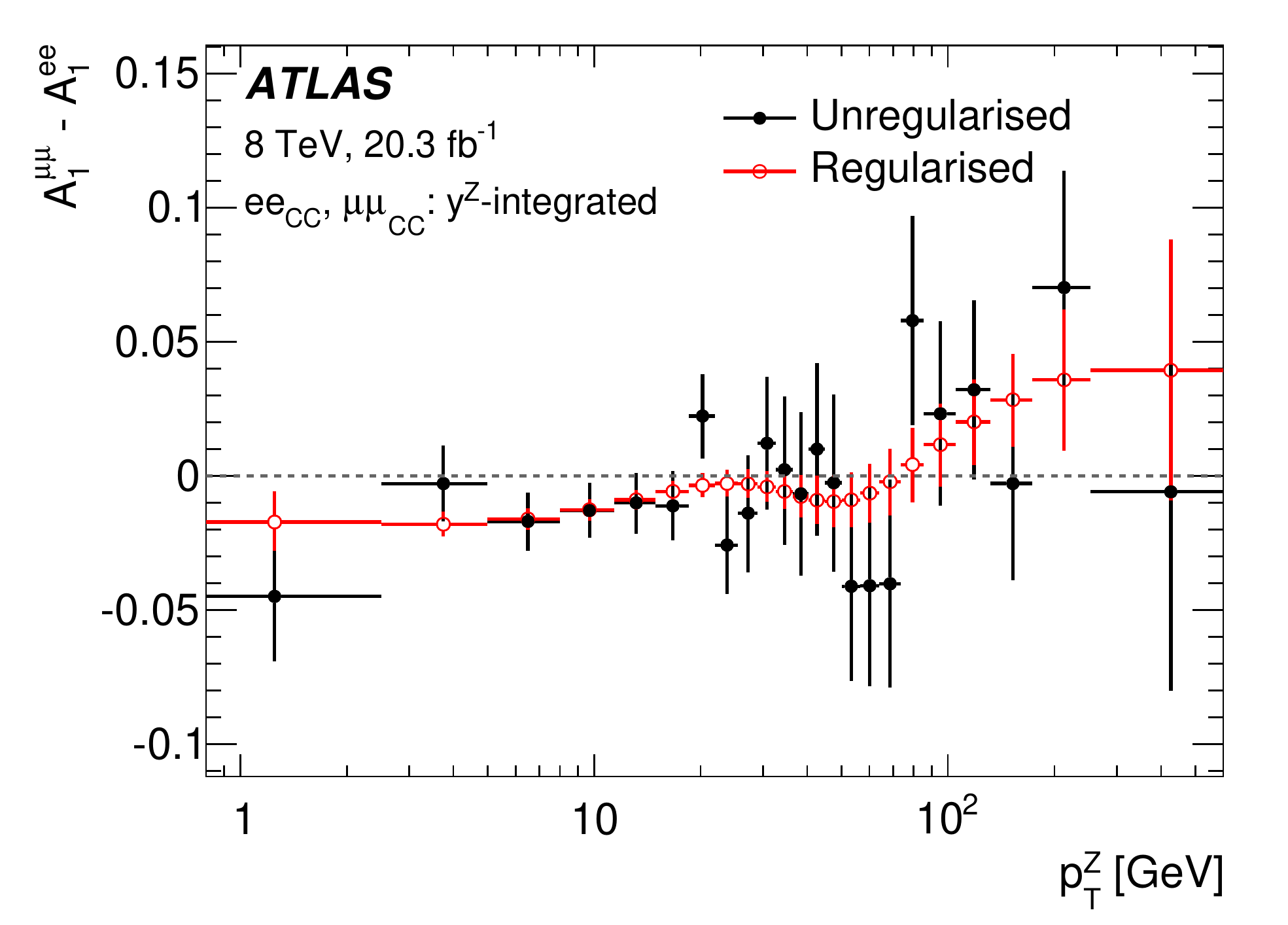}
    \includegraphics[width=7.3cm,angle=0]{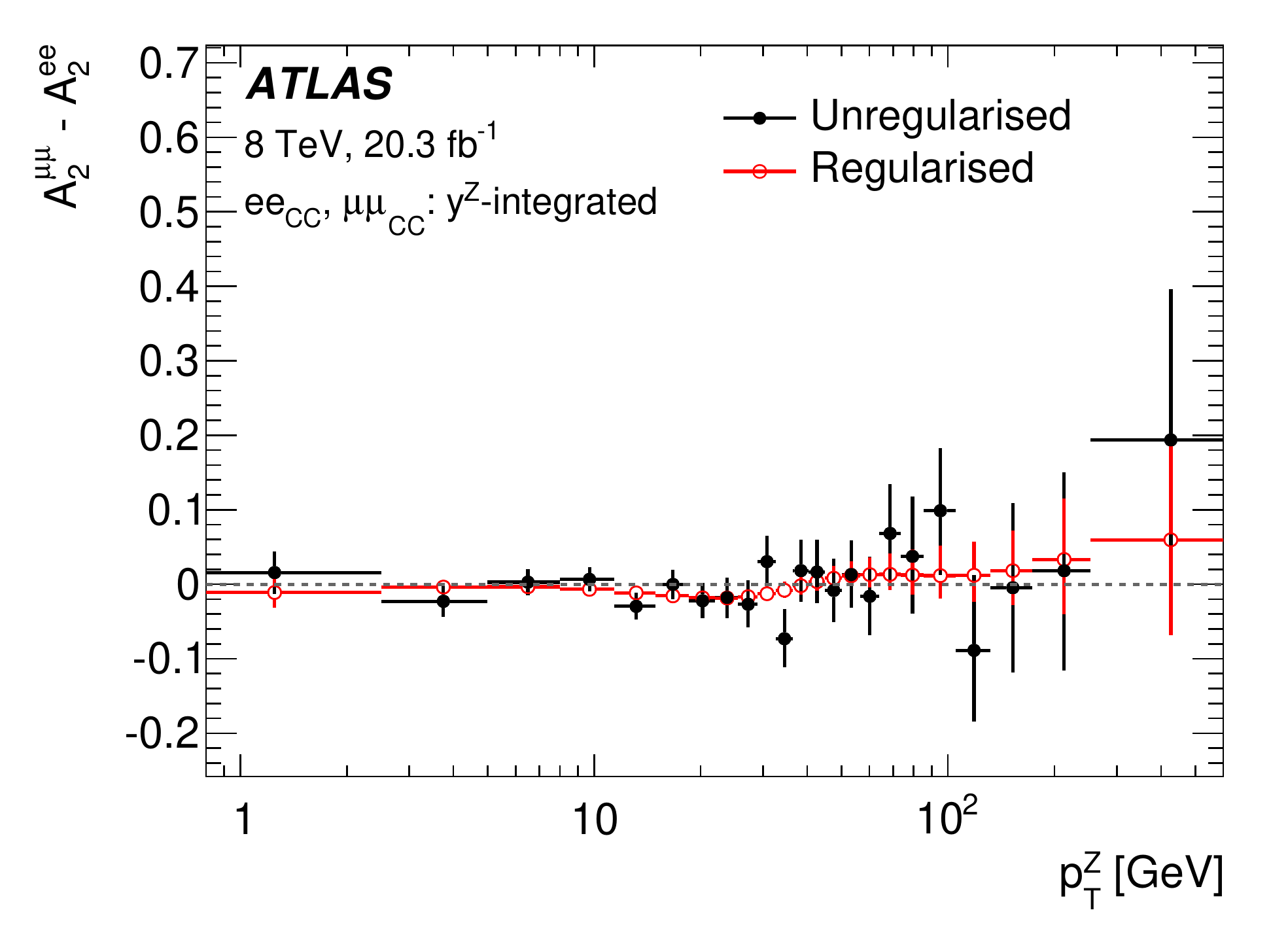}
    \includegraphics[width=7.3cm,angle=0]{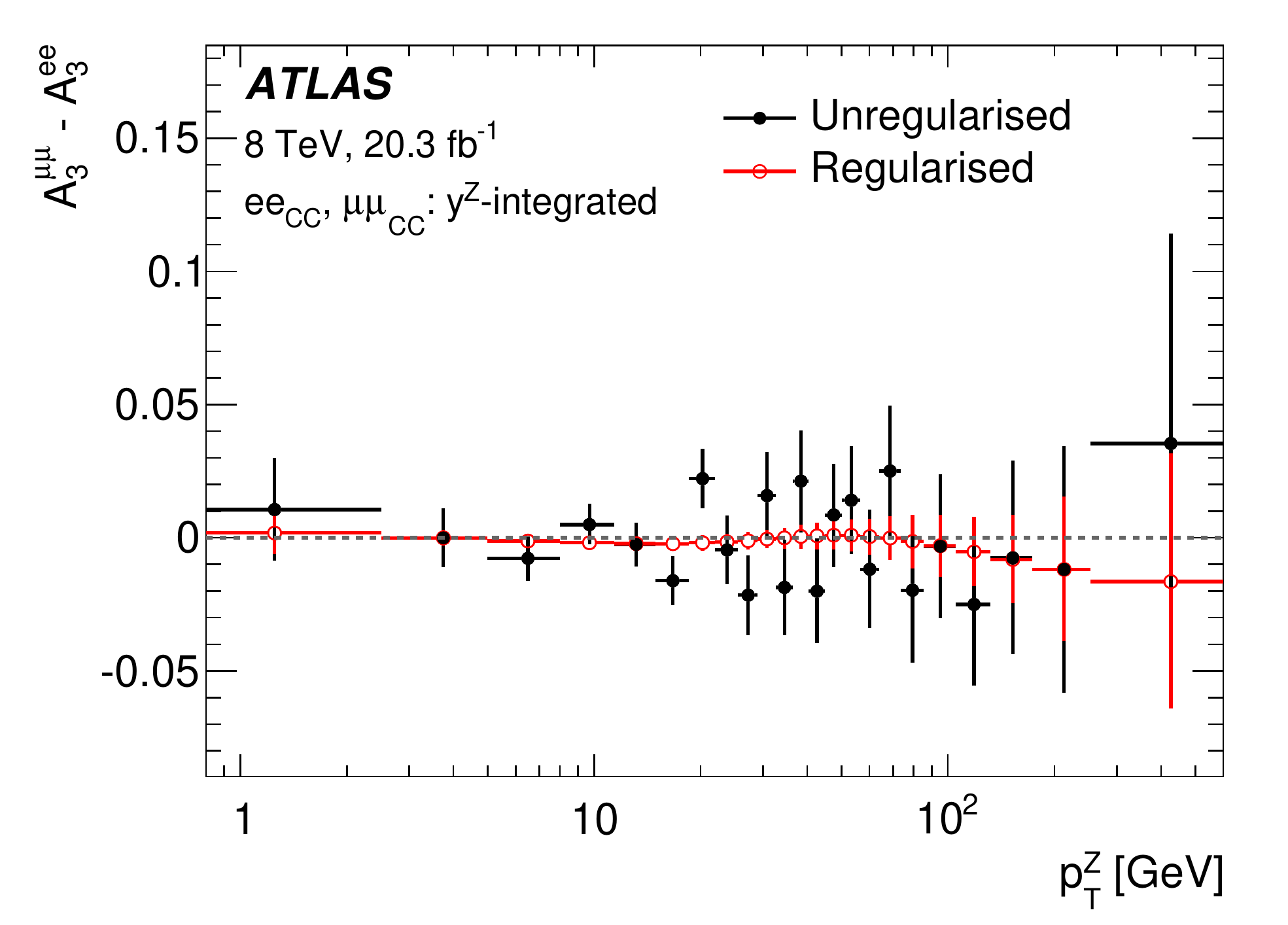}
    \includegraphics[width=7.3cm,angle=0]{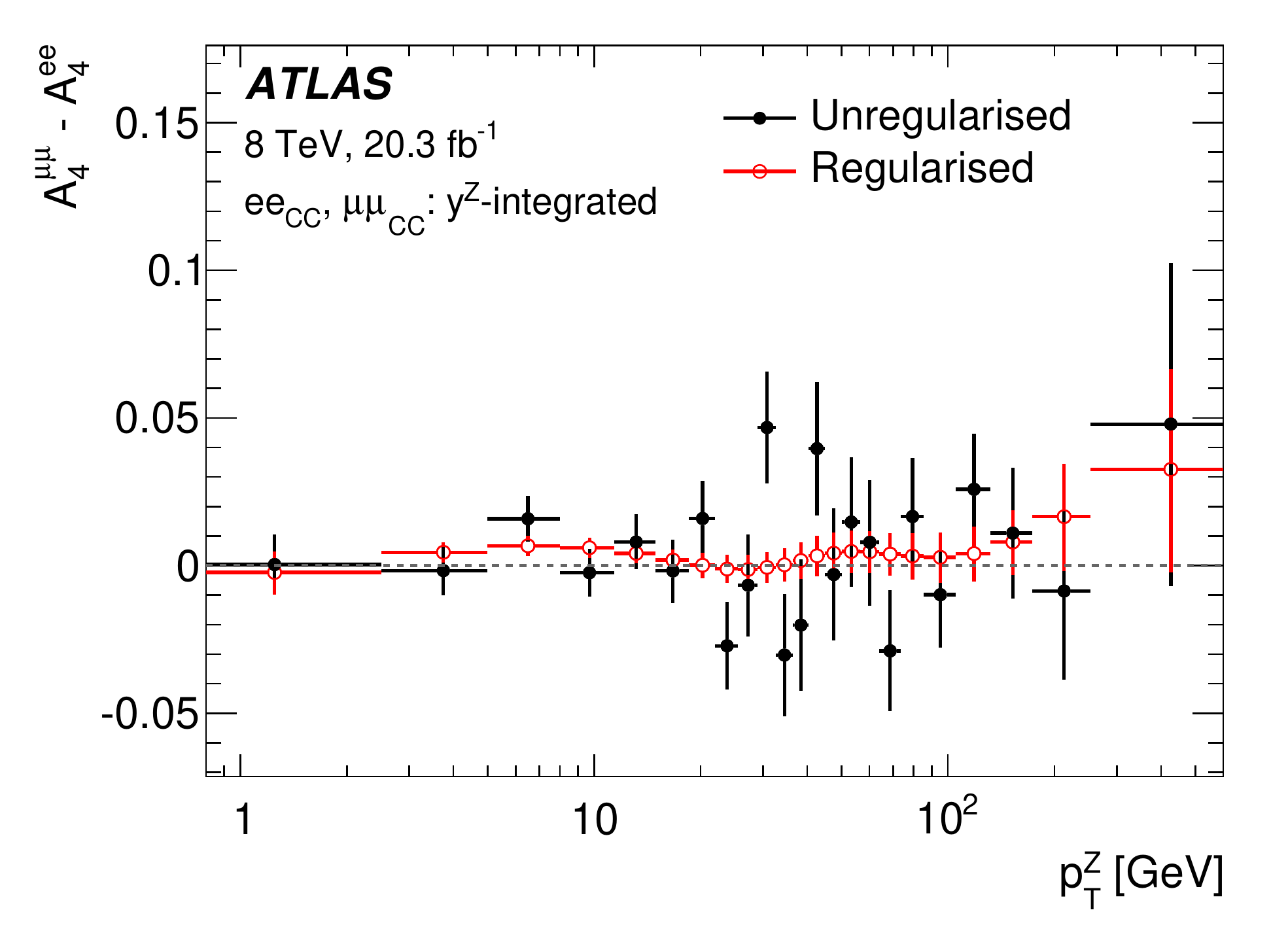}
    \includegraphics[width=7.3cm,angle=0]{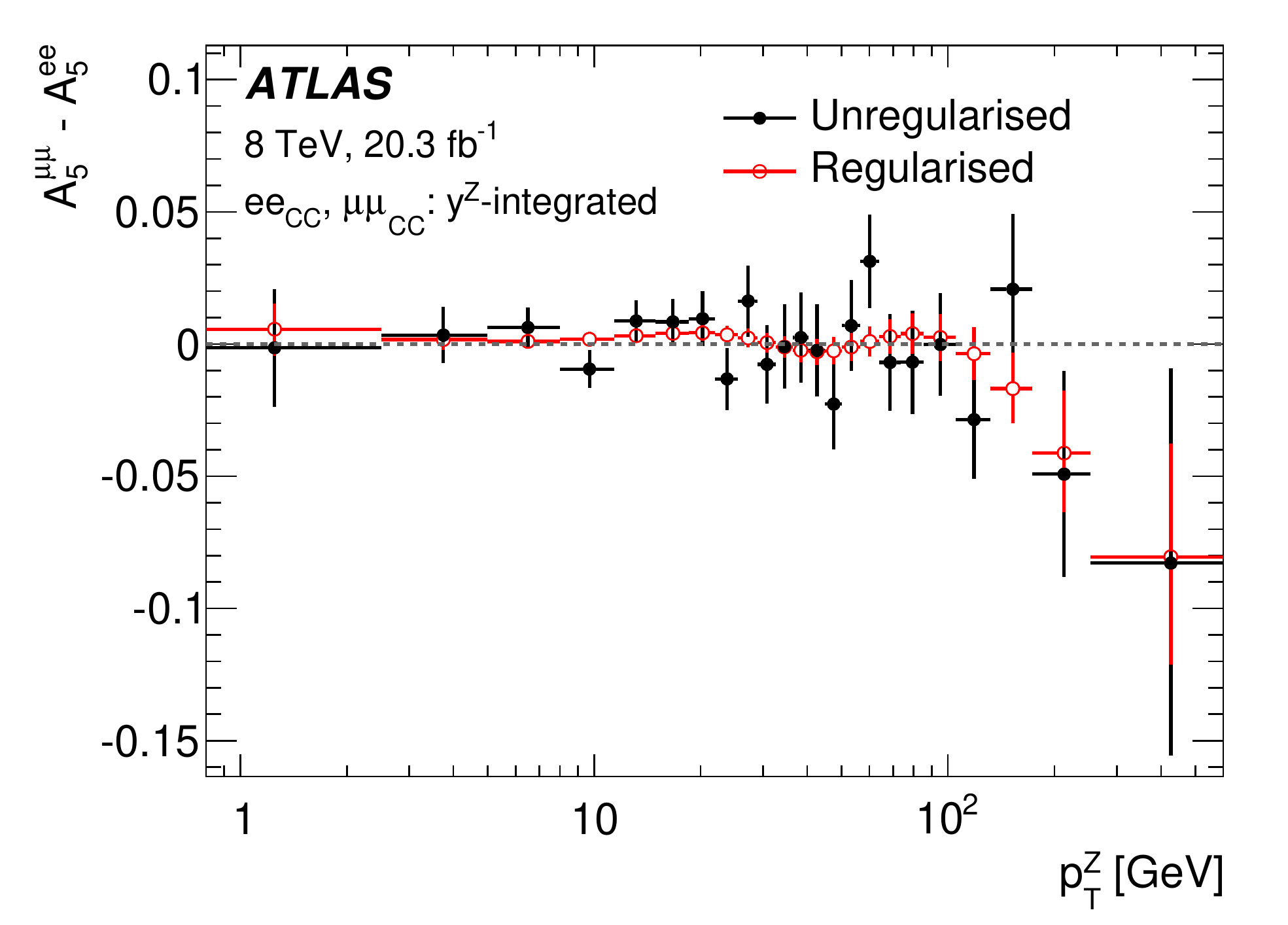}
    \includegraphics[width=7.3cm,angle=0]{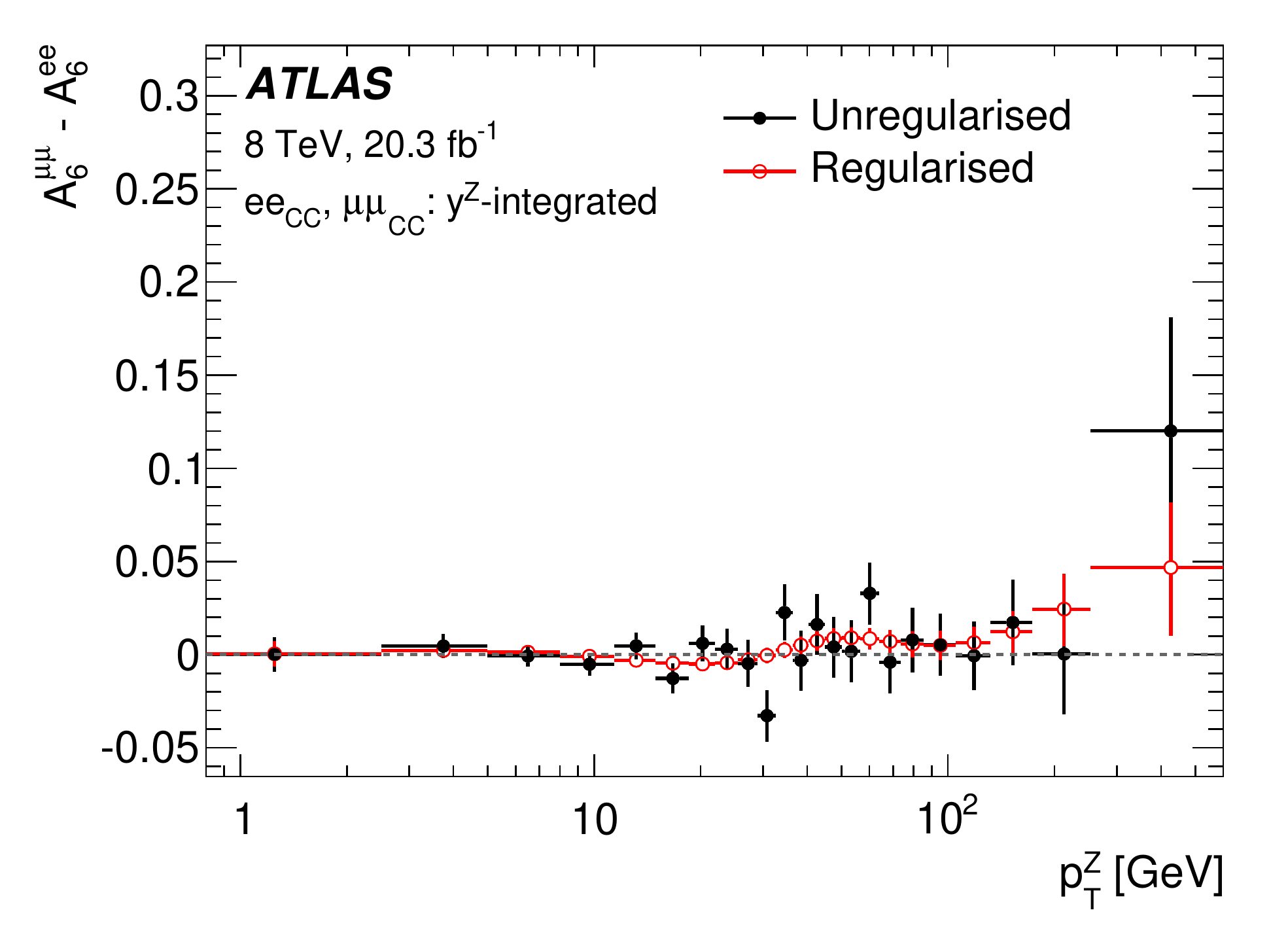}
    \includegraphics[width=7.3cm,angle=0]{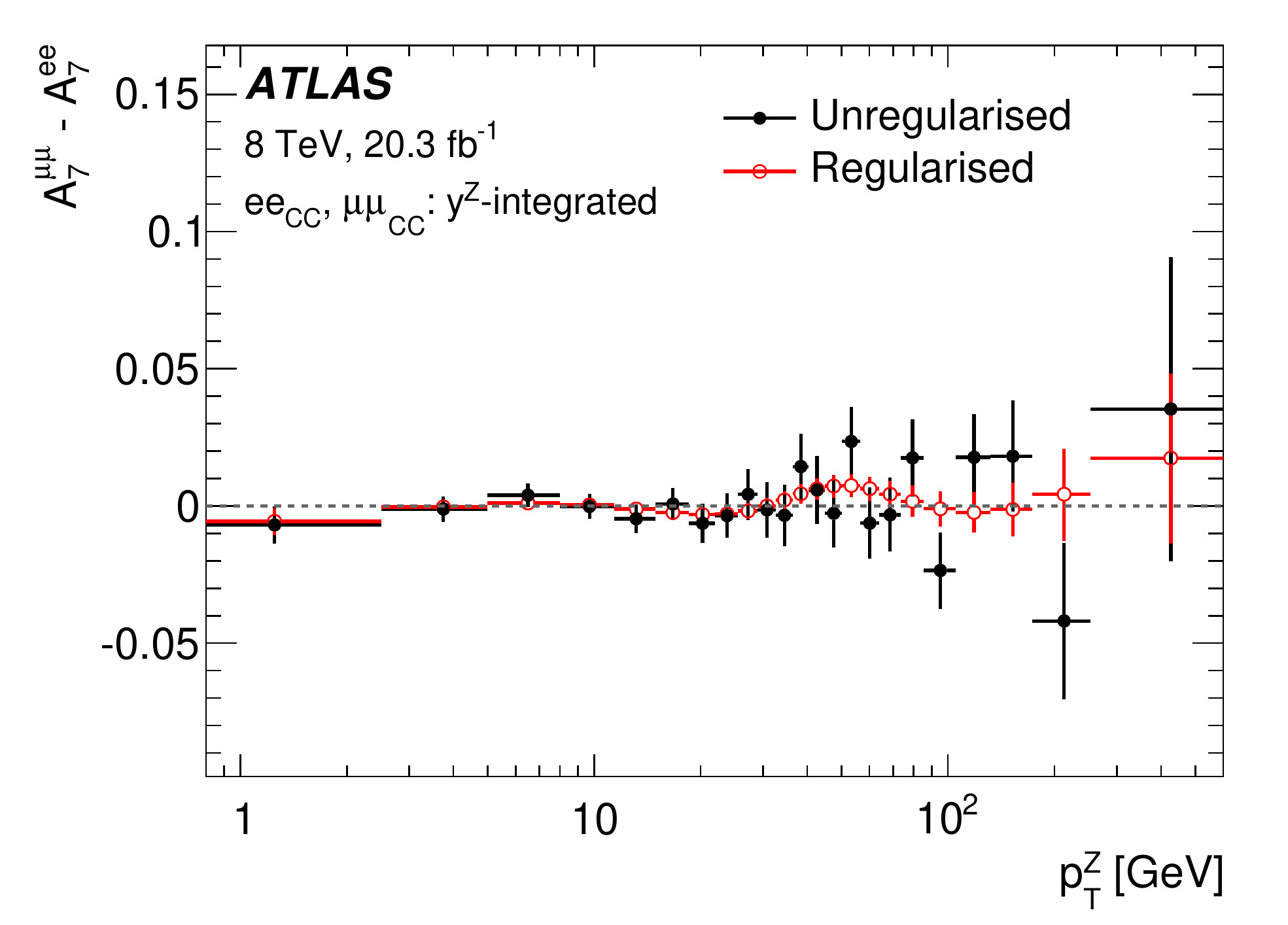}
}
\end{center}
\caption{Differences between the measured angular coefficients in the $\mu\mu_{\text{CC}}$ and $ee_{\text{CC}}$ channels, shown as a function of~\ptz\ from top left to bottom right, for all measured coefficients in the $\yz$-integrated configuration. The full (open) circles represent the measured differences before (after) regularisation. The error bars represent the total uncertainty in the measurements.
\label{Fig:DiffeeCC-mumuCC}}
\end{figure}

\subsection{Results in the individual and combined channels}
\label{sec:baseline_results}

The measurements represent the full set of $\yz$-integrated coefficients, including the difference $A_{0}-A_{2}$, as a function of $\ptz$, as well as the $\yz$-dependent coefficients as a function of $\ptz$ in the available $\yz$ bins. The combination of the $ee_{\text{CC}}$ and $\mu\mu_{\text{CC}}$ channels is used for the $\yz$-integrated measurements and the measurements in the first two $\yz$ bins, while the $ee_{\text{CF}}$ channel is used for the measurements in the last $\yz$ bin. A summary of these measurements is tabulated in Tables~\ref{Tab:test}--\ref{Tab:test2} for three representative $\ptz$ bins. Figure~\ref{Fig:measuredAi} shows the $\yz$-integrated measurements for all $\Ai$ and overlays of the $\yz$-dependent $\Ai$ in each accessible $\yz$ bin. 
The $A_1$ and $A_6$ measurements are missing from the third $\yz$ bin since they are inaccessible in the projections used in the $ee_{\text{CF}}$ channel (see Section~\ref{sec:likelihood}). Also, a measurement of $A_0-A_2$ is missing in this bin since $A_0$ and $A_2$ are accessible in different projections. Complete tables can be found in Appendix~\ref{sec:additional_results} along with additional figures in~$\yz$ bins. Similarly to the regularised $\Delta A_{ij}$ measurements, there is a large degree of correlation from bin to bin. This, coupled with statistical fluctuations, can lead to correlated deviations in the spectra, for example near $\ptz~=~40$~GeV for $A_{4}$ in the $2<|\yz|<3.5$ bin, and for $A_1$ in the $0<|\yz|<1$ bin. Visually, the coefficients $A_{5,6,7}$ all show a trend towards non-zero positive values in the region with $\ptz$ around~$100$~GeV.

\begin{figure}
  \begin{center}
{
    \includegraphics[width=7.5cm,angle=0]{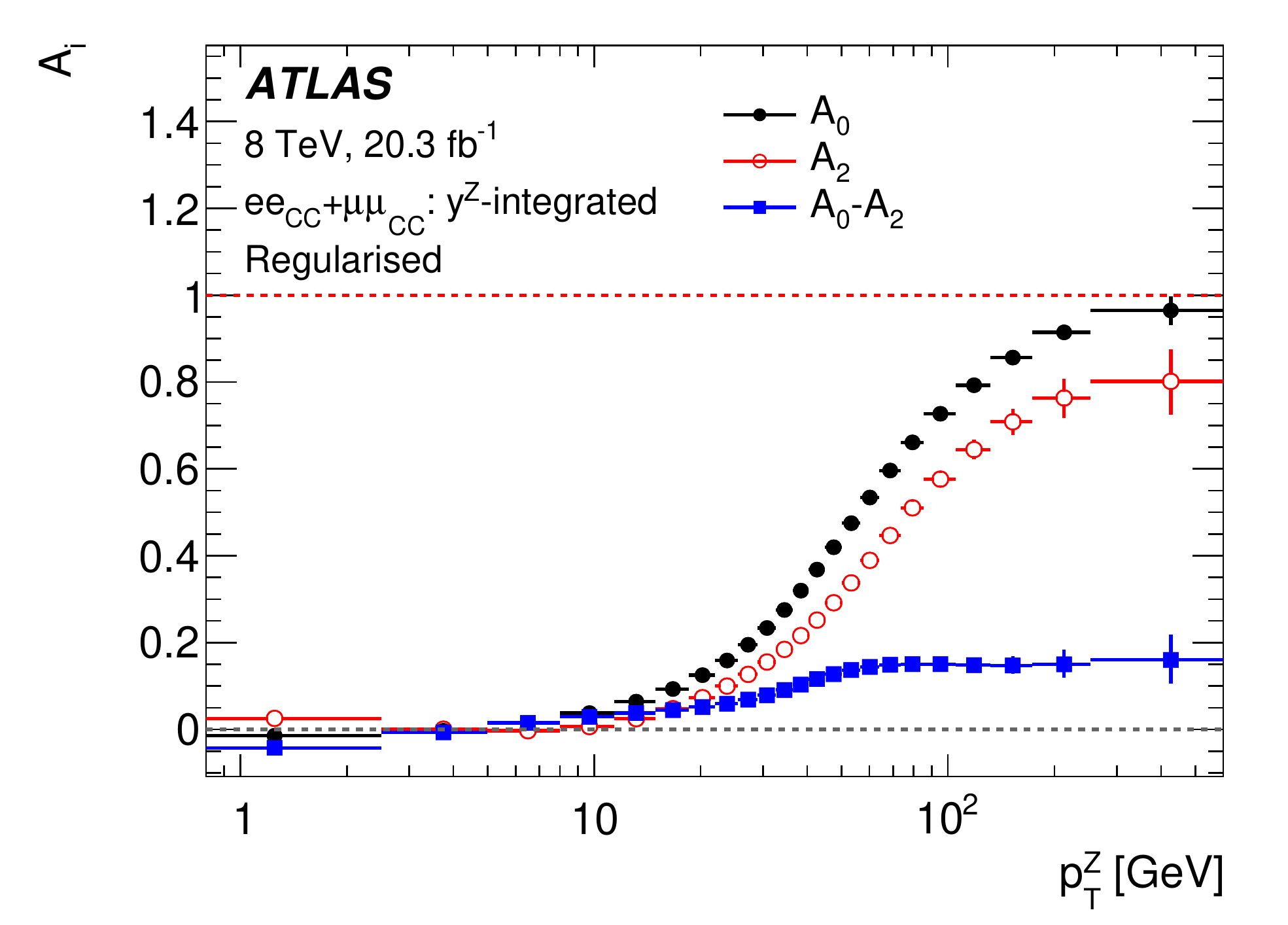}
    \includegraphics[width=7.5cm,angle=0]{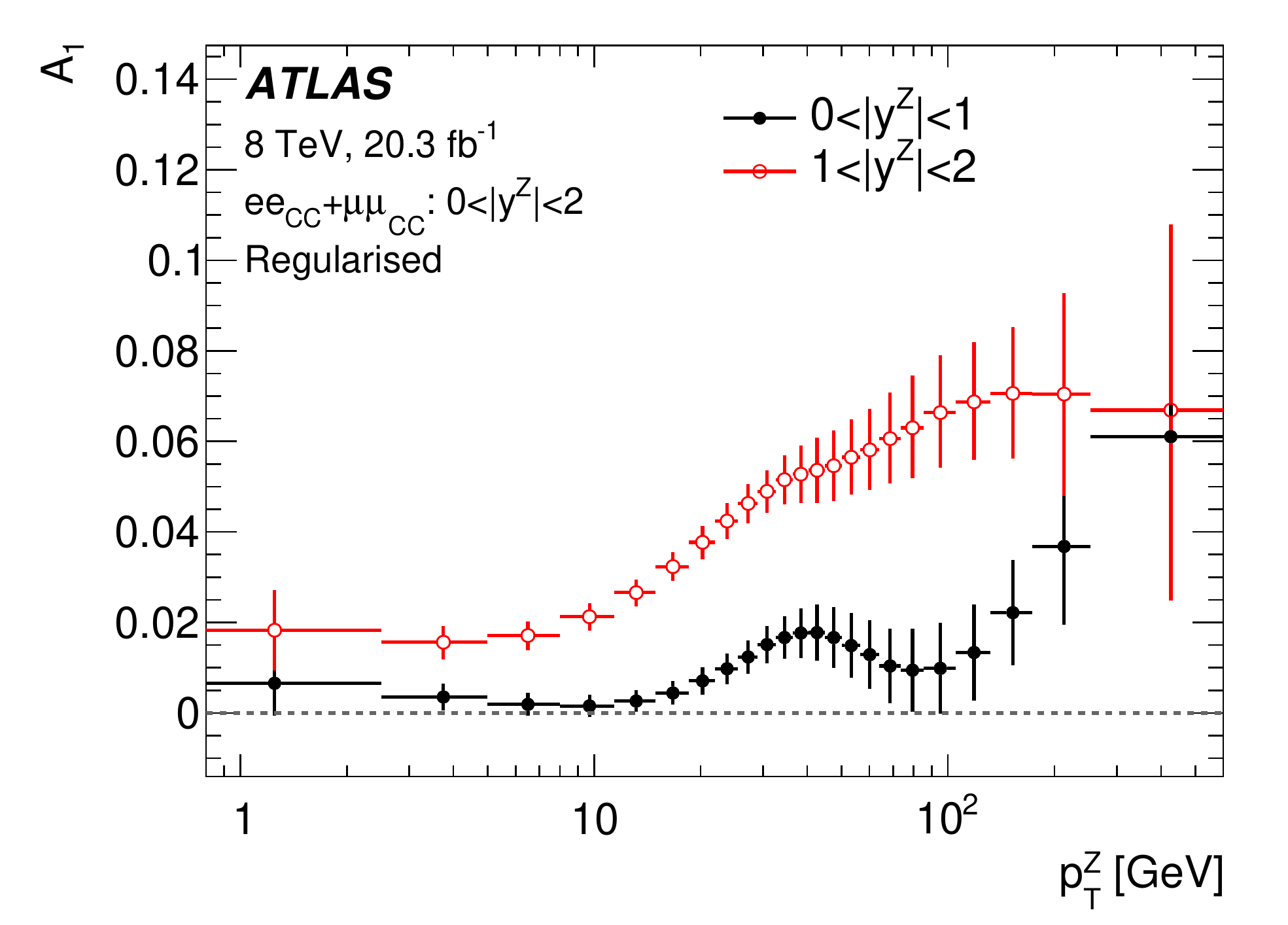}
    \includegraphics[width=7.5cm,angle=0]{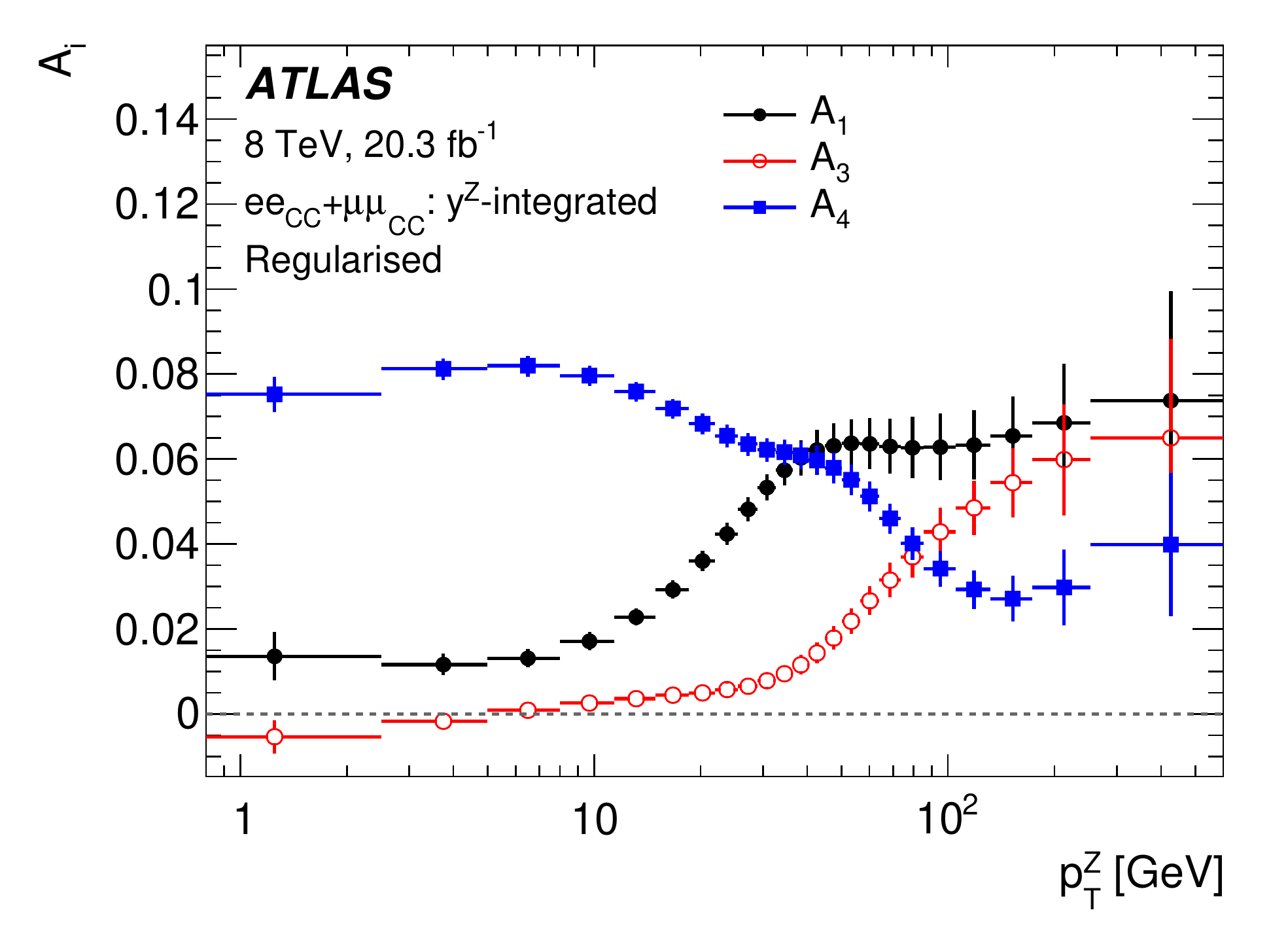}
    \includegraphics[width=7.5cm,angle=0]{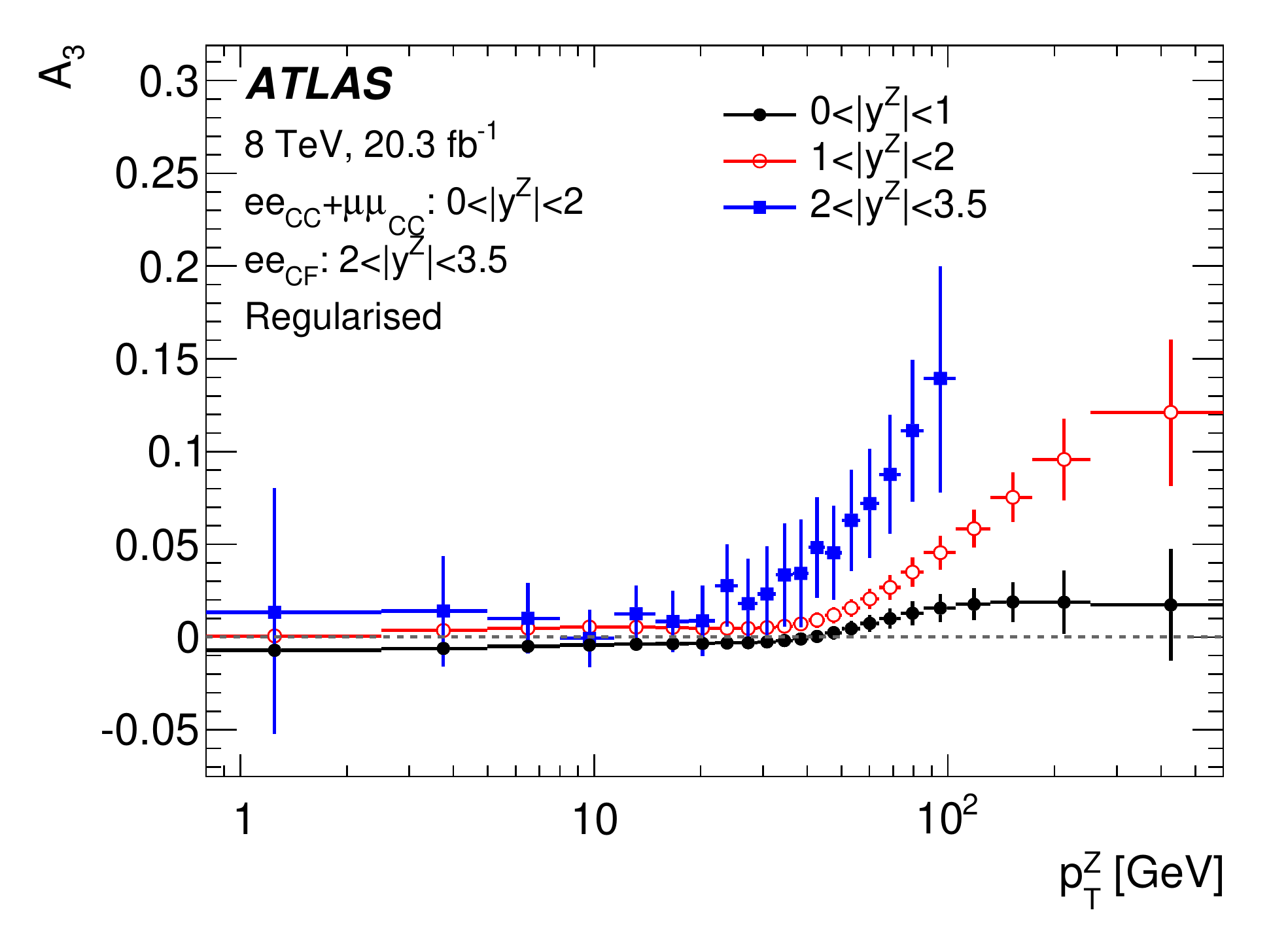}
    \includegraphics[width=7.5cm,angle=0]{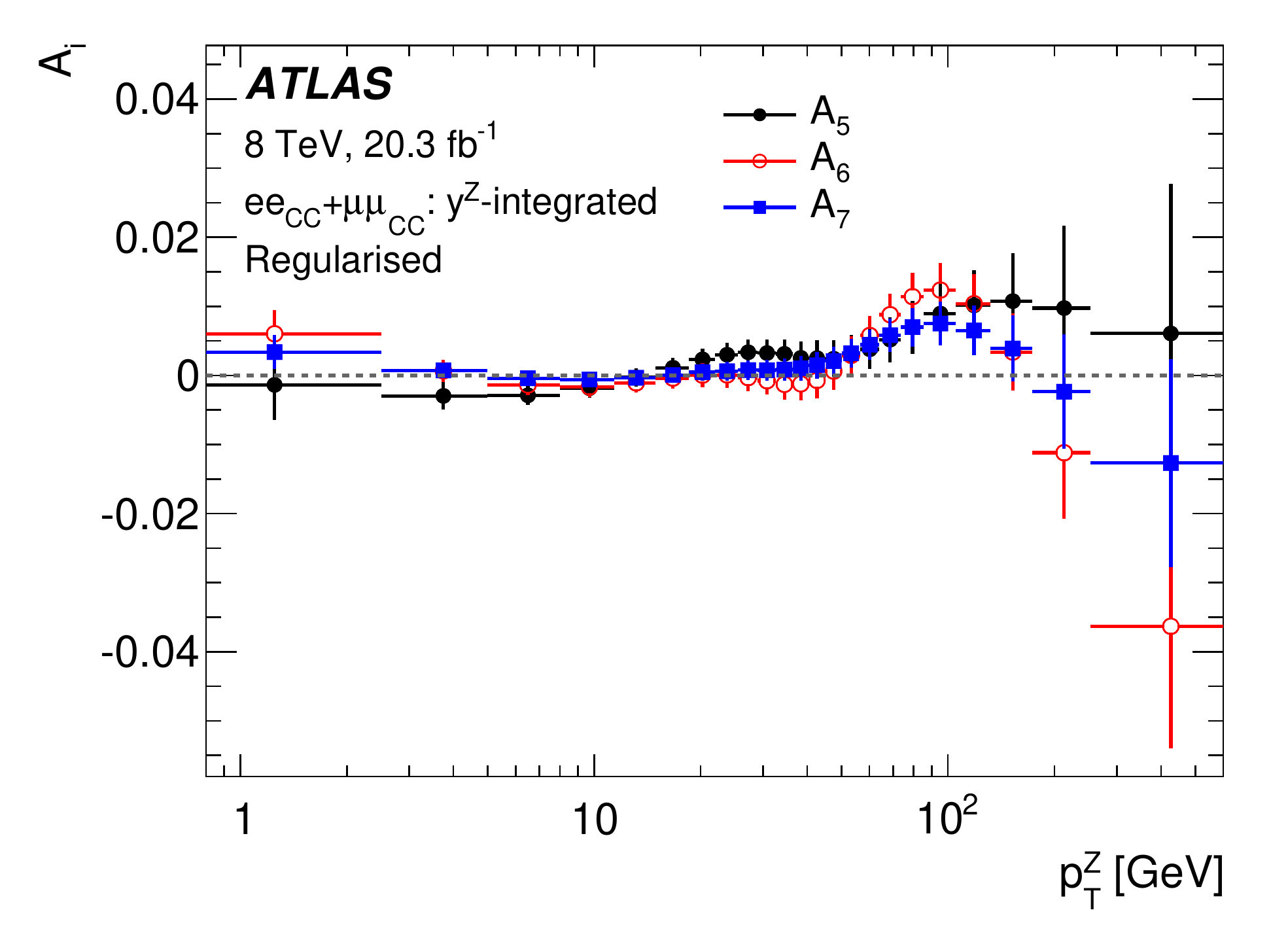}
    \includegraphics[width=7.5cm,angle=0]{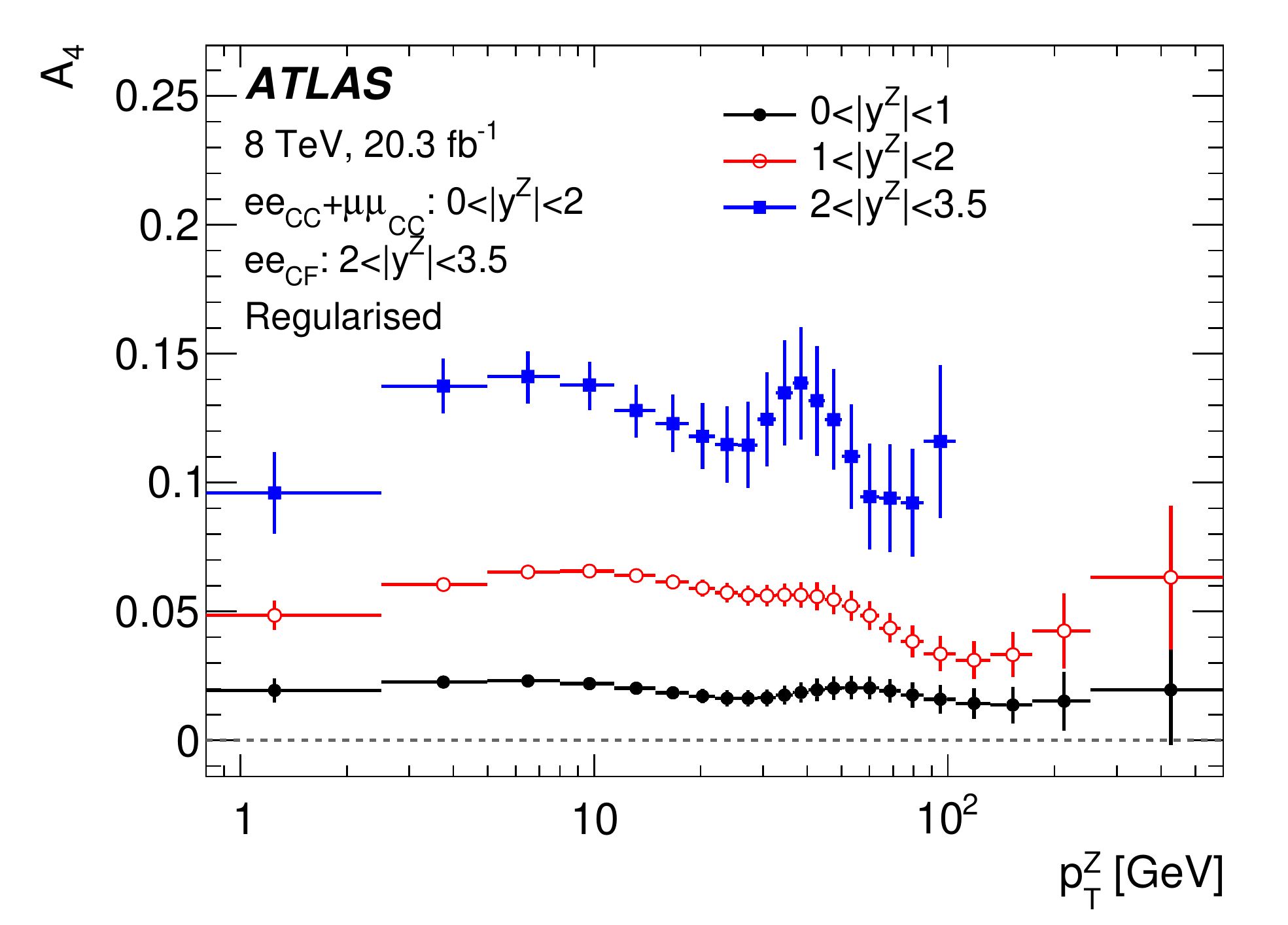}
}
\end{center}
\caption{Measurements of the angular coefficients in the $\yz$-integrated and $\yz$-binned configurations versus $\ptz$. Among the $\yz$-integrated configurations, are shown $A_{0,2}$ and $A_{0}-A_{2}$ (top left), $A_{1,3,4}$ (middle left), and $A_{5,6,7}$ (bottom left). The $\yz$-binned $\Ai$ are overlayed in each accessible $\yz$ bin for $A_{1}$ (top right), $A_{3}$ (middle right), and $A_{4}$ (bottom right). The error bars represent the total uncertainty in the measurements. 
\label{Fig:measuredAi}}
\end{figure}

\subsection{Cross-checks}
\label{sec:cross-checks}

Several cross-checks are performed to ensure that the fit is of good quality and that the underlying theoretical assumptions are valid to the extent of the precision of the analysis. 

The signal MC distributions are reweighted to the full set of measured parameters. An event-by-event weight is calculated as a ratio using the right-hand side of~Eq.~(\ref{Eq:master2}): the numerator uses the measured parameters, and the denominator uses the reference values in the MC~simulation. Distributions are obtained after applying this reweighting; the $\costhetacs$ and $\phics$ distributions integrated in $\yz$ are shown in Fig.~\ref{Fig:rw_dists}, along with their bin-by-bin pulls, obtained by combining the statistical and systematic uncertainties. Overall, the data and MC~simulation agree well. One observes significant pulls near~$\costhetacs=0$ for the $ee_{\text{CF}}$~channel, but the number of events in this region is very small and its impact on the coefficient measurements is negligible.

\begin{figure}
  \begin{center}
{
    \includegraphics[width=7.5cm,angle=0]{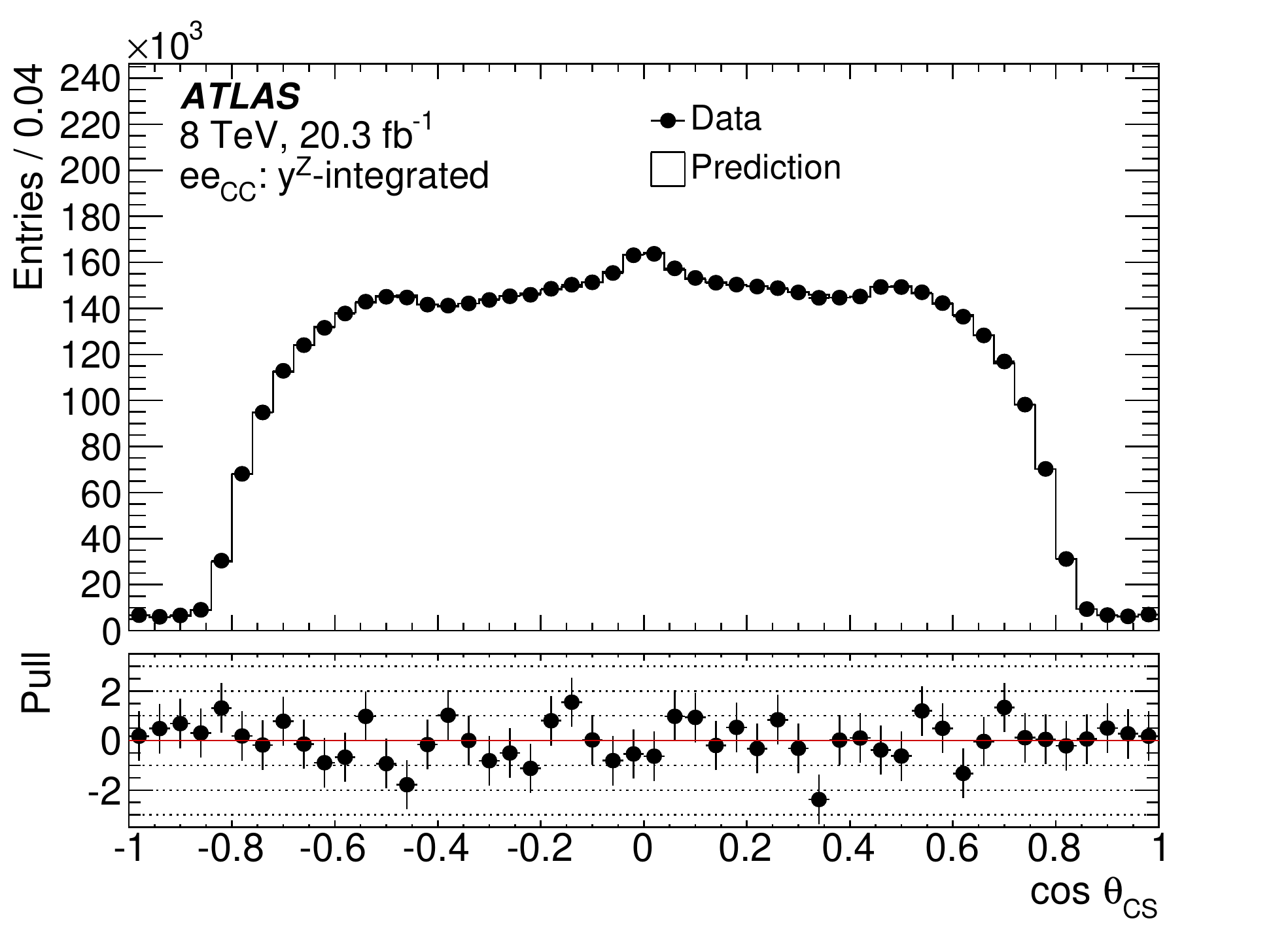}
    \includegraphics[width=7.5cm,angle=0]{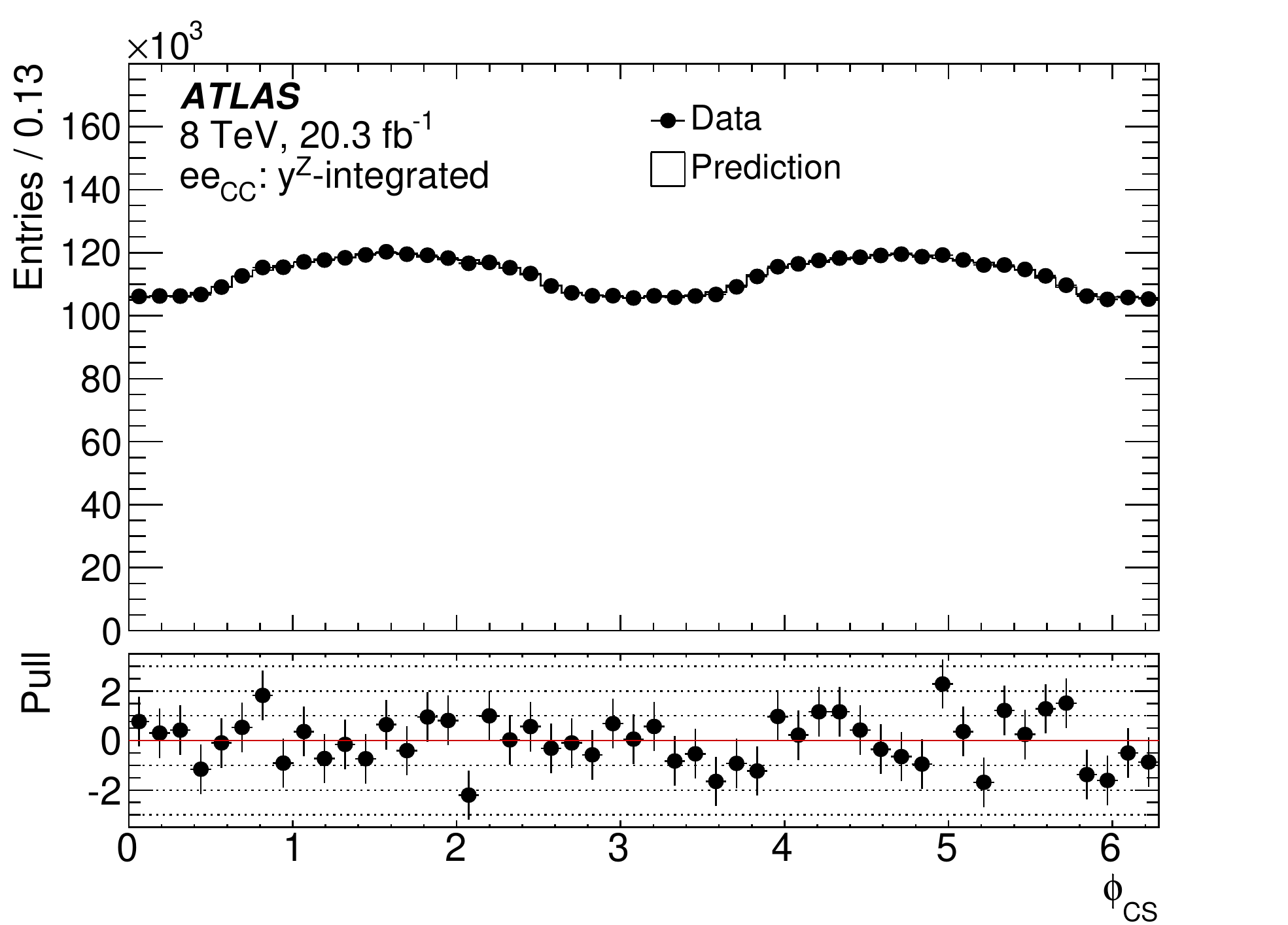}

    \includegraphics[width=7.5cm,angle=0]{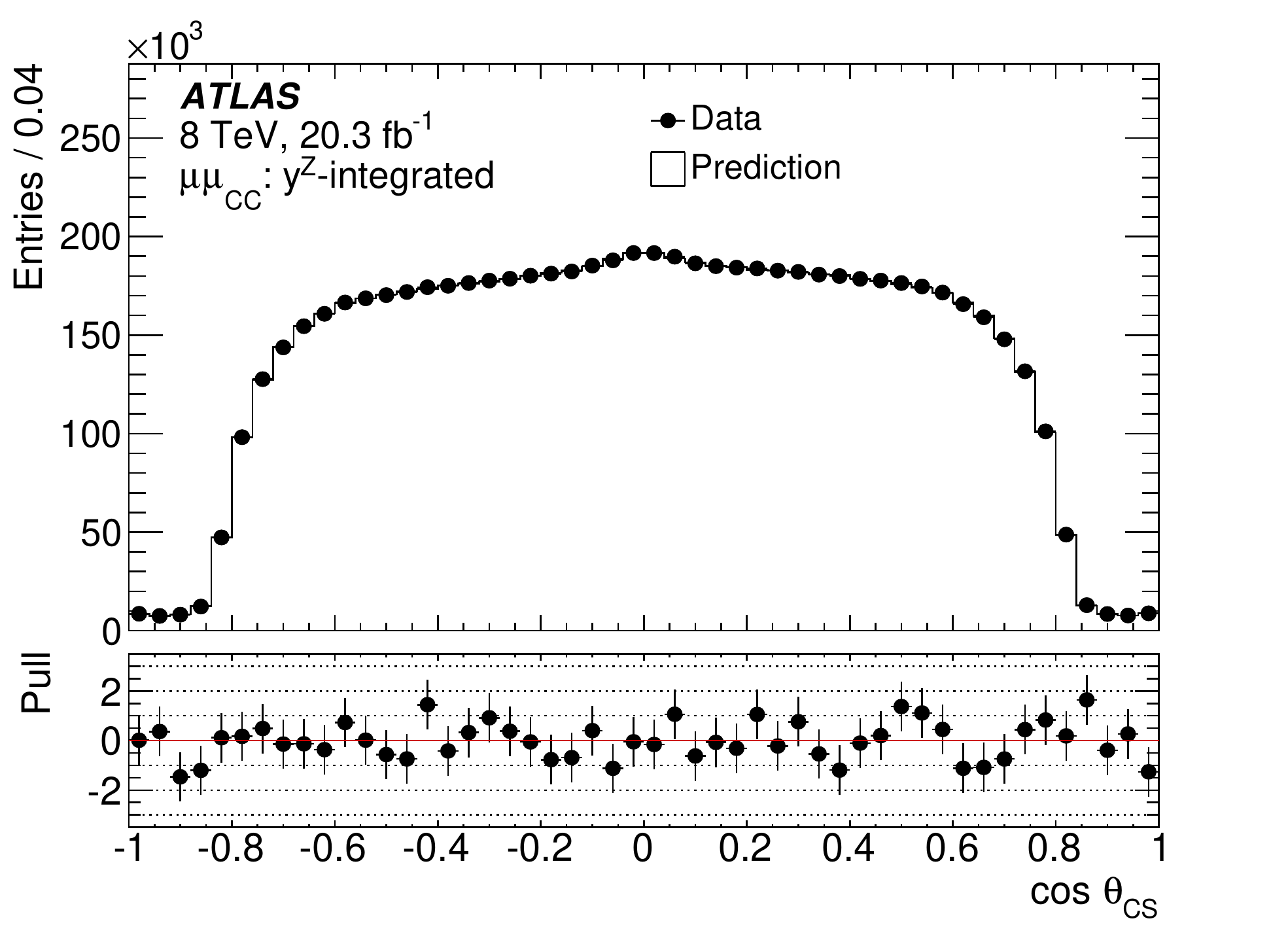}
    \includegraphics[width=7.5cm,angle=0]{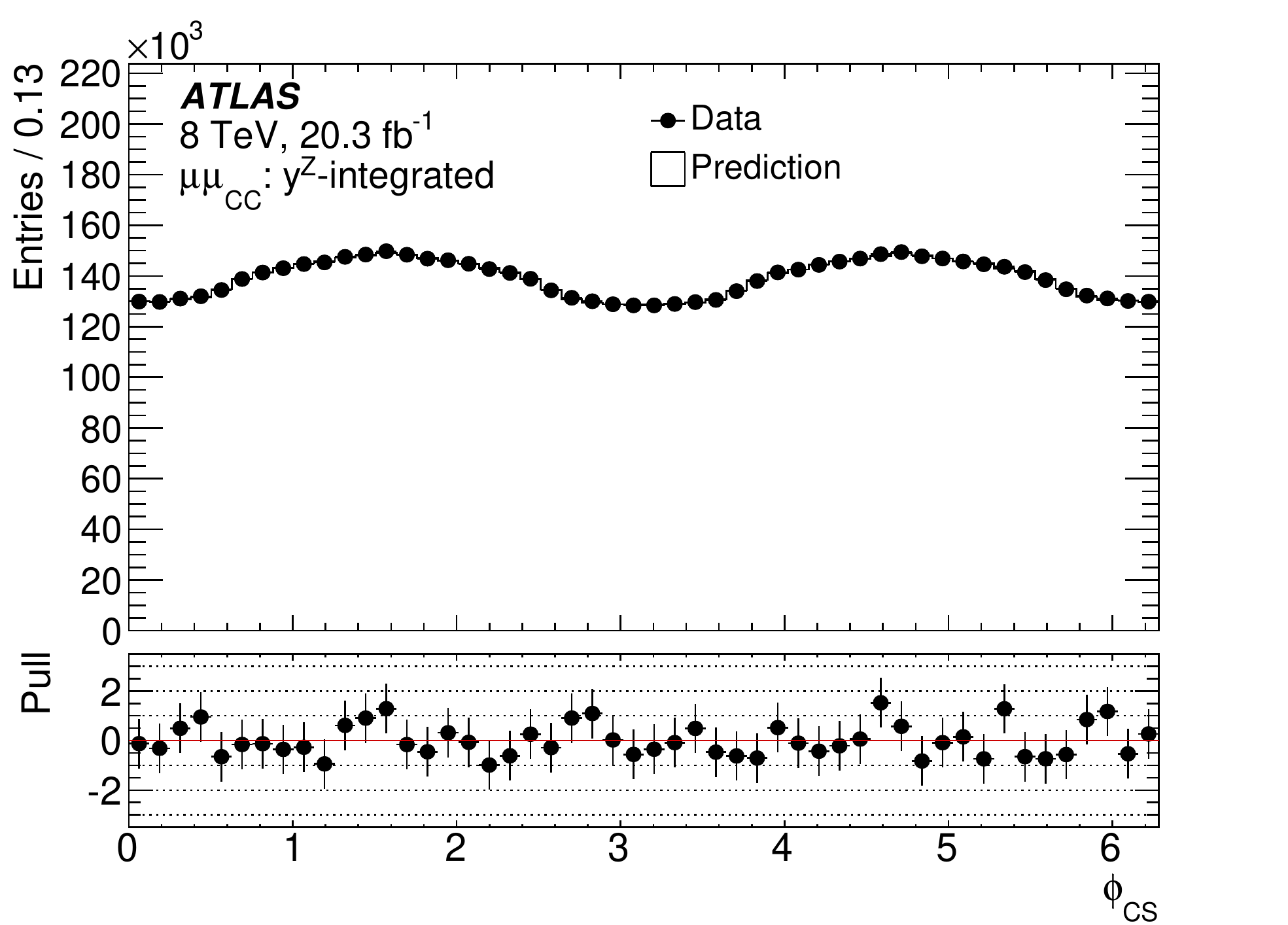}

    \includegraphics[width=7.5cm,angle=0]{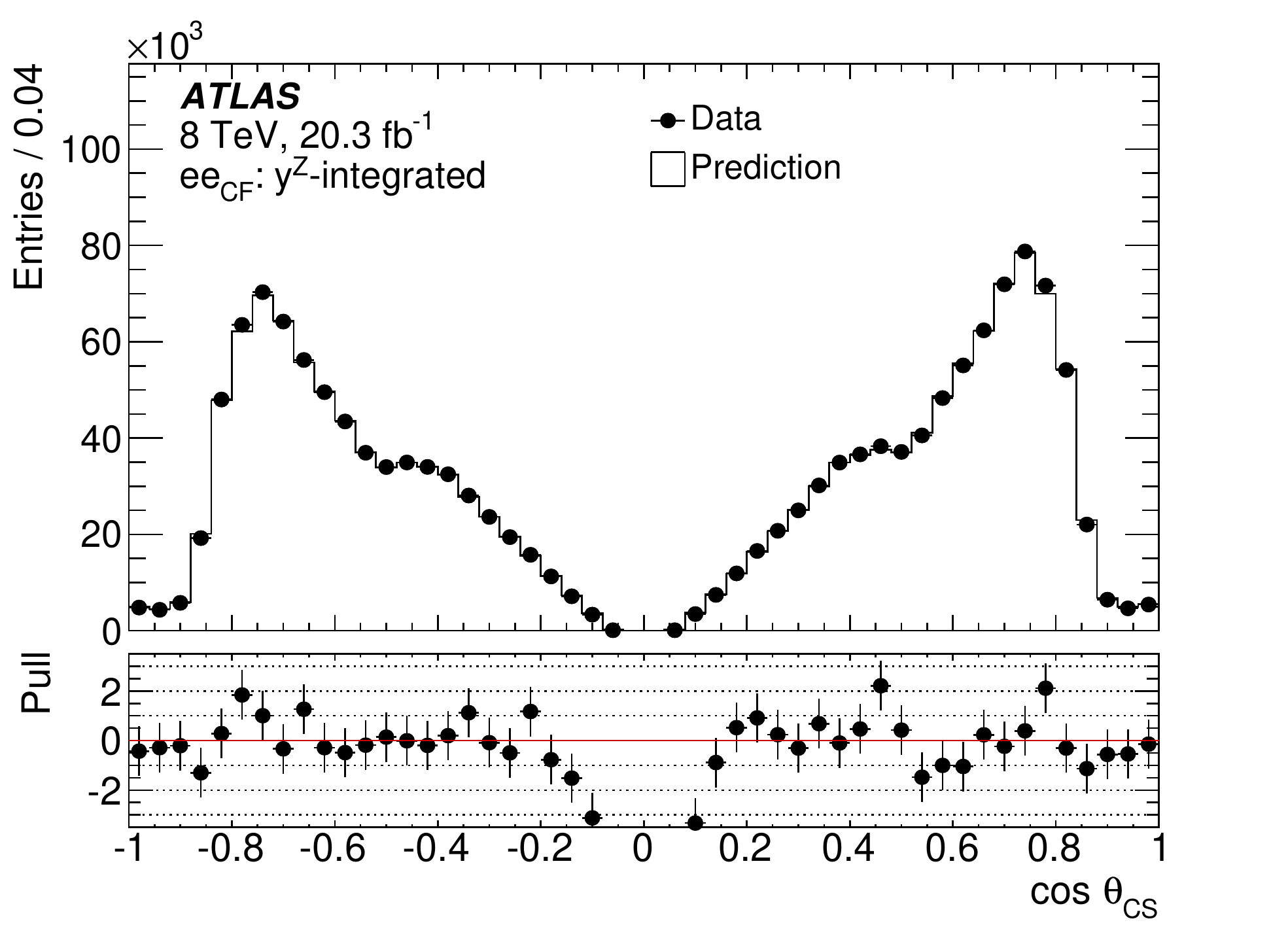}
    \includegraphics[width=7.5cm,angle=0]{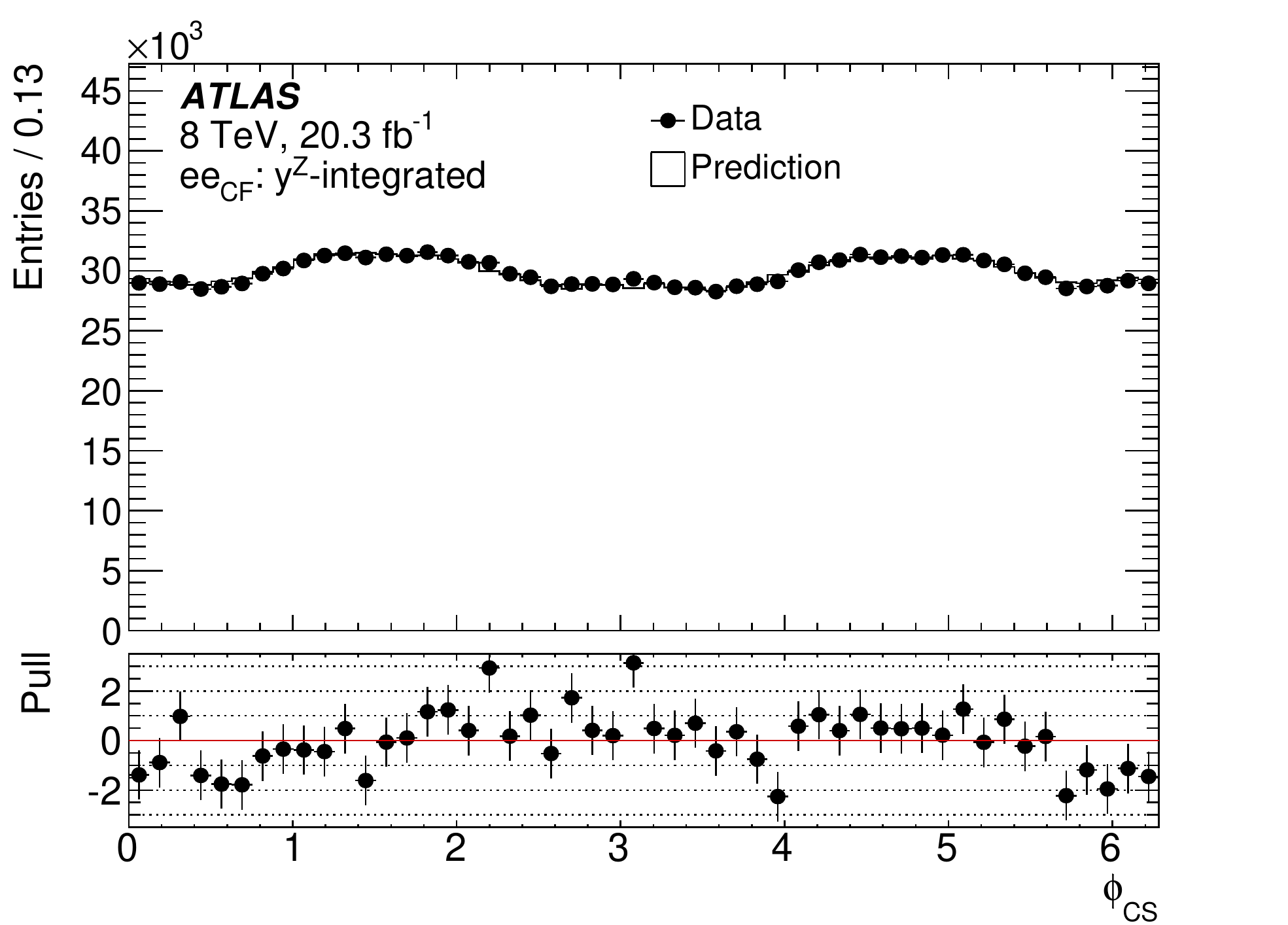}
}
\end{center}
\caption{ Reweighted $\costhetacs$ (left) and $\phics$ (right) distribution integrated over $\yz$ in the $ee_{\text{CC}}$ (top), $\mu\mu_{\text{CC}}$ (middle), and $ee_{\text{CF}}$ (bottom) channels and the corresponding pulls of the distributions after reweighting them predictions to data. The pulls are computed using the full statistical and systematic uncertainties. Two points in the bottom-left pull plot near $\costhetacs=0$ fall below the range shown, but the number of events in these two bins is very small.
\label{Fig:rw_dists}}
\end{figure}

After reweighting the signal MC~events to the measured parameters, the global fit quality is evaluated by computing the $\chi^{2}$ of the data with respect to the sum of expected events in each bin used in the likelihood fit. This test takes into account data statistical and MC~statistical uncertainties, but not other systematic effects. The resulting $\chi^{2}$~values for all channels are consistent with expectations across all $\yz$~configurations.

\begin{table}
\caption{ Summary of the measured coefficients in the $ee_{\text{CC}}+\mu\mu_{\text{CC}}$ $\yz$-integrated channel at low (5--8~GeV), mid (22--25.5~GeV), and high (132--173~GeV) $\ptz$ \label{Tab:test}. The uncertainties are given as $\pm \mathrm{(stat.)} \pm \mathrm{(syst.)}$.}
\begin{center}
\begin{tabular}{l||c|c|c}
\hline
\multicolumn{4}{c}{|$\yz$|-integrated measurements}\\
\hline
 $\ptz$ range [GeV] & $A_{0}$ & $A_{2}$ & $A_{0}-A_{2}$ \\
\hline
 5.0--8.0   & 0.015 $\pm$ 0.002 $\pm$ 0.007 & --0.003 $\pm$ 0.003 $\pm$ 0.003 & 0.018 $\pm$ 0.003 $\pm$ 0.007 \\
 22.0--25.5 & 0.159 $\pm$ 0.003 $\pm$ 0.007 &  0.100 $\pm$ 0.003 $\pm$ 0.003 & 0.059 $\pm$ 0.005 $\pm$ 0.006 \\
 132--173   & 0.856 $\pm$ 0.008 $\pm$ 0.008 &  0.708 $\pm$ 0.022 $\pm$ 0.020 & 0.148 $\pm$ 0.019 $\pm$ 0.011 \\
 \hline
\hline
 $\ptz$ range [GeV] & $A_{1}$ & $A_{3}$ & $A_{4}$ \\
\hline
 5.0--8.0   & 0.013 $\pm$ 0.002 $\pm$ 0.002 & 0.001 $\pm$ 0.001 $\pm$ 0.001 & 0.082 $\pm$ 0.001 $\pm$ 0.002 \\
 22.0--25.5 & 0.042 $\pm$ 0.002 $\pm$ 0.002 & 0.006 $\pm$ 0.001 $\pm$ 0.001 & 0.065 $\pm$ 0.002 $\pm$ 0.002 \\
 132--173   & 0.065 $\pm$ 0.008 $\pm$ 0.005 & 0.054 $\pm$ 0.007 $\pm$ 0.003 & 0.027 $\pm$ 0.005 $\pm$ 0.002 \\
 \hline
\hline
 $\ptz$ range [GeV] & $A_{5}$ & $A_{6}$ & $A_{7}$ \\
\hline
 5.0--8.0   & --0.002 $\pm$ 0.001 $\pm$ 0.001 & --0.001 $\pm$ 0.001 $\pm$ 0.001 & 0.000 $\pm$ 0.001 $\pm$ 0.001 \\
 22.0--25.5 &  0.003 $\pm$ 0.002 $\pm$ 0.001 &  0.000 $\pm$ 0.002 $\pm$ 0.001 & 0.001 $\pm$ 0.001 $\pm$ 0.001 \\
 132--173   &  0.011 $\pm$ 0.006 $\pm$ 0.003 &  0.003 $\pm$ 0.005 $\pm$ 0.002 & 0.004 $\pm$ 0.004 $\pm$ 0.002 \\
 \hline
\end{tabular}
\end{center}
\end{table}
\begin{table}
\caption{ Summary of the measured coefficients in the $ee_{\text{CC}}+\mu\mu_{\text{CC}}$ channel for the two bins $0<|\yz|<1$ and $1<|\yz|<2$ and in the $ee_{\text{CF}}$ channel for the $2<|\yz|<3.5$ bin at low (5--8~GeV), mid (22--25.5~GeV), and high (132--173~GeV) $\ptz$ \label{Tab:test2}. The uncertainties are given as $\pm \mathrm{(stat.)} \pm \mathrm{(syst.)}$.}
\begin{center}
\begin{tabular}{l||c|c|c}
\hline
\multicolumn{4}{c}{|$\yz$|-binned measurements}\\
\hline
\multicolumn{4}{c}{$A_{1}$}\\
\hline
 $\ptz$ range [GeV] & $0<|y^{Z}|<1$ & $1<|y^{Z}|<2$ & $2<|y^{Z}|<3.5$ \\
\hline
 5.0--8.0   & 0.002 $\pm$ 0.002 $\pm$ 0.001 & 0.017 $\pm$ 0.002 $\pm$ 0.002 &   \\
 22.0--25.5 & 0.010 $\pm$ 0.003 $\pm$ 0.002 & 0.042 $\pm$ 0.003 $\pm$ 0.002 &   \\
 132--173   & 0.022 $\pm$ 0.010 $\pm$ 0.006 & 0.071 $\pm$ 0.013 $\pm$ 0.007 &   \\
 \hline
\hline
\multicolumn{4}{c}{$A_{3}$}\\
\hline
 $\ptz$ range [GeV] & $0<|y^{Z}|<1$ & $1<|y^{Z}|<2$ & $2<|y^{Z}|<3.5$ \\
\hline
 5.0--8.0  & --0.005 $\pm$ 0.001 $\pm$ 0.001 & 0.005 $\pm$ 0.002 $\pm$ 0.001 &  0.005 $\pm$ 0.002 $\pm$ 0.001 \\
 22.0--25.5 & --0.003 $\pm$ 0.002 $\pm$ 0.001 & 0.005 $\pm$ 0.002 $\pm$ 0.001 &  0.005 $\pm$ 0.002 $\pm$ 0.001 \\
 132--173   &  0.019 $\pm$ 0.010 $\pm$ 0.004 & 0.075 $\pm$ 0.012 $\pm$ 0.006 &   \\
 \hline
\hline
\multicolumn{4}{c}{$A_{4}$}\\
\hline
 $\ptz$ range [GeV] & $0<|y^{Z}|<1$ & $1<|y^{Z}|<2$ & $2<|y^{Z}|<3.5$ \\
\hline
 5.0--8.0   & 0.023 $\pm$ 0.002 $\pm$ 0.001 & 0.065 $\pm$ 0.002 $\pm$ 0.001 &  0.065 $\pm$ 0.002 $\pm$ 0.001 \\
 22.0--25.5 & 0.016 $\pm$ 0.003 $\pm$ 0.001 & 0.057 $\pm$ 0.003 $\pm$ 0.002 &  0.057 $\pm$ 0.003 $\pm$ 0.002 \\
 132--173   & 0.014 $\pm$ 0.006 $\pm$ 0.003 & 0.033 $\pm$ 0.008 $\pm$ 0.004 &   \\
 \hline
\end{tabular}
\end{center}
\end{table}

The best-fit values of each nuisance parameter along with their post-fit constraints are checked. Most parameters have a fit value close to zero with a constraint close to unity. It was also checked that the regularisation procedure does not significantly change the best-fit value or post-fit constraint of the nuisance parameters.

Finally, the degree to which the data follow the nine-$\poli$ polynomial decomposition is tested by checking for the presence of higher-order~$\poli$ in the data. The original nine~$\poli$ are up to second order in spherical harmonics. The template-building methodology described in Section~\ref{sec:templates} is extended to have more than nine~$\poli$ by using third- and fourth-order spherical harmonics, corresponding to 16~additional~$\poli$. One additional $\poli$~template is fitted at a time. The higher-order coefficients are found to be compatible with zero using a $\chi^2$ test as in Section~\ref{sec:compatibility}, leading to the conclusion that any possible breaking of the nine~$\poli$ polynomial decomposition is beyond the sensitivity of the analysis.

\section{Comparisons with theory}
\label{sec:comparisons}

In this section, the measurements are compared to the most precise fixed-order calculations currently available. They probe the dynamics of perturbative~QCD, including the presence of higher-order corrections, and explore the effects from the $V-A$ structure of \Zboson-boson couplings. These comparisons are made with both the $\yz$-integrated and $\yz$-binned measurements. For the $\yz$-integrated measurements and for the $0 <|\yz| < 1$ and $1 < |\yz| < 2$ bins, the combined $ee_{\text{CC}}$ and $\mu\mu_{\text{CC}}$ measurements are used, while the $ee_{\text{CF}}$ measurements are used for the $2 < |\yz| < 3.5$ bin. In all cases, the regularised uncertainties described in~Section~\ref{sec:Methodology} are used for the data. The measurements are also compared to various event generators, in particular to probe different parton-shower models and event-generator implementations.

The overlays of the $\yz$-integrated measurements are shown in Figs.~\ref{Fig::comp-overlays-a}--\ref{Fig::comp-overlays-c} for all coefficients. The calculations from \DYNNLO are shown at NNLO for~$\ptz > 2.5$~GeV with their uncertainties computed as a sum in quadrature of statistical, QCD~scale, and PDF uncertainties, as described in~Section~\ref{sec:theory}. The \POWHEG+~\MINLO predictions, which are shown only including statistical uncertainties, were obtained using the $Z+\mathrm{jet}$~process at~NLO~\cite{arXiv:1206.3572} over the full $\ptz$~range. Owing to numerical issues in the phase-space integration, the \POWHEG+~\MINLO results show fluctuations beyond their statistical uncertainties. The formal accuracy of both calculations is the same, namely~$\mathcal{O}(\alpha_{\text{s}})$ for the predictions of the $A_i$~coefficients as a function of~$\ptz$. The left-hand plots in these figures illustrate the behaviour of each coefficient as a function of~$\ptz$, while the right-hand plots, in which the data measurements are used as a reference, show to which extent the various theoretical predictions agree with the data. In the very first~$\ptz$ bin, $\phics$ has poor resolution and therefore suffers from larger measurement uncertainties. This is reflected in the deviation from the prediction in $A_2$, for example, which is derived primarily from $\phics$.

The predictions from the \DYNNLO and \POWHEG+~\MINLO calculations agree with the data within uncertainties for most coefficients. The striking exception is the $A_2$~coefficient, which rises more slowly as $\ptz$ increases in the data than in the calculations. The data confirm that the Lam--Tung relation ($A_0-A_2=0$) does not hold at~$\mathcal{O}(\alpha^{2}_{\text{s}})$. For~$\ptz > 50$~GeV, significant deviations from zero, almost a factor of two larger than those predicted by the calculations, are observed. Since the impact of the PDF~uncertainties on the calculations is very small, these deviations must be due to higher-order QCD effects. 

In the case of the $A_{5,6,7}$~coefficients, the trend towards non-zero values at high~$\ptz$ discussed in~Section~\ref{sec:results} is also compatible with that from the predictions, although it is at the limit of the sensitivity for both the data and the calculations. As shown in~Fig.~\ref{Fig::comp-overlays-c} and also in~Table~\ref{Tab:PredSummaryMGV}, the predictions from~\DYNNLO\ suggest that the values of the $A_{5,6,7}$~coefficients should be at the level of 0.005 at high values of~$\ptz$. A test is performed to quantify the significance of the deviation from zero (see Appendix~\ref{sec:app_a567}). A signed $\chi^2$ test statistic is defined based on the tail probability of each individual measurement, taking into account the correlations between the parameters in bins of~$\ptz$. An ensemble test is performed to compute the observed and expected significance of all three coefficients together, where pseudo-data from~\DYNNLO\ is used for the expected value. This test gives an observed (expected) significance of~3.0~(3.2)~standard deviations.

The measurements of the $A_1$, $A_3$, and $A_4$~coefficients in the three $\yz$~bins (only the first two bins are available for the $A_1$~coefficient) are compared to the predictions in~Figs.~\ref{Fig::comp-overlays-ybin1}--\ref{Fig::comp-overlays-ybin3}. Overall, the predictions and the data agree for all three $\yz$~bins. These coefficients are the only ones that display any significant $\yz$~dependence and it is interesting to note that, for high values of~$\ptz$, the~$A_1$ and $A_3$~coefficients increase as $\yz$~increases. 
As explained in~Section~\ref{sec:intro} and detailed in~Appendix~\ref{appendix:theory}, at low values of~$\ptz$, the measured value of the $A_4$~coefficient can be directly related to the Weinberg angle~$\sin^2\theta_{\text{W}}$~\cite{Aad:2015uau}. The strong dependence of the value of the $A_4$~coefficient on~$|\yz|$ is, however, mostly a consequence of the approximation made for the interacting quark direction in the CS~reference frame on an event-by-event basis. The impact of this approximation 
decreases at higher values of~$|\yz|$, and, as a result, the measured and expected values of the $A_4$~coefficient increase, as can be seen in~Figs.~\ref{Fig::comp-overlays-ybin1}~--~\ref{Fig::comp-overlays-ybin3}.

The effect of the parton-shower modelling and matching scheme on the reference angular coefficients is explored in~Fig.~\ref{Fig::comp-overlay-powheg}, which shows a comparison of the measurements of $A_0$, $A_1$, $A_2$, and $A_0 - A_2$ with \DYNNLO at NLO and NNLO, \POWHEGBOX (without parton shower), and with the same process in~\POWHEGBOX with the parton shower simulated with \PYTHIA8 (\POWHEGBOX+~\PYTHIA8) and~\HERWIG (\POWHEGBOX+~\HERWIG). The predictions from \DYNNLO at~NLO and \POWHEGBOX\ without parton shower, which are formally at the same level of accuracy, agree for~$A_1$ and~$A_2$. For the $A_2$~coefficient, which is the most sensitive one to higher-order corrections, adding the parton-shower simulation to the \POWHEGBOX\Zboson-boson production process brings the predictions closer to \DYNNLO at NNLO. This is consistent with the assumption that the parton-shower model emulates higher-order effects, although the discrepancy between the measurements and the parton-shower models is larger than that with~\DYNNLO at~NNLO. The $A_0$~coefficient has an unexpected offset of~$-0.025$ at low values of~$\ptz$ in the \POWHEGBOX implementation. This effect is also reflected in the predictions for~$A_0 - A_2$ and has been corrected in the more recent version of \POWHEGBOX~(v2.1) used in this paper for the $Z+\mathrm{jet}$ predictions with \POWHEG+~\MINLO~\cite{arXiv:0409146,Frixione:2007vw,Powheg1,Powheg2}. The predictions from~\DYNNLO at~NLO and~NNLO agree well with the data measurements for the $A_0$~coefficient, but overestimate the rise of the $A_2$~coefficient at higher values of~$\ptz$, as discussed above. Finally, it is interesting to note that, whereas the agreement between \PYTHIA8 and \HERWIG is good for most of the coefficients, the $A_1$~coefficient displays significant differences between the two predictions over most of the $\ptz$~range. Although this might be ascribed to differences between the parton-shower model and matching schemes at intermediate values of~$\ptz$, it is somewhat surprising to observe large differences for the highest values of~$\ptz$.

Figure~\ref{Fig::comp-overlay-sherpa} shows a comparison of the measurements of $A_0$, $A_2$ and $A_0 - A_2$ with \SHERPA~1.4 (up to five jets at~LO) and \SHERPA~2.1~\cite{Gleisberg:2008ta,Hoeche:2009rj,Gleisberg:2008fv,Schumann:2007mg}. The effect of simulating \SHERPA~2.1 events (up to two jets at~NLO and up to five jets at~LO for higher jet multiplicities) is explicitly shown. None of the configurations correctly predict the behaviour of $A_0$ or $A_2$. The \SHERPA~2.1 version follows the data more closely than the \SHERPA~1.4 version. In addition, in all versions except \SHERPA~2.1 with two jets, significant higher-order polynomial behaviour was found to be present. This is probably due to the matrix-element matching scheme used in the event generator for the calculation of the $Z+n$-jet process for $n>2$.

\begin{figure}[hp]
  \begin{center}                               
{
    \includegraphics[width=7.5cm,angle=0]{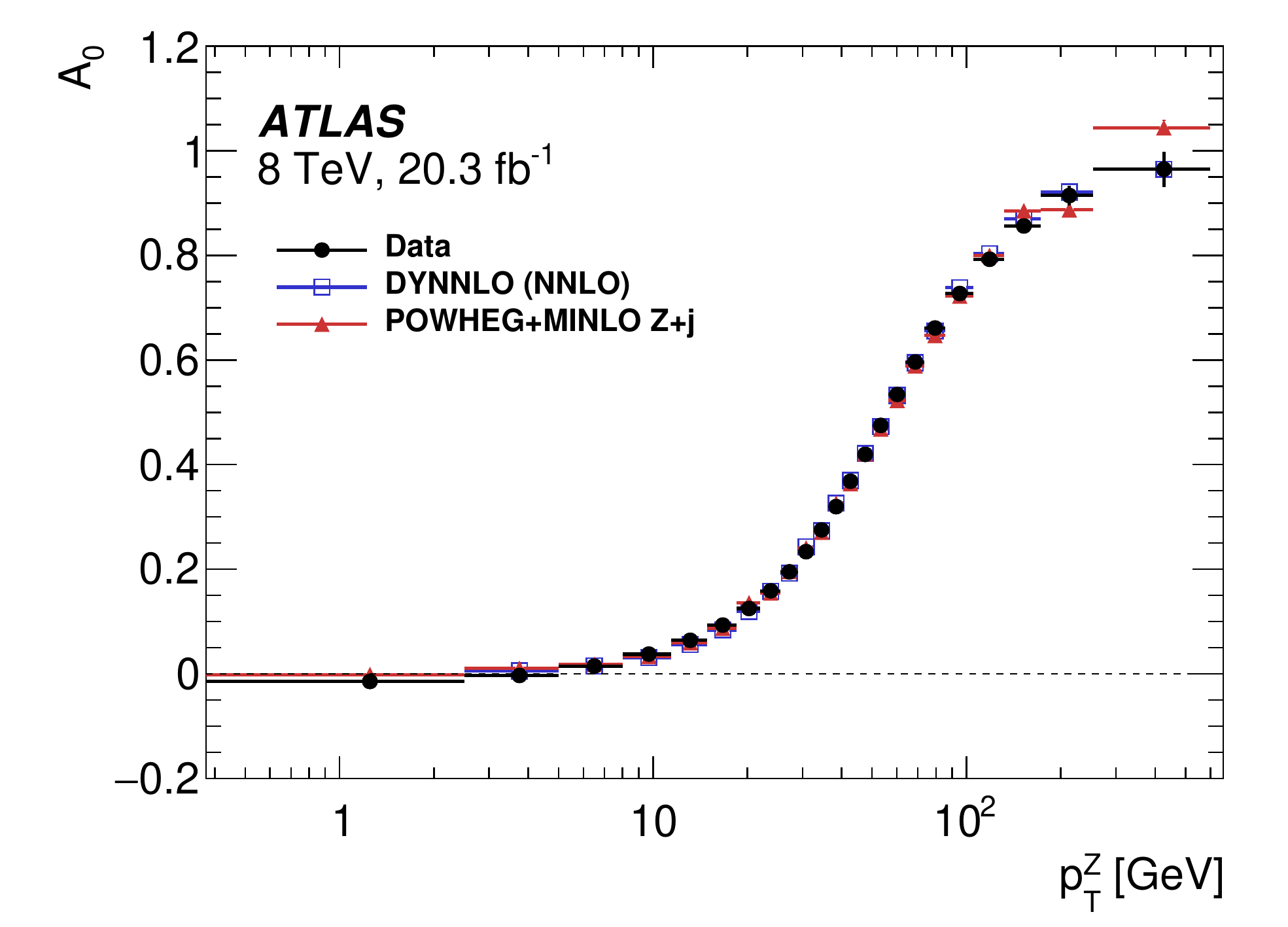}
    \includegraphics[width=7.5cm,angle=0]{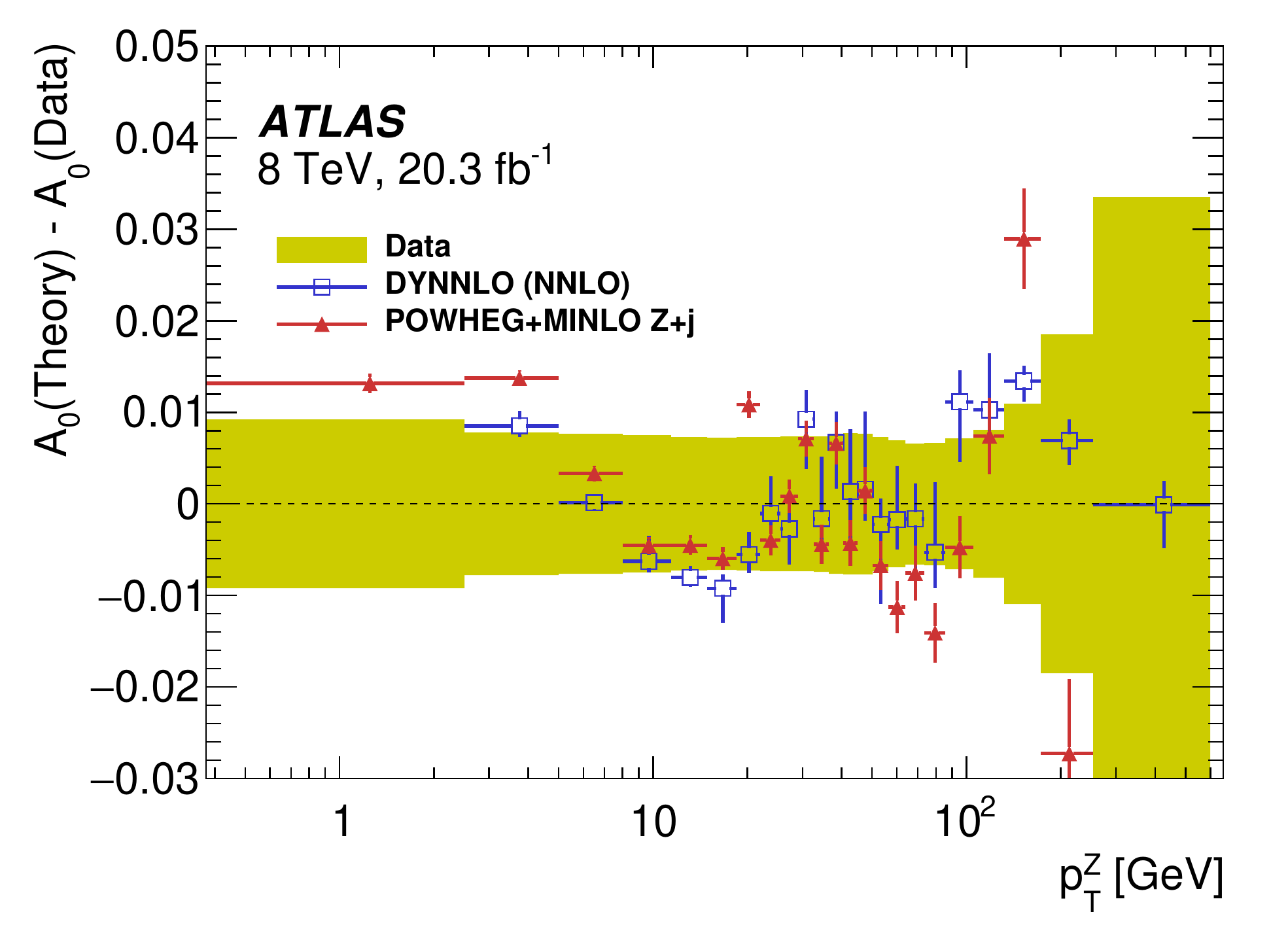}
}
{
    \includegraphics[width=7.5cm,angle=0]{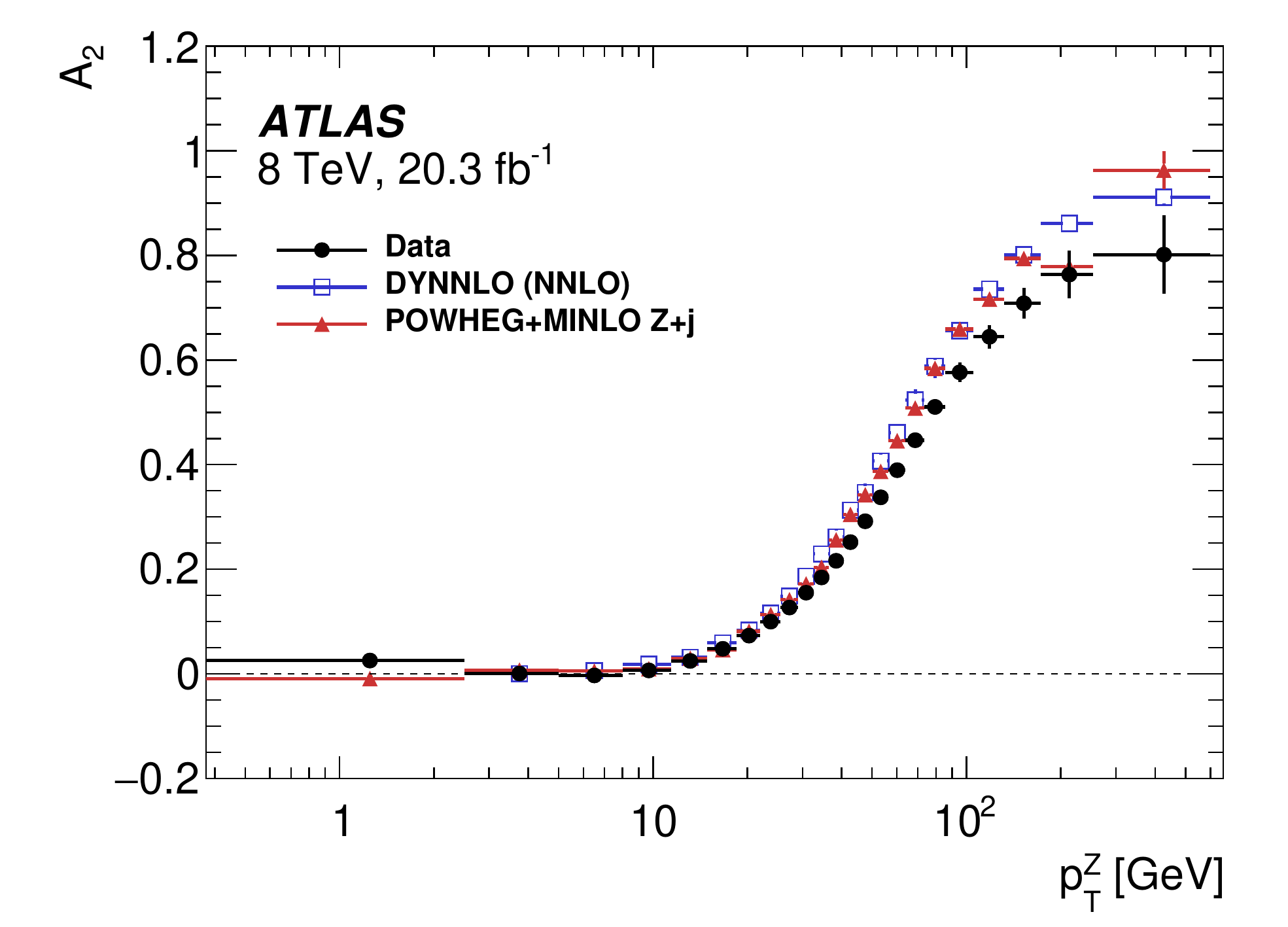}
    \includegraphics[width=7.5cm,angle=0]{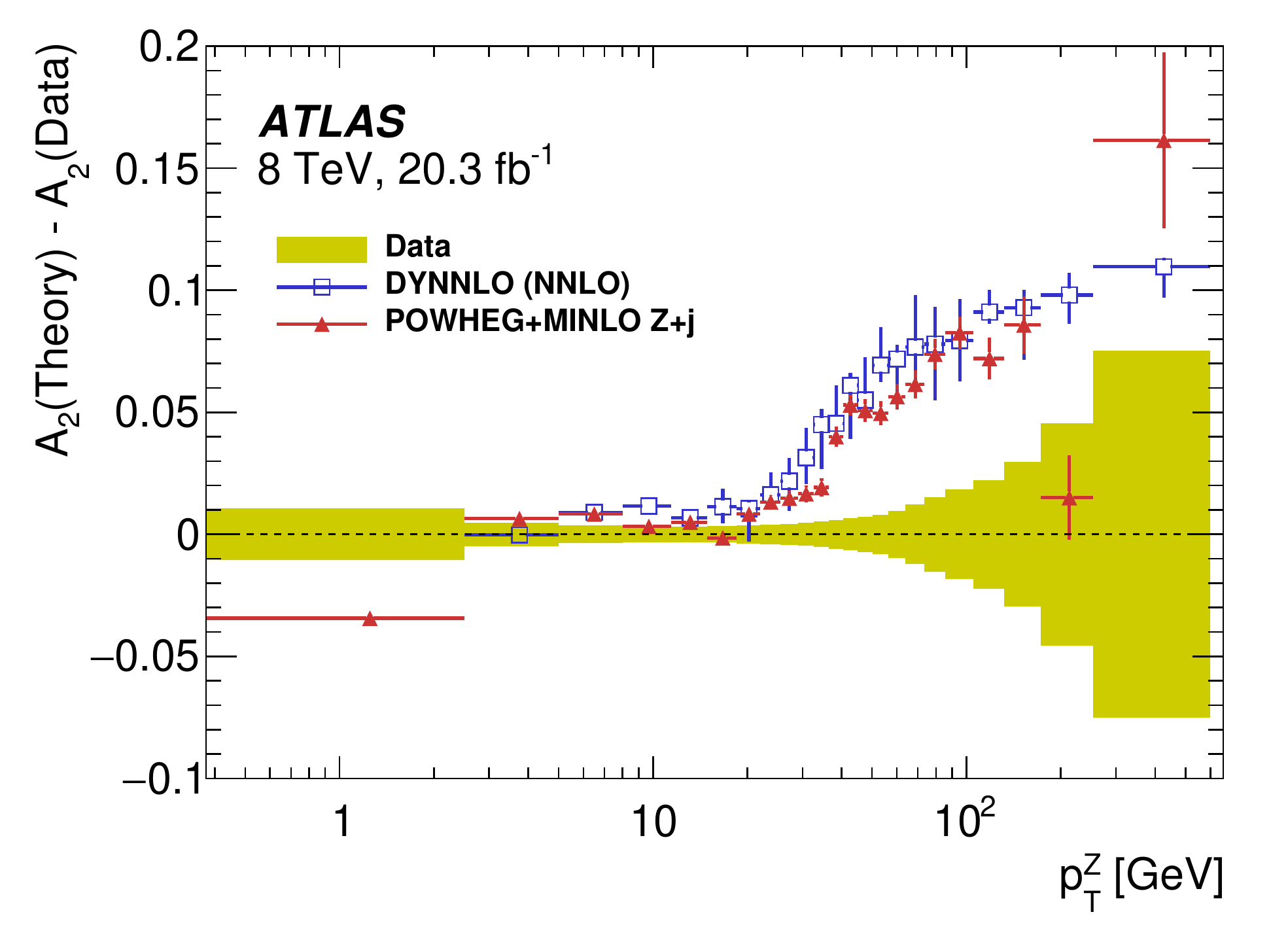}
}
{
    \includegraphics[width=7.5cm,angle=0]{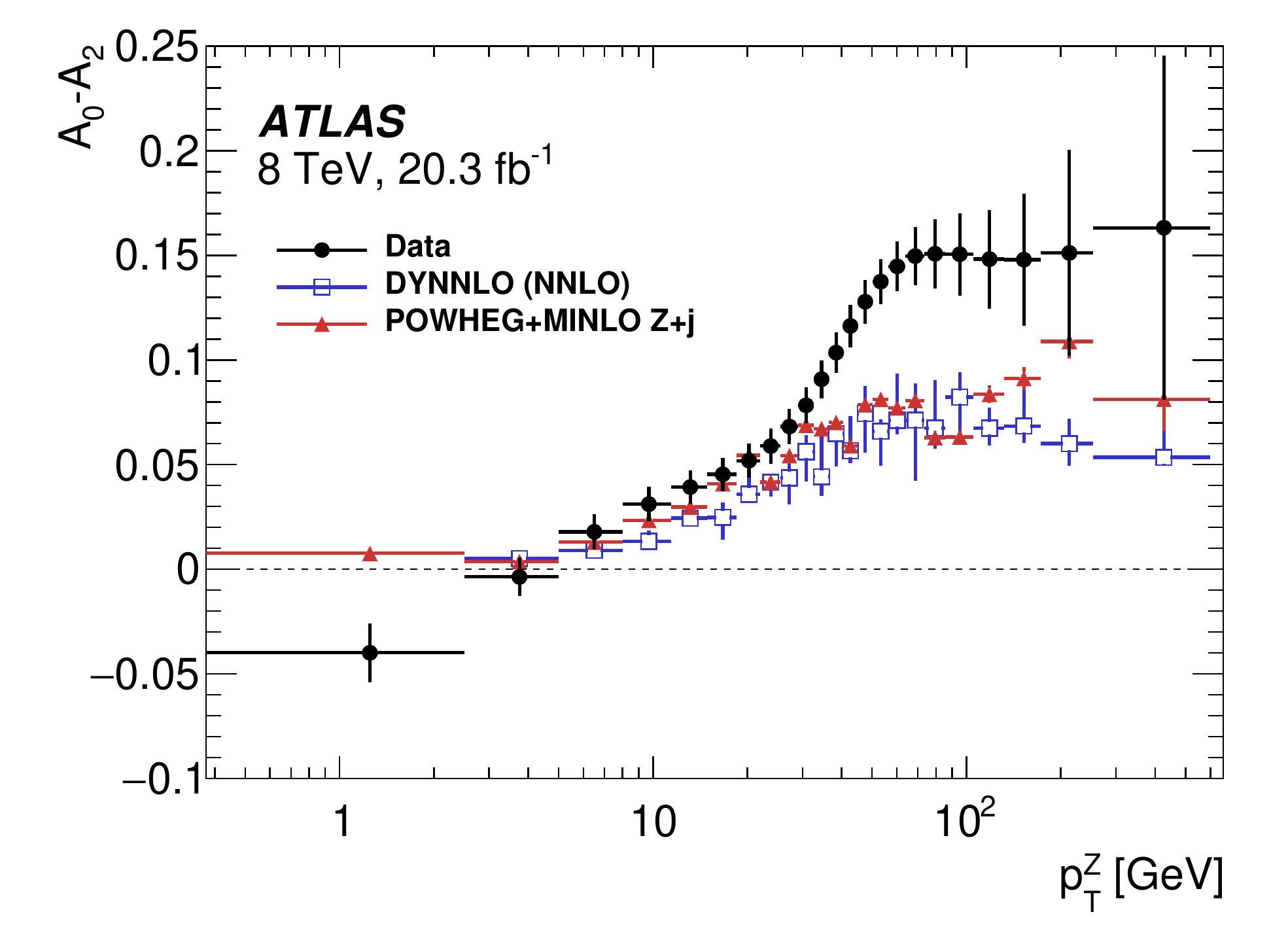}
    \includegraphics[width=7.5cm,angle=0]{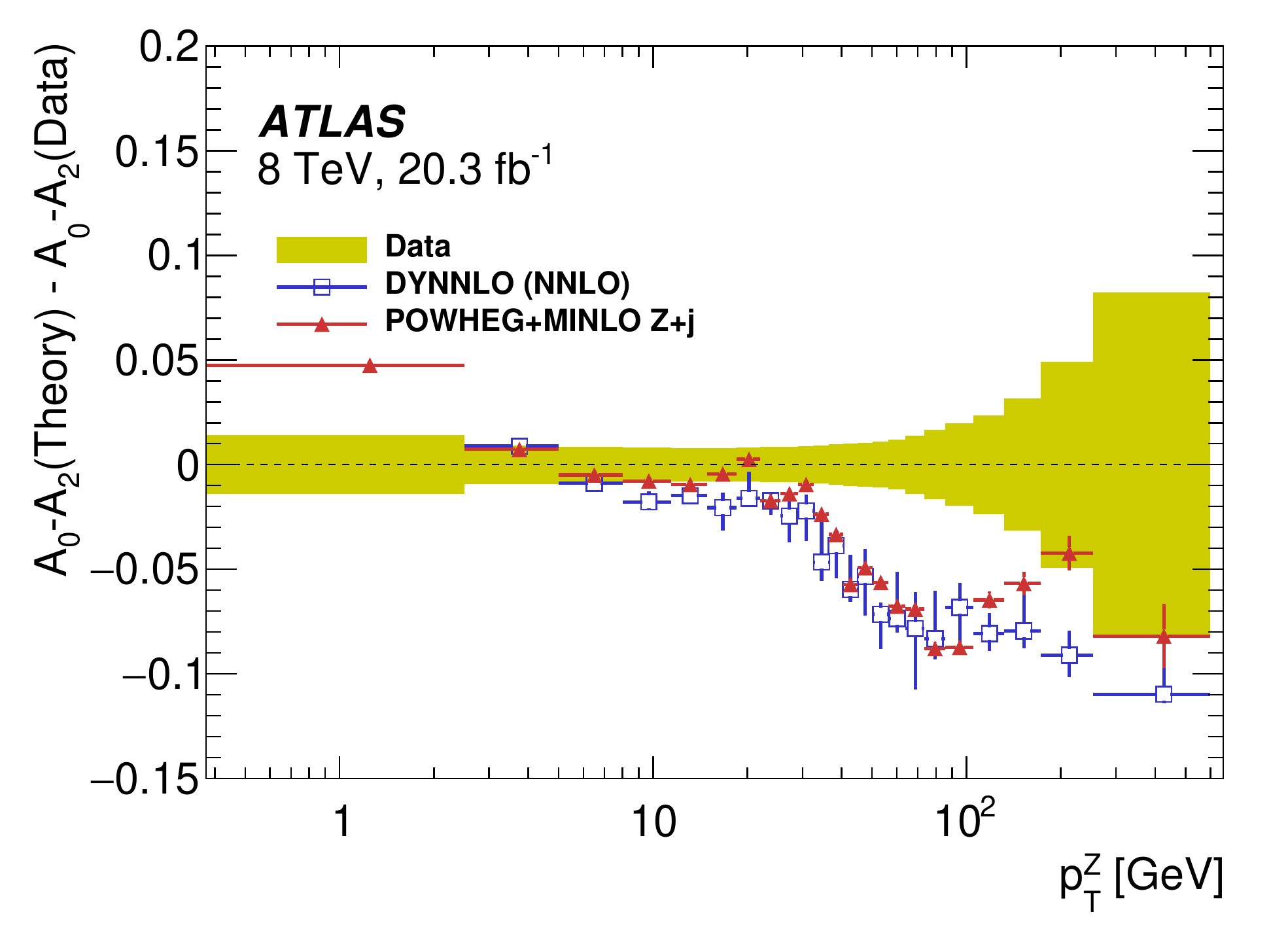}
}
\end{center}
\caption{ 
Distributions of the angular coefficients $A_0$ (top), $A_2$ (middle) and $A_0-A_2$ (bottom) as a function of~$\ptz$. The results from the $\yz$-integrated measurements are compared to the \DYNNLO\ and \POWHEG \MINLO predictions (left).
The differences between the two calculations and the data are also shown (right), with the shaded band around zero representing the total uncertainty in the measurements. The error bars for the calculations show the total uncertainty for \DYNNLO, but only the statistical uncertainties for \POWHEG \MINLO (see text). 
\label{Fig::comp-overlays-a} }
\end{figure}

\begin{figure}[hp]
  \begin{center}                               
{
    \includegraphics[width=7.5cm,angle=0]{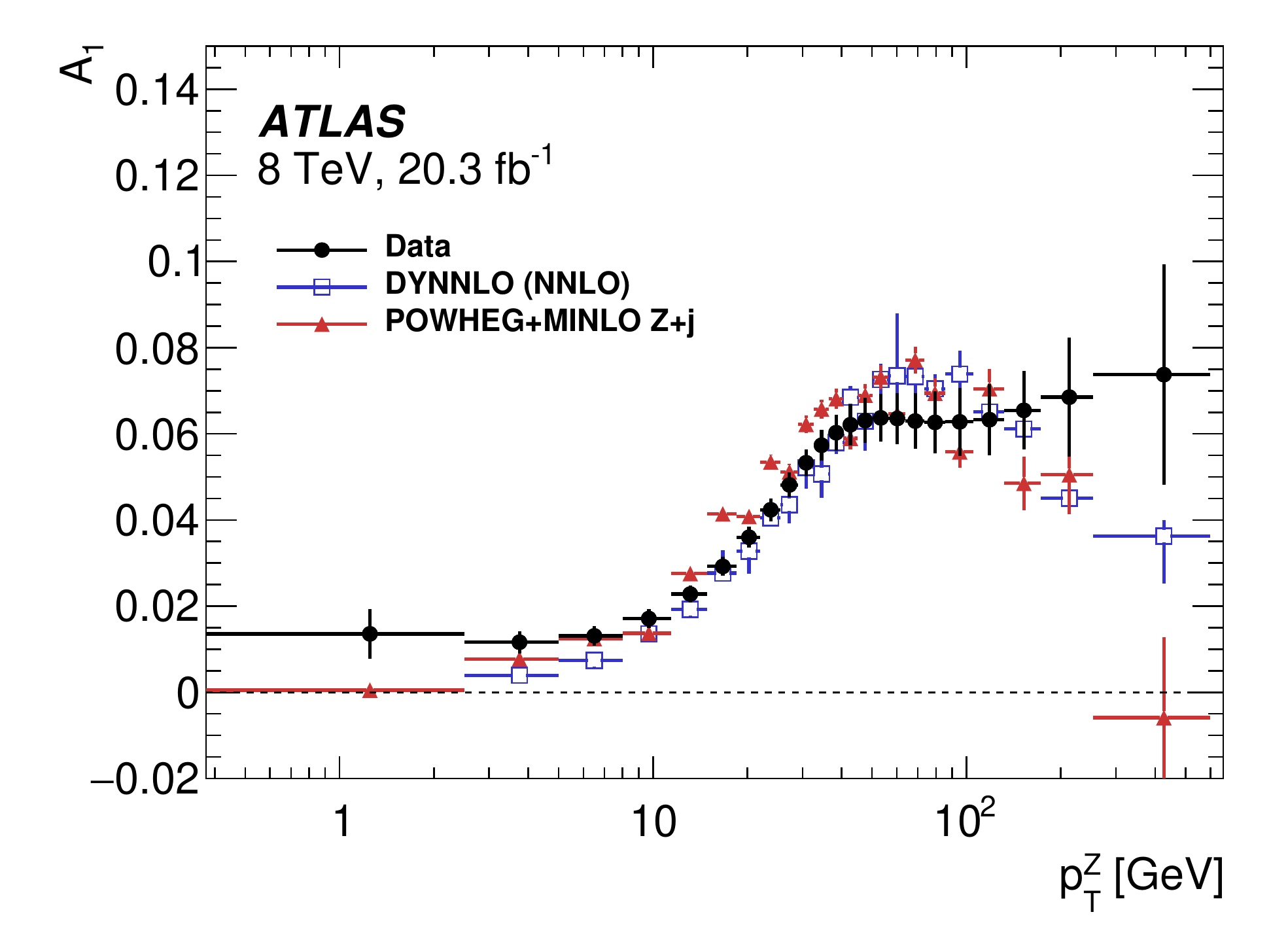}
    \includegraphics[width=7.5cm,angle=0]{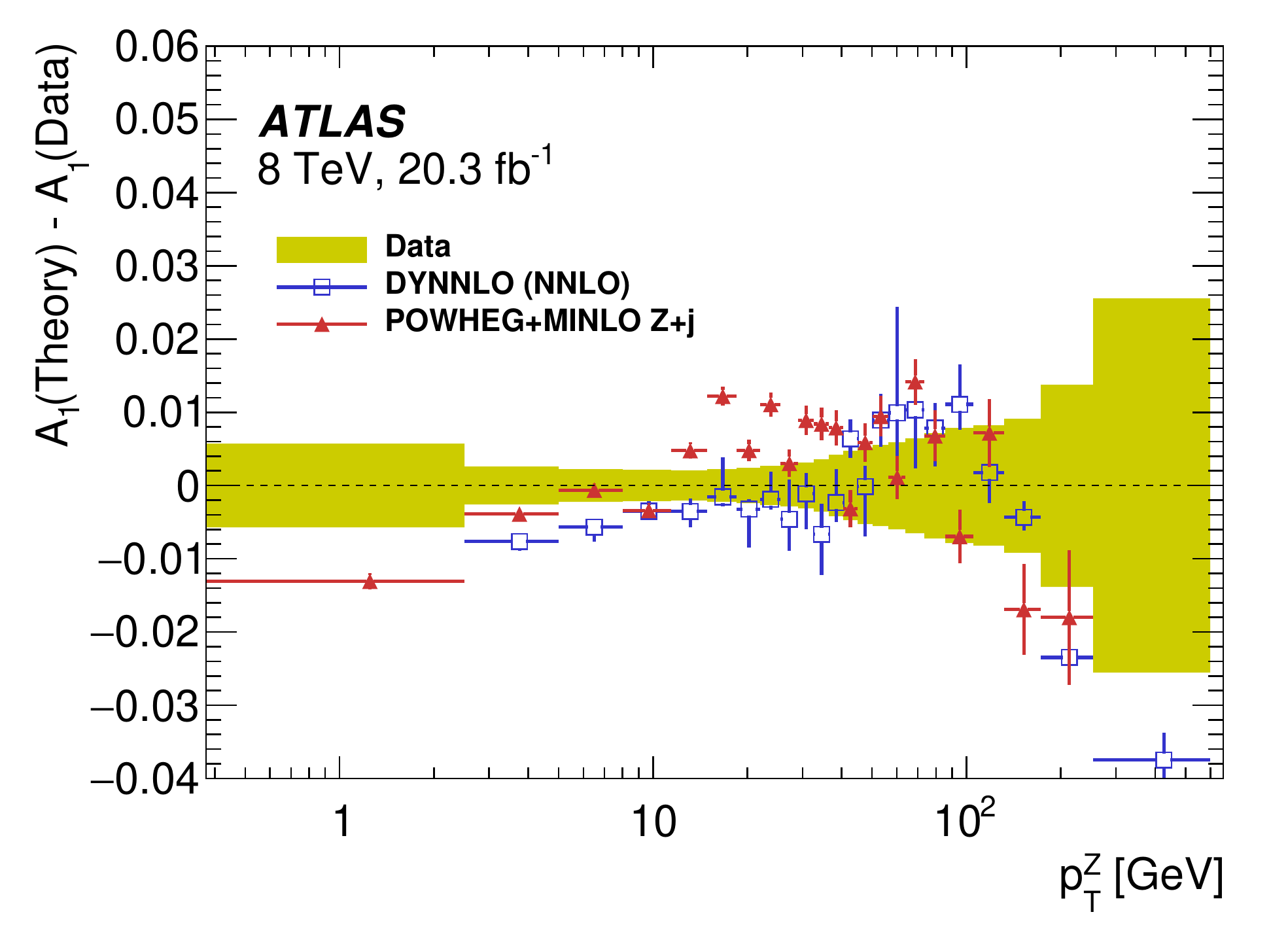}
}
{
    \includegraphics[width=7.5cm,angle=0]{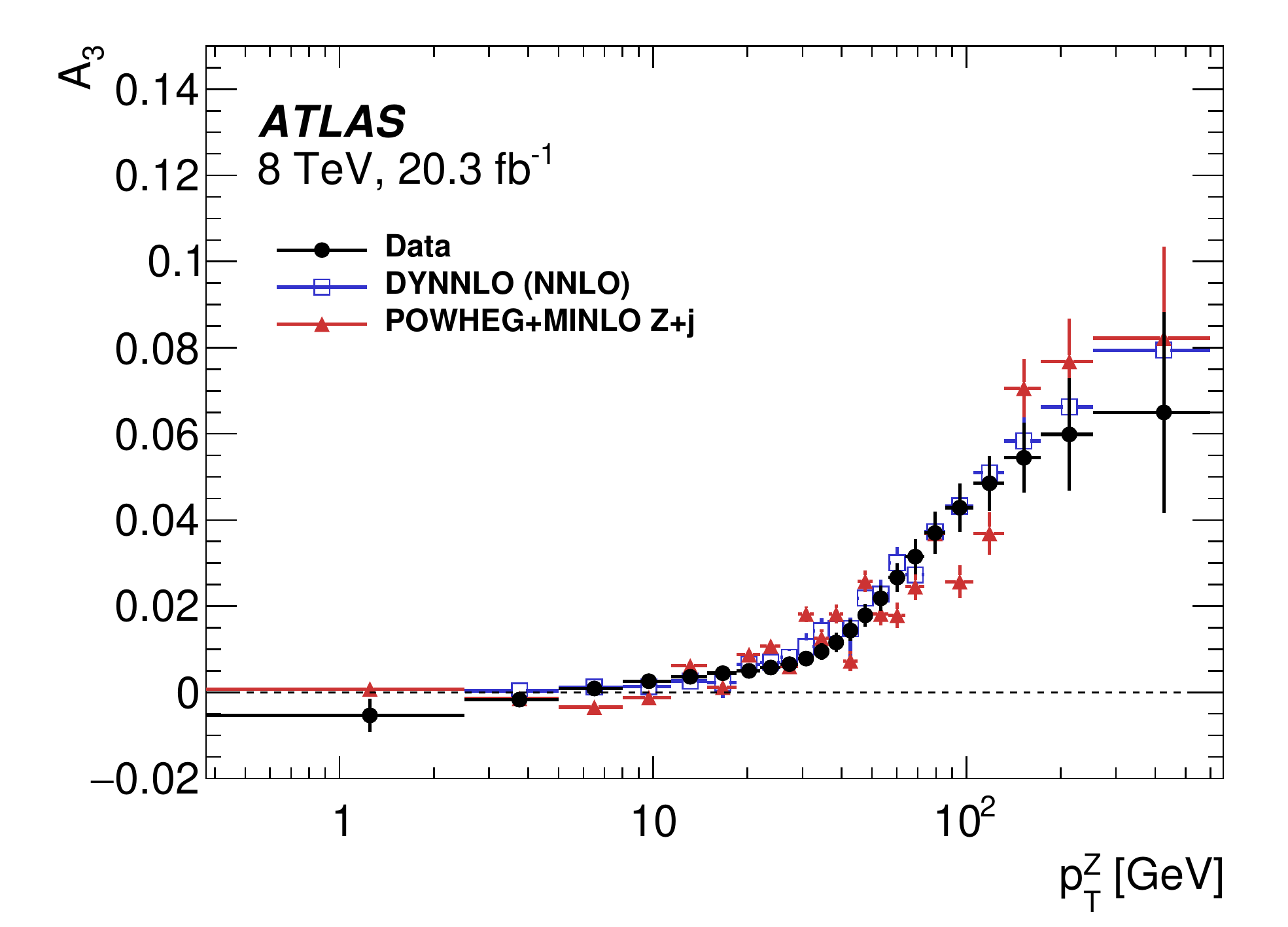}
    \includegraphics[width=7.5cm,angle=0]{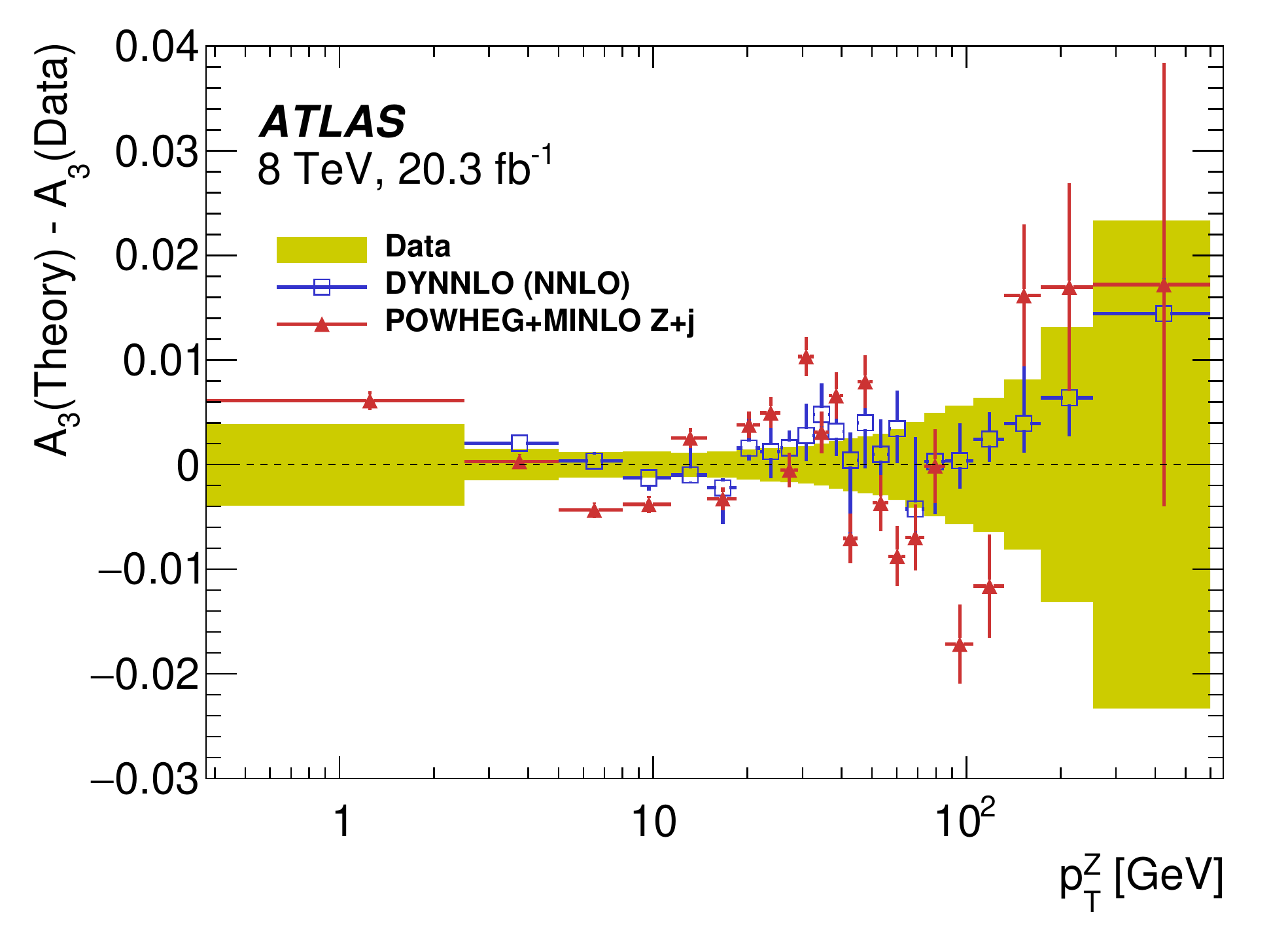}
}
{
    \includegraphics[width=7.5cm,angle=0]{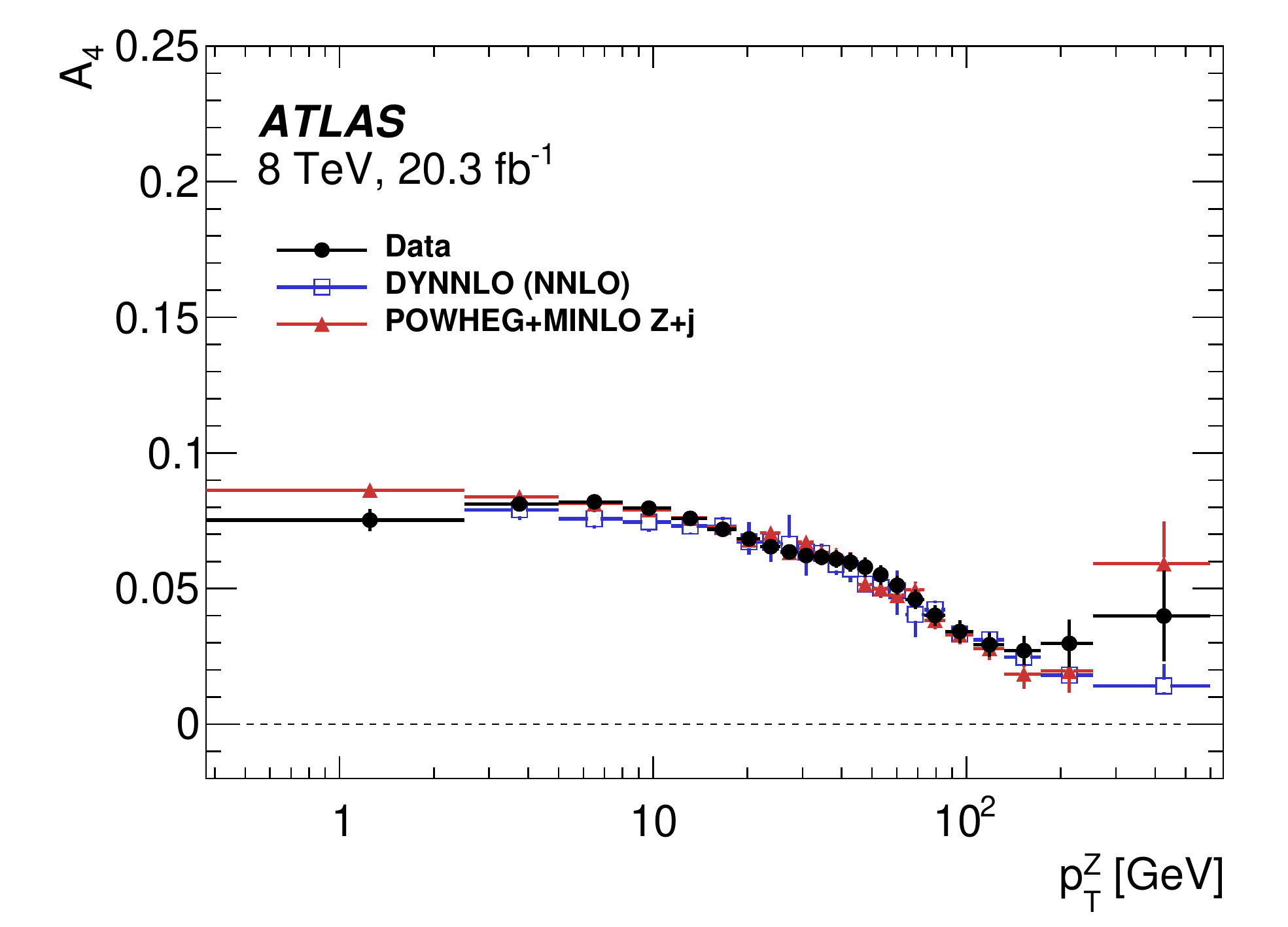}
    \includegraphics[width=7.5cm,angle=0]{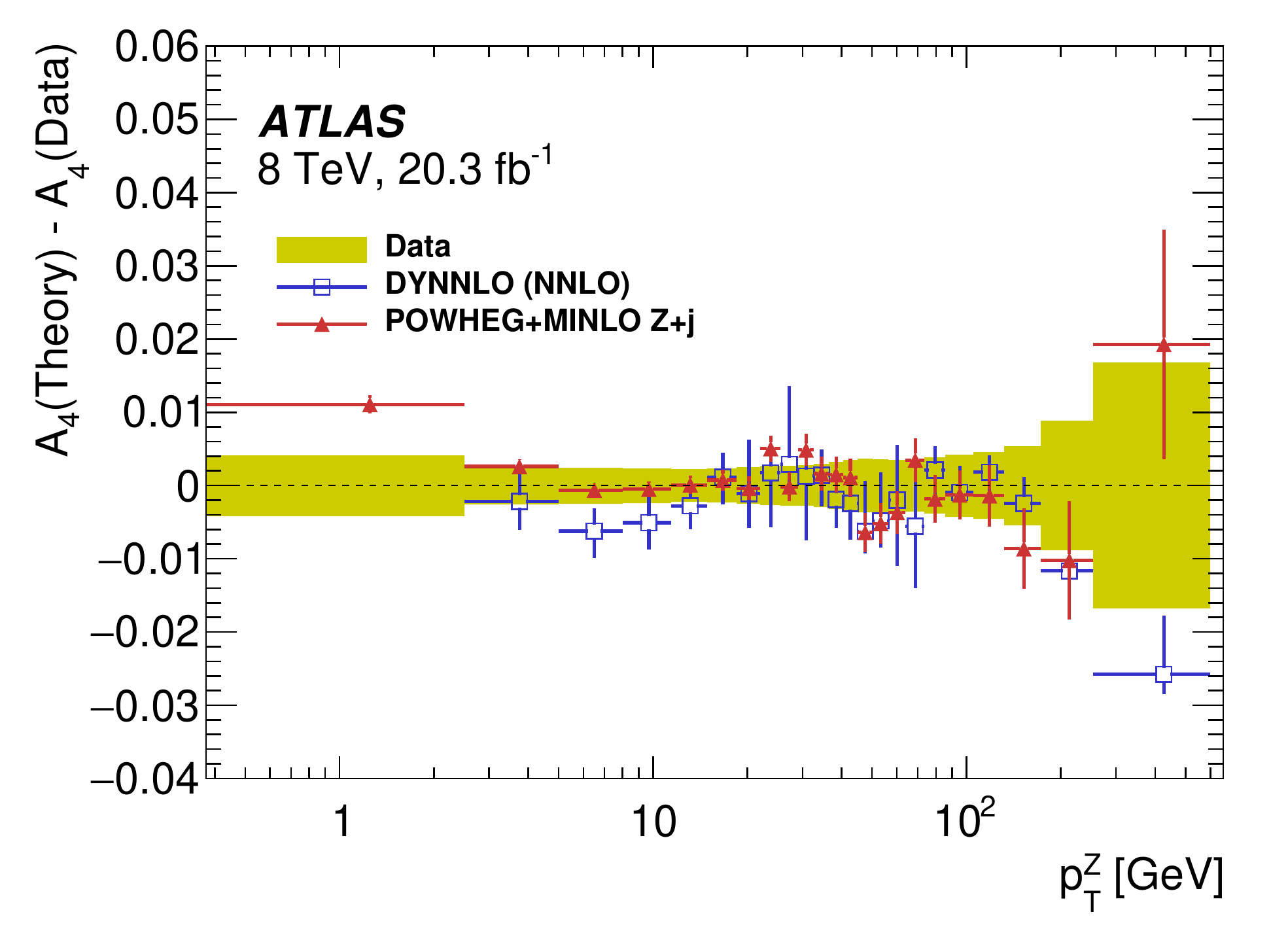}
}
\end{center}
\caption{ 
Distributions of the angular coefficients $A_1$ (top), $A_3$ (middle) and $A_4$ (bottom) as a function of~$\ptz$. The results from the $\yz$-integrated measurements are compared to the \DYNNLO\ and \POWHEG \MINLO predictions (left).
The differences between the two calculations and the data are also shown (right), with the shaded band around zero representing the total uncertainty in the measurements. The error bars for the calculations show the total uncertainty for \DYNNLO, but only the statistical uncertainties for \POWHEG \MINLO (see text). 
\label{Fig::comp-overlays-b} }
\end{figure}

\begin{figure}[hp]
  \begin{center}                               
{
    \includegraphics[width=7.5cm,angle=0]{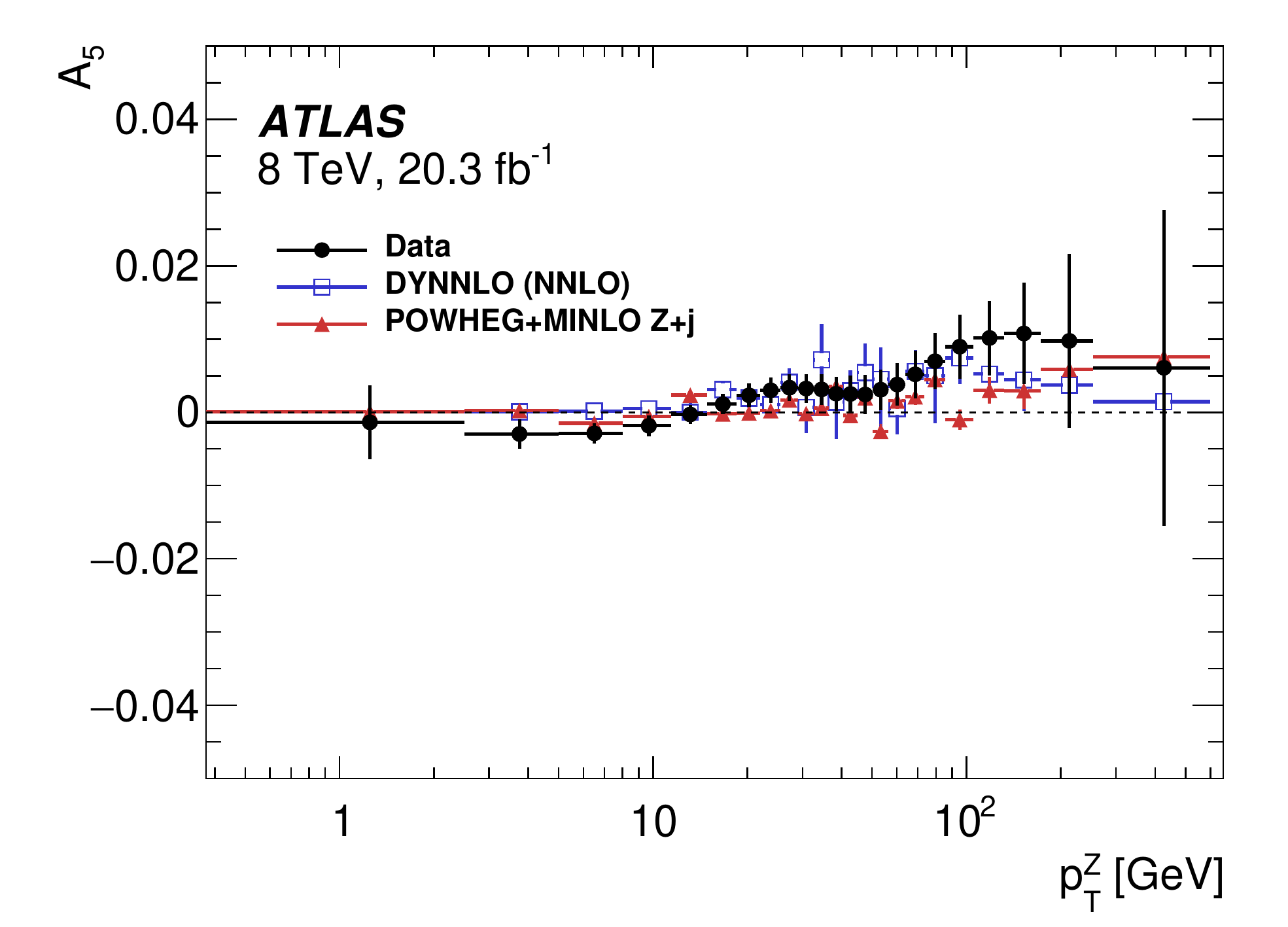}
    \includegraphics[width=7.5cm,angle=0]{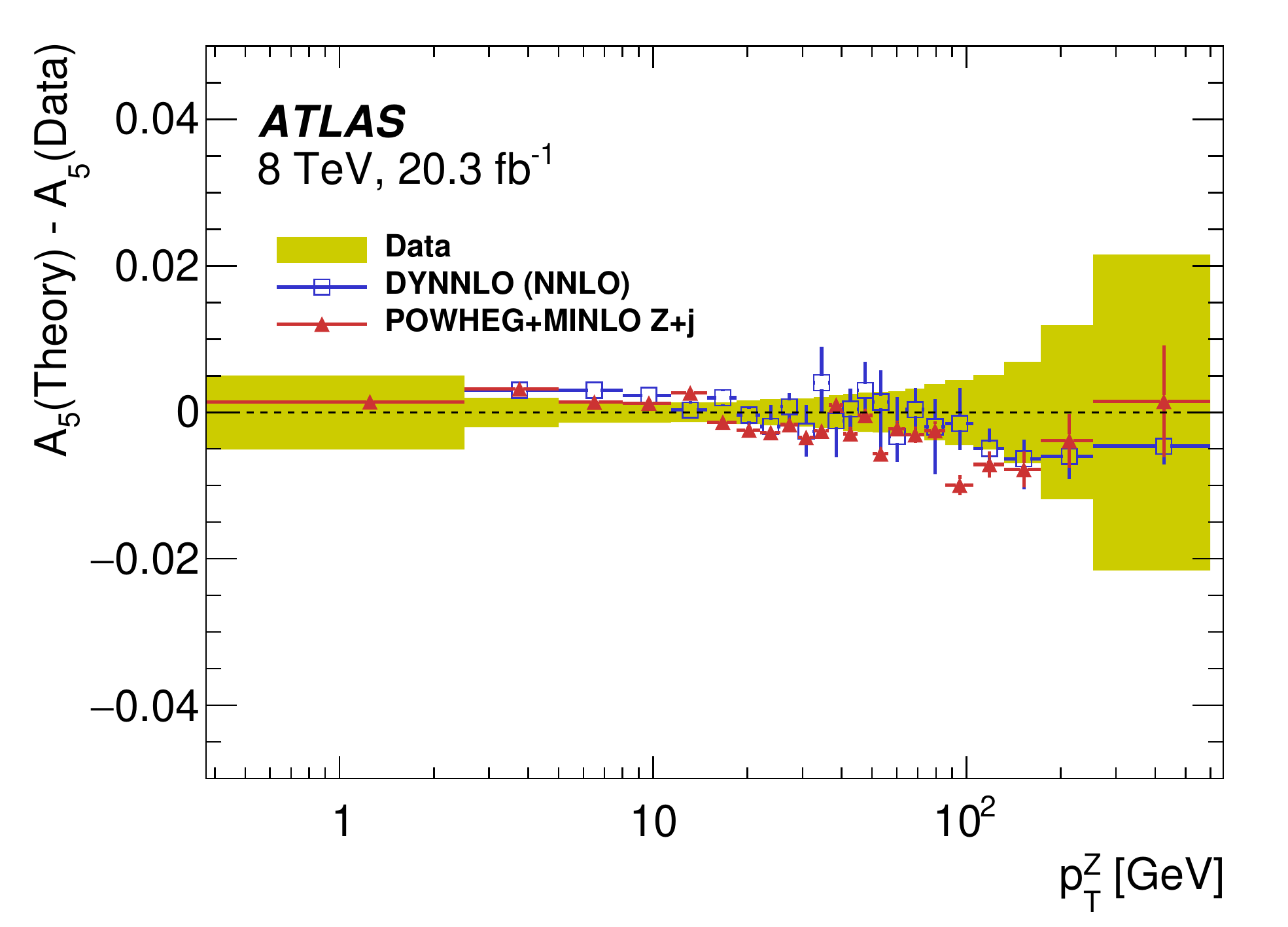}
}
{
    \includegraphics[width=7.5cm,angle=0]{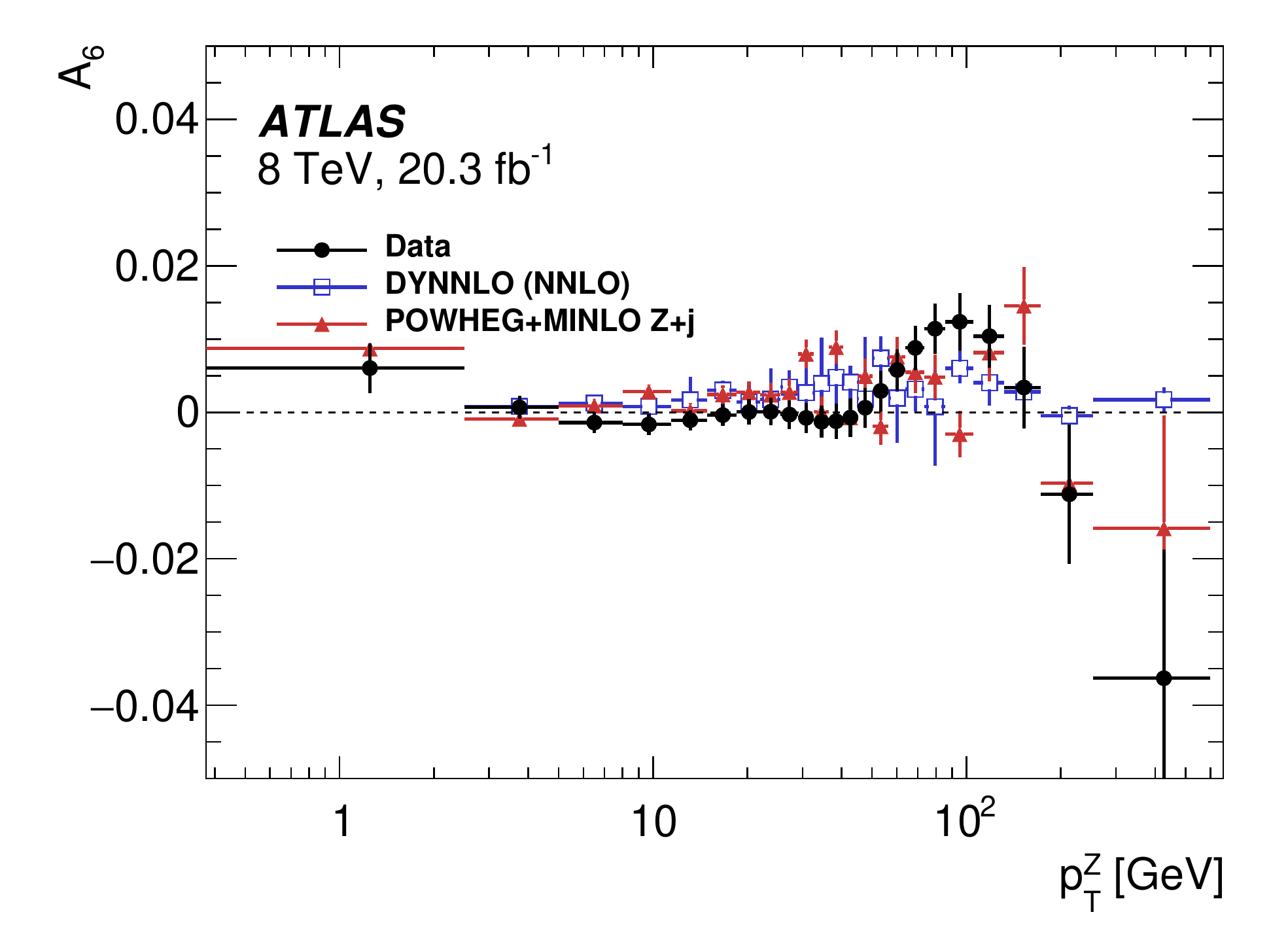}
    \includegraphics[width=7.5cm,angle=0]{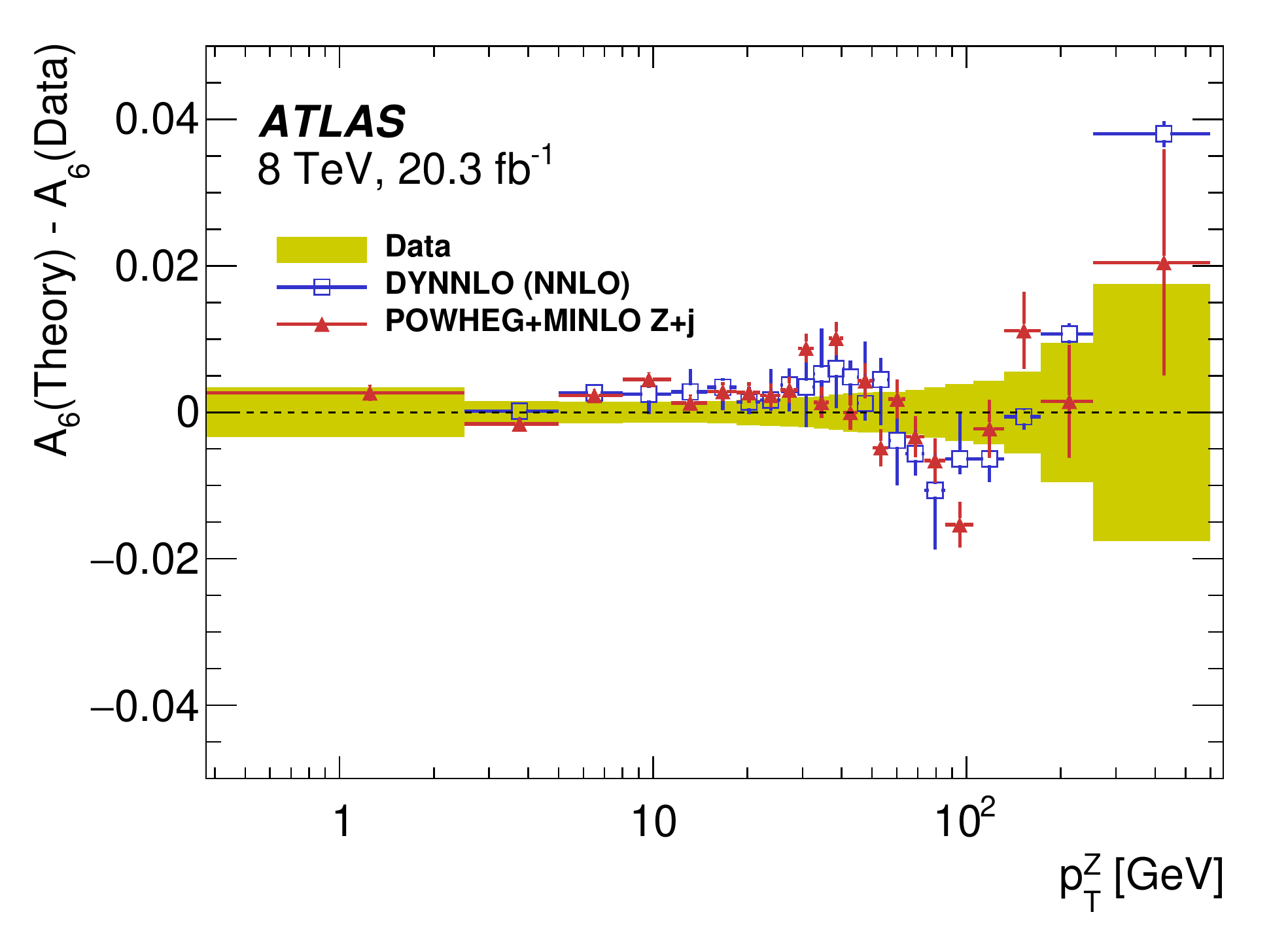}
}
{
    \includegraphics[width=7.5cm,angle=0]{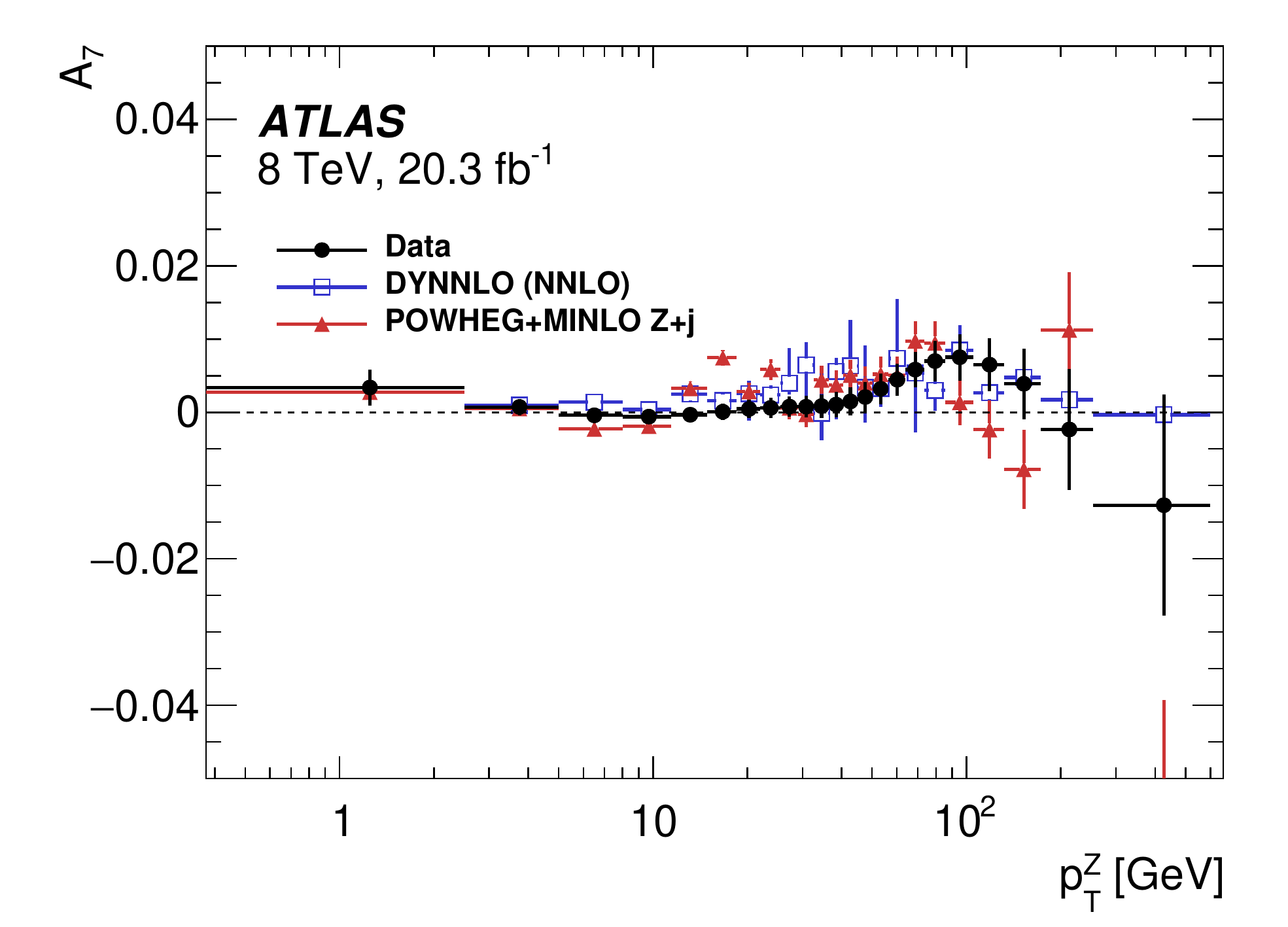}
    \includegraphics[width=7.5cm,angle=0]{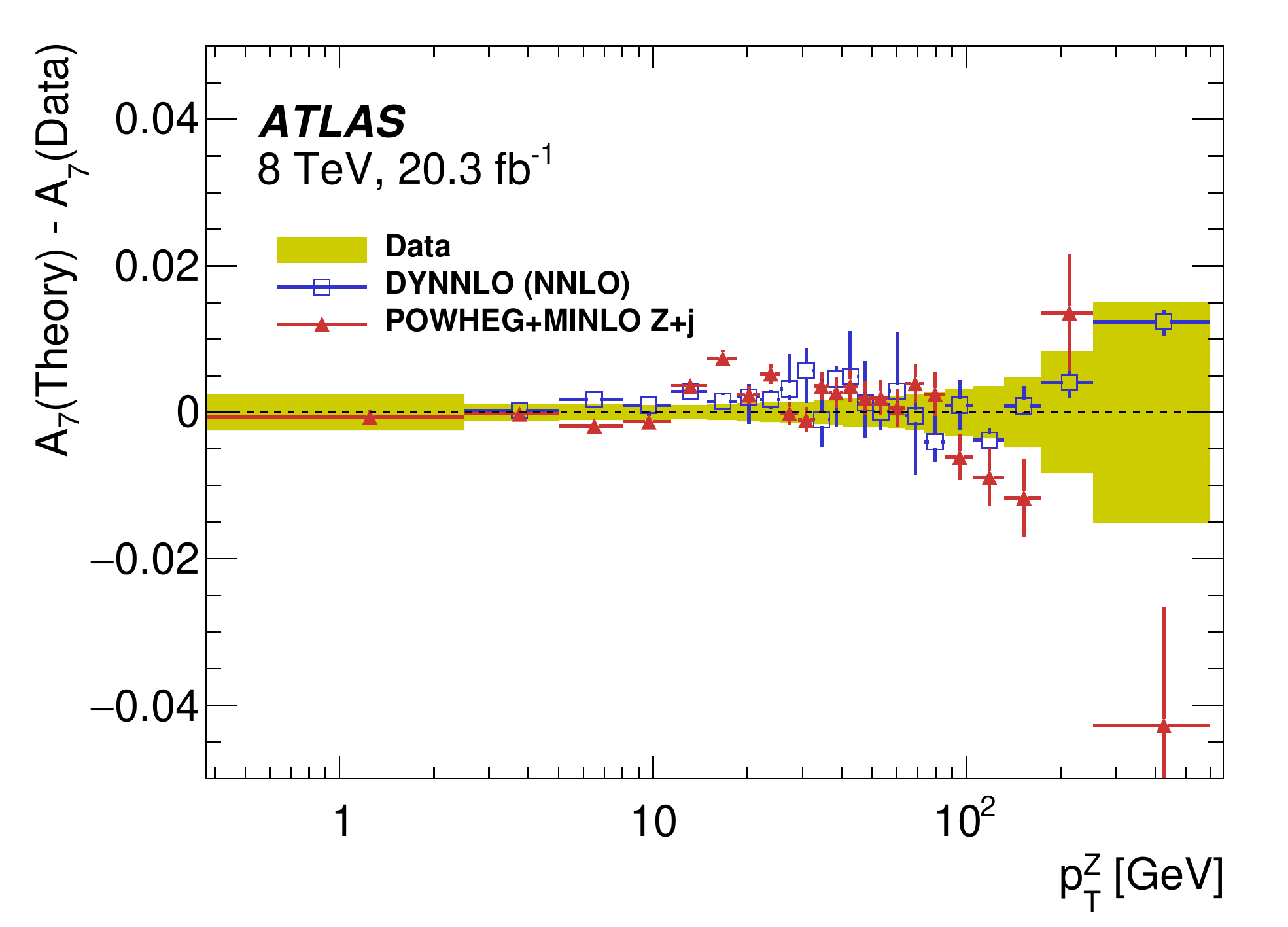}
}
\end{center}
\caption{
Distributions of the angular coefficients $A_5$ (top), $A_6$ (middle) and $A_7$ (bottom) as a function of~$\ptz$. The results from the $\yz$-integrated measurements are compared to the \DYNNLO\ and \POWHEG \MINLO predictions (left).
The differences between the two calculations and the data are also shown (right), with the shaded band around zero representing the total uncertainty in the measurements. The error bars for the calculations show the total uncertainty for \DYNNLO, but only the statistical uncertainties for \POWHEG \MINLO (see text). 
\label{Fig::comp-overlays-c} }
\end{figure}

\begin{figure}[hp]
  \begin{center}
{
    \includegraphics[width=7.5cm,angle=0]{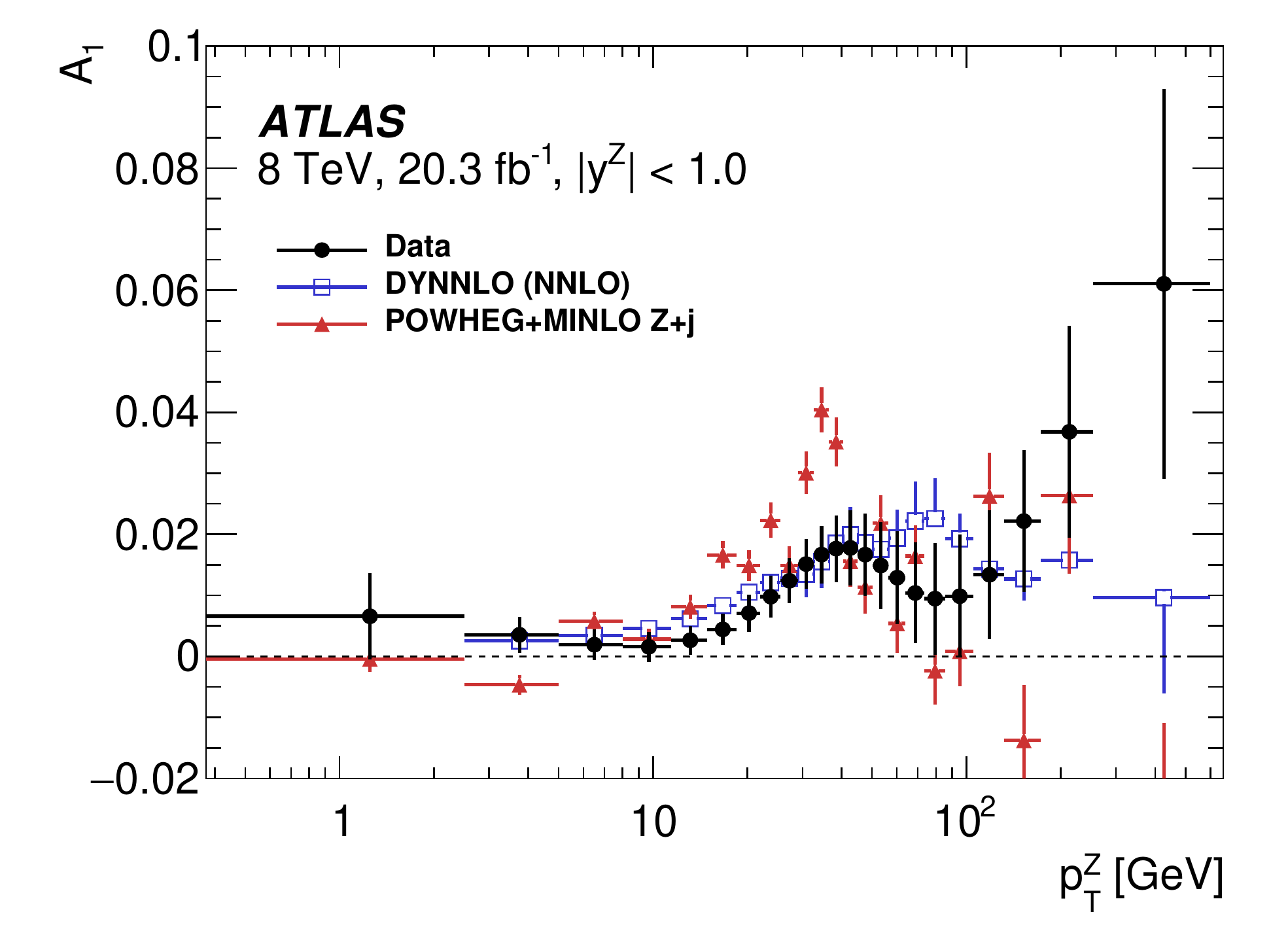}
    \includegraphics[width=7.5cm,angle=0]{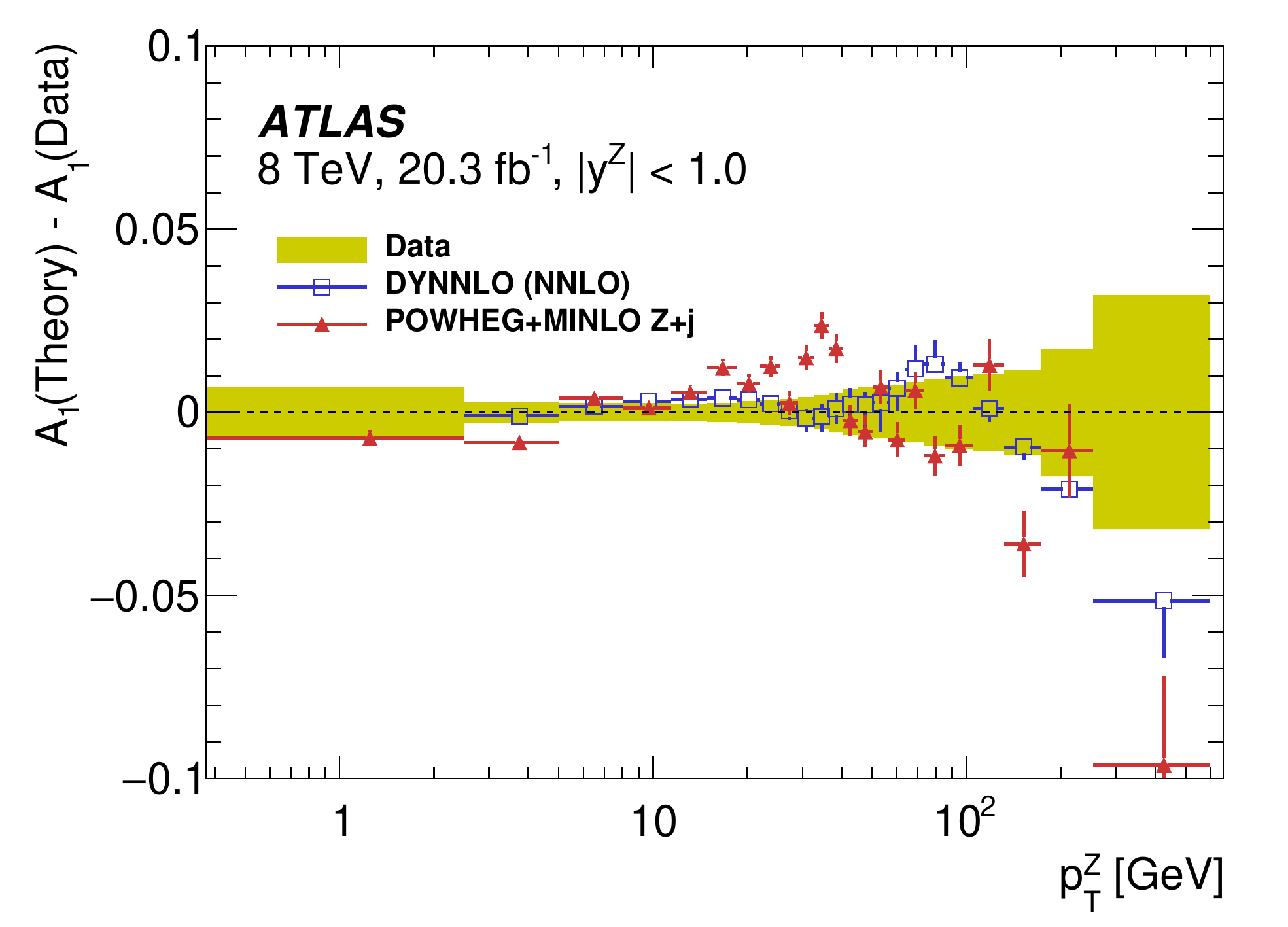}
}
{
    \includegraphics[width=7.5cm,angle=0]{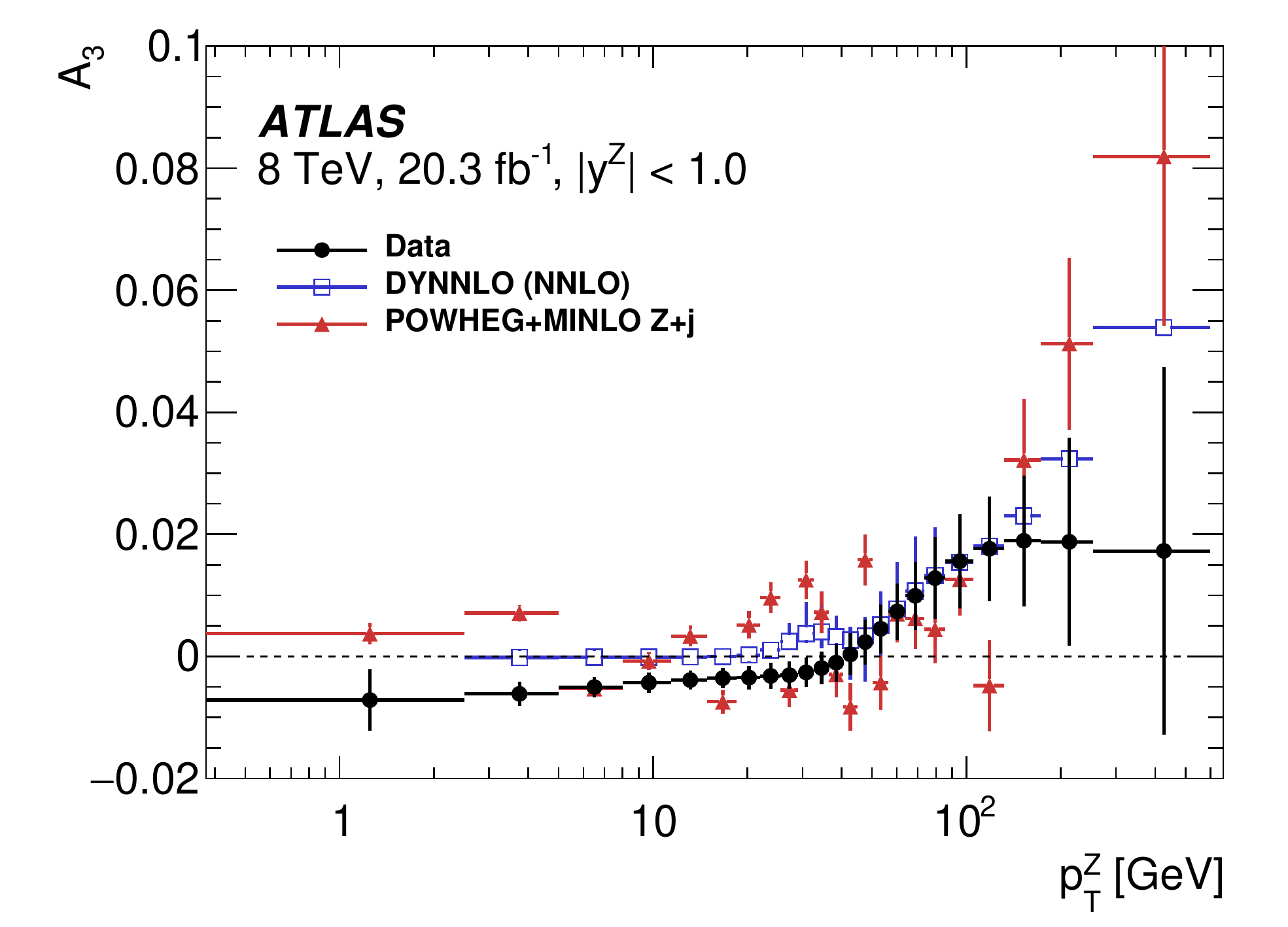}
    \includegraphics[width=7.5cm,angle=0]{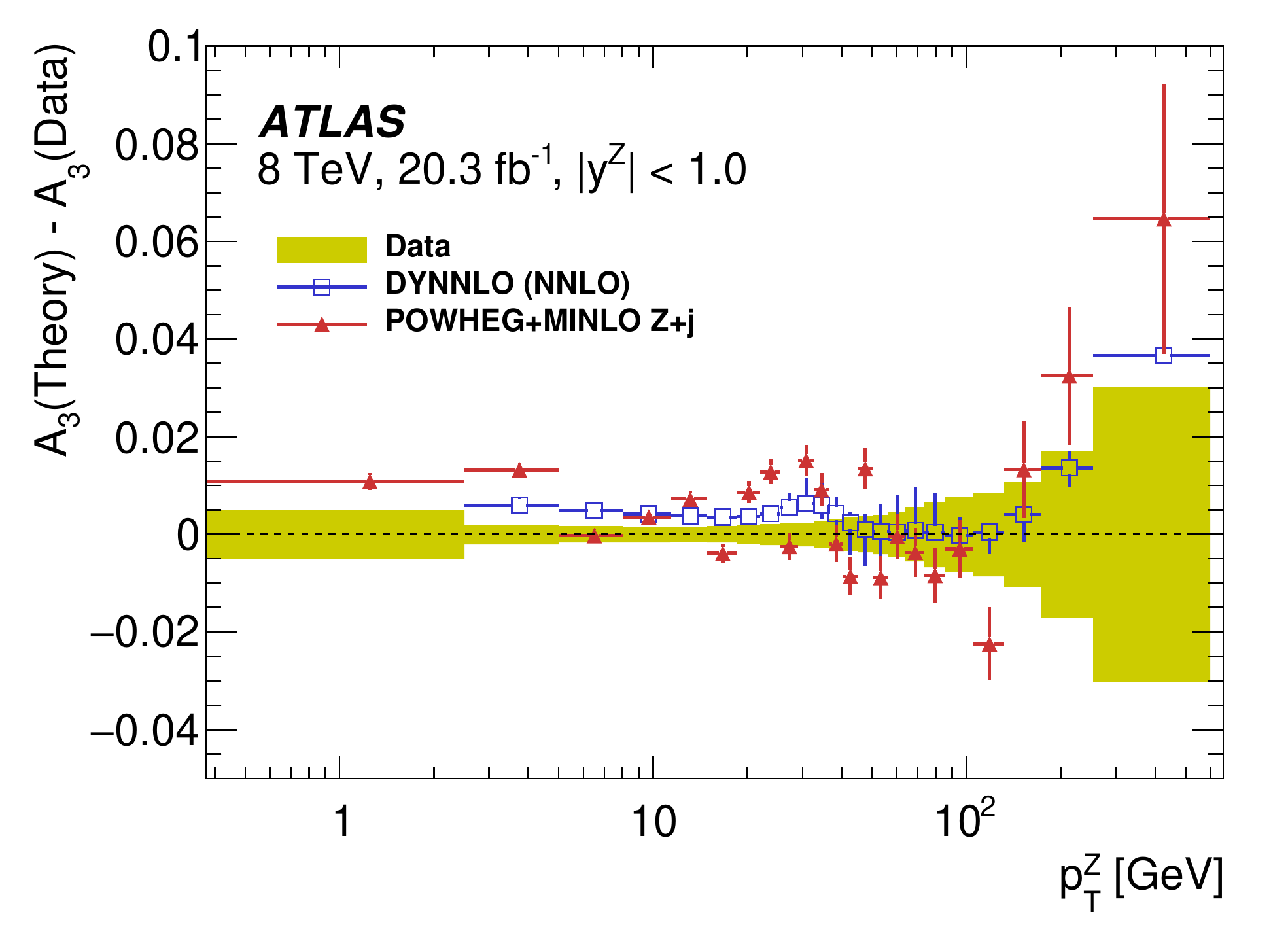}
}
{
    \includegraphics[width=7.5cm,angle=0]{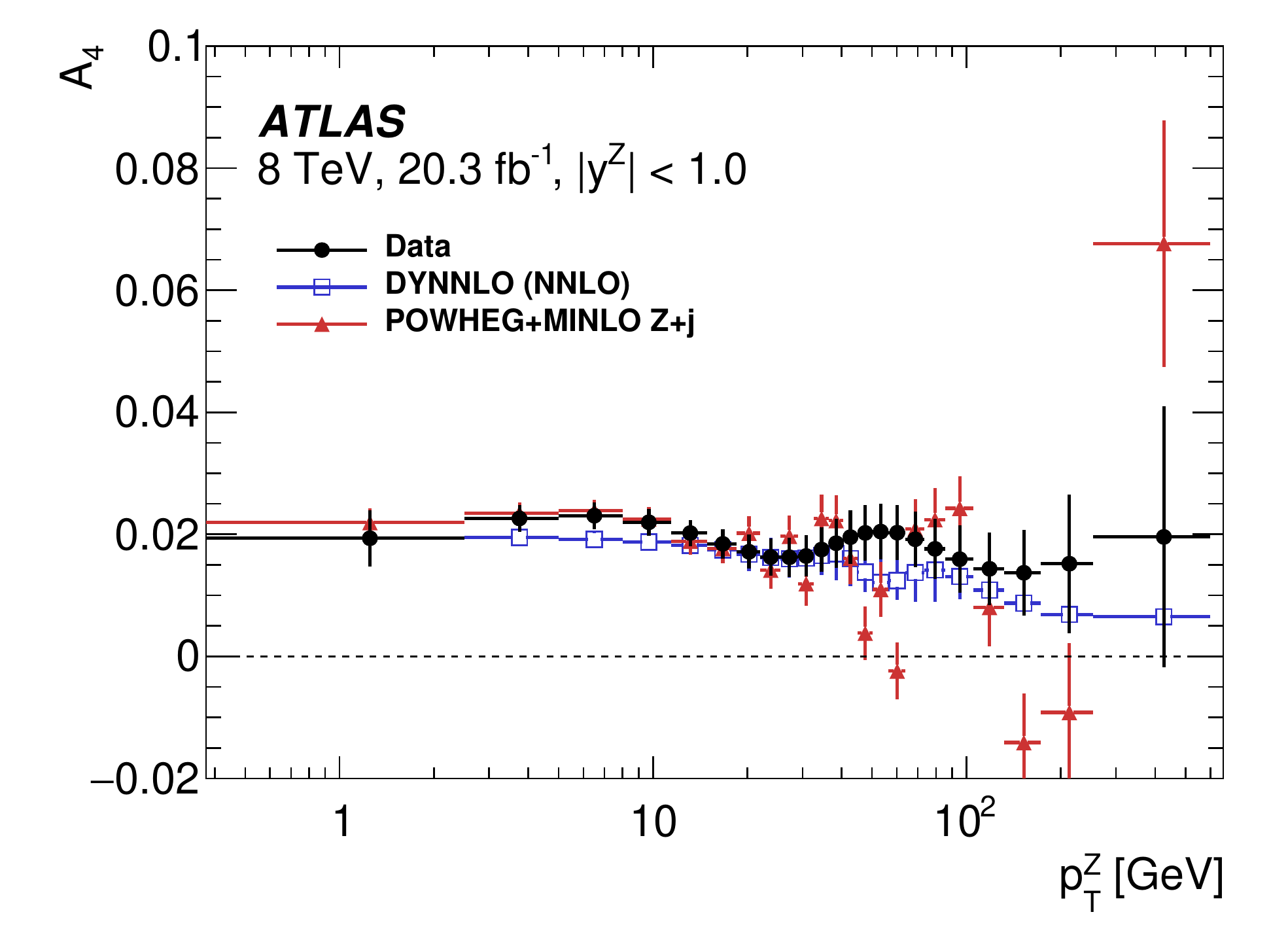}
    \includegraphics[width=7.5cm,angle=0]{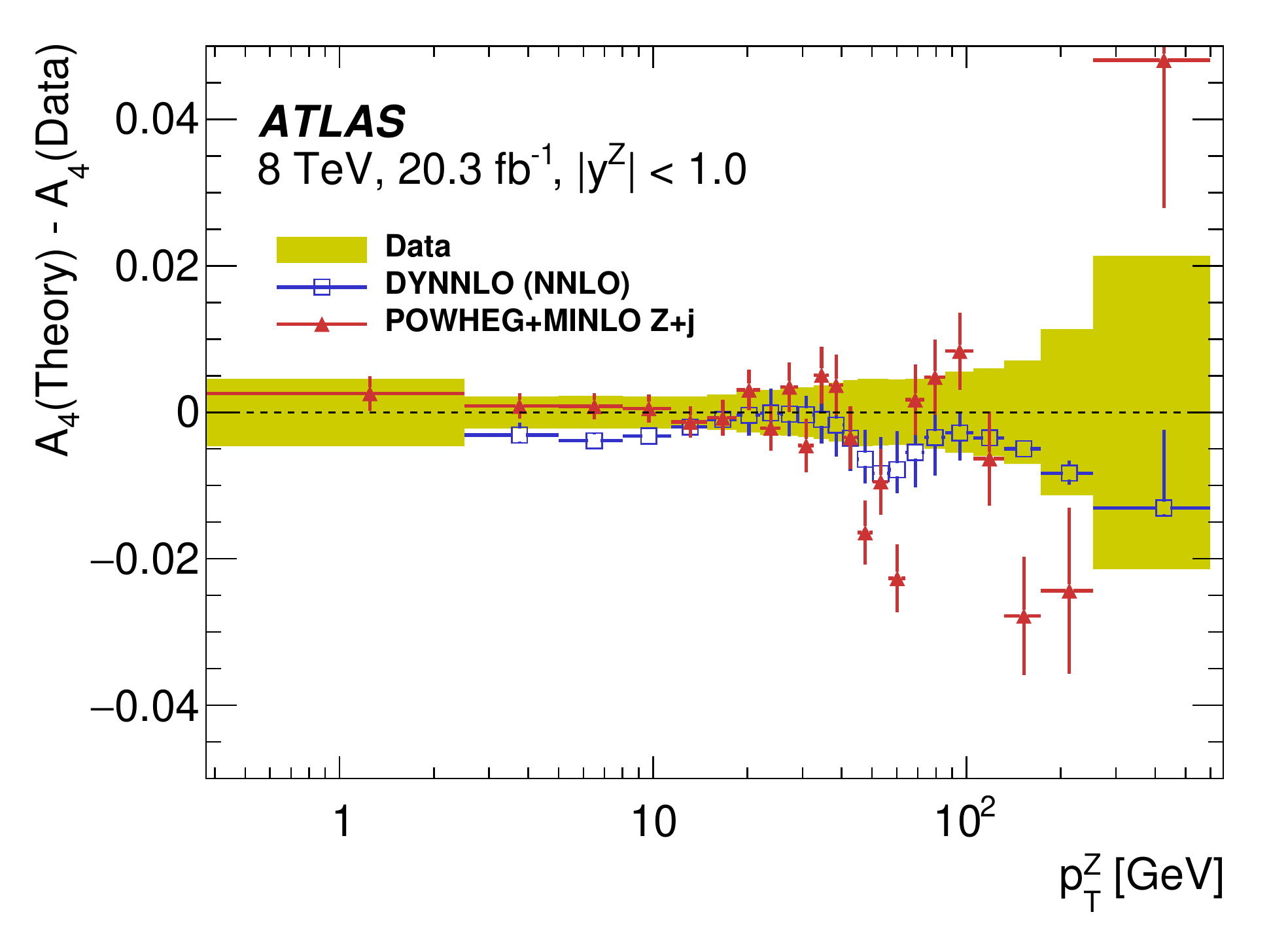}
}
\end{center}
\caption{
Distributions of the angular coefficients $A_1$ (top), $A_3$ (middle) and $A_4$ (bottom) as a function of~$\ptz$ for~$0 < |\yz| < 1$. The results from the measurements are compared to the \DYNNLO\ and \POWHEG \MINLO predictions (left).
The differences between the two calculations and the data are also shown (right), with the shaded band around zero representing the total uncertainty in the measurements. The error bars for the calculations show the total uncertainty for \DYNNLO, but only the statistical uncertainties for \POWHEG \MINLO (see text). 
\label{Fig::comp-overlays-ybin1} }
\end{figure}

\begin{figure}[hp]
  \begin{center}                               
{
    \includegraphics[width=7.5cm,angle=0]{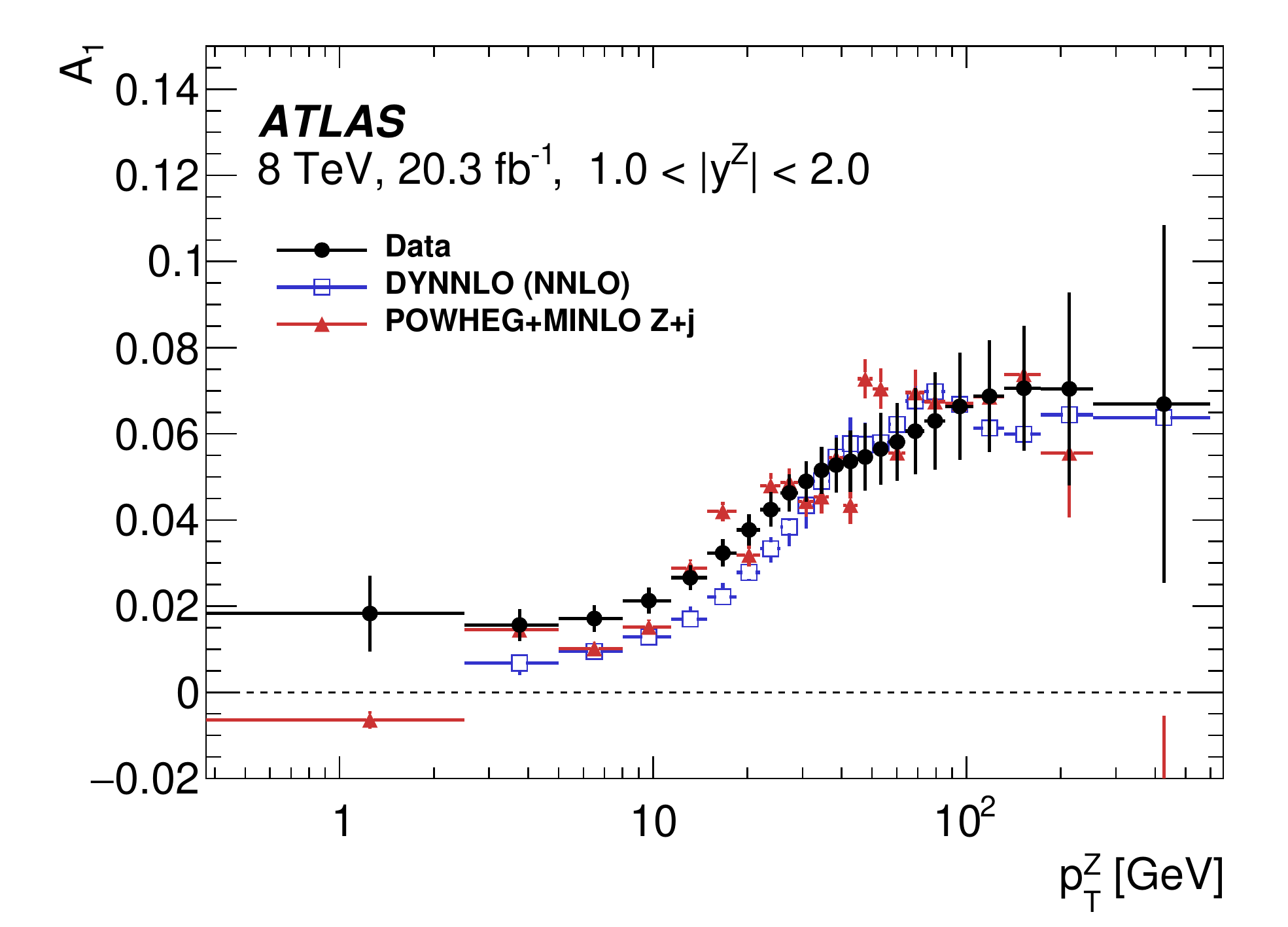}
    \includegraphics[width=7.5cm,angle=0]{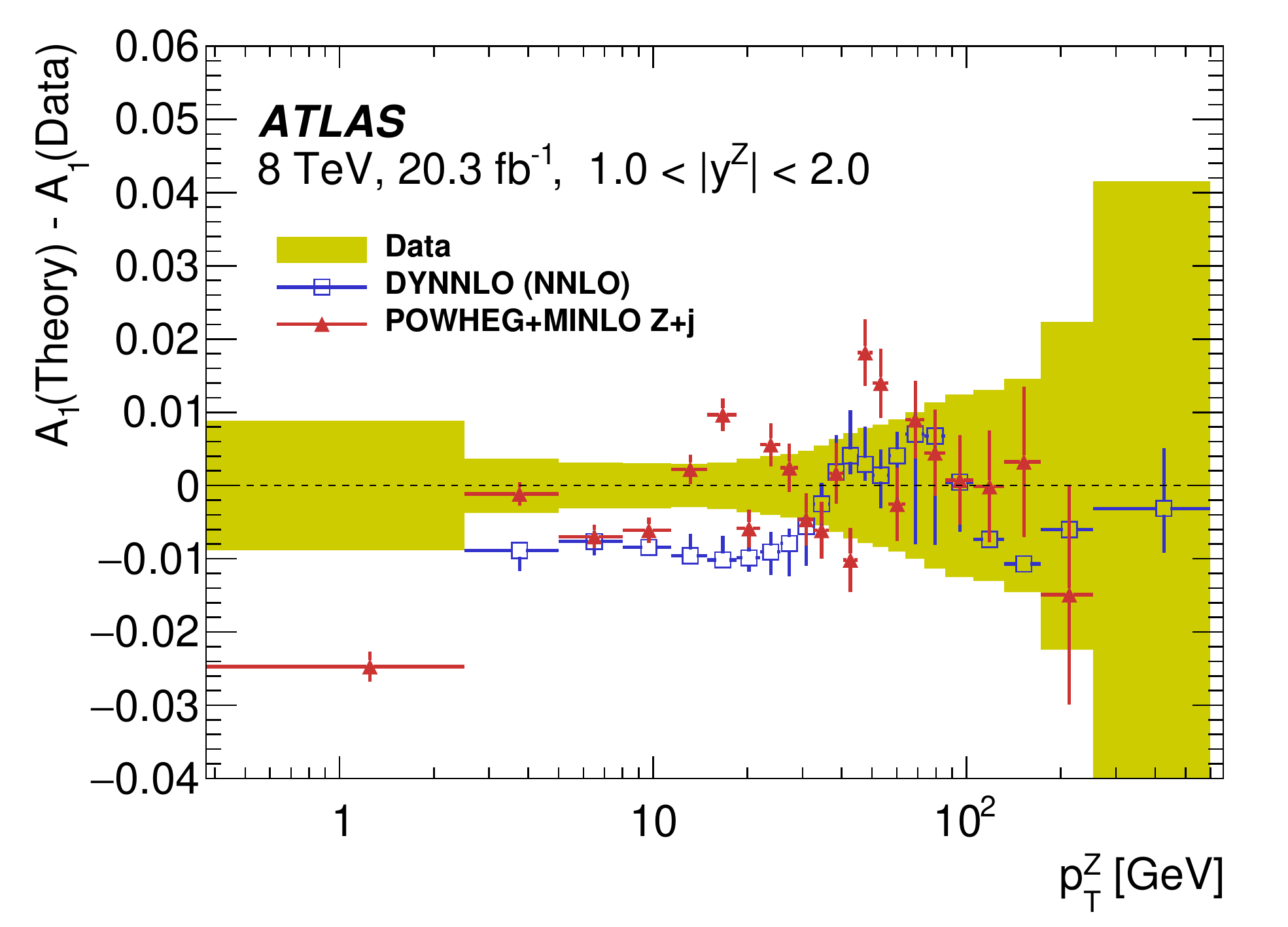}
}
{
    \includegraphics[width=7.5cm,angle=0]{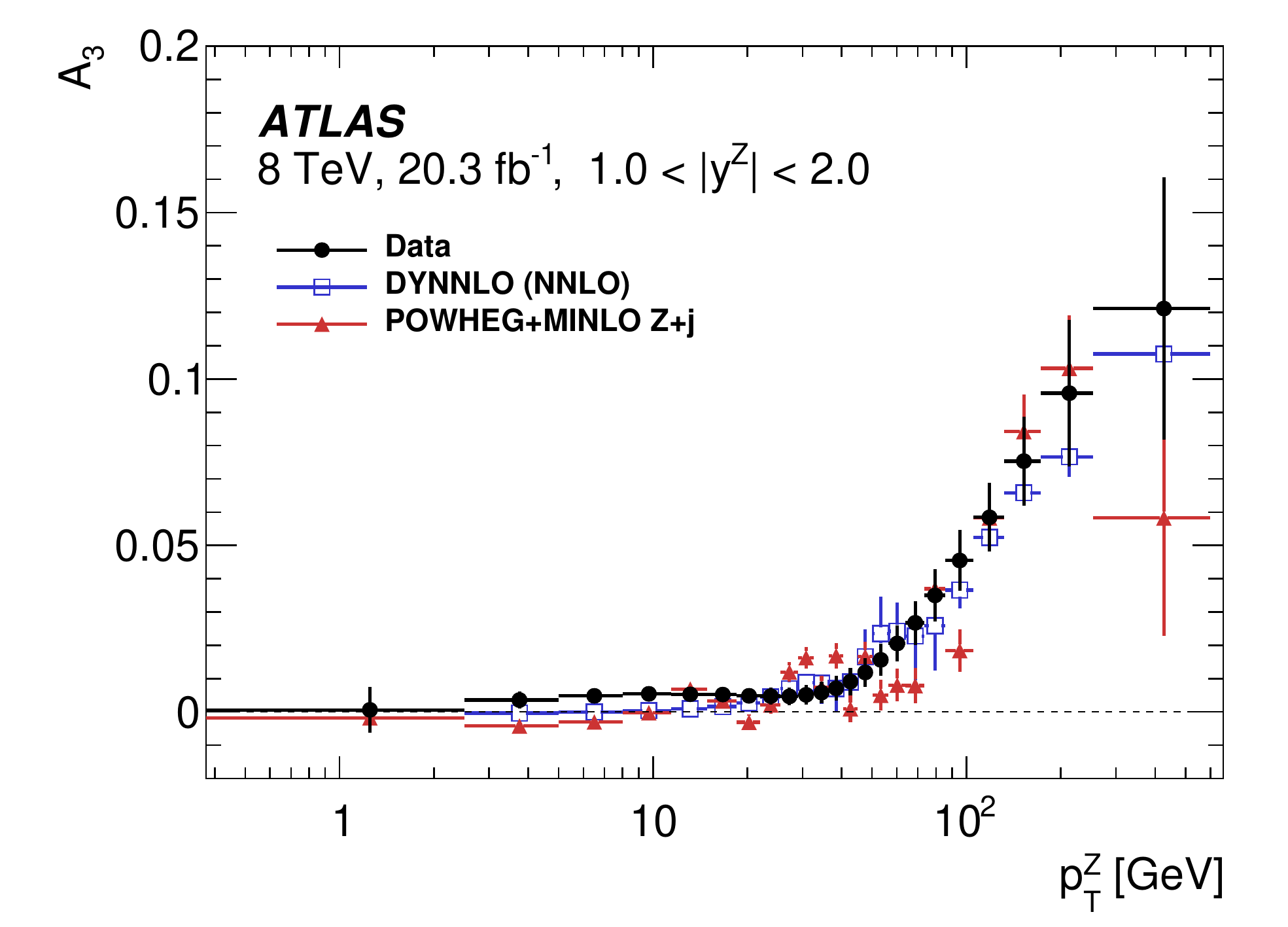}
    \includegraphics[width=7.5cm,angle=0]{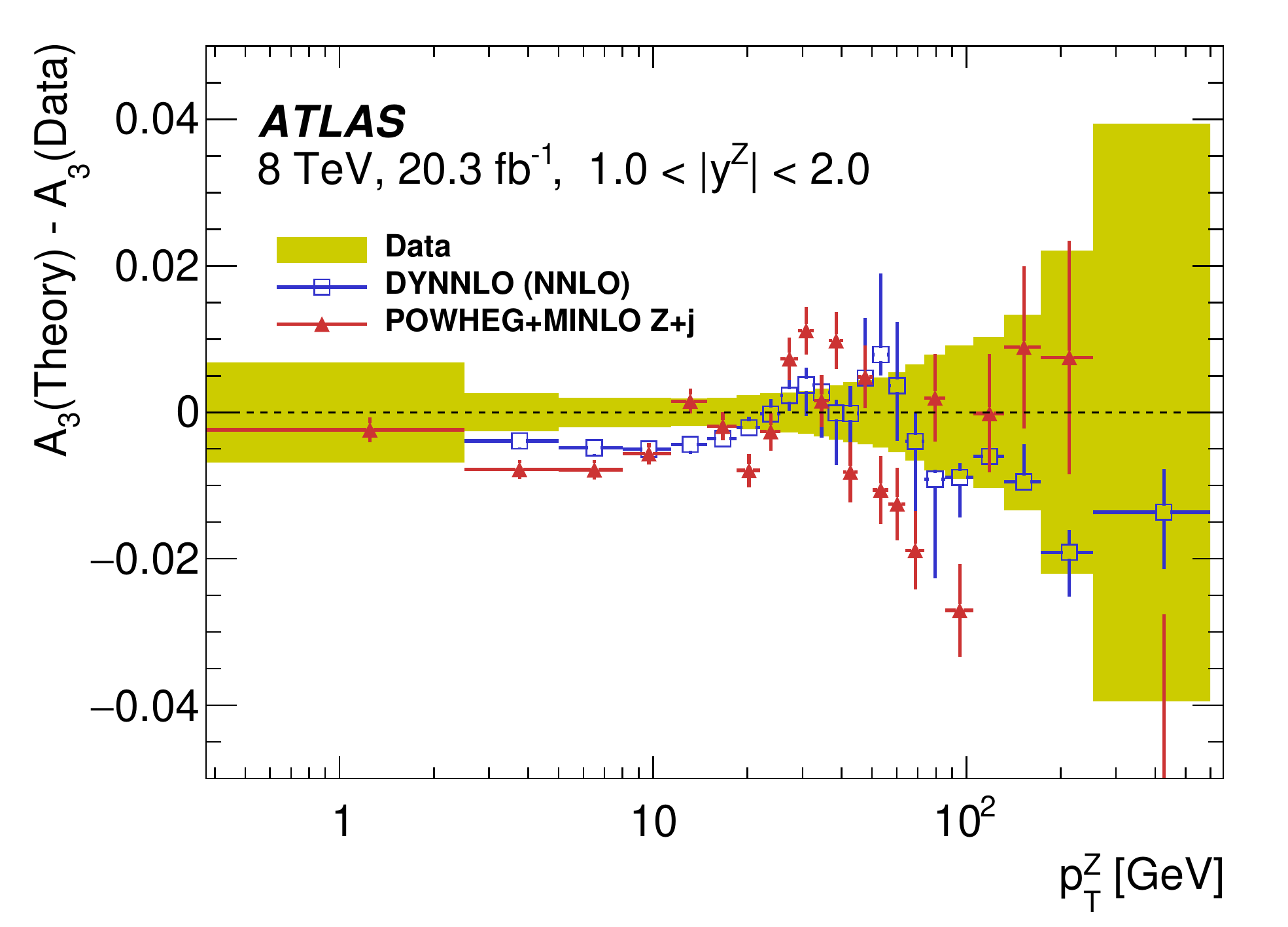}\\
}
{
    \includegraphics[width=7.5cm,angle=0]{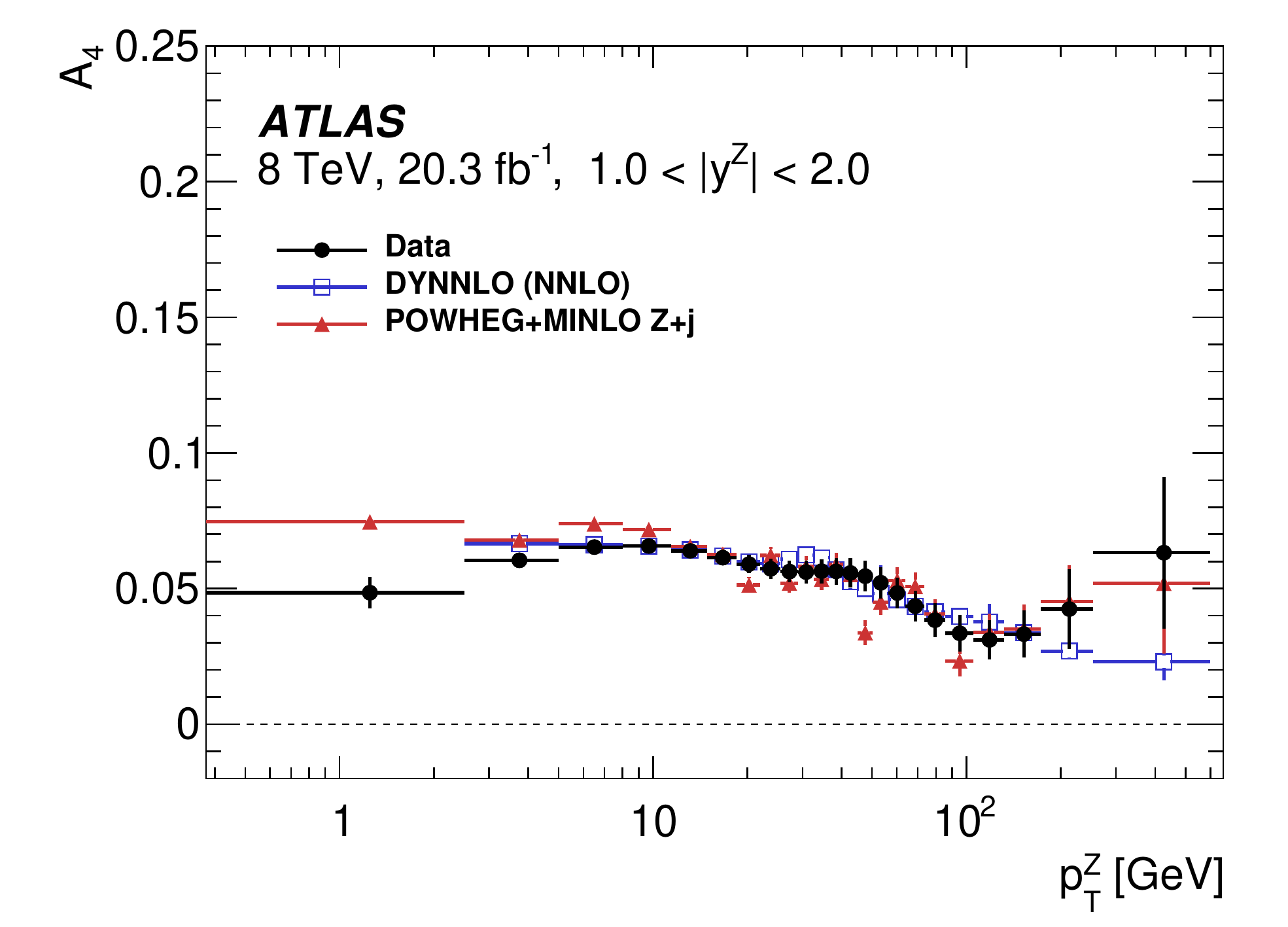}
    \includegraphics[width=7.5cm,angle=0]{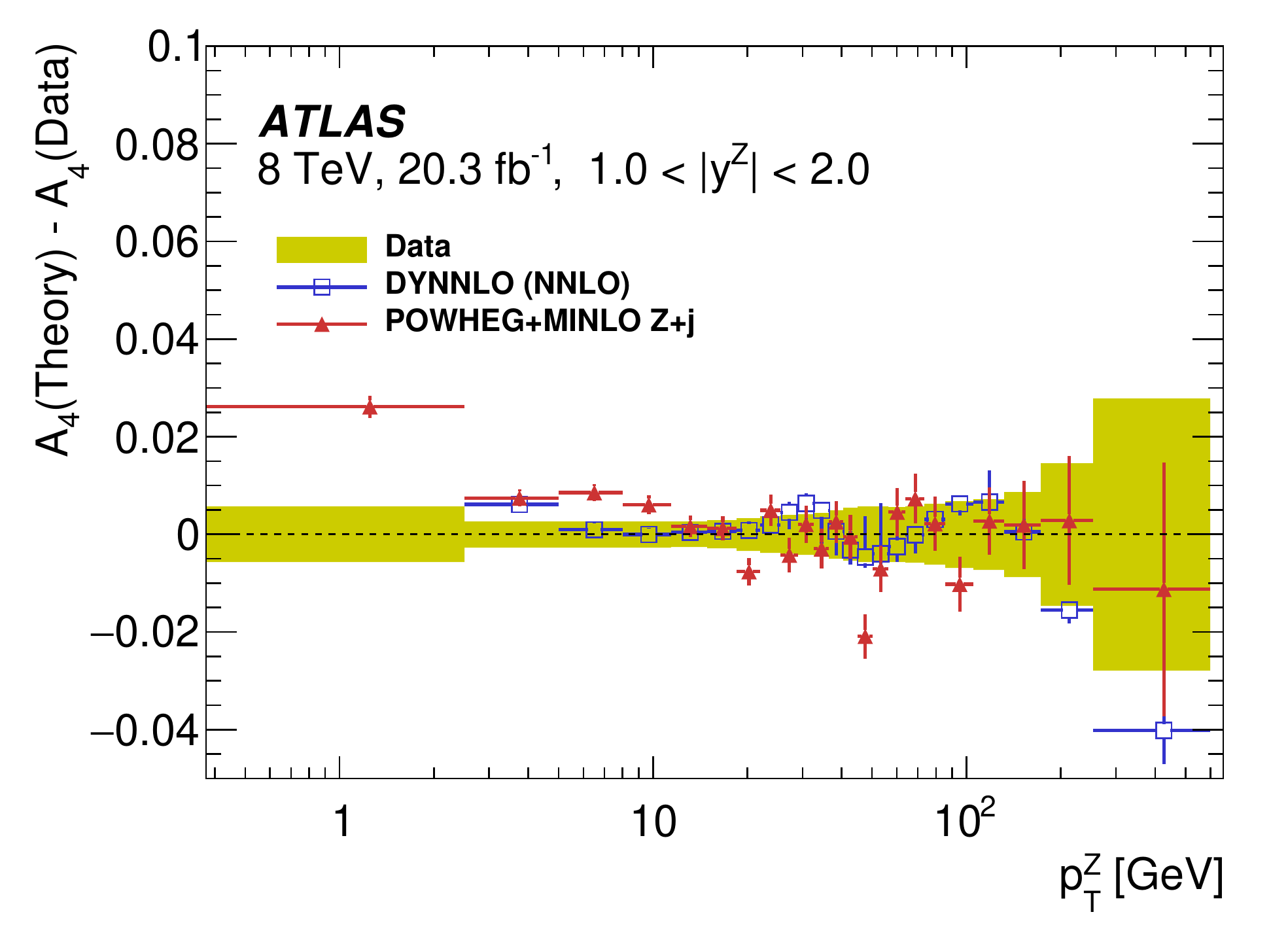}
}
\end{center}
\caption{
Distributions of the angular coefficients $A_1$ (top), $A_3$ (middle) and $A_4$ (bottom) as a function of~$\ptz$ for~$1 < |\yz| < 2$. The results from the measurements are compared to the \DYNNLO\ and \POWHEG \MINLO predictions (left).
The differences between the two calculations and the data are also shown (right), with the shaded band around zero representing the total uncertainty in the measurements. The error bars for the calculations show the total uncertainty for \DYNNLO, but only the statistical uncertainties for \POWHEG \MINLO (see text). 
\label{Fig::comp-overlays-ybin2} }
\end{figure}

\begin{figure}[hp]
  \begin{center}                               
{
    \includegraphics[width=7.5cm,angle=0]{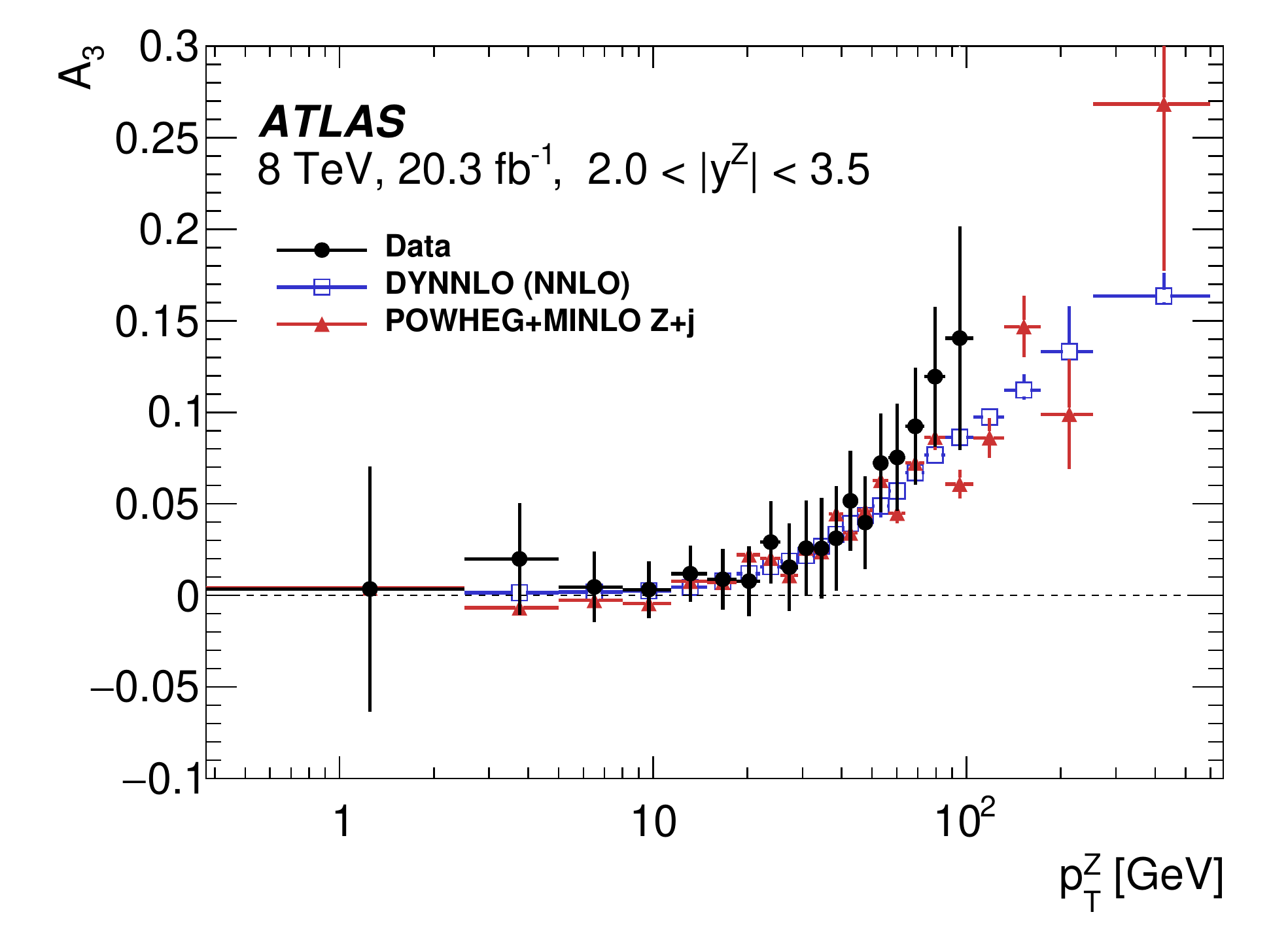}
    \includegraphics[width=7.5cm,angle=0]{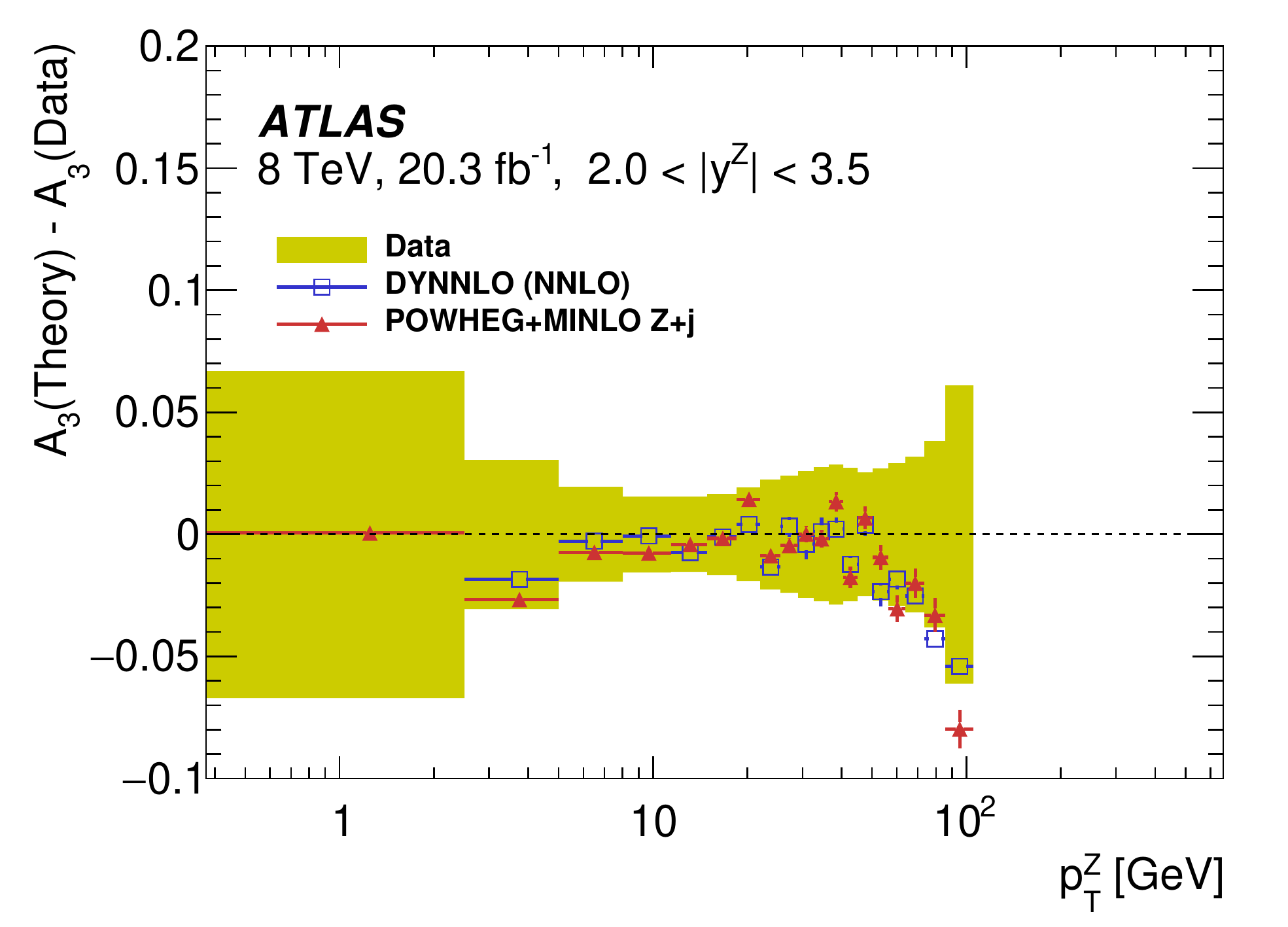}
}
{
    \includegraphics[width=7.5cm,angle=0]{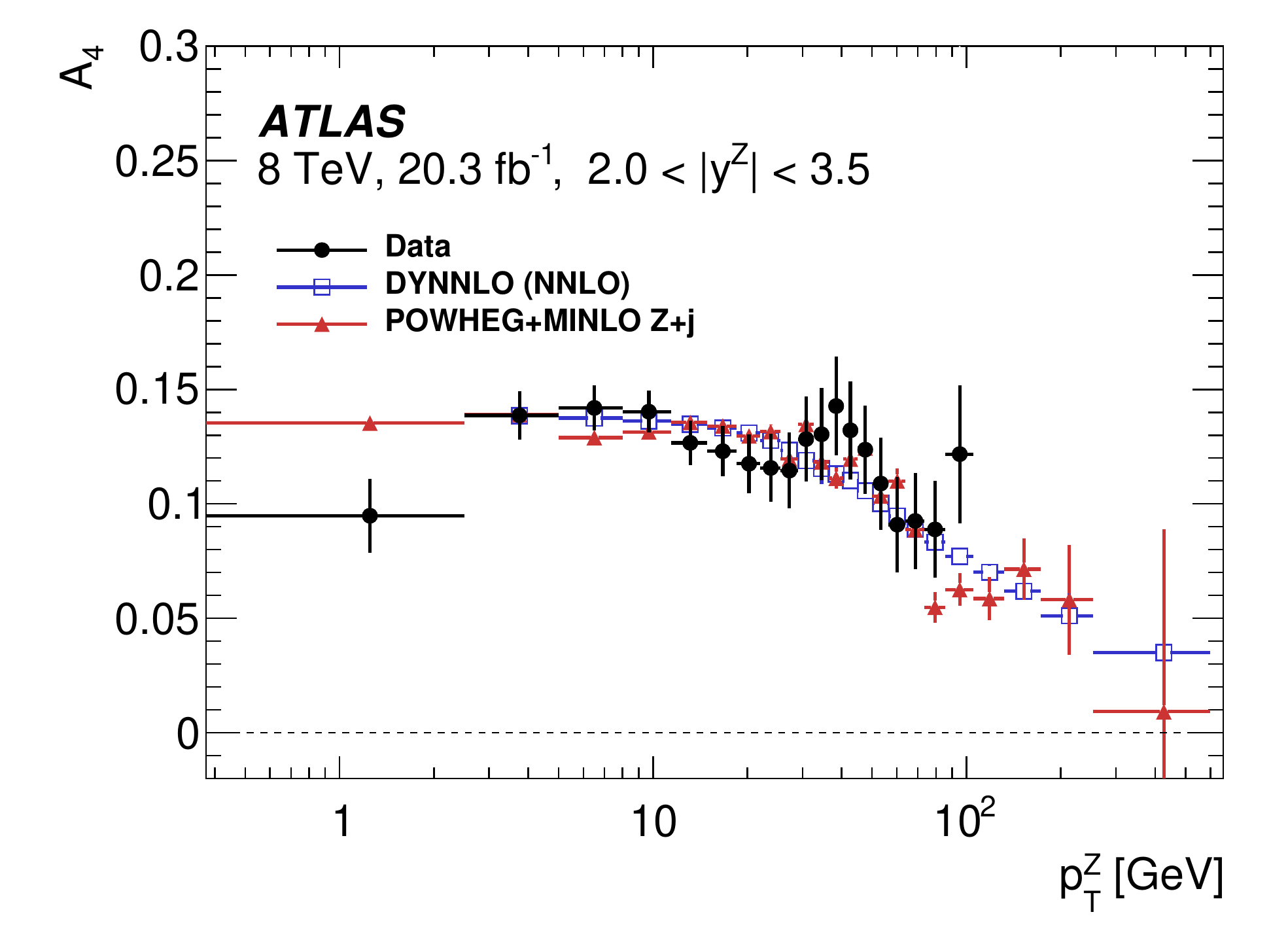}
    \includegraphics[width=7.5cm,angle=0]{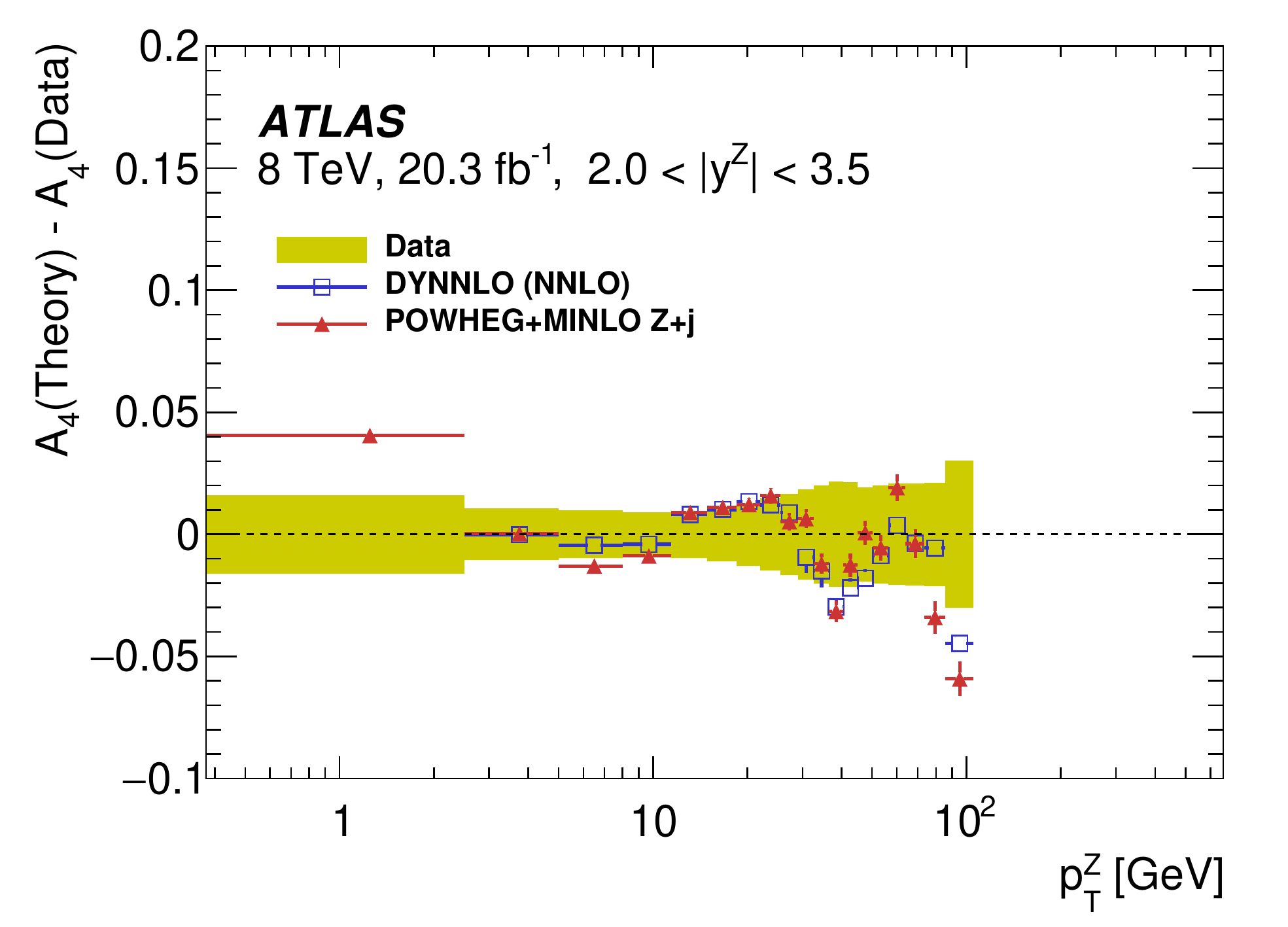}
}
\end{center}
\caption{ 
Distributions of the angular coefficients $A_3$ (top) and $A_4$ (bottom) as a function of~$\ptz$ for~$2 < |\yz| < 3.5$. The results from the measurements are compared to the \DYNNLO\ and \POWHEG \MINLO predictions (left).
The differences between the two calculations and the data are also shown (right), with the shaded band around zero representing the total uncertainty in the measurements. The error bars for the calculations show the total uncertainty for \DYNNLO, but only the statistical uncertainties for \POWHEG \MINLO (see text). 
\label{Fig::comp-overlays-ybin3} }
\end{figure}

\begin{figure}[p]
\vspace{-4mm}
  \begin{center}                               
{
    \includegraphics[width=7.5cm,angle=0]{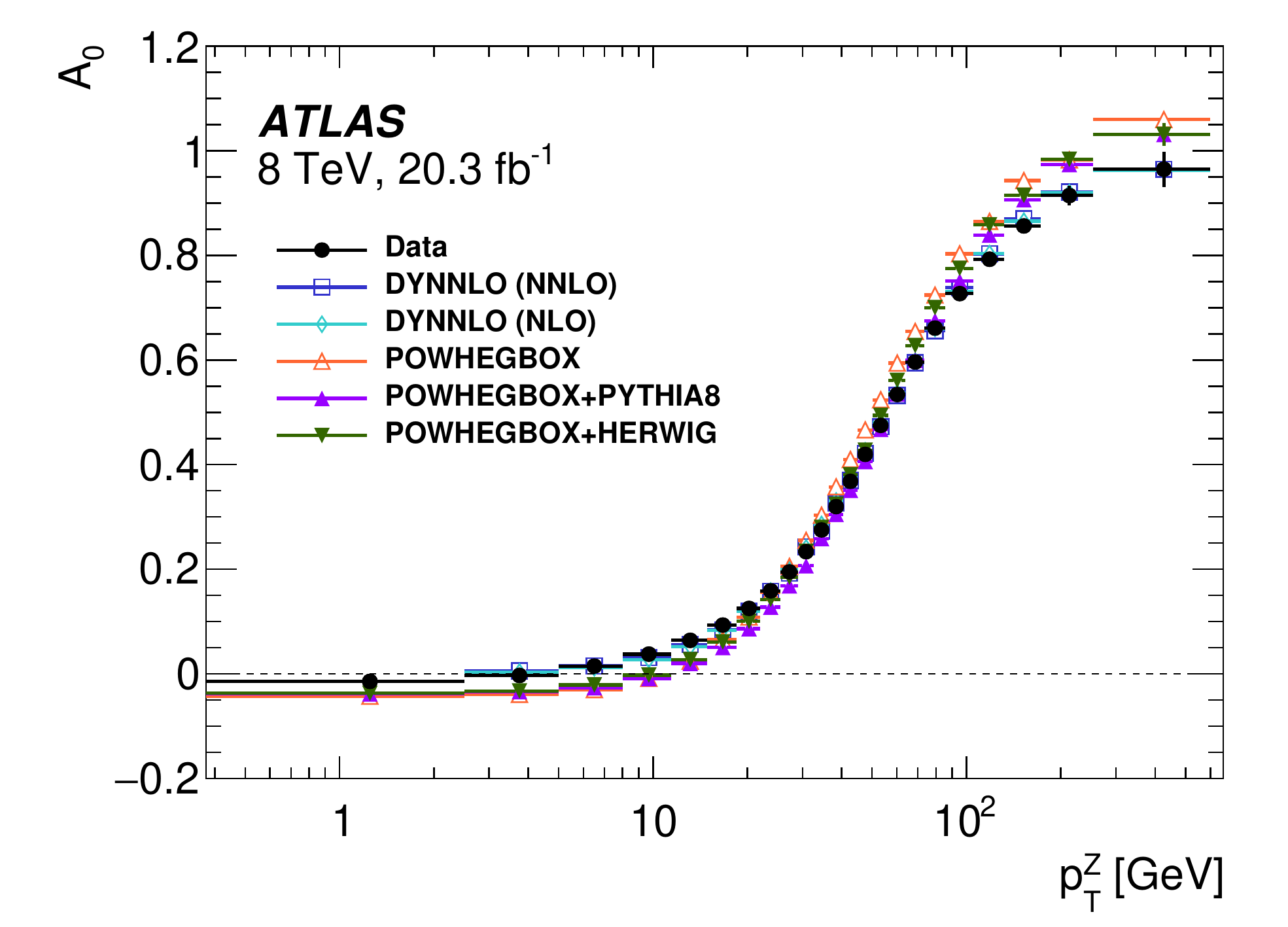}
    \includegraphics[width=7.5cm,angle=0]{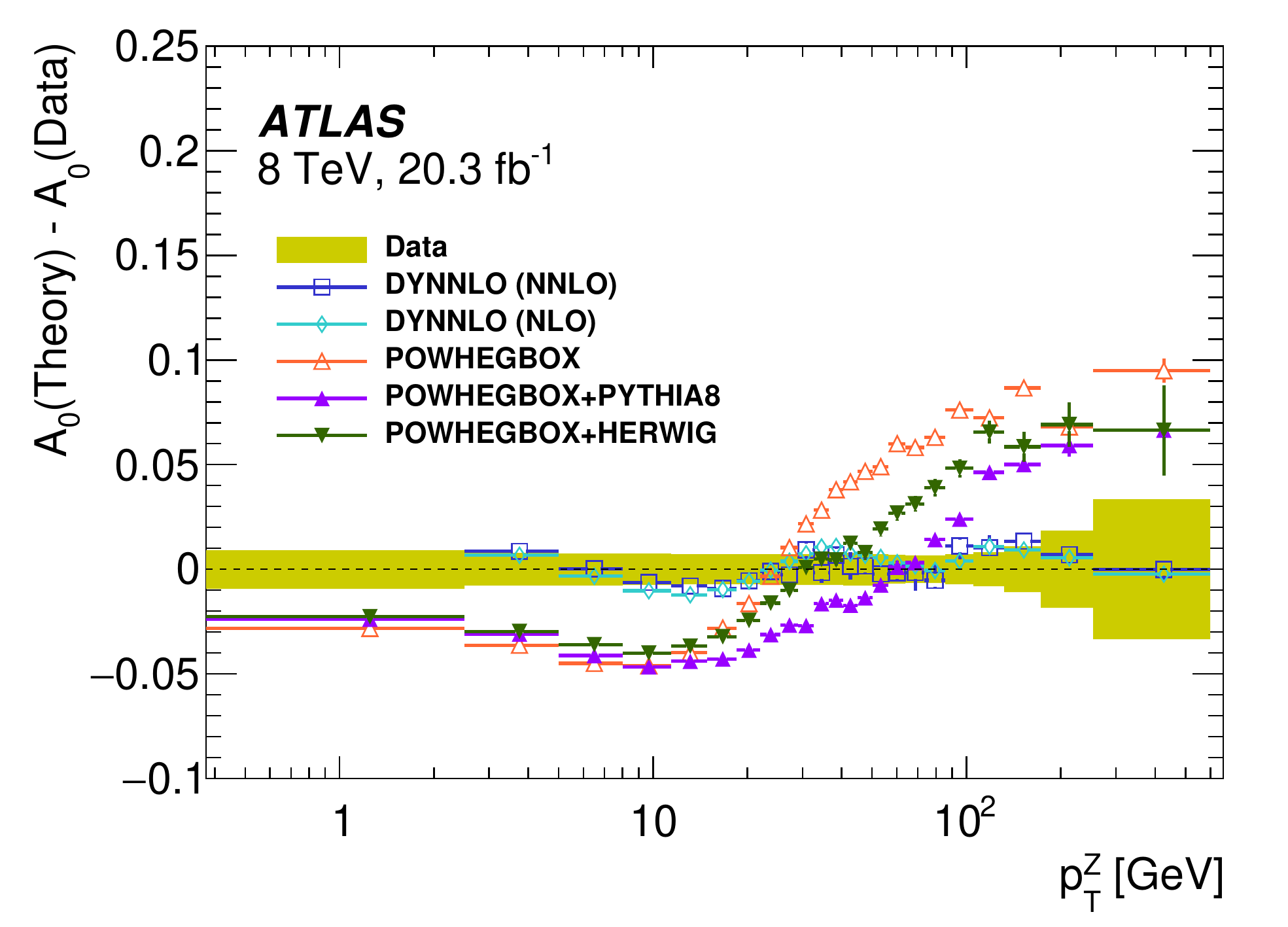}
    \includegraphics[width=7.5cm,angle=0]{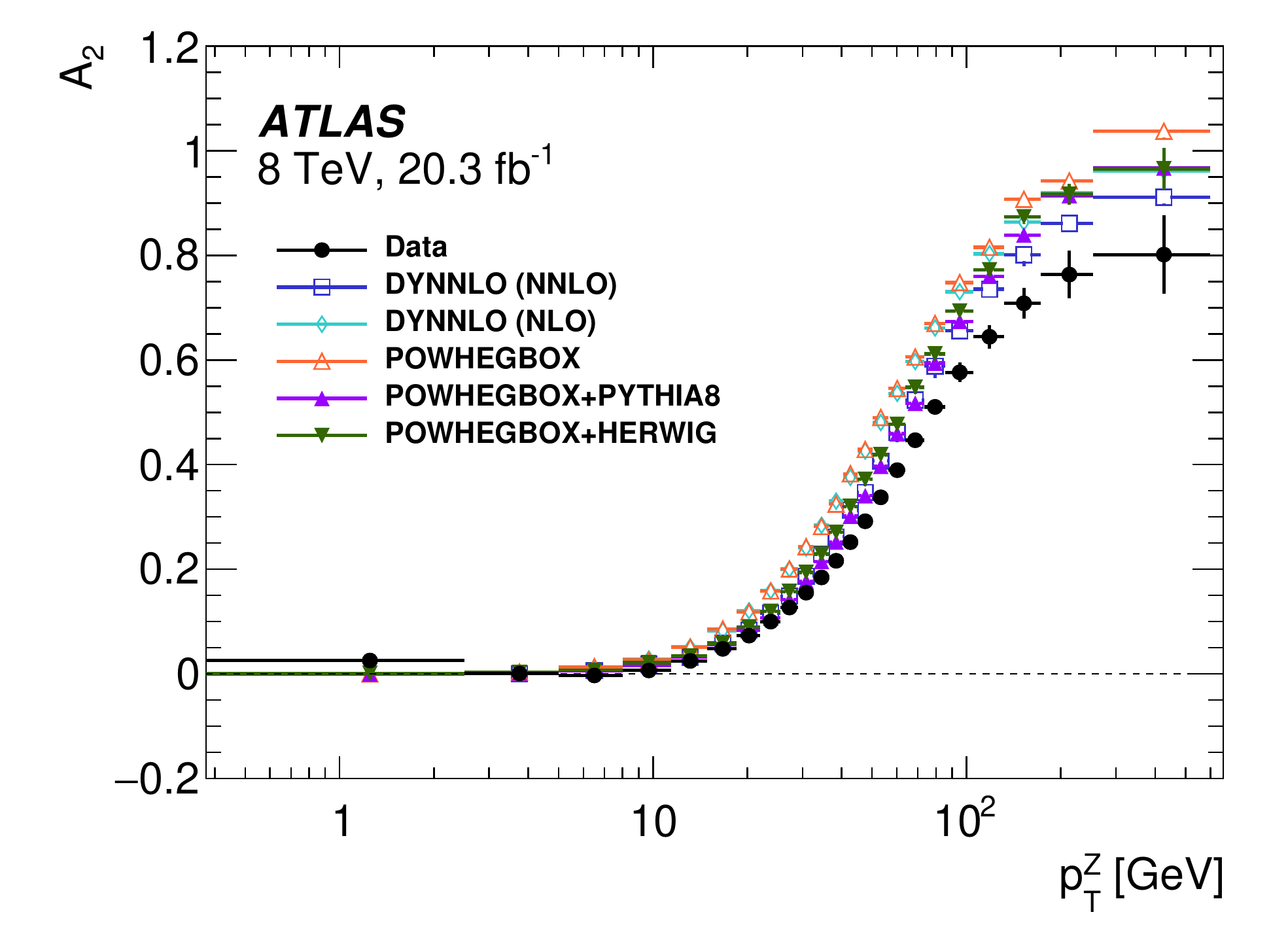}
    \includegraphics[width=7.5cm,angle=0]{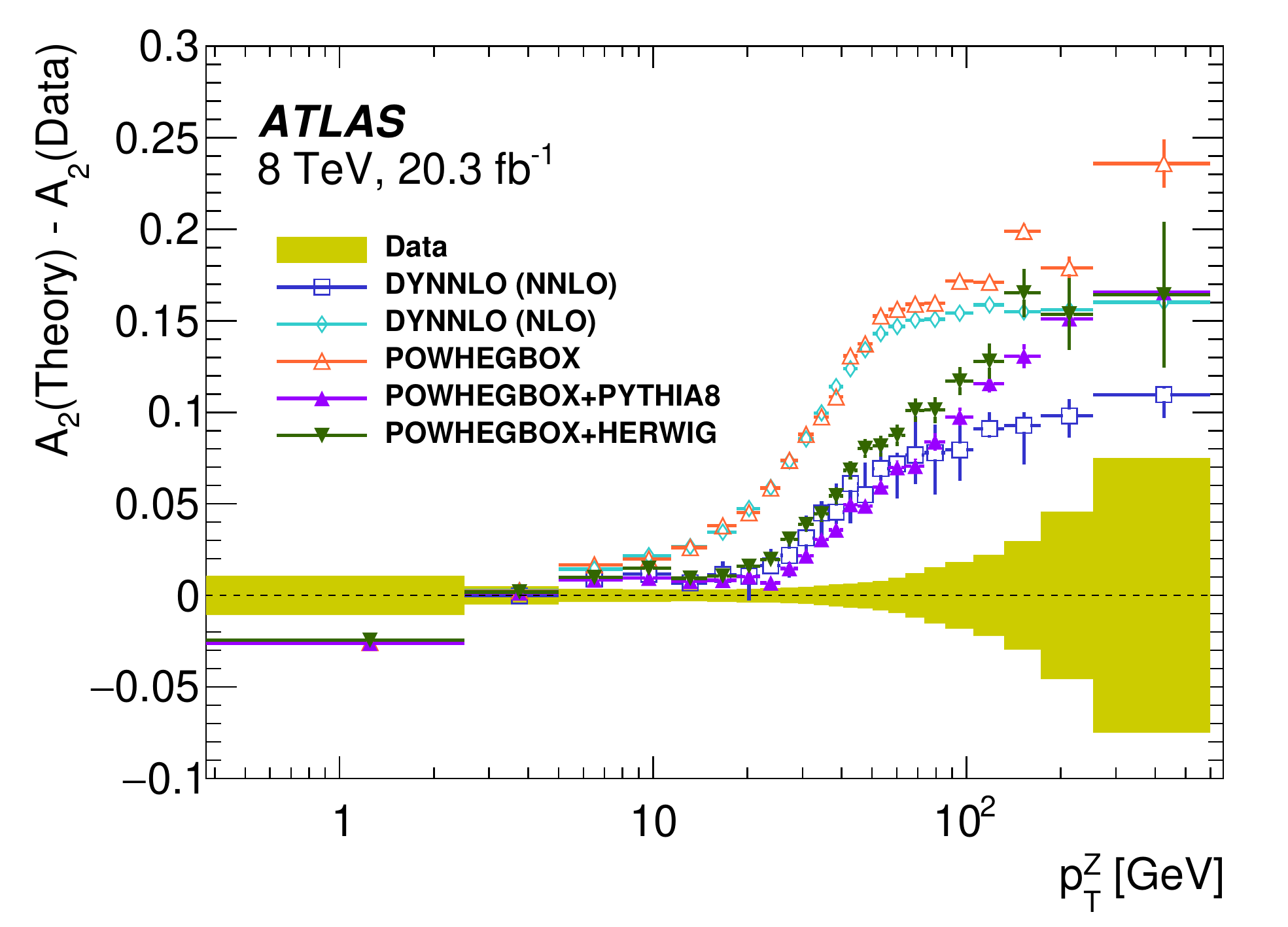}
    \includegraphics[width=7.5cm,angle=0]{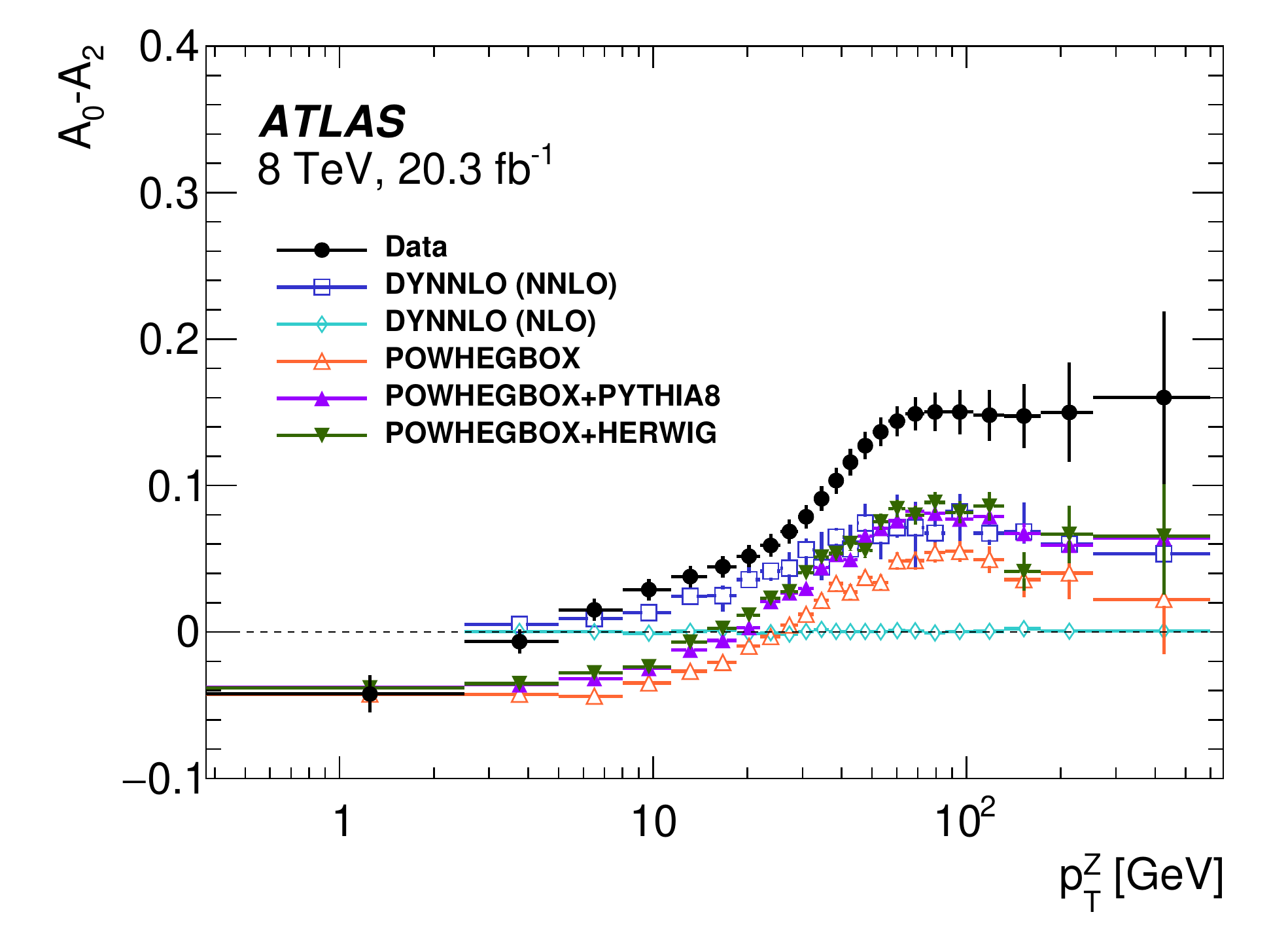}
    \includegraphics[width=7.5cm,angle=0]{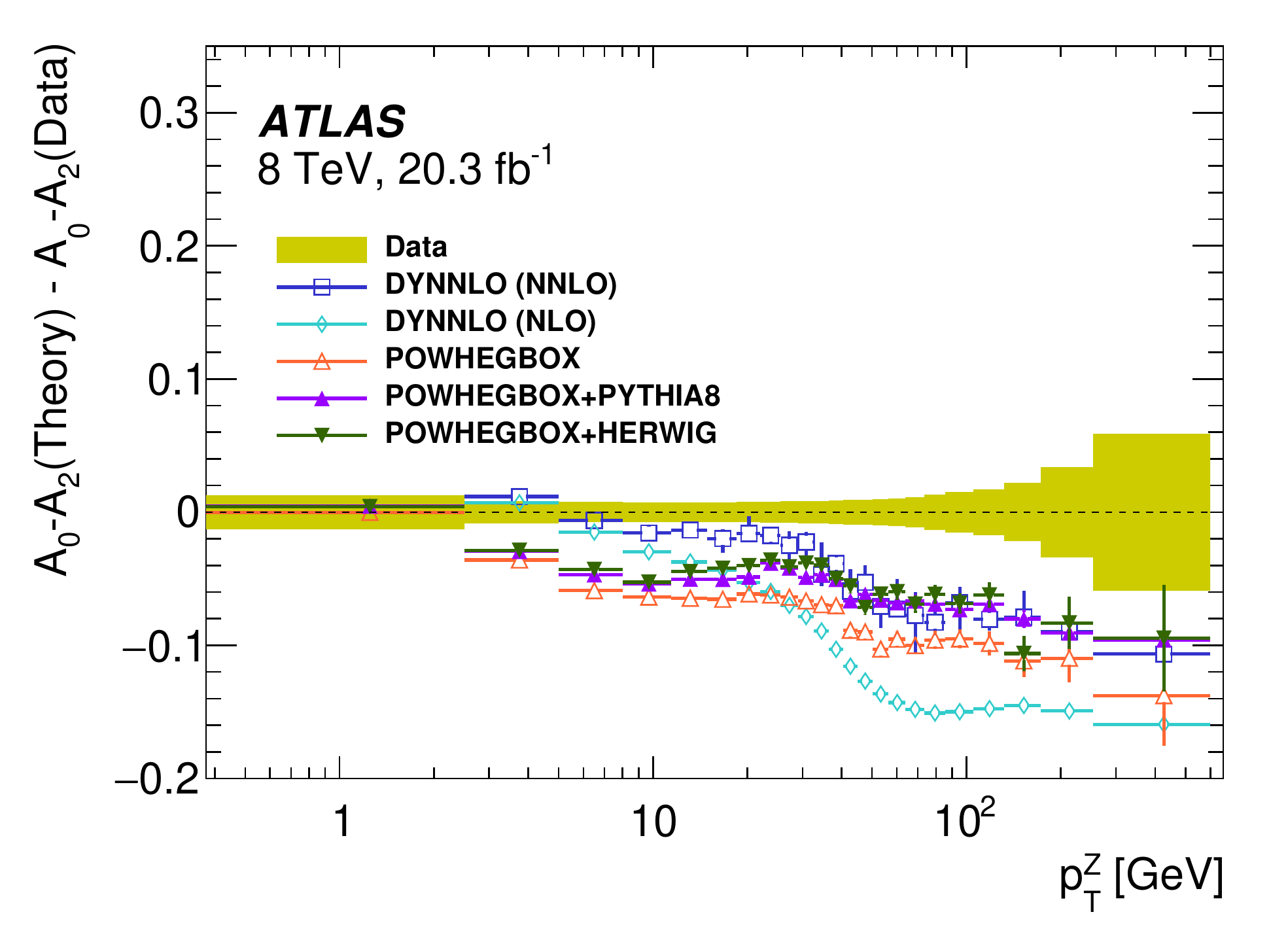}
    \includegraphics[width=7.5cm,angle=0]{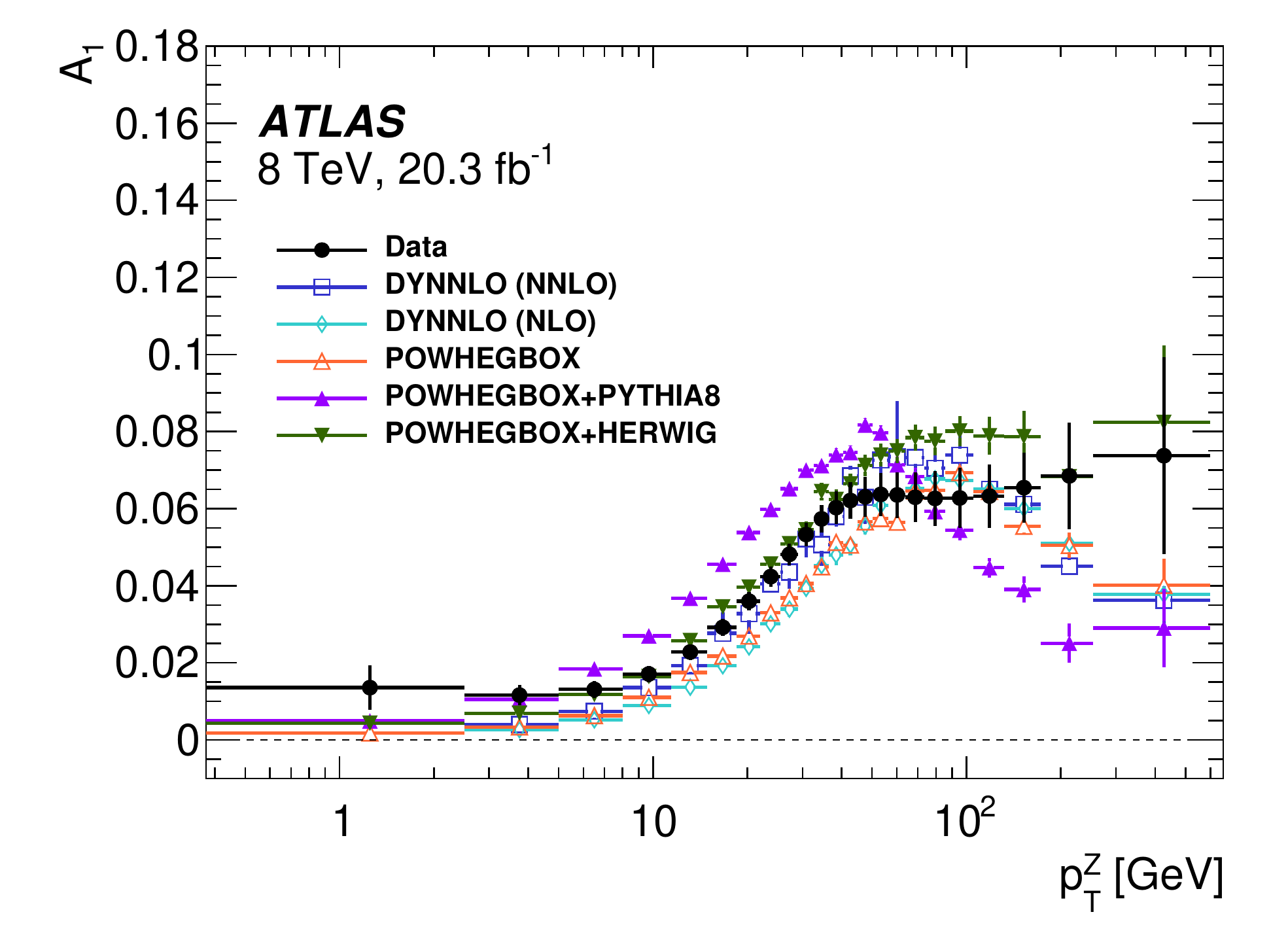}
    \includegraphics[width=7.5cm,angle=0]{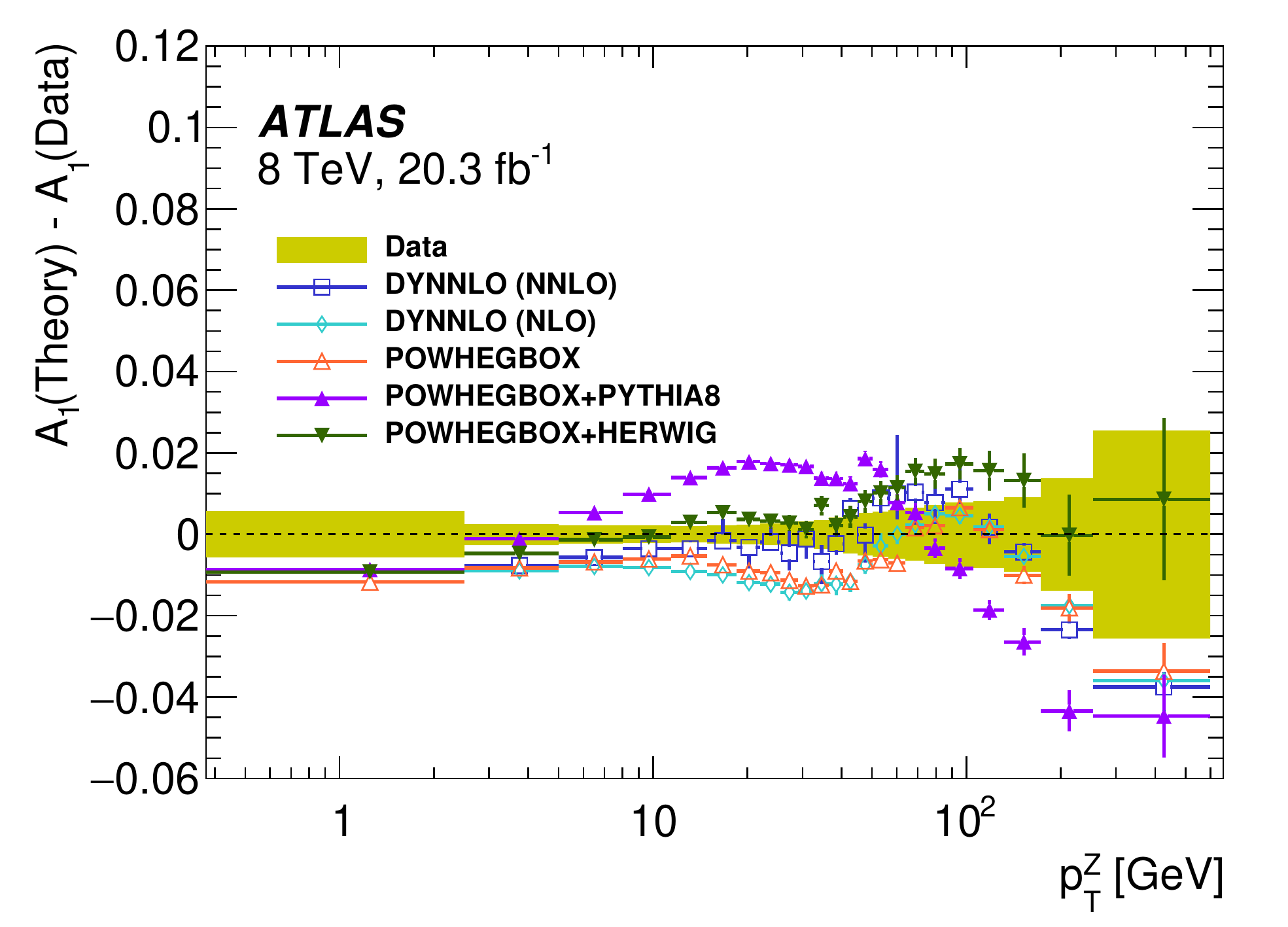}

}
\end{center}
\vspace{-8mm}
\caption{ 
Distributions of the angular coefficients $A_0$, $A_2$, $A_0-A_2$ and $A_1$ (from top to bottom) as a function of~$\ptz$. The results from the $\yz$-integrated measurements are compared to the \DYNNLO\ predictions at~NLO and at~NNLO, as well as to those from $\POWHEGBOX +~\PYTHIA$8 and $\POWHEGBOX +~\HERWIG$~(left). The differences between the calculations and the data are also shown (right), with the shaded band around zero representing the total uncertainty in the measurements. The error bars for the calculations show the total uncertainty for \DYNNLO, but only the statistical uncertainties for \POWHEGBOX. 
\label{Fig::comp-overlay-powheg} }
\end{figure}

\begin{figure}[p]
  \begin{center}                               
{
    \includegraphics[width=7.5cm,angle=0]{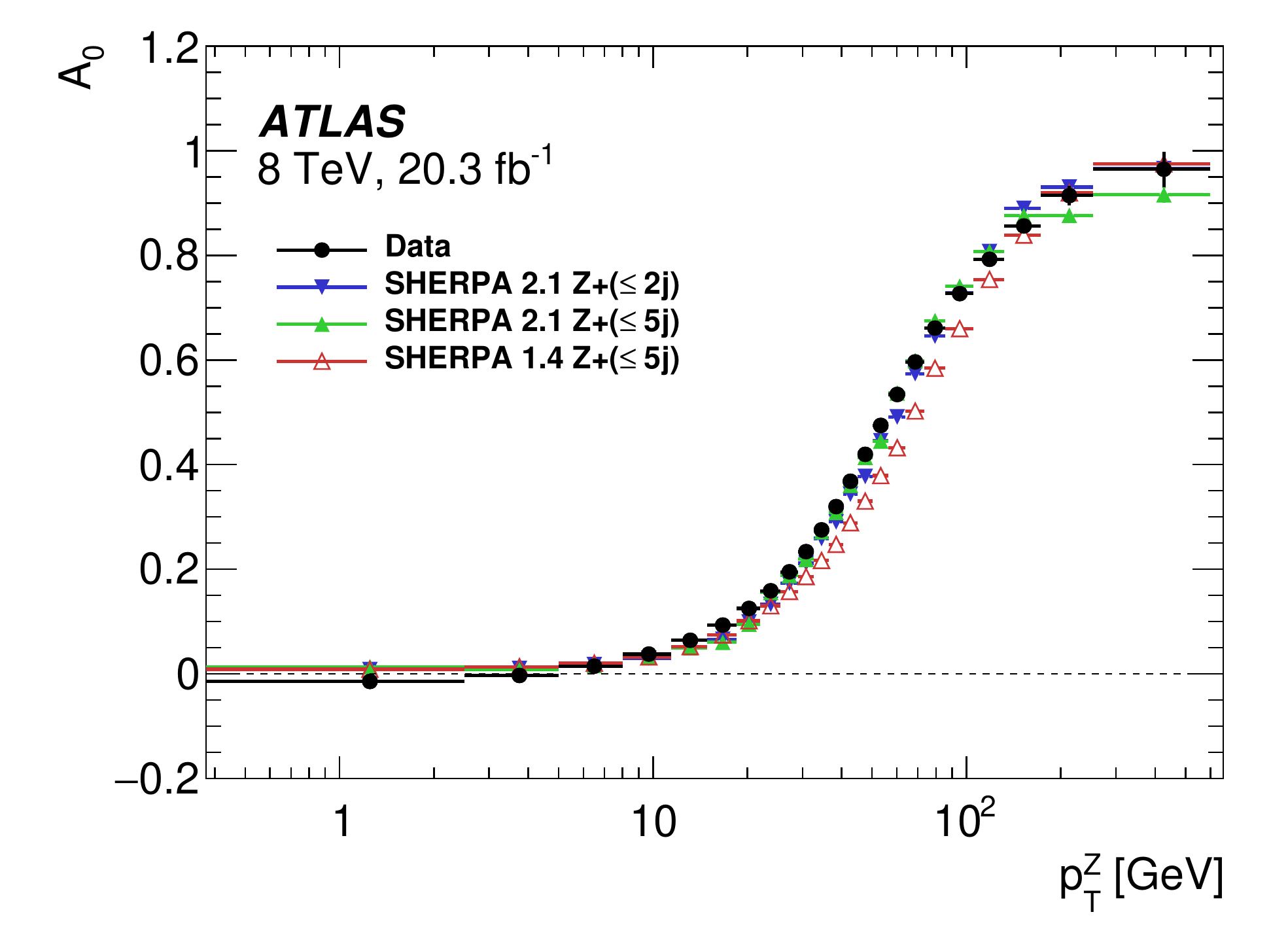}
    \includegraphics[width=7.5cm,angle=0]{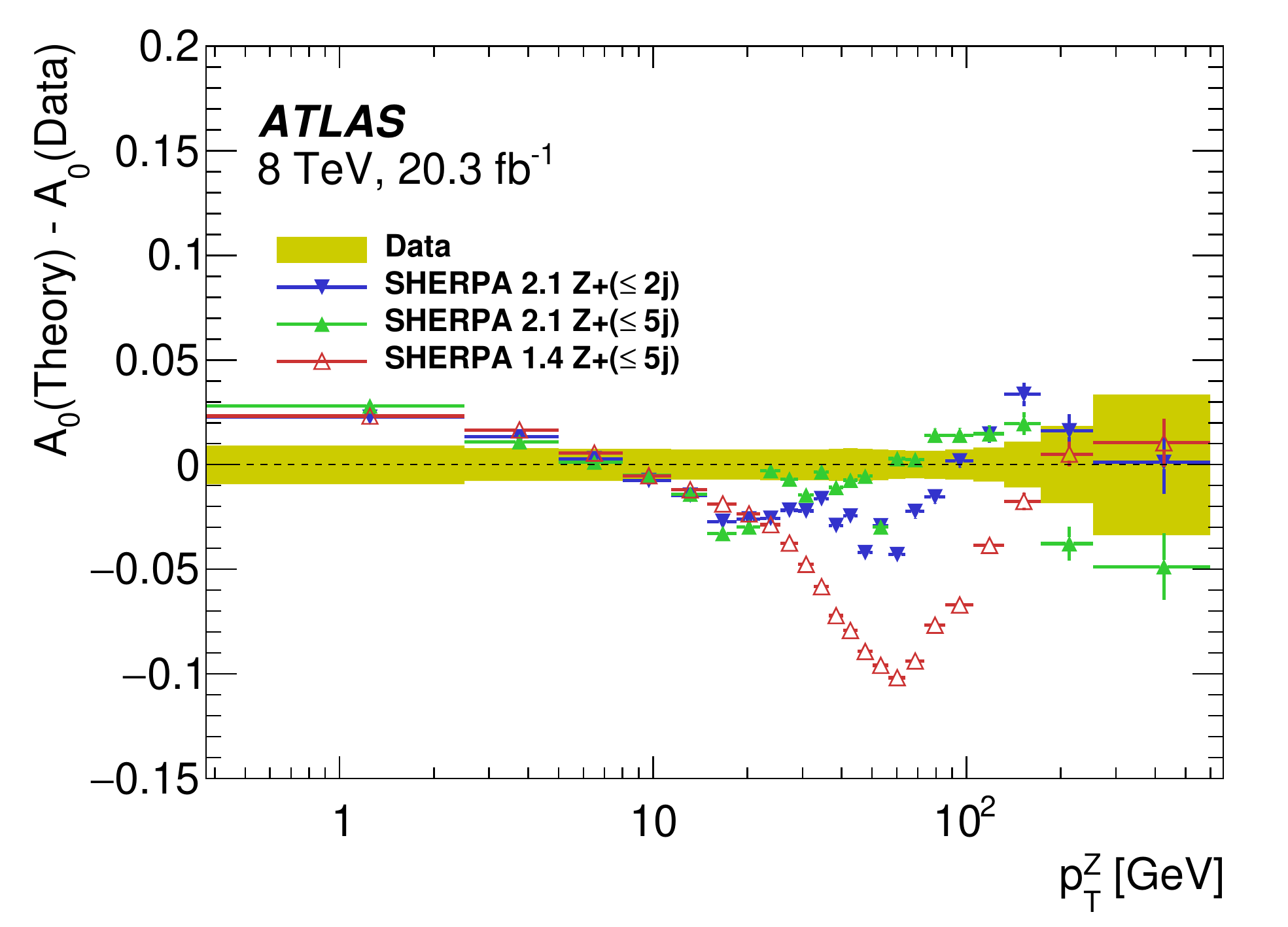}
    \includegraphics[width=7.5cm,angle=0]{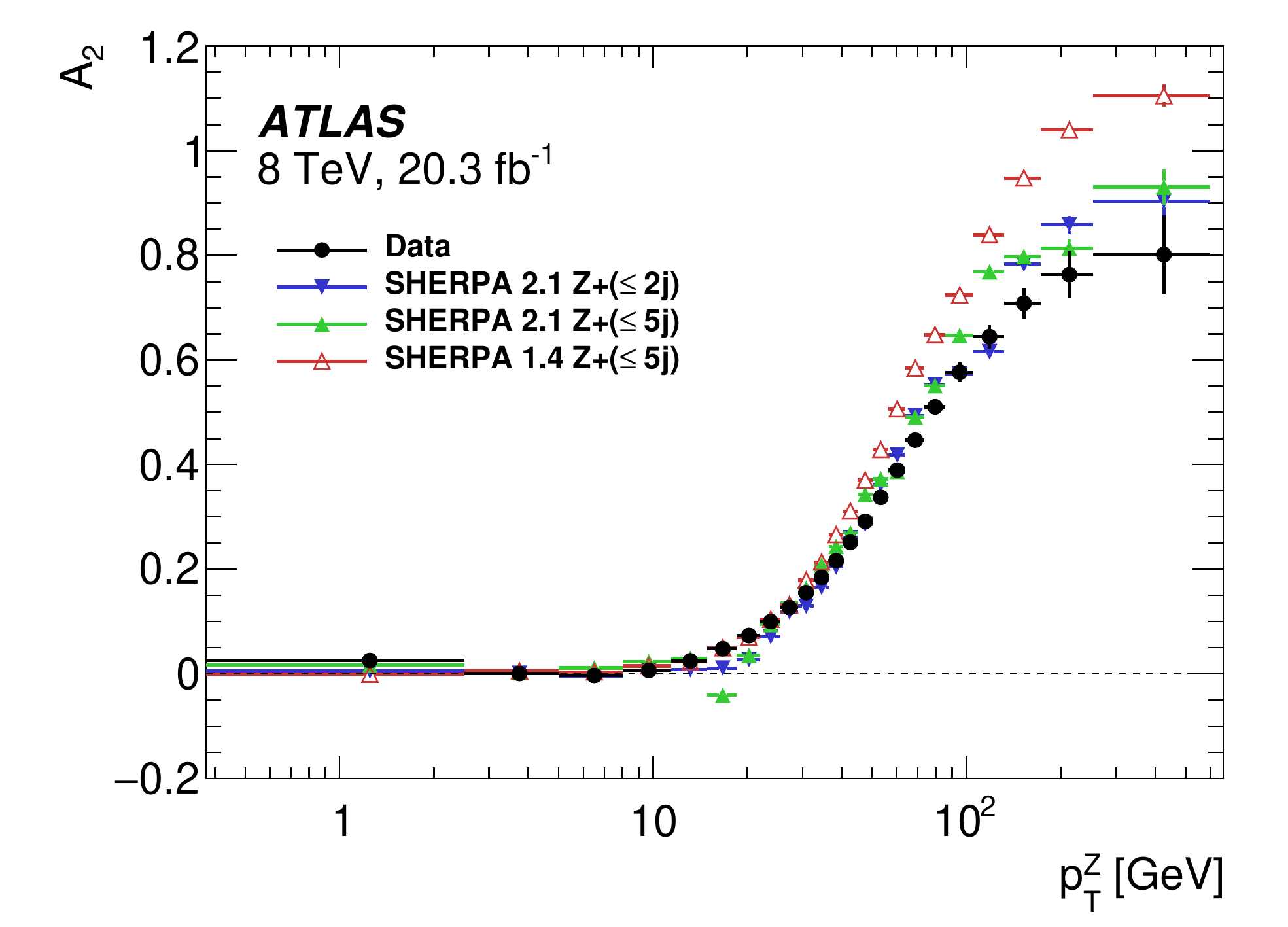}
    \includegraphics[width=7.5cm,angle=0]{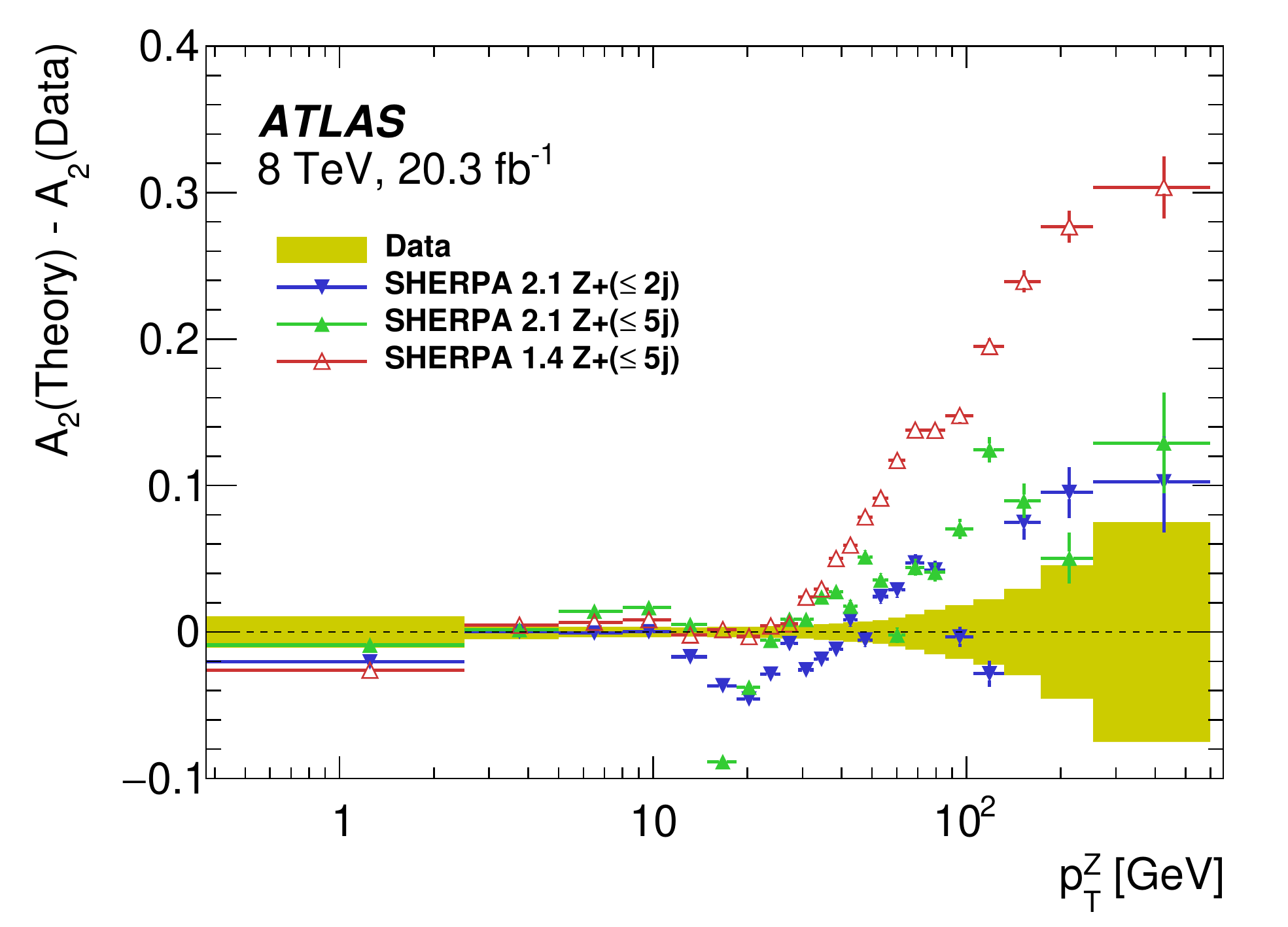}
    \includegraphics[width=7.5cm,angle=0]{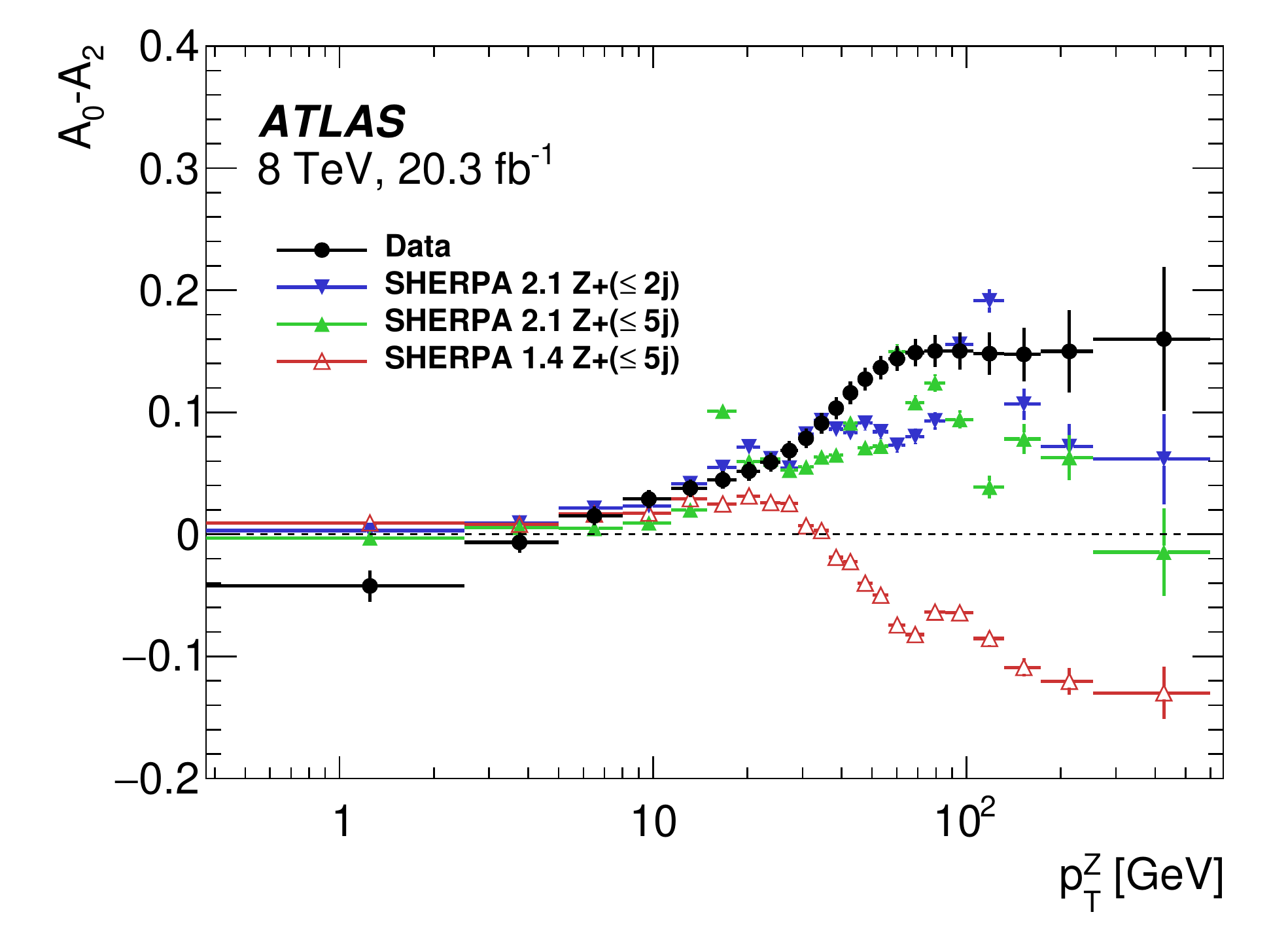}
    \includegraphics[width=7.5cm,angle=0]{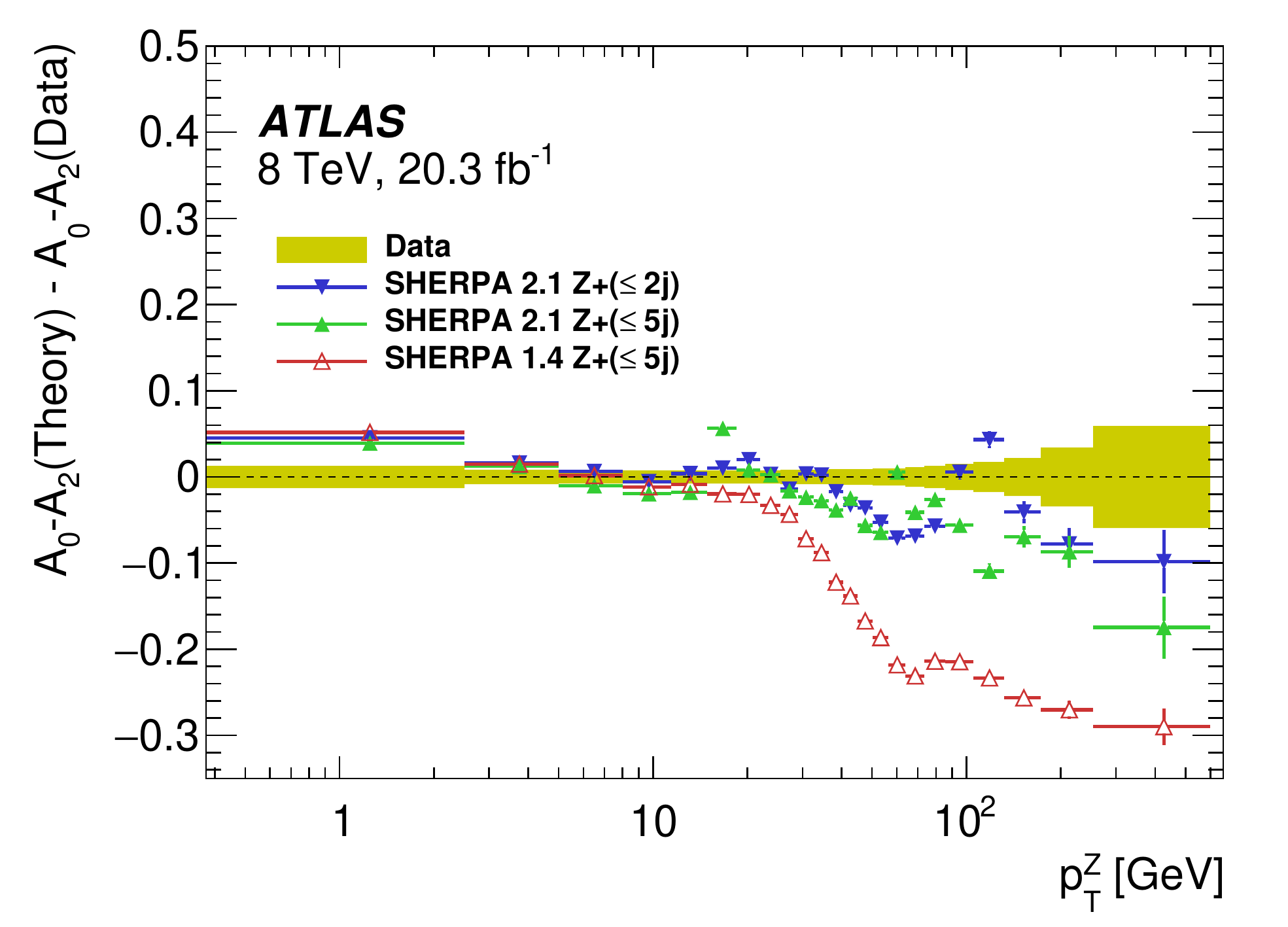}

}
\end{center}
\caption{  
Distributions of the angular coefficients $A_0$ (top), $A_2$ (middle) and $A_0-A_2$ (bottom) as a function of~$\ptz$. The results from the $\yz$-integrated measurements are compared to various predictions from the \SHERPA~event generator~(left). The differences between the calculations and the data are also shown (right), with the shaded band around zero representing the total uncertainty in the measurements. The error bars for the \SHERPA~predictions represent only the statistical uncertainties. 
\label{Fig::comp-overlay-sherpa} }
\end{figure}



\clearpage
\section{Summary}
\label{sec:conclusions}

This paper presents a precise set of measurements of the $Z$-boson production dynamics in the $Z$-boson pole region, through the angular distributions of the leptons. The data analysed correspond to 20.3~\ifb of $pp$ collisions at~$\sqrt{s} = 8$~TeV, collected by the~ATLAS detector at the CERN~LHC. The measurements are obtained as a function of~$\ptz$, integrated over~$\yz$ and in bins of~$\yz$, covering almost the full range of~$\yz$ spanned by $Z$-boson production at~$\sqrt{s} = 8$~TeV. This is made possible by exploiting the decomposition of the production cross-section into nine terms, where in each term the angular coefficients that encapsulate the production dynamics are factorised from the decay dynamics described by angular polynomials. Templates of the nine polynomials folded to the detector level are fitted to the data to extract the angular coefficients in the full phase space of the \Zboson~boson.

Over most of the phase space, the measurements that are obtained from samples of electron and muon pairs covering respectively the ranges $0 <  |\yz| < 3.5$ and~$0 <  |\yz| < 2.5$ are limited only by statistical uncertainties in the data. These uncertainties are small and range from~0.002 at low~$\ptz$ to~0.008 at~$\ptz~=$~150~GeV. The experimental systematic uncertainties are much smaller in almost all cases. The theory systematic uncertainties are minimised through the template-building procedure, such that the PDF~uncertainties, which are the dominant source of theoretical uncertainties, are below~0.004 in all cases.

The measurements, when compared to theoretical calculations and to predictions from MC~generators, are precise enough to probe QCD corrections beyond the formal accuracy of the calculations. A significant deviation from the $\mathcal{O}(\alpha^{2}_{\text{s}})$ predictions from~DYNNLO is observed for $A_{0}-A_{2}$, indicating that higher-order QCD corrections are required to describe the data. Evidence at the~3$\sigma$~level is found for non-zero $A_{5,6,7}$ coefficients, consistent with expectations from~DYNNLO at~$\mathcal{O}(\alpha^{2}_{\text{s}})$. The measurements also provide discrimination between various event generators, in particular in terms of the related implementation of different parton-shower models.

The measurements of the~$A_i$ coefficients,  in particular through the correlation of the angular distributions with the lepton transverse momentum distributions, are thus an important ingredient to the next steps in precision measurements of electroweak parameters at the LHC, both for the effective weak mixing angle $\sin^2\theta_{\text{W}}$ and for the \Wboson-boson mass.

\clearpage
\section*{Acknowledgements}


We thank CERN for the very successful operation of the LHC, as well as the
support staff from our institutions without whom ATLAS could not be
operated efficiently.

We acknowledge the support of ANPCyT, Argentina; YerPhI, Armenia; ARC, Australia; BMWFW and FWF, Austria; ANAS, Azerbaijan; SSTC, Belarus; CNPq and FAPESP, Brazil; NSERC, NRC and CFI, Canada; CERN; CONICYT, Chile; CAS, MOST and NSFC, China; COLCIENCIAS, Colombia; MSMT CR, MPO CR and VSC CR, Czech Republic; DNRF and DNSRC, Denmark; IN2P3-CNRS, CEA-DSM/IRFU, France; GNSF, Georgia; BMBF, HGF, and MPG, Germany; GSRT, Greece; RGC, Hong Kong SAR, China; ISF, I-CORE and Benoziyo Center, Israel; INFN, Italy; MEXT and JSPS, Japan; CNRST, Morocco; FOM and NWO, Netherlands; RCN, Norway; MNiSW and NCN, Poland; FCT, Portugal; MNE/IFA, Romania; MES of Russia and NRC KI, Russian Federation; JINR; MESTD, Serbia; MSSR, Slovakia; ARRS and MIZ\v{S}, Slovenia; DST/NRF, South Africa; MINECO, Spain; SRC and Wallenberg Foundation, Sweden; SERI, SNSF and Cantons of Bern and Geneva, Switzerland; MOST, Taiwan; TAEK, Turkey; STFC, United Kingdom; DOE and NSF, United States of America. In addition, individual groups and members have received support from BCKDF, the Canada Council, CANARIE, CRC, Compute Canada, FQRNT, and the Ontario Innovation Trust, Canada; EPLANET, ERC, FP7, Horizon 2020 and Marie Sk{\l}odowska-Curie Actions, European Union; Investissements d'Avenir Labex and Idex, ANR, R{\'e}gion Auvergne and Fondation Partager le Savoir, France; DFG and AvH Foundation, Germany; Herakleitos, Thales and Aristeia programmes co-financed by EU-ESF and the Greek NSRF; BSF, GIF and Minerva, Israel; BRF, Norway; Generalitat de Catalunya, Generalitat Valenciana, Spain; the Royal Society and Leverhulme Trust, United Kingdom.

The crucial computing support from all WLCG partners is acknowledged
gratefully, in particular from CERN and the ATLAS Tier-1 facilities at
TRIUMF (Canada), NDGF (Denmark, Norway, Sweden), CC-IN2P3 (France),
KIT/GridKA (Germany), INFN-CNAF (Italy), NL-T1 (Netherlands), PIC (Spain),
ASGC (Taiwan), RAL (UK) and BNL (USA) and in the Tier-2 facilities
worldwide.

\appendix
\part*{Appendix}


\section{Theoretical formalism}
\label{appendix:theory}

Following the notation in~Ref.~\cite{arXiv9406381}, the lepton--hadron correlations in the $pp\rightarrow Z\rightarrow\ell\ell$ process are described by the contraction of the lepton tensor $L_{\mu \nu}$ with the parton-level hadron tensor $H^{\mu \nu}$. The tensor~$L_{\mu \nu}$ acts as an analyser of the structure of $H^{\mu \nu}$, which carries effective information about the polarisation of the \Zboson~boson mediating the interaction. The angular dependence can be elucidated by introducing nine helicity density matrix elements

\begin{equation}
  H_{m m'} = \epsilon^*_{\mu}(m) H^{\mu  \nu} \epsilon_{\nu}(m')
\end{equation}

where $m, m' = +,0,-$ and 

\begin{equation} 
  \epsilon_{\mu}(\pm) = \frac{1}{\sqrt{2}} (0; \pm 1, -i, 0), \ \ \ \ \epsilon_{\mu} (0) = (0;0,0,1)
\end{equation}

are the polarisation vectors for the \Zboson~boson, defined with respect to its rest frame. 
The angular dependence of the differential cross-section can be written as:

\begin{equation}
 \label{Eq:master1}
  \frac{ \mathrm{d}\sigma}{\mathrm{d}\ptz~\mathrm{d}\yz~\mathrm{d}\mz~\mathrm{d}\cos\theta~\mathrm{d}\phi} = 
     \displaystyle\sum_{ \alpha  \epsilon M} g^{ \alpha }( \theta,  \phi)
     \frac{3}{ 16 \pi} \frac{\mathrm{d} \sigma ^{\alpha }}{ \mathrm{d}\ptz~\mathrm{d}\yz~\mathrm{d}\mz}, \ \ \ \  M = \{ U+L, L, T, I, P, A, 7, 8, 9\},
\end{equation}

where the $ g^{ \alpha }( \theta,  \phi)$ are second-order harmonic polynomials, multiplied by normalisation constants. The helicity cross-sections $\sigma ^{\alpha }$ are linear combinations of the helicity density matrix elements $H_{m m'}$:

\begin{equation}
\label{Eq:helicity_elements}
\begin{array}{l l l}
   \sigma^{U+L} &\propto &H_{00} + H_{++} + H_{--} \\
   \sigma^{L} &\propto &H_{00} \\
   \sigma^{T} &\propto &1/2 (H_{+-} + H_{-+}) \\
   \sigma^{I} &\propto &1/4 ( H_{+0} + H_{0+} - H_{-0} - H_{0-} ) \\
   \sigma^{P} &\propto &H_{++} - H_{--}   \\
   \sigma^{A} &\propto &1/4 ( H_{+0} + H_{0+} + H_{-0} + H_{0-}) \\
   \sigma^{7} &\propto &-i/2 ( H_{+-} - H_{-+}) \\
   \sigma^{8} &\propto &-i/4 ( H_{+0} - H_{0+} + H_{-0} - H_{0-}) \\
   \sigma^{9} &\propto &-i/4 ( H_{+0} - H_{0+} - H_{-0} + H_{0-}) . \\
\end{array}
\end{equation}

The unpolarised cross-section is denoted historically by $\sigma^{U+L}$, whereas  $\sigma^{L, T, I, P, A, 7, 8, 9 }$ characterise the $Z$-boson polarisation. Respectively, these are the contributions to the \Zboson-boson cross-section from longitudinally and transversely polarised states, transverse--longitudinal interference, etc., as described in~Ref.~\cite{Mirkes92}.

The individual helicity cross-sections depend on the coupling coefficients of the $Z$~boson as follows:

\begin{equation}
\label{Eq:VA_structure}
\begin{array}{l l l}
   \sigma^{U+L, L, T, I }  & \propto& (v_{\ell}^2 + a_{\ell}^2)(v_{\ell}^2 + a_{q}^2) \nonumber \\ 
   \sigma^{P, A }  & \propto &v_{\ell} a_{\ell} v_q a_q \nonumber \\ 
   \sigma^{7,8 }  & \propto &(v_{\ell}^2 + a_{\ell}^2) ( v_q  a_q) \nonumber \\
   \sigma^{9 }  &  \propto & v_{\ell} a_{\ell} ( v_q^2 + a_q^2), \\
\end{array}
\end{equation}

where $v_q(v_{\ell})$ and $a_q (a_{\ell})$ denote the vector and axial-vector coupling of the \Zboson~boson to the quarks (leptons). The cross-sections $\sigma^{U+L, L, T, I, 9 }$ receive contributions from the parity-conserving component of the hadron tensor, while the others, $\sigma^{P, A, 7, 8 }$, are proportional to the parity-violating component of ~$H^{\mu  \nu}$. However, the angular polynomials $g^{P, A, 9 }( \theta,  \phi)$ are parity-violating as well, so contributions to the angular distributions from $\sigma^{U+L, L, T, I, P, A }$ are parity-conserving.

It is standard notation to factorise out the unpolarised cross-section, and to present the five-dimensional differential cross-section as an expansion in harmonic polynomials $\poli(\cos\theta,\phi)$ and dimensionless angular coefficients $A_{0-7}$, which represent ratios of helicity cross-sections with respect to the unpolarised one, as follows: 

 \begin{align}
   A_0 &= 2 \mathrm{d} \sigma^{L} /  \mathrm{d} \sigma^{U+L} \nonumber \\ 
   A_1 &= 2  \sqrt{2} \mathrm{d} \sigma^{I} /  \mathrm{d} \sigma^{U+L} \nonumber \\
   A_2 &= 4  \mathrm{d} \sigma^{T} /  \mathrm{d} \sigma^{U+L} \nonumber \\ 
   A_3 &= 4  \sqrt{2} \mathrm{d} \sigma^{A} /  \mathrm{d} \sigma^{U+L} \nonumber \\
   A_4 &= 2  \mathrm{d} \sigma^{P} /  \mathrm{d} \sigma^{U+L} \\
   A_5 &= 2  \mathrm{d} \sigma^{7} /  \mathrm{d} \sigma^{U+L} \nonumber \\
   A_6 &= 2  \sqrt{2}  \mathrm{d} \sigma^{8} /  \mathrm{d} \sigma^{U+L} \nonumber \\
   A_7 &= 4  \sqrt{2}  \mathrm{d} \sigma^{9} /  \mathrm{d} \sigma^{U+L}. \nonumber 
  \end{align}

This leads to~Eq.~(\ref{Eq:master2}), as discussed in~Section~\ref{sec:intro}.

\section{Additional Templates}
\label{sec:add_templates}

To expand upon the one-dimensional templates shown in Section~\ref{sec:Methodology}, the two-dimensional versions are shown here. The dimension corresponding to migrations in $\ptll$ is integrated over. Figures~\ref{Fig:templates2D_Ai_1}--\ref{Fig:templates2D_Ai_3} show each analytical polynomial for each corresponding coefficient along with the templated versions in three representative $\ptz$ bins after acceptance and selection requirements. The differences between the analytical polynomials and their templates reflect primarily the effect of the acceptance shape in the angular variables, and to a lesser extent resolution effects.

\begin{figure}
  \begin{center}
{
    \includegraphics[width=5cm,angle=0]{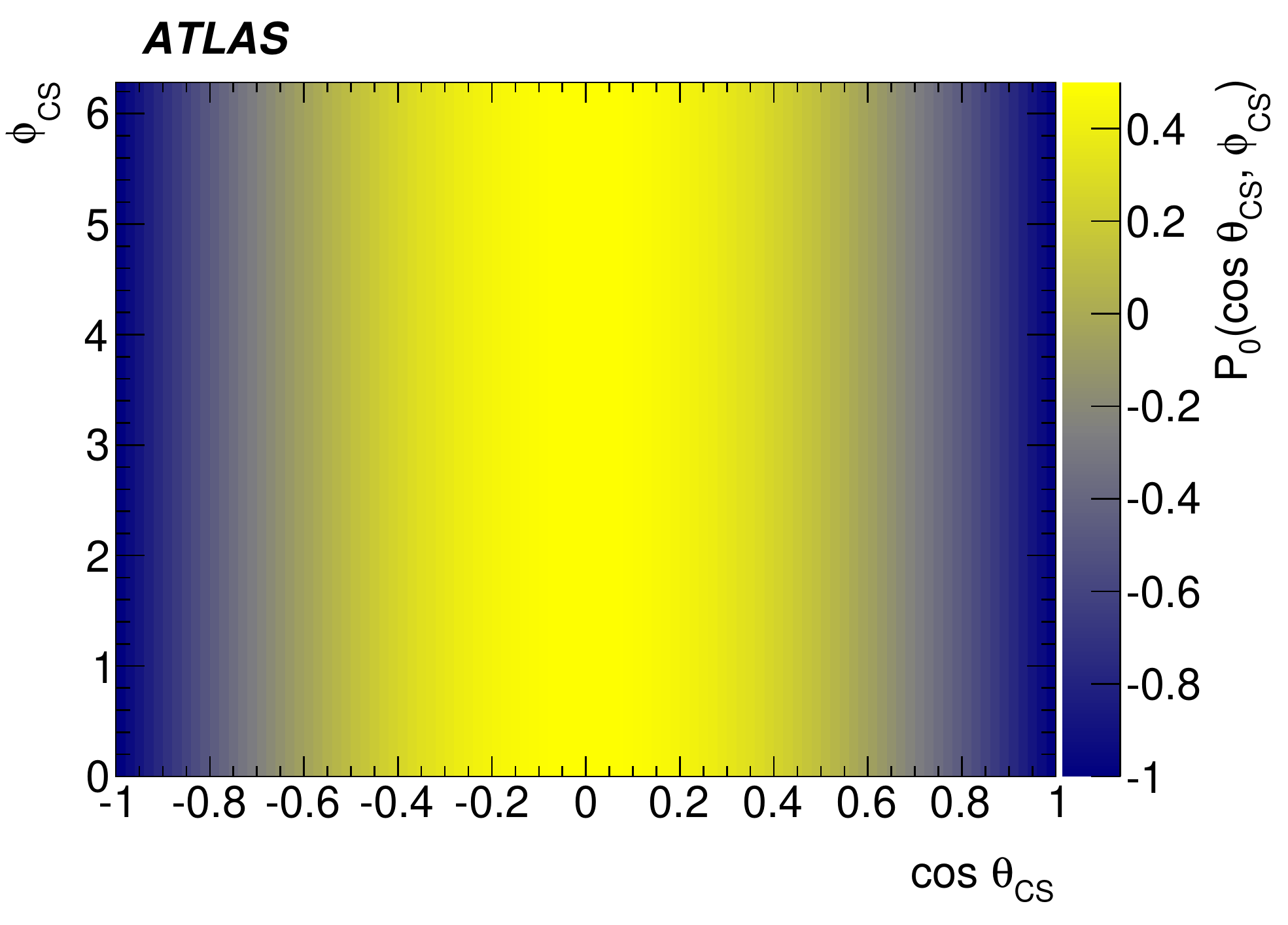}
    \includegraphics[width=5cm,angle=0]{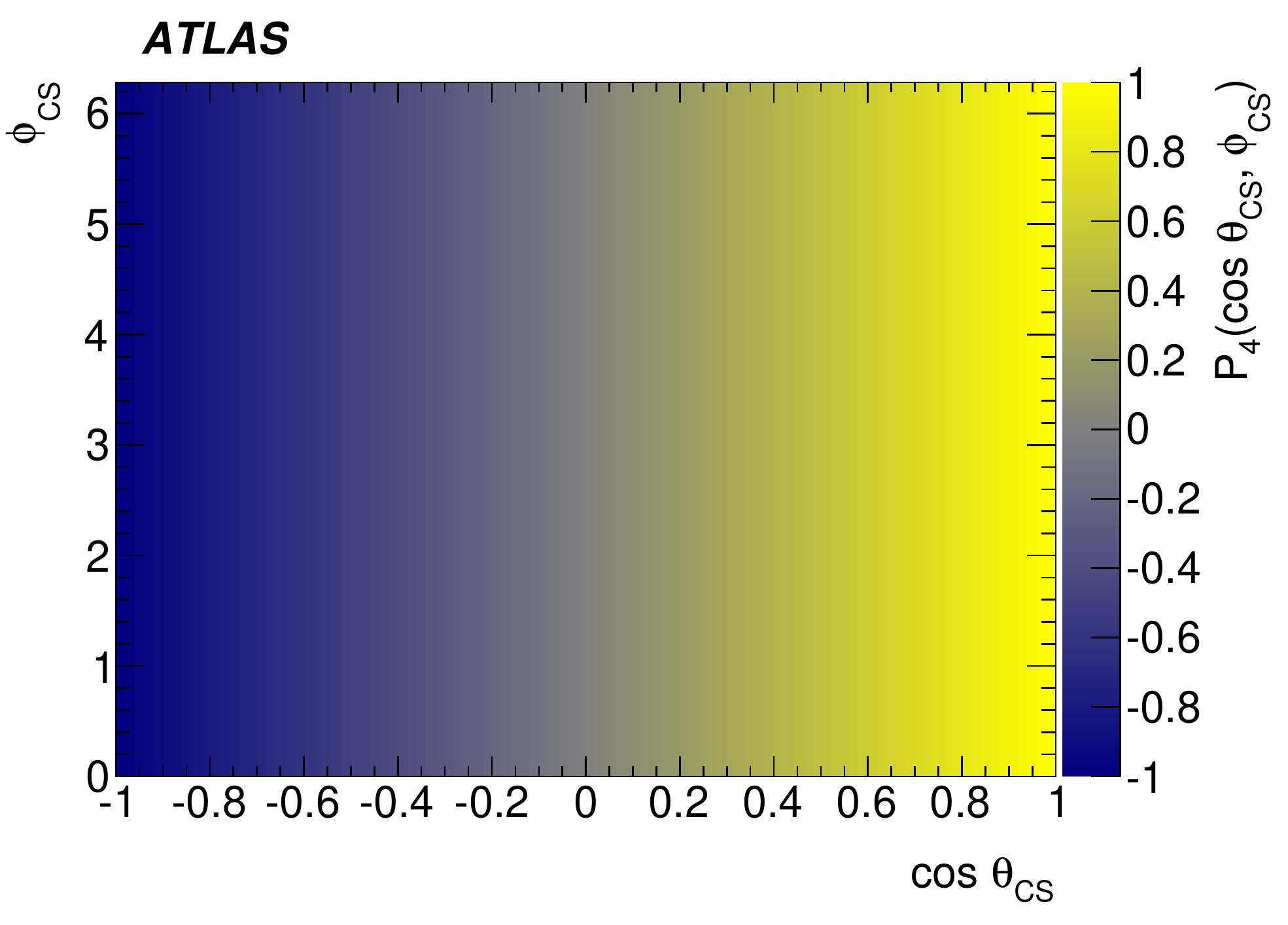}
    \includegraphics[width=5cm,angle=0]{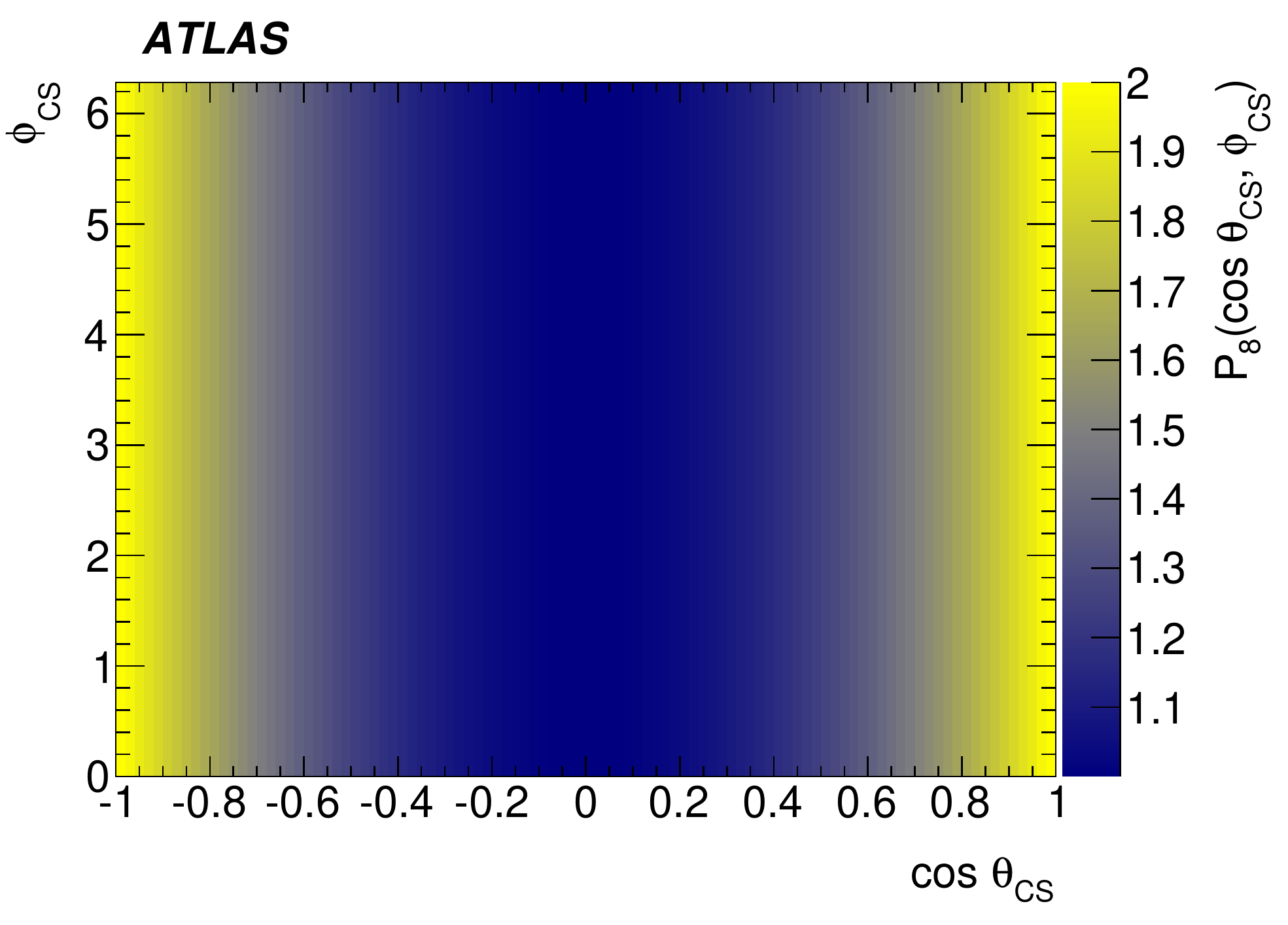}

    \includegraphics[width=5cm,angle=0]{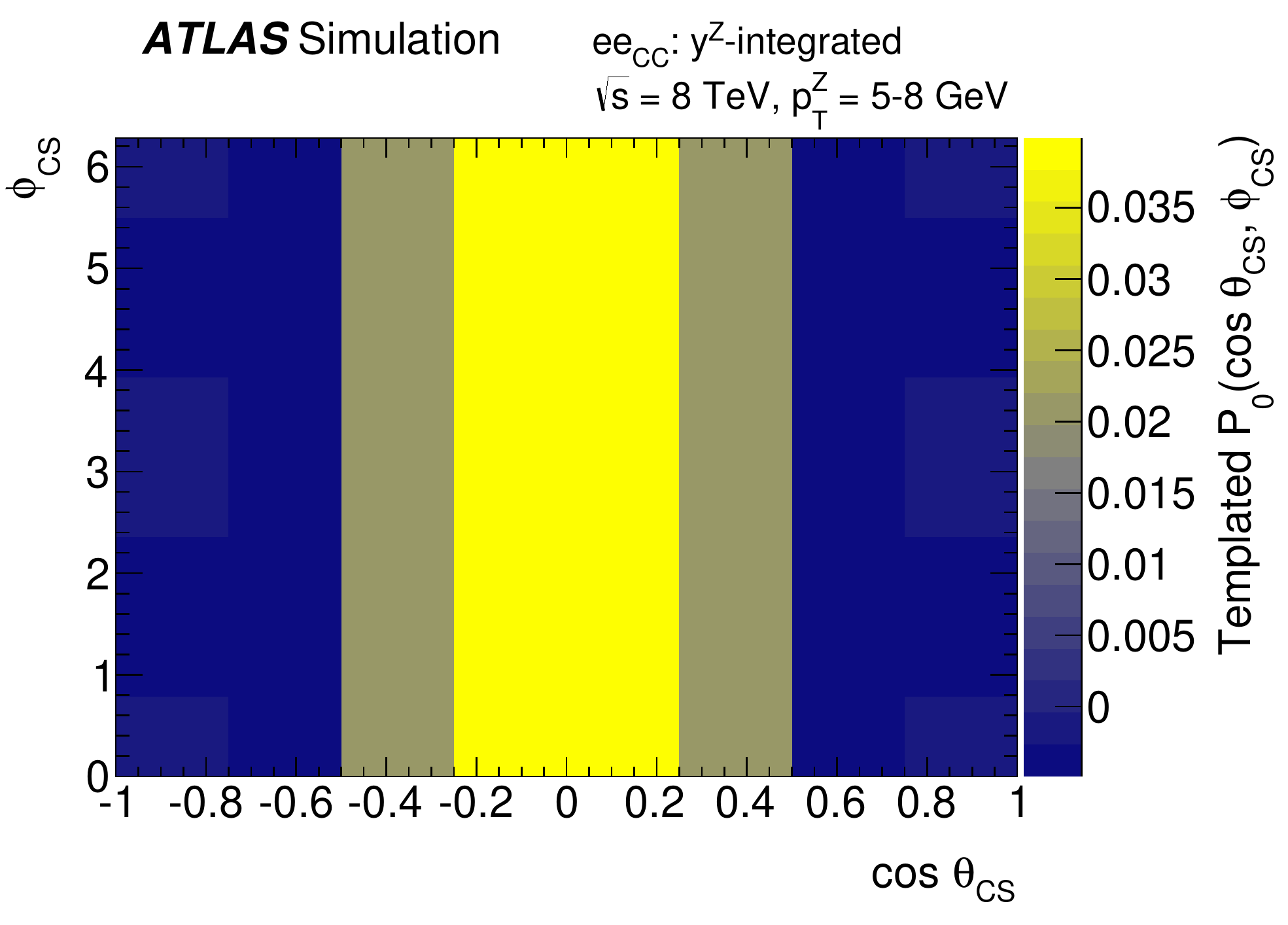}
    \includegraphics[width=5cm,angle=0]{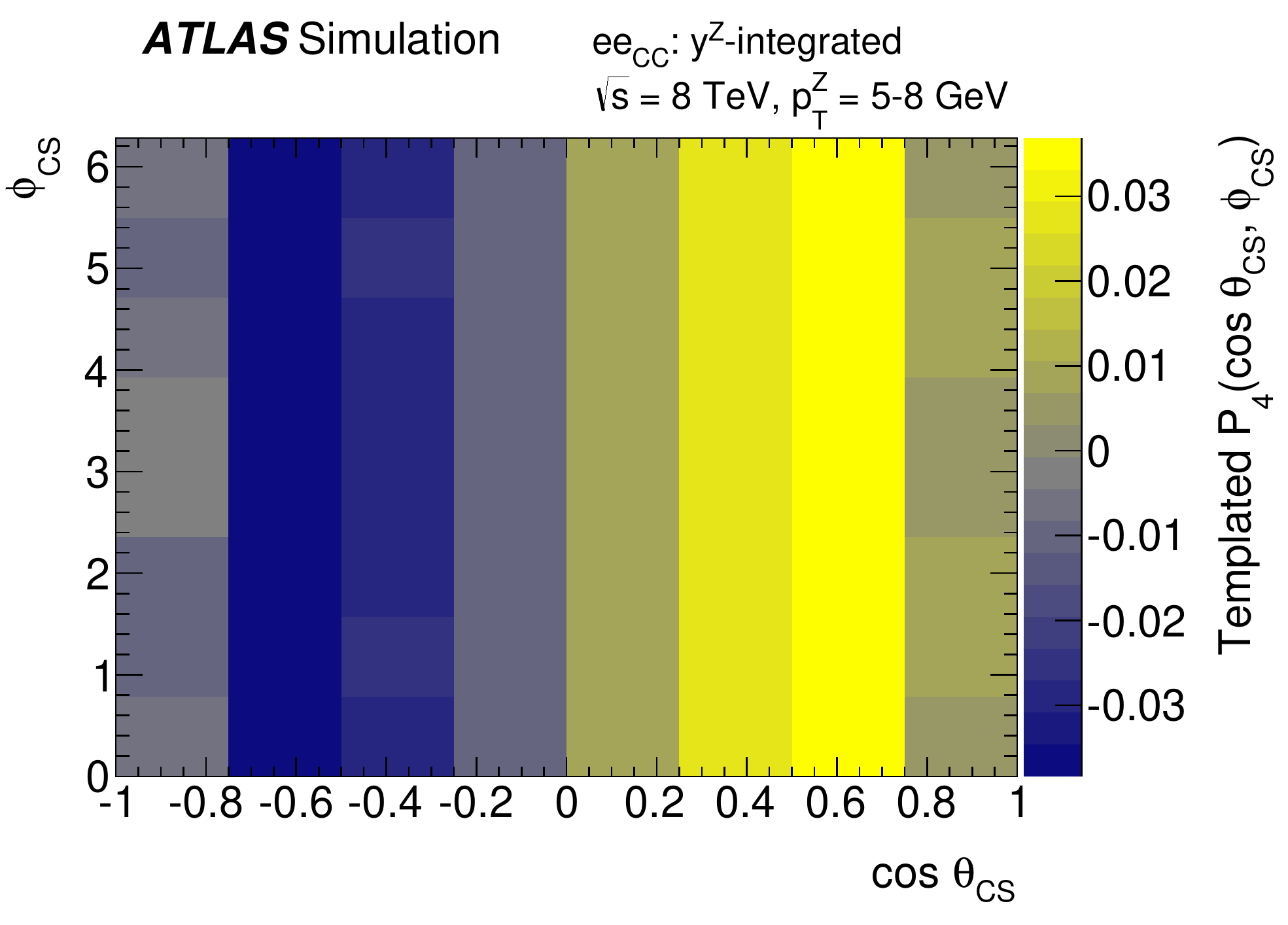}
    \includegraphics[width=5cm,angle=0]{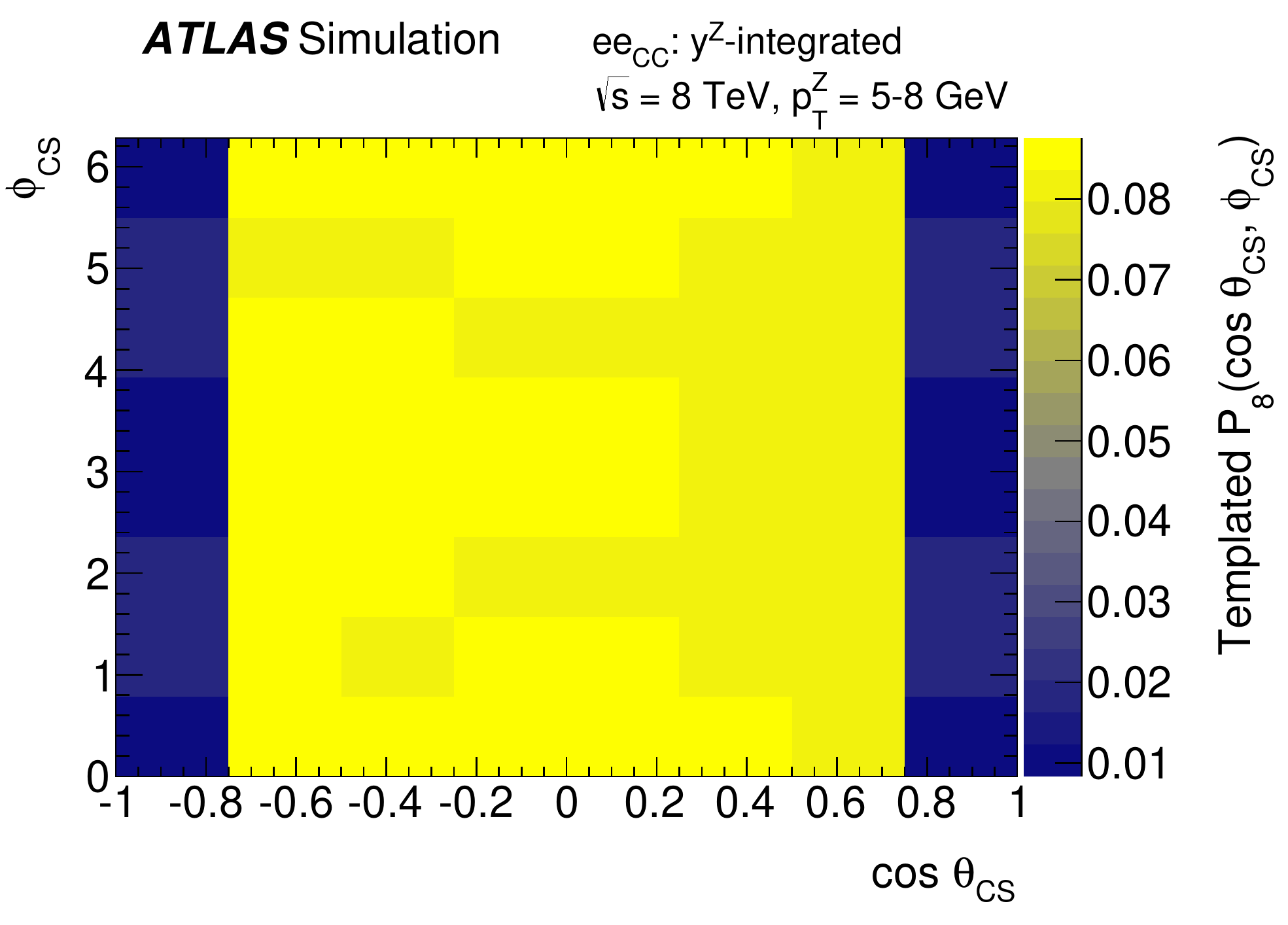}

    \includegraphics[width=5cm,angle=0]{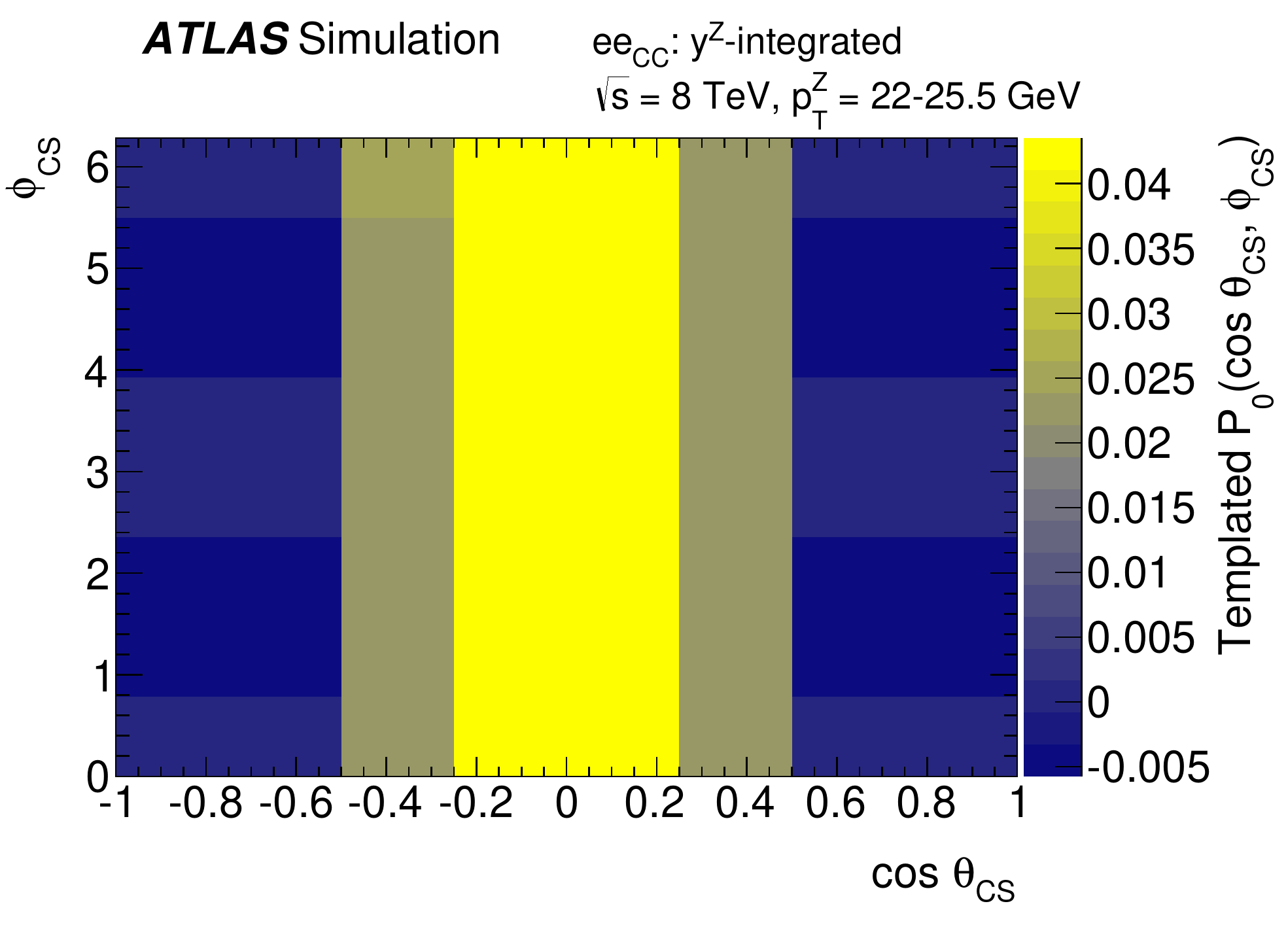}
    \includegraphics[width=5cm,angle=0]{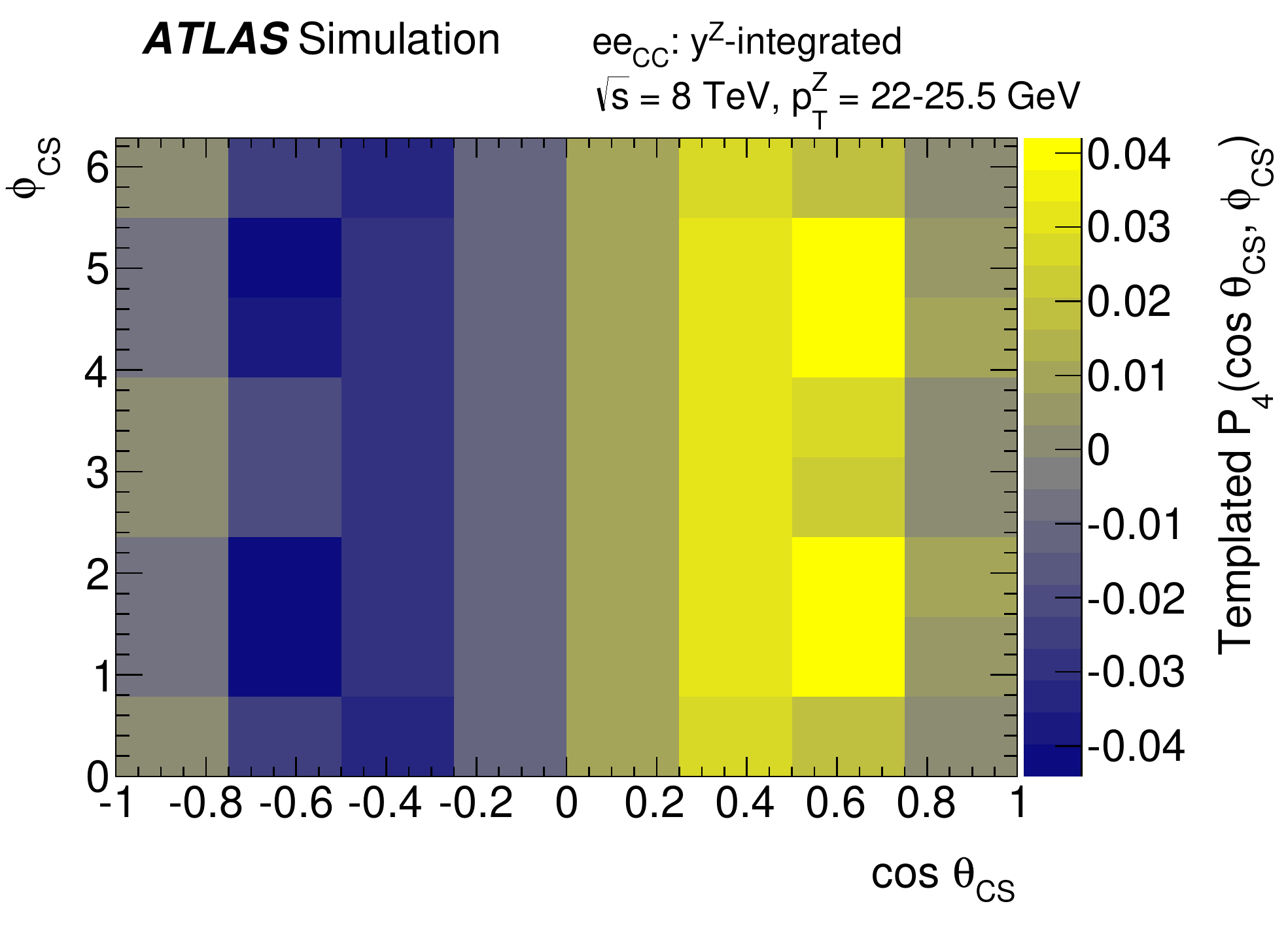}
    \includegraphics[width=5cm,angle=0]{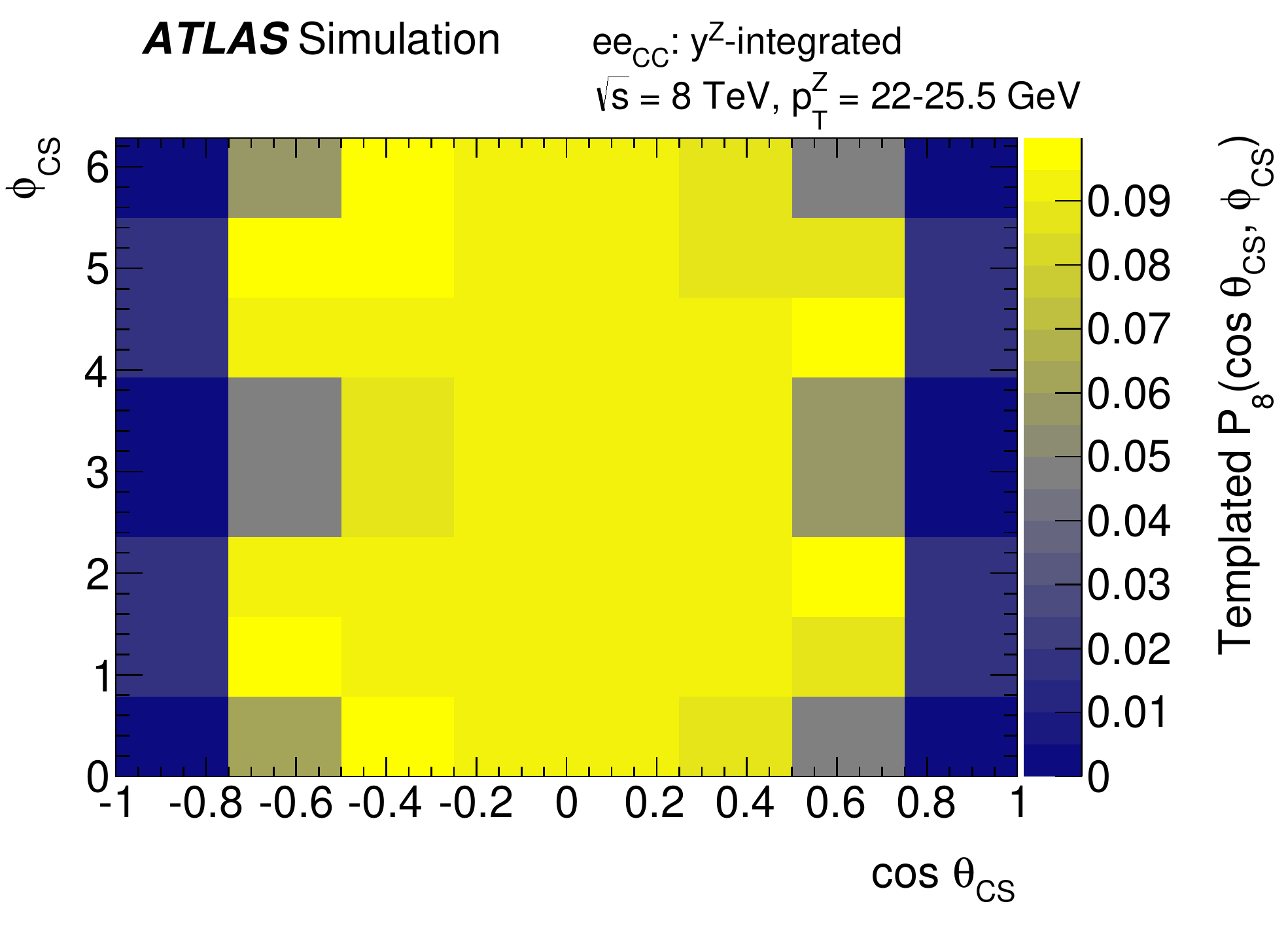}

    \includegraphics[width=5cm,angle=0]{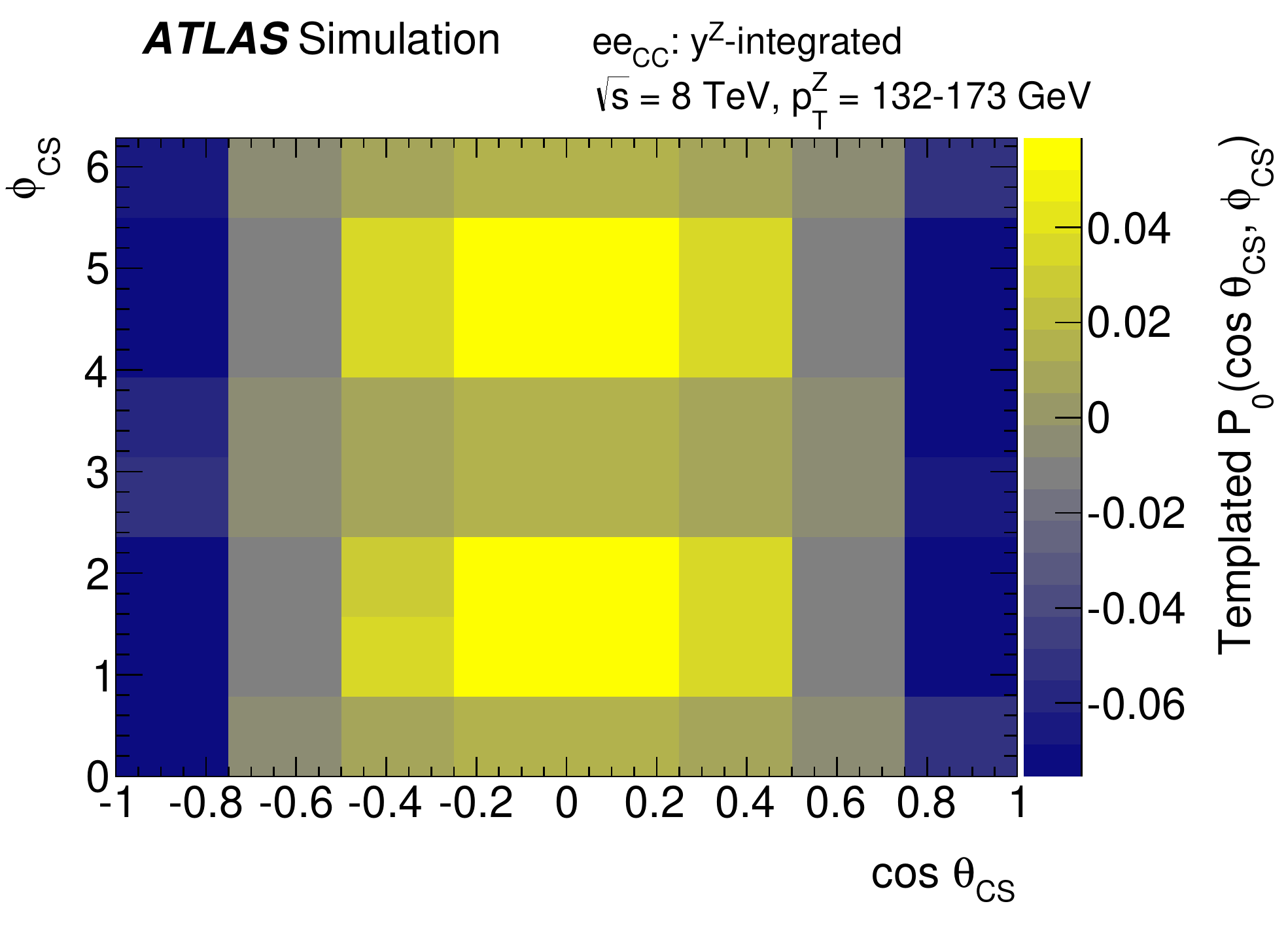}
    \includegraphics[width=5cm,angle=0]{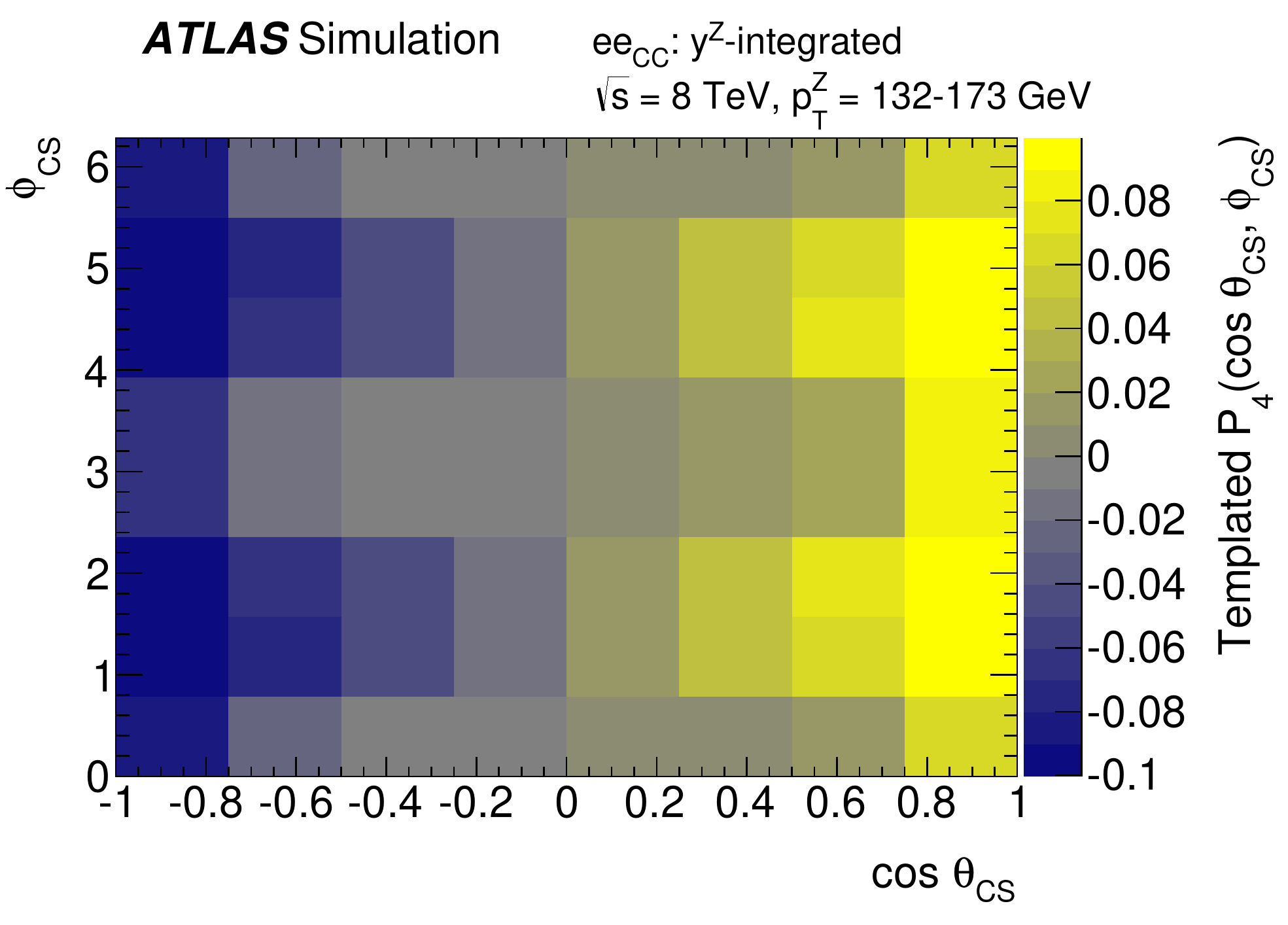}
    \includegraphics[width=5cm,angle=0]{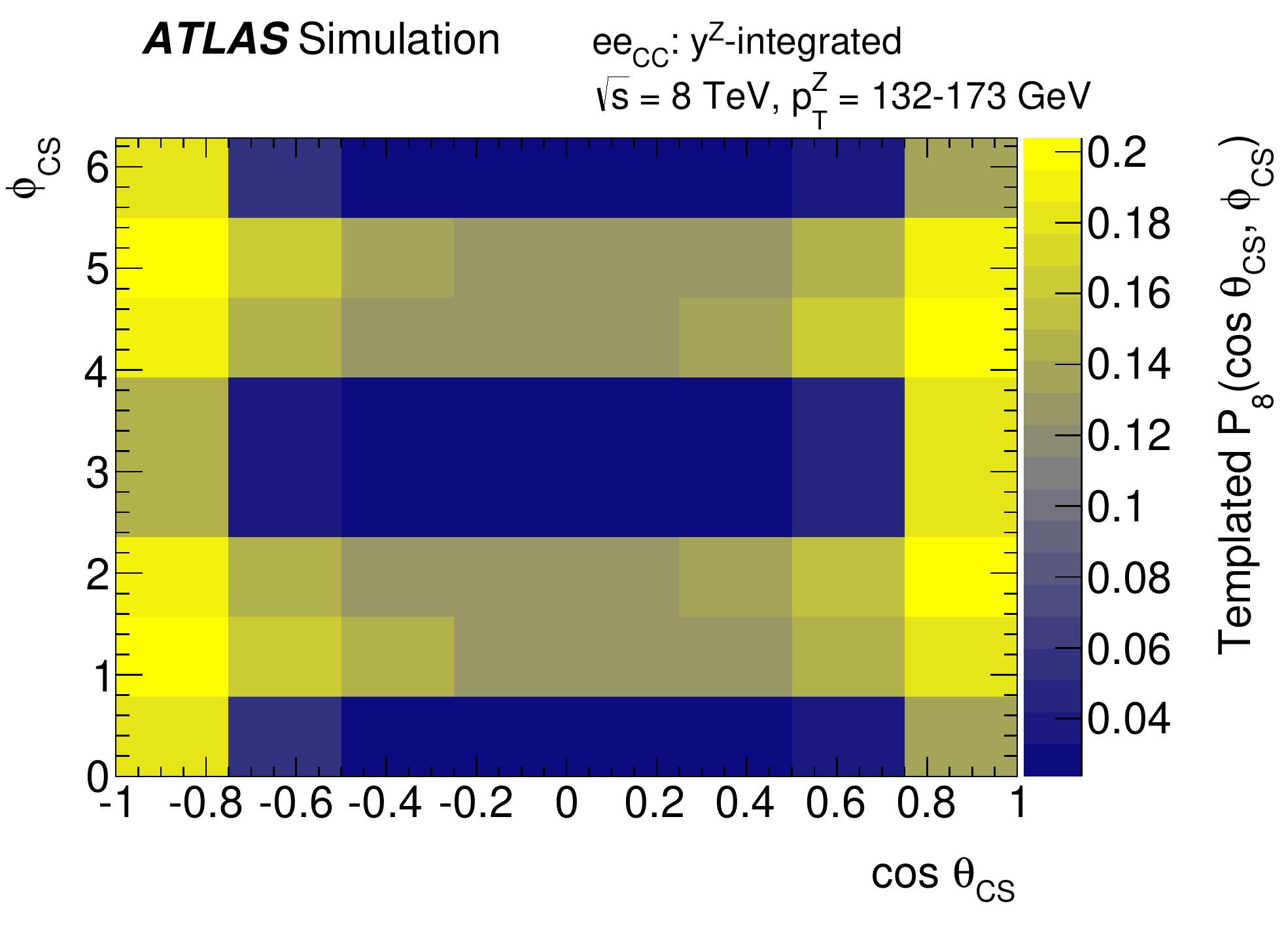}

}
\end{center}
\vspace{-5mm}
\caption{Shapes of the polynomials $P_{0,4,8}$ as a function of~$\costhetacs$ and $\phics$ (top). Below these are the templated polynomials for the $\yz$-integrated $ee_{CC}$~events at low~($5-8$~GeV), medium~($22-25.5$~GeV), and high~($132-173$~GeV) values of~$\ptz$.
\label{Fig:templates2D_Ai_1}}
\end{figure}

\begin{figure}
  \begin{center}
{
    \includegraphics[width=5cm,angle=0]{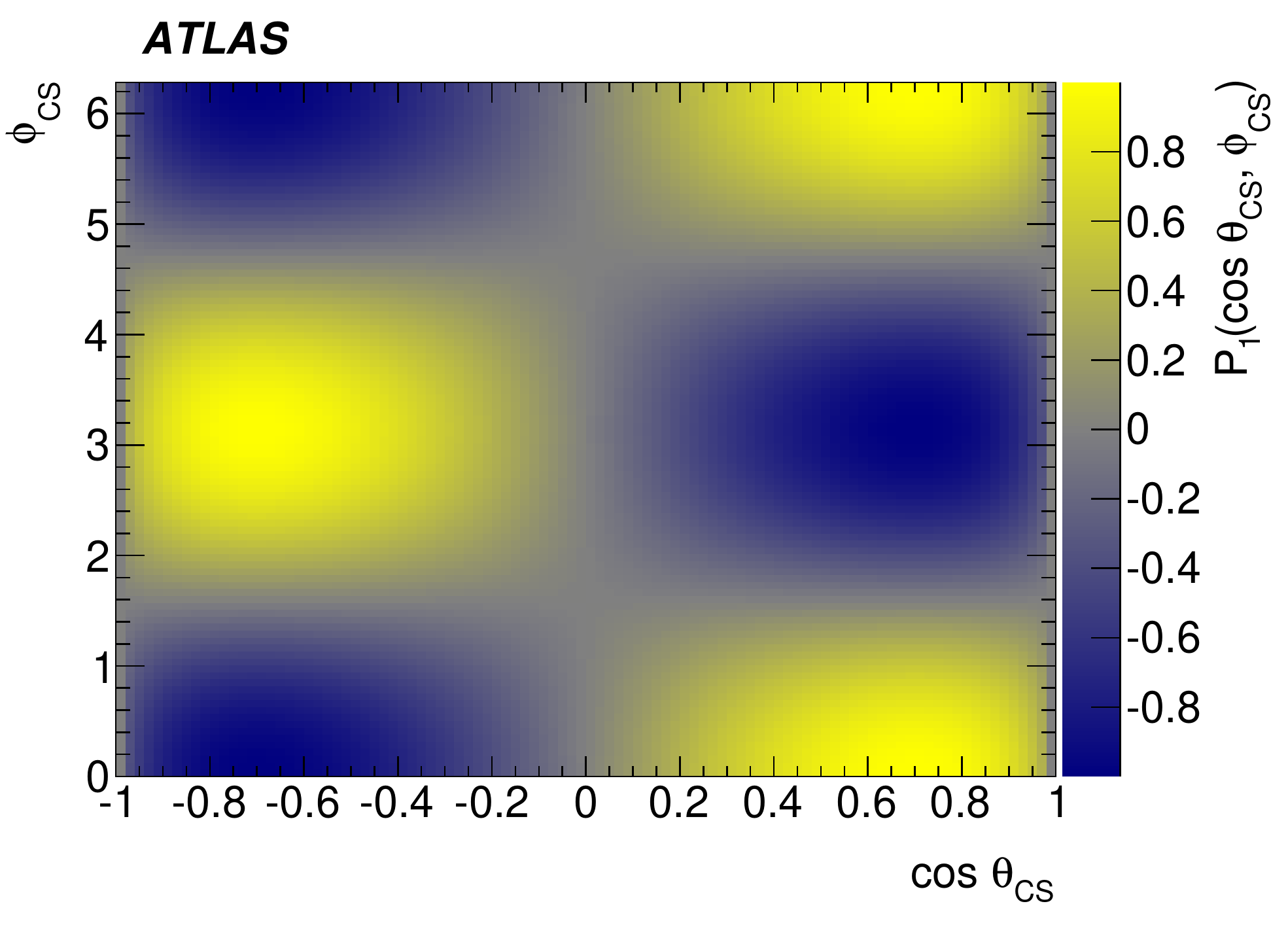}
    \includegraphics[width=5cm,angle=0]{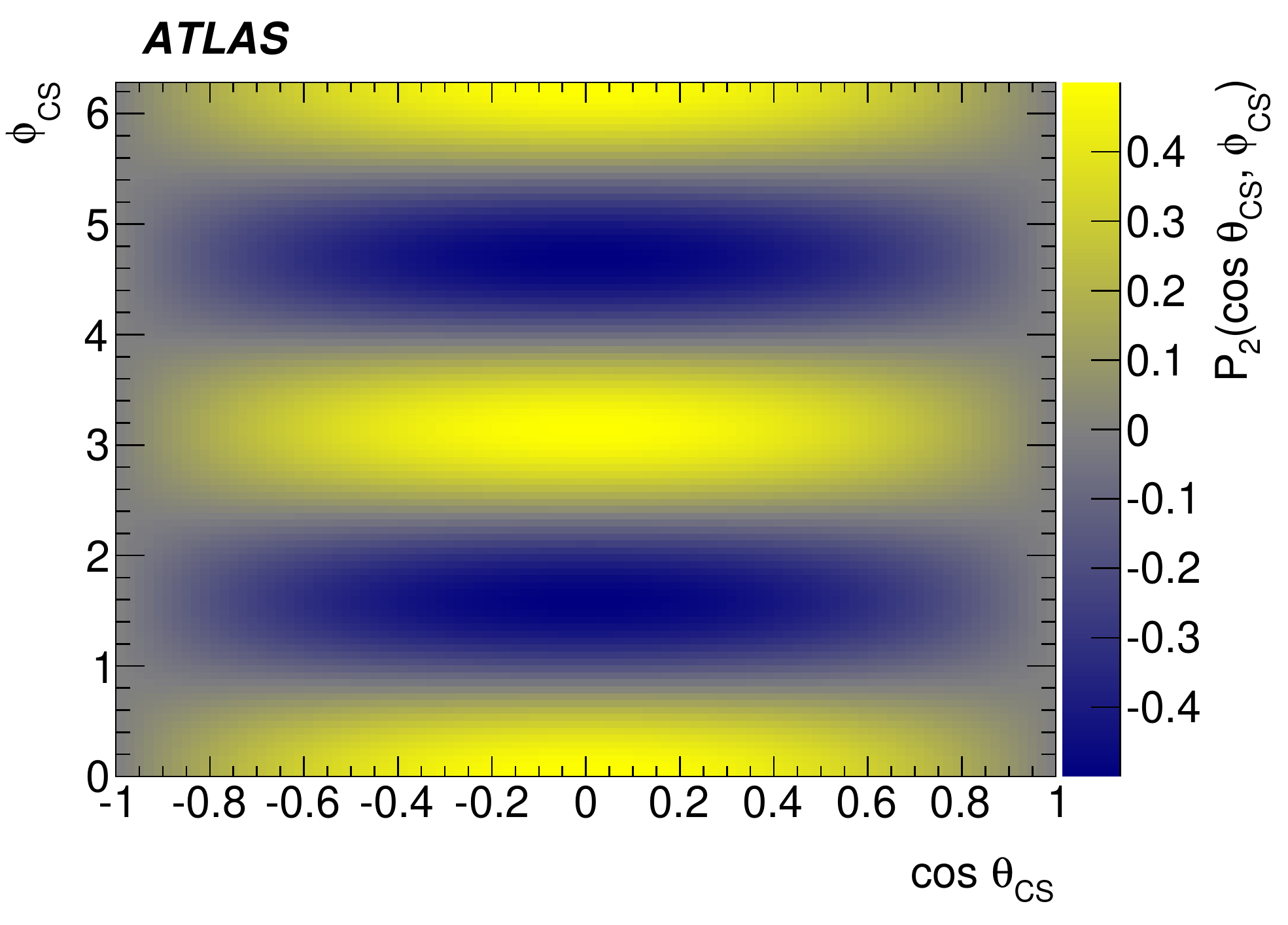}
    \includegraphics[width=5cm,angle=0]{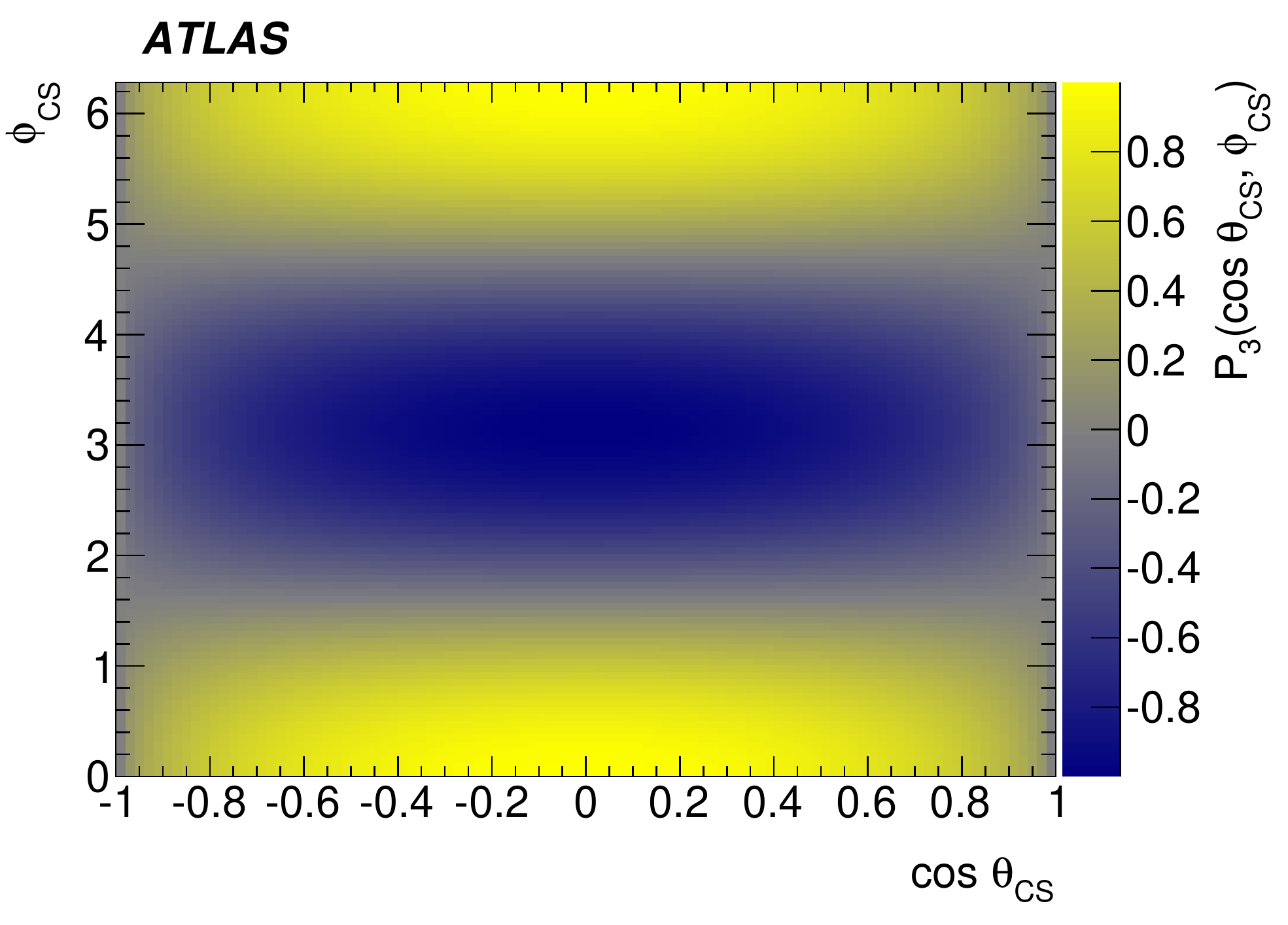}

    \includegraphics[width=5cm,angle=0]{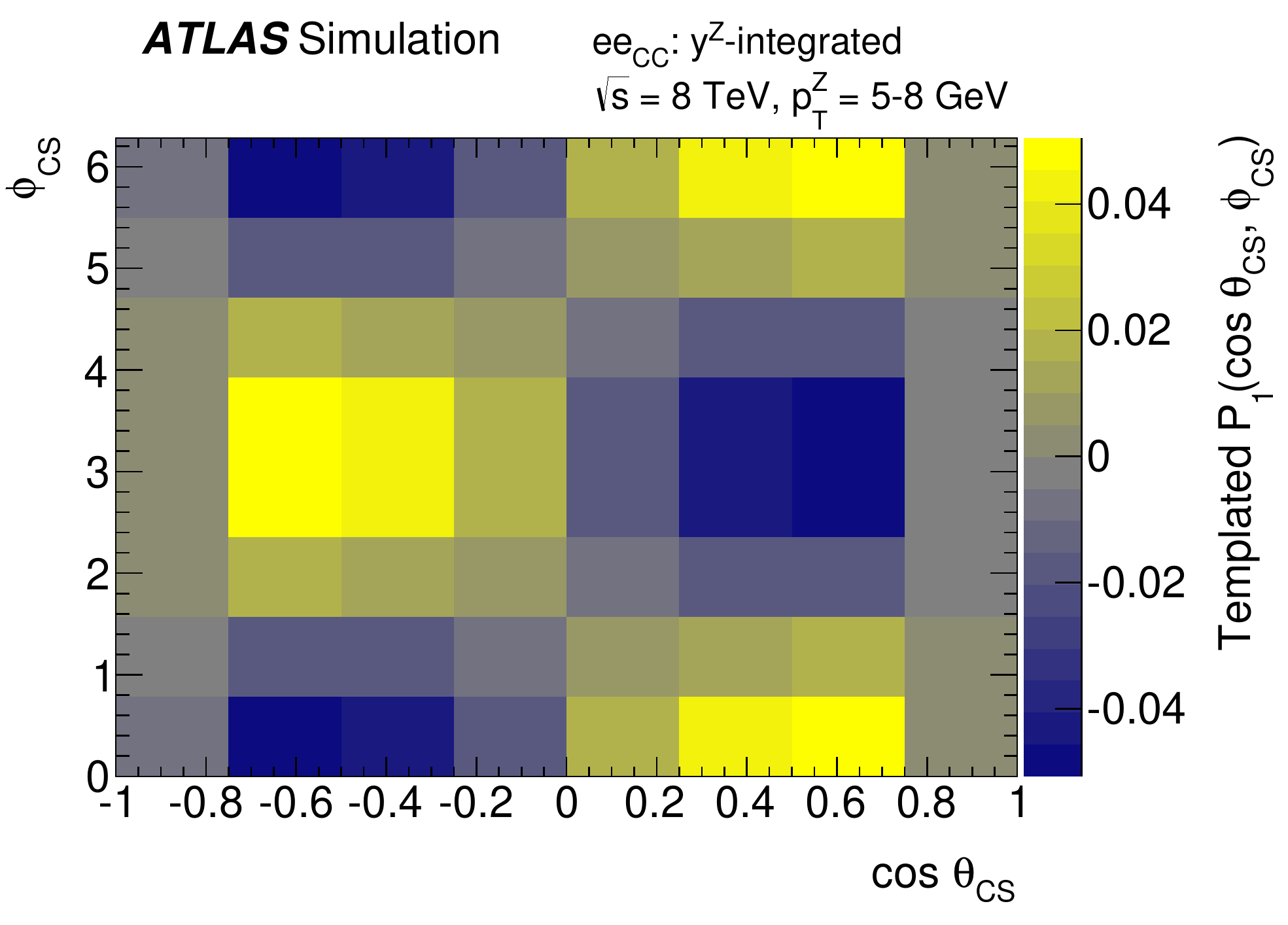}
    \includegraphics[width=5cm,angle=0]{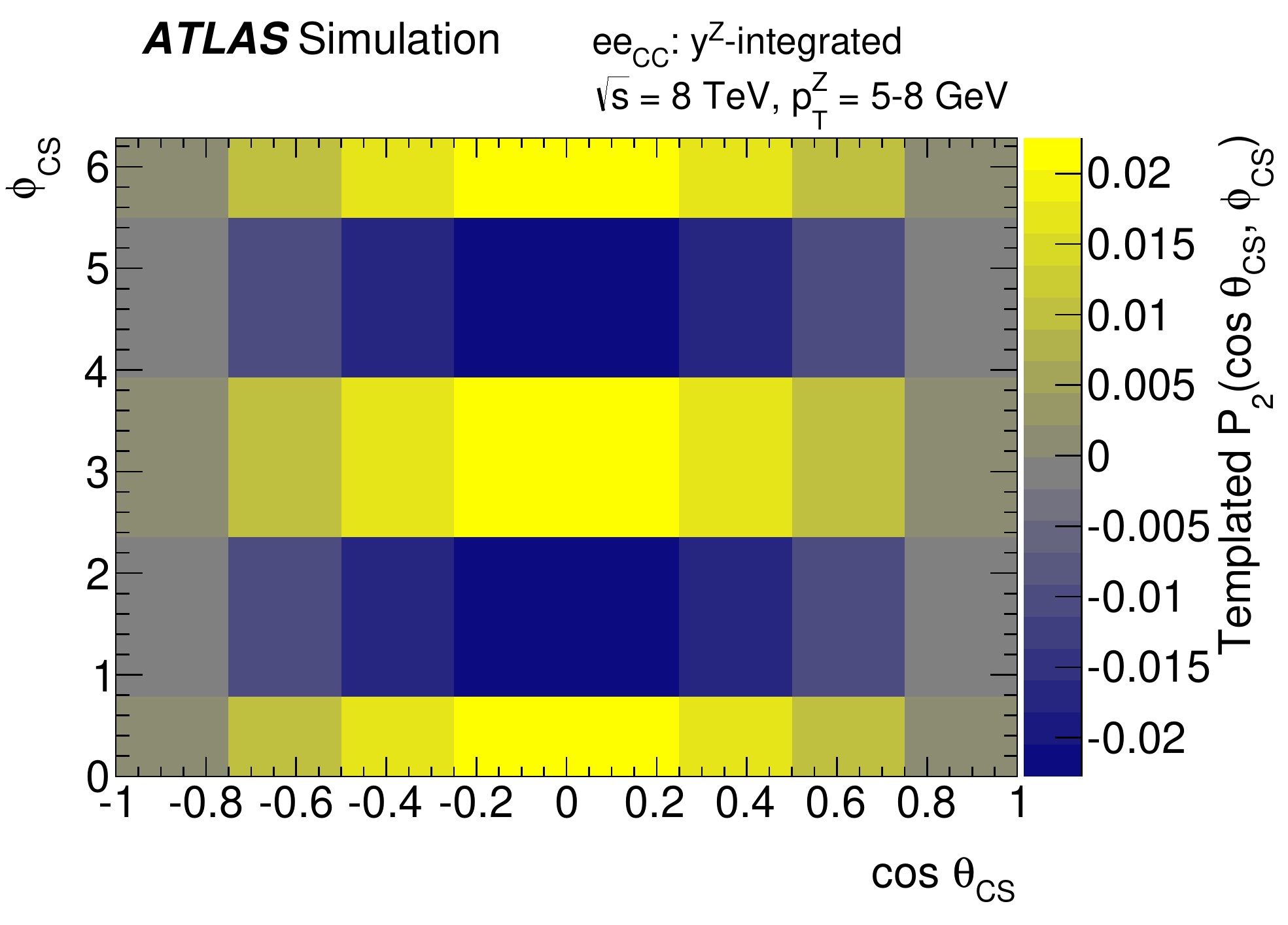}
    \includegraphics[width=5cm,angle=0]{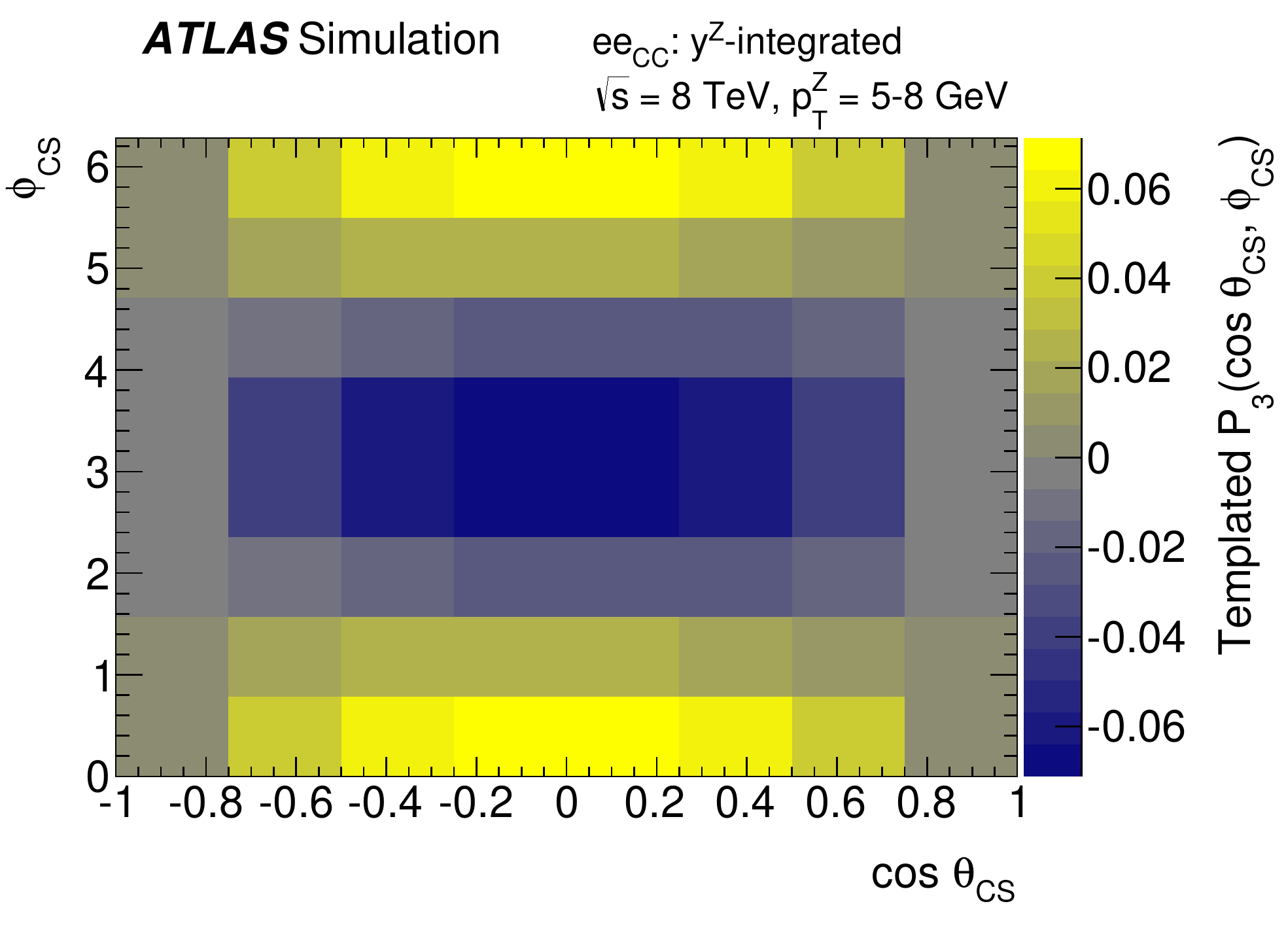}

    \includegraphics[width=5cm,angle=0]{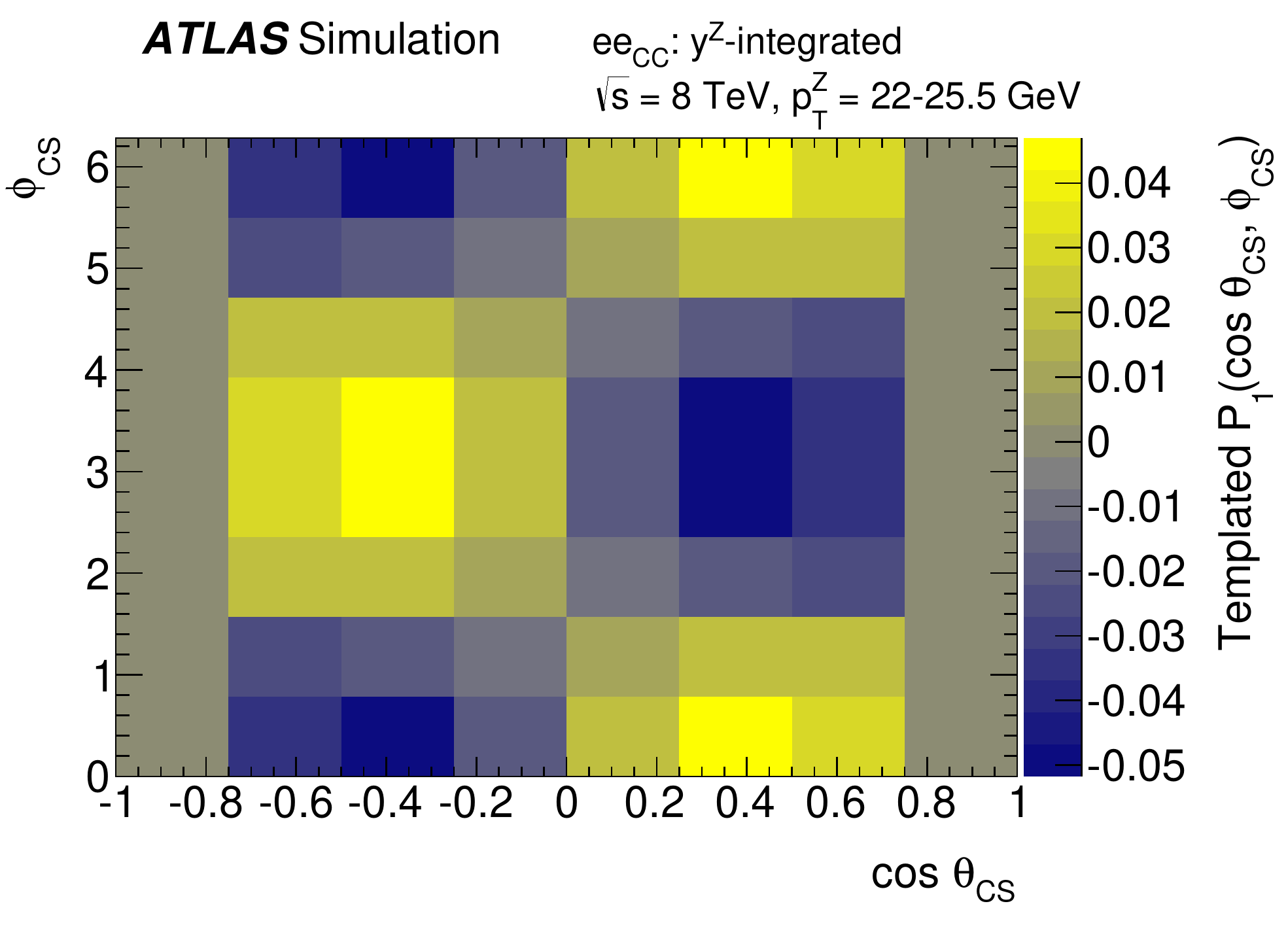}
    \includegraphics[width=5cm,angle=0]{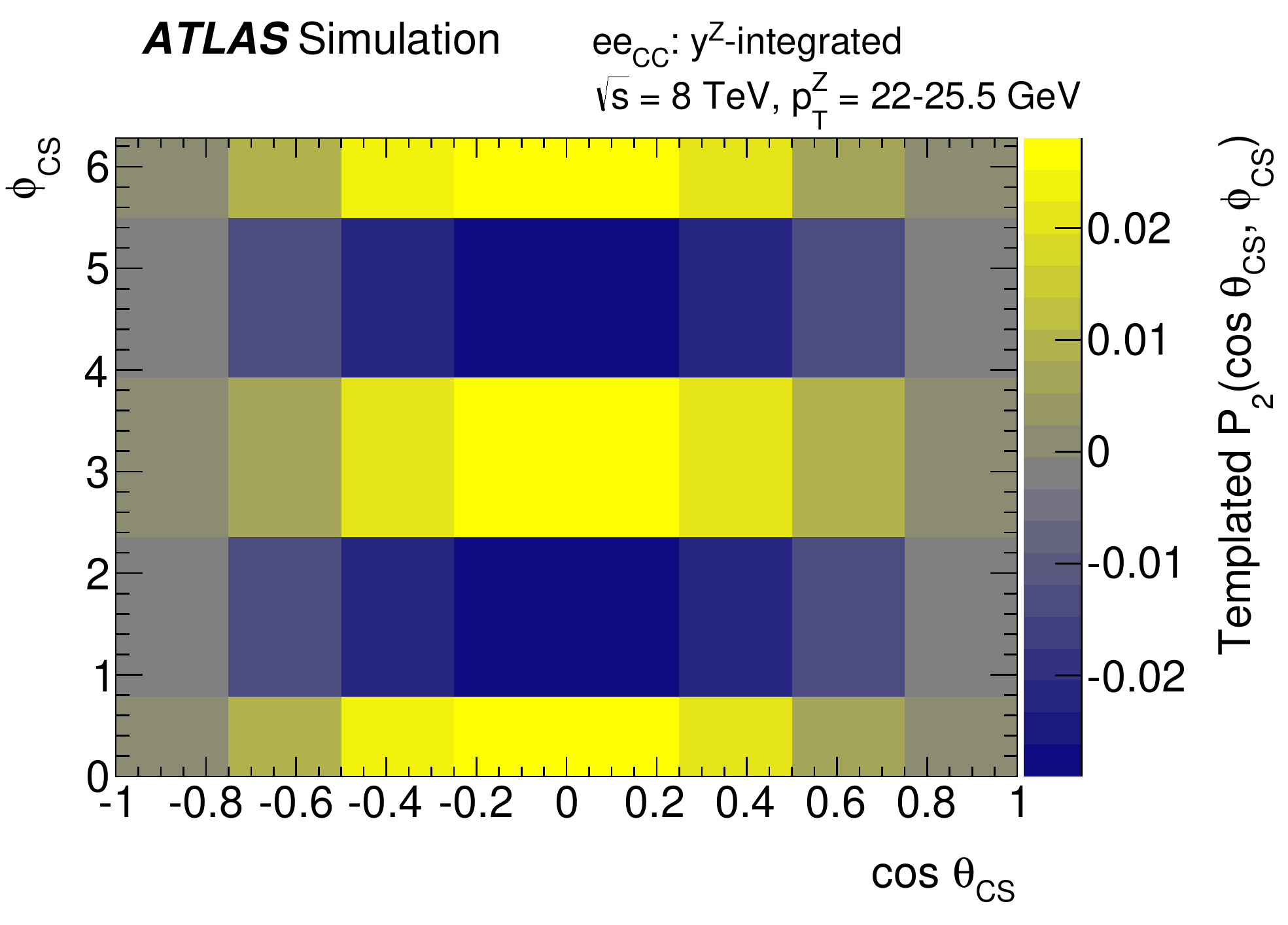}
    \includegraphics[width=5cm,angle=0]{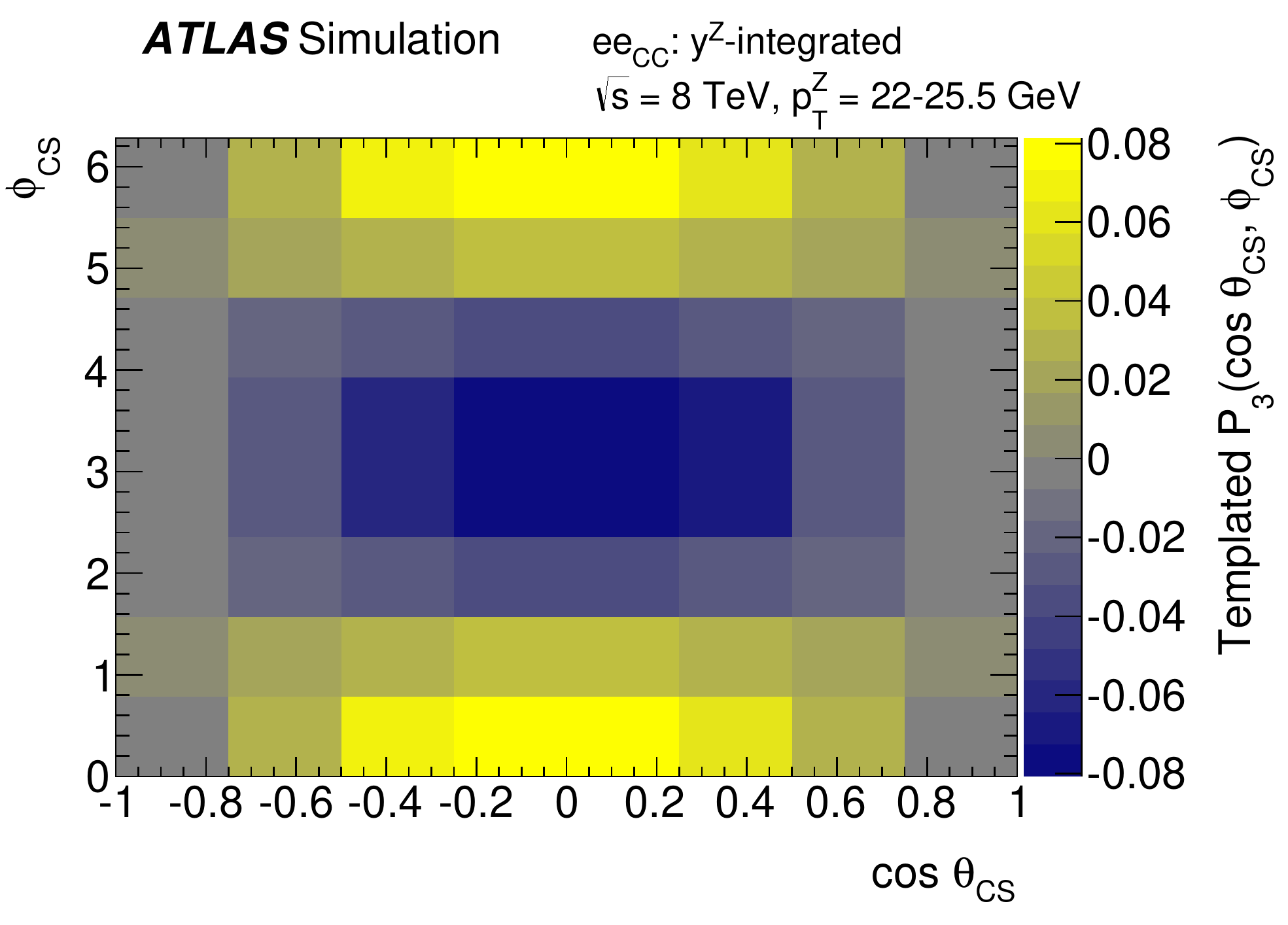}

    \includegraphics[width=5cm,angle=0]{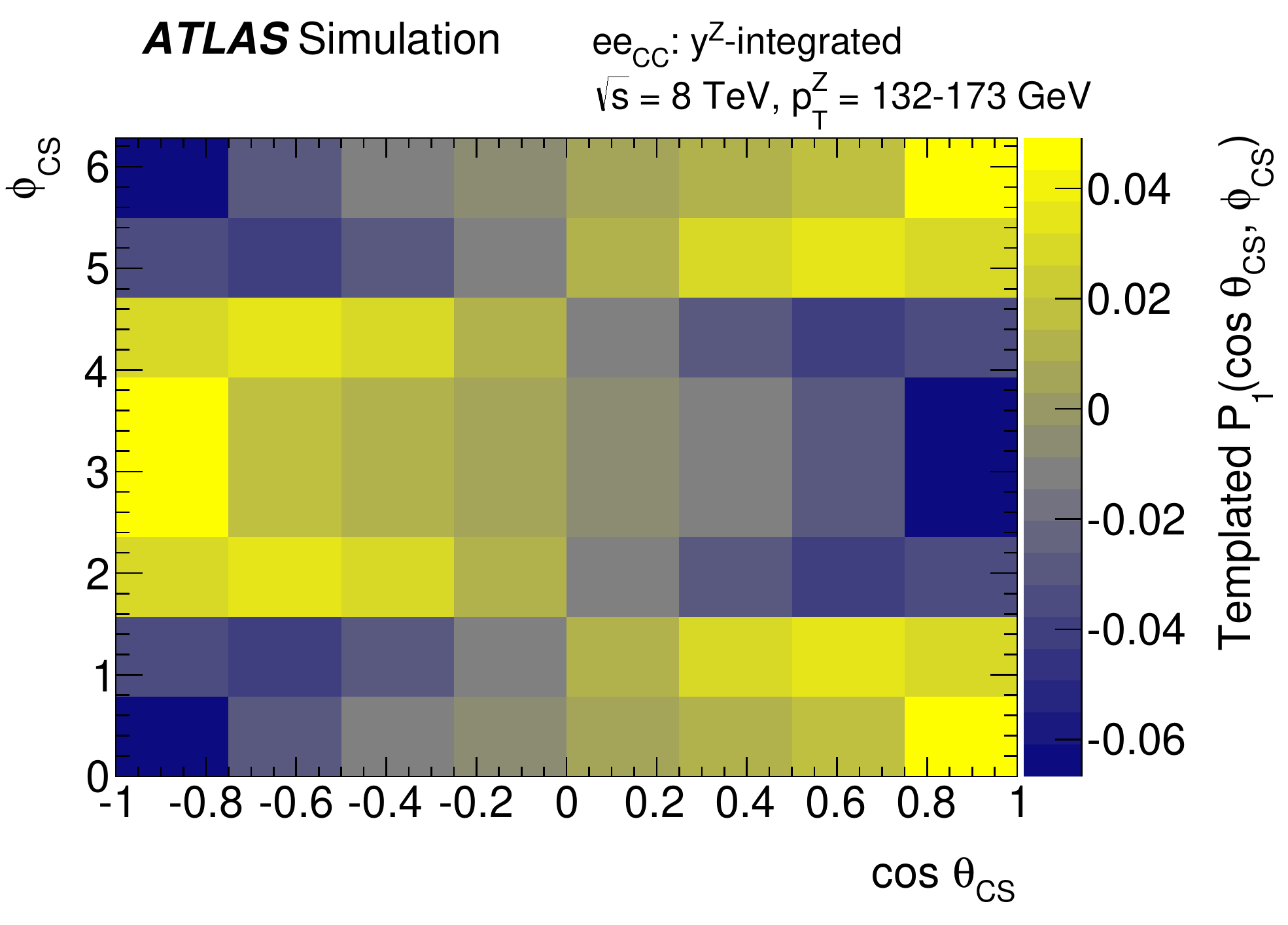}
    \includegraphics[width=5cm,angle=0]{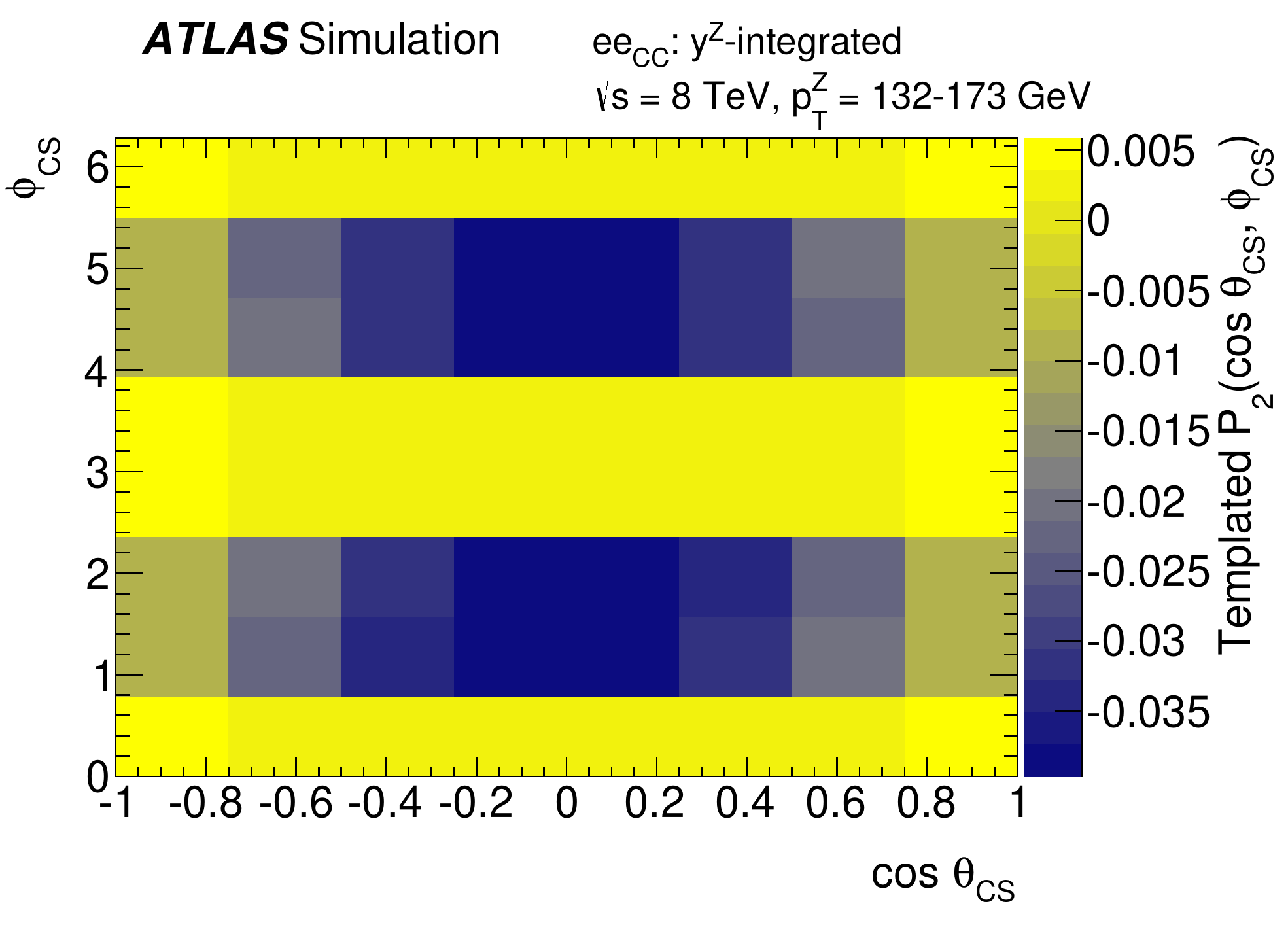}
    \includegraphics[width=5cm,angle=0]{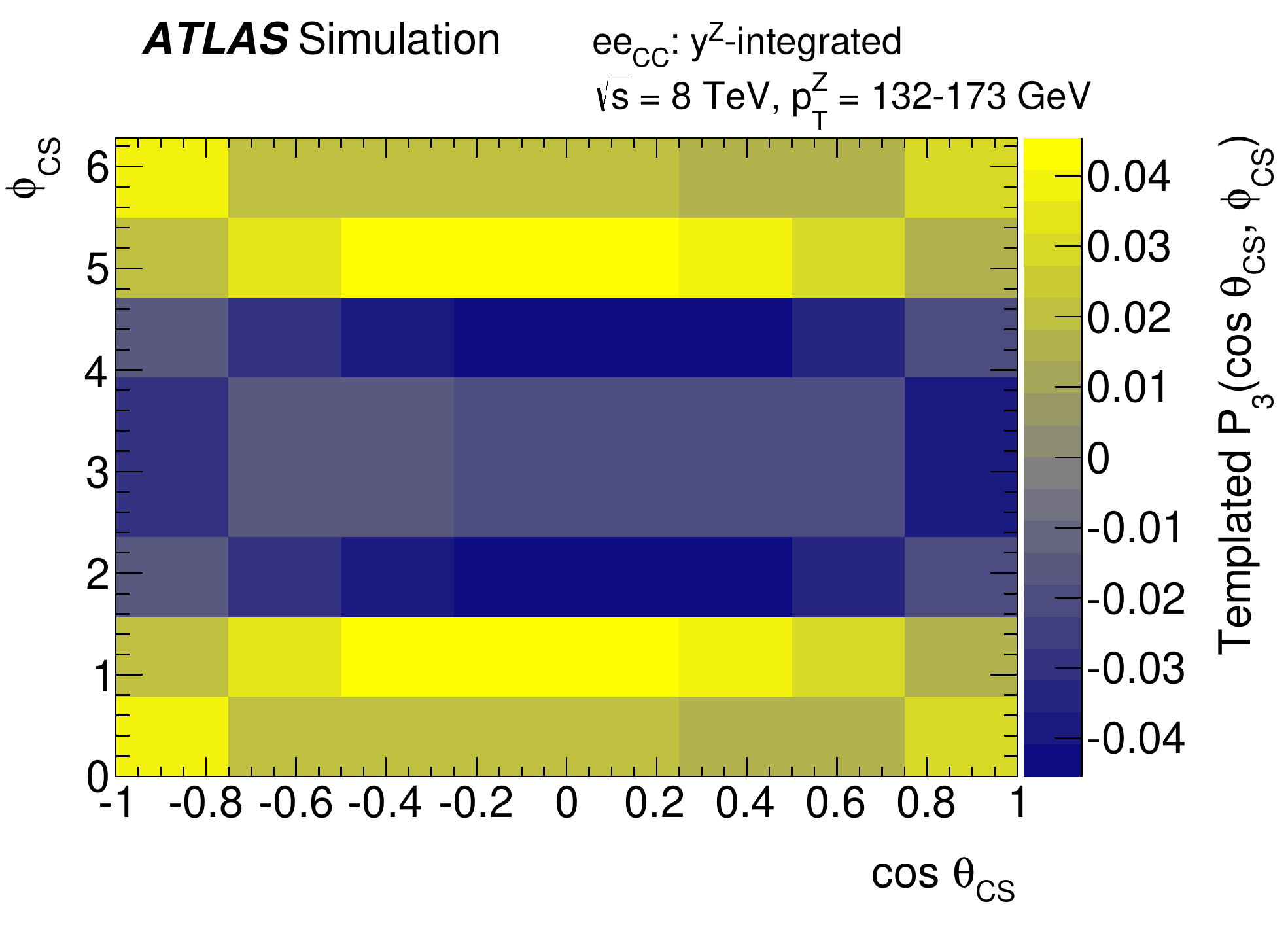}

}
\end{center}
\vspace{-5mm}
\caption{Shapes of the polynomials $P_{1,2,3}$ as a function of~$\costhetacs$ and $\phics$ (top). Below these are the templated polynomials for the $\yz$-integrated $ee_{CC}$~events at low~($5-8$~GeV), medium~($22-25.5$~GeV), and high~($132-173$~GeV) values of~$\ptz$.
\label{Fig:templates2D_Ai_2}}
\end{figure}

\begin{figure}
  \begin{center}
{
    \includegraphics[width=5cm,angle=0]{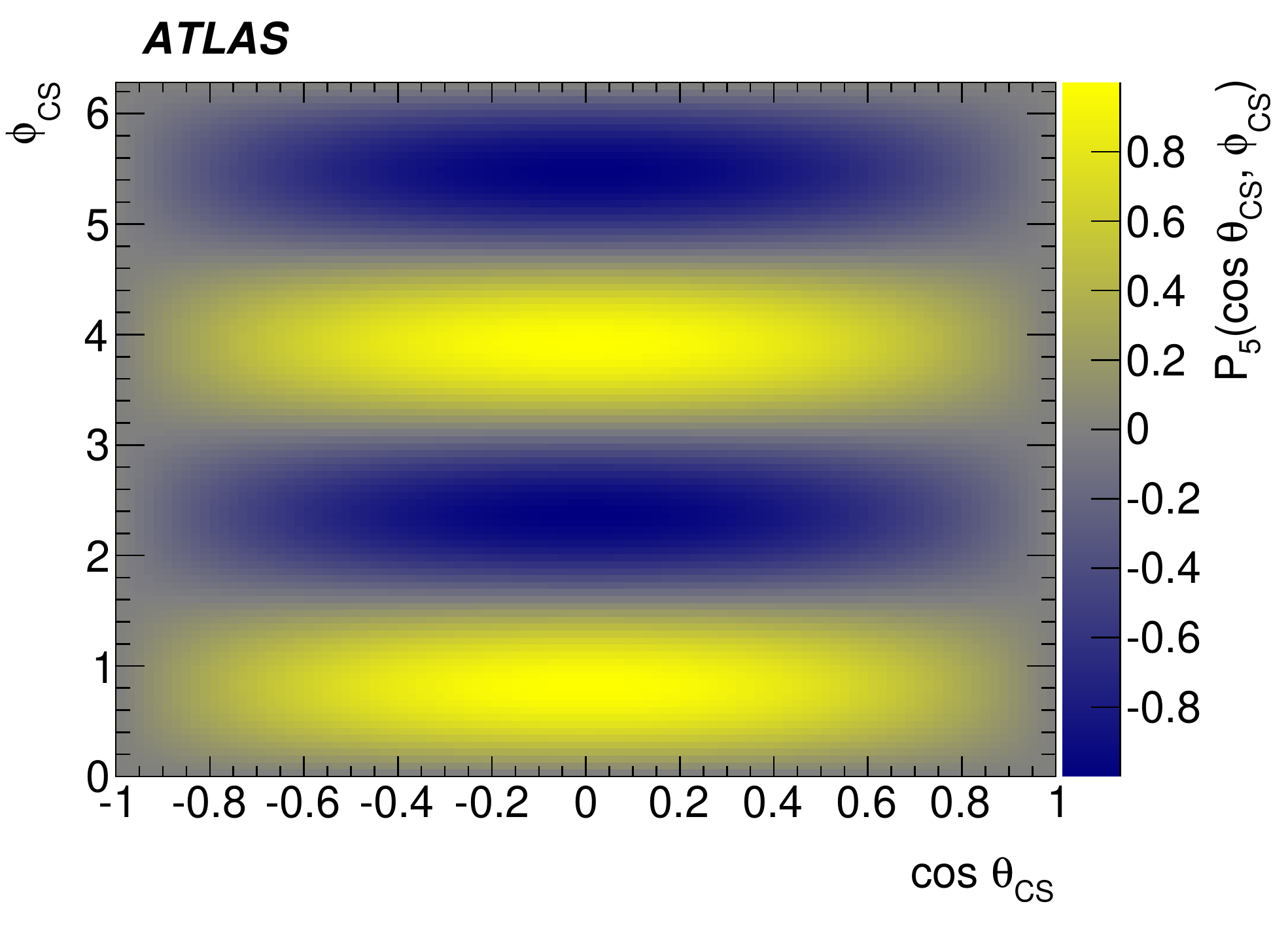}
    \includegraphics[width=5cm,angle=0]{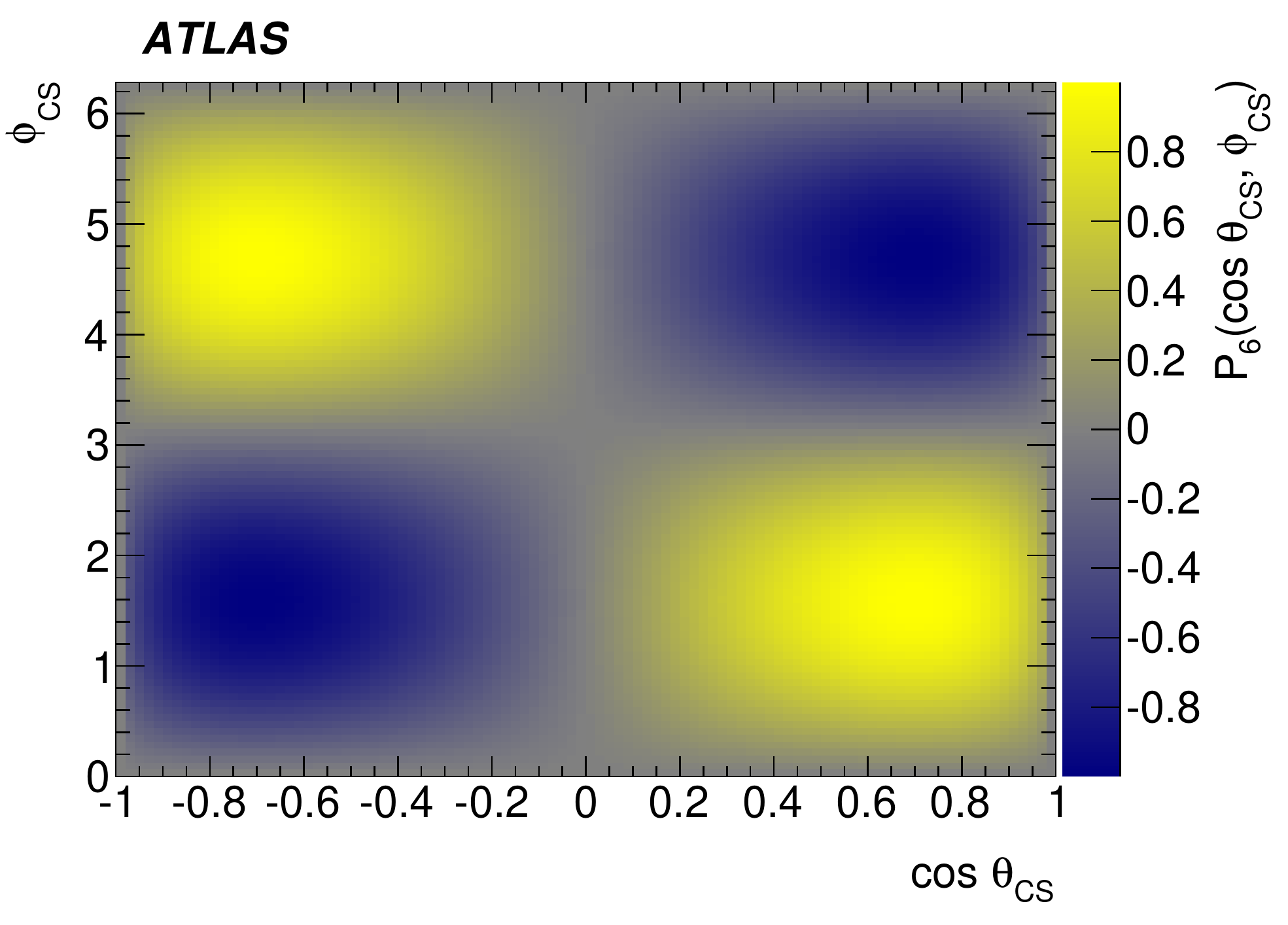}
    \includegraphics[width=5cm,angle=0]{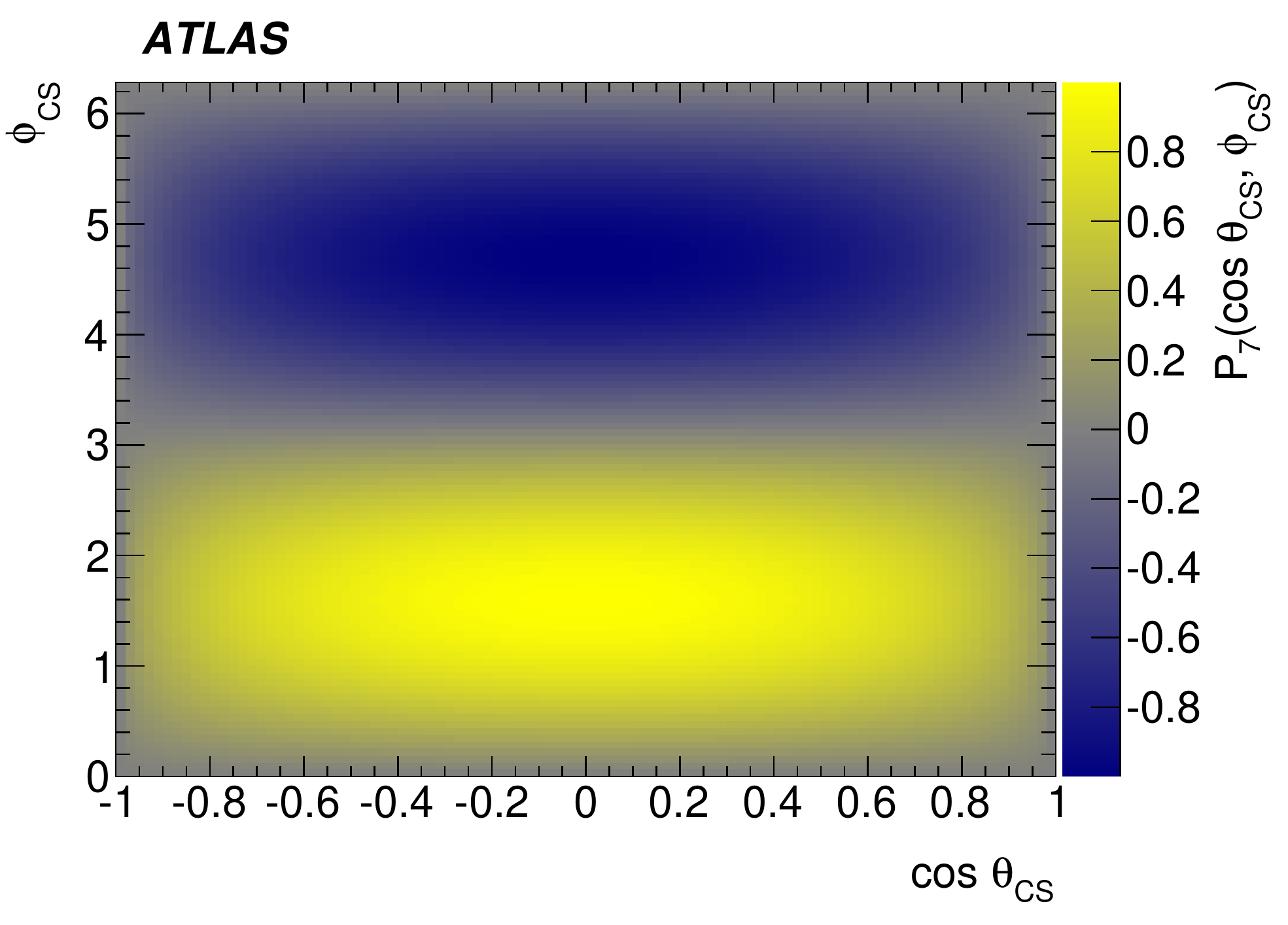}

    \includegraphics[width=5cm,angle=0]{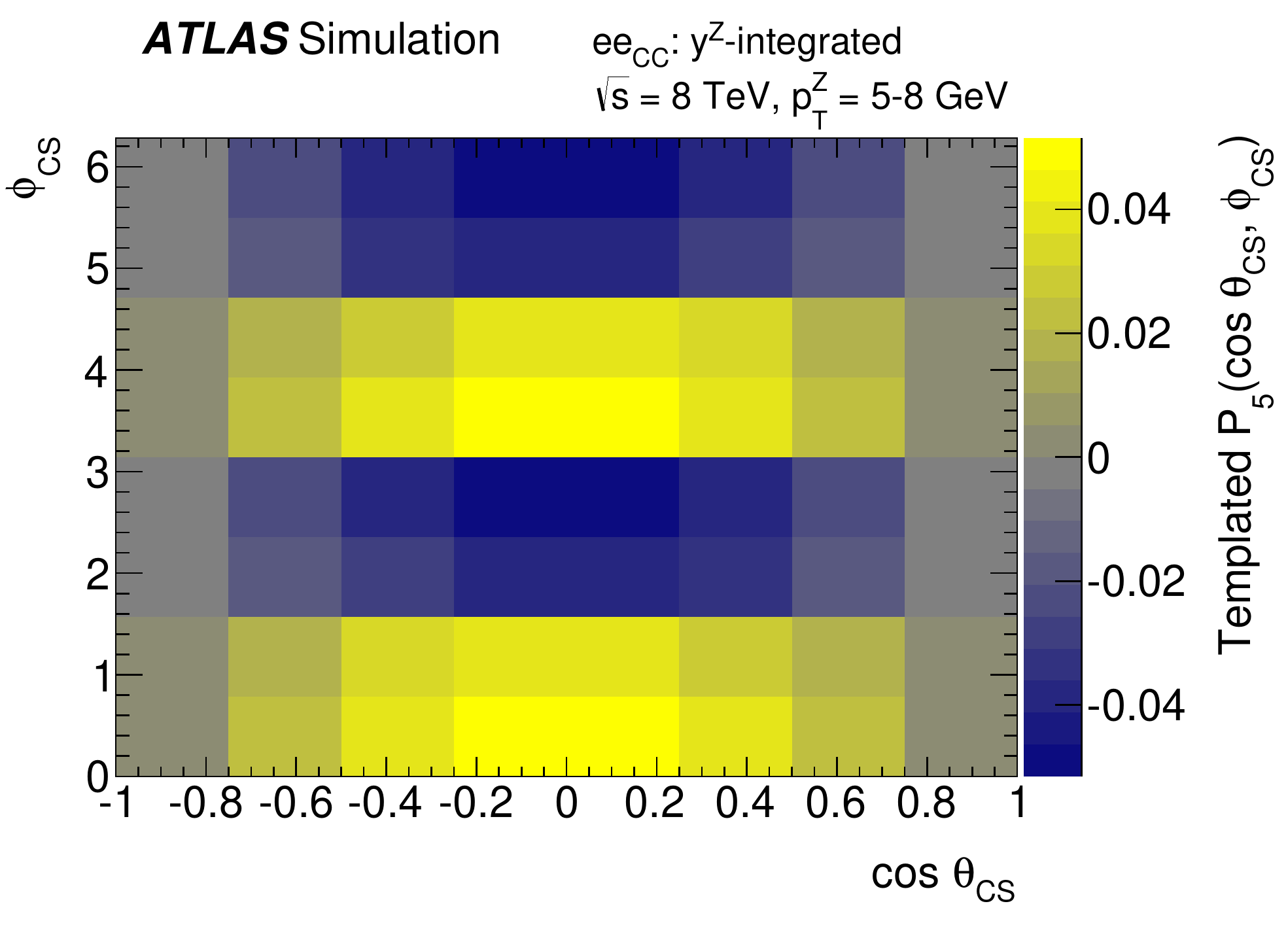}
    \includegraphics[width=5cm,angle=0]{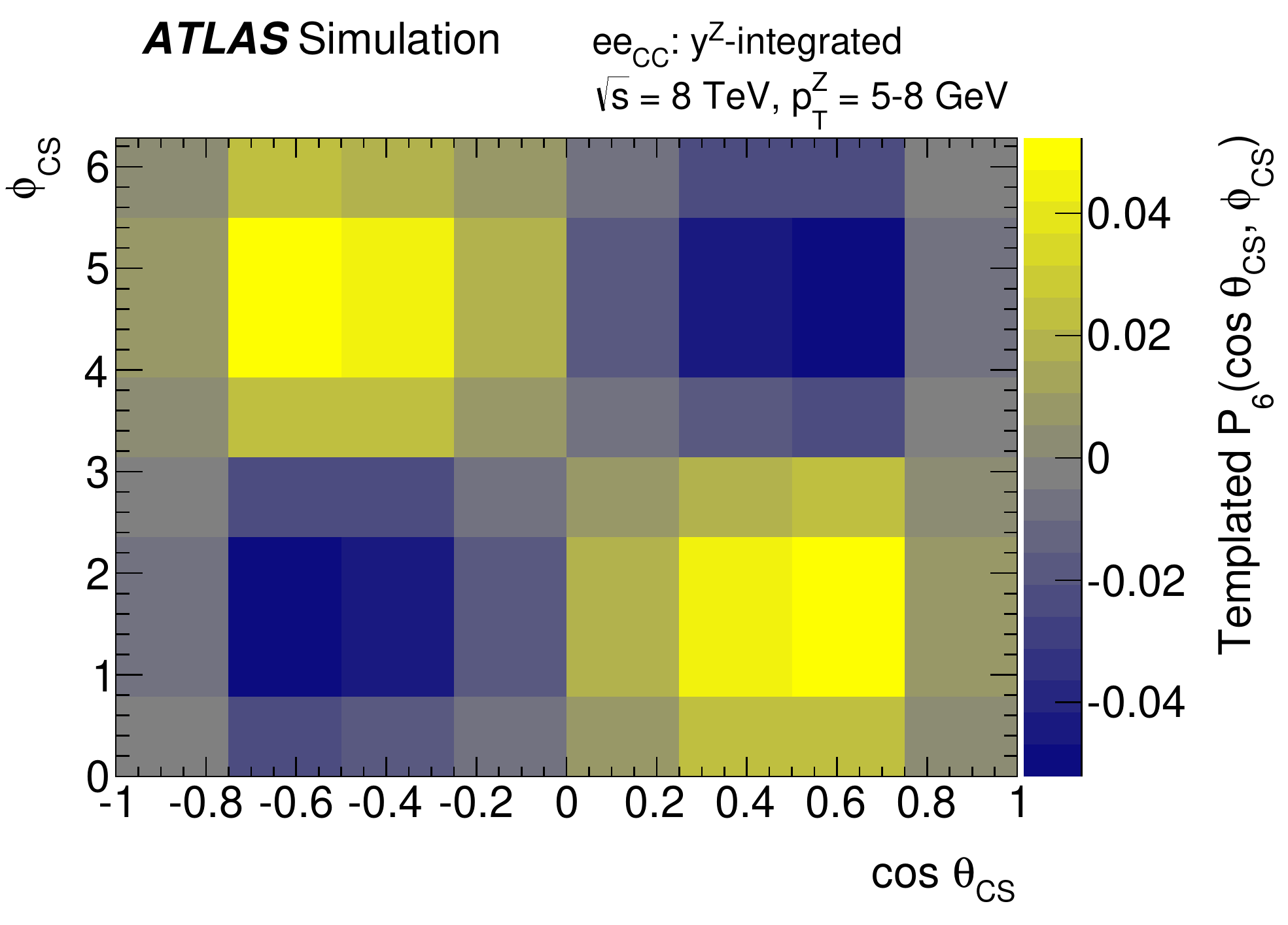}
    \includegraphics[width=5cm,angle=0]{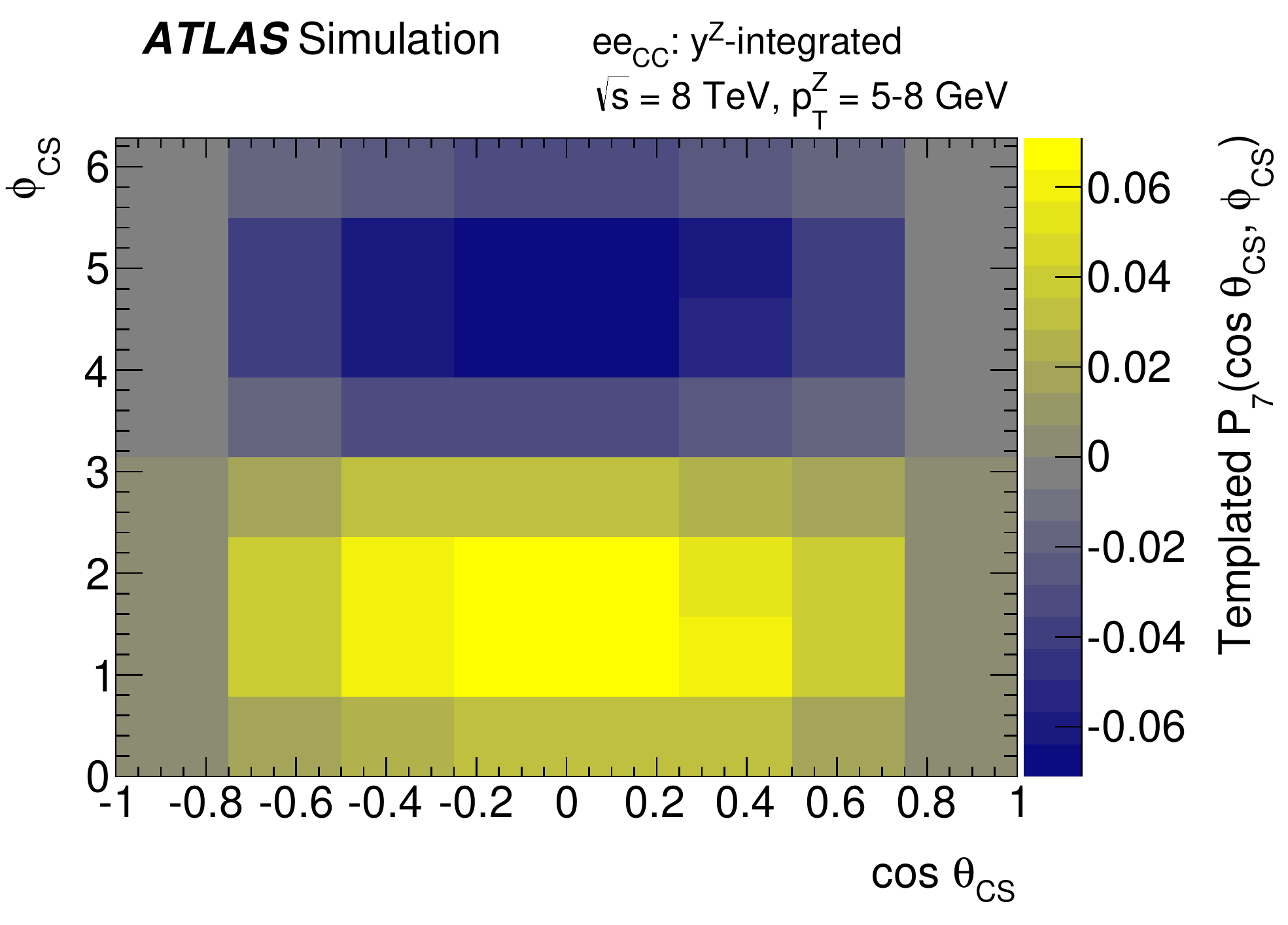}

    \includegraphics[width=5cm,angle=0]{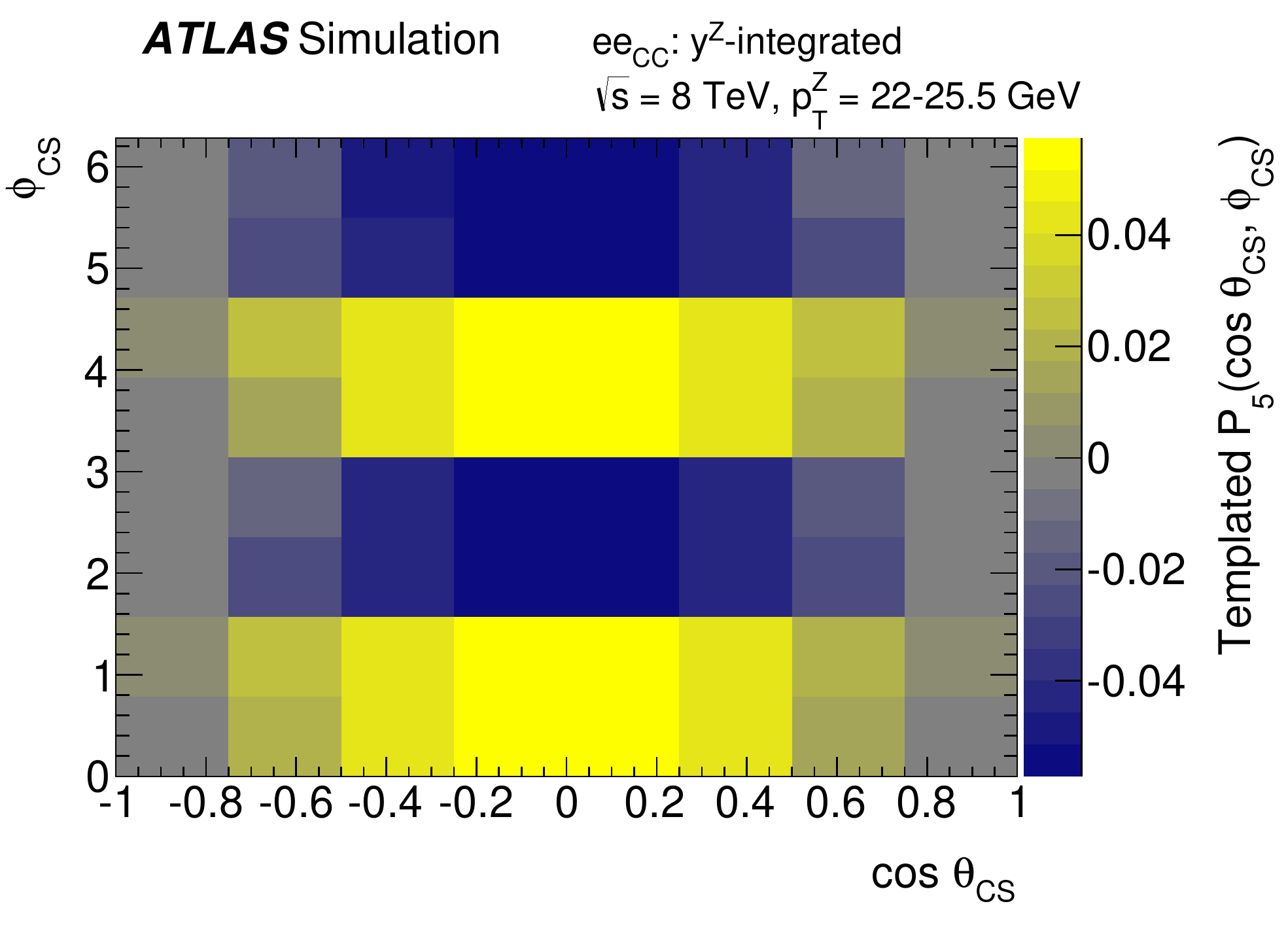}
    \includegraphics[width=5cm,angle=0]{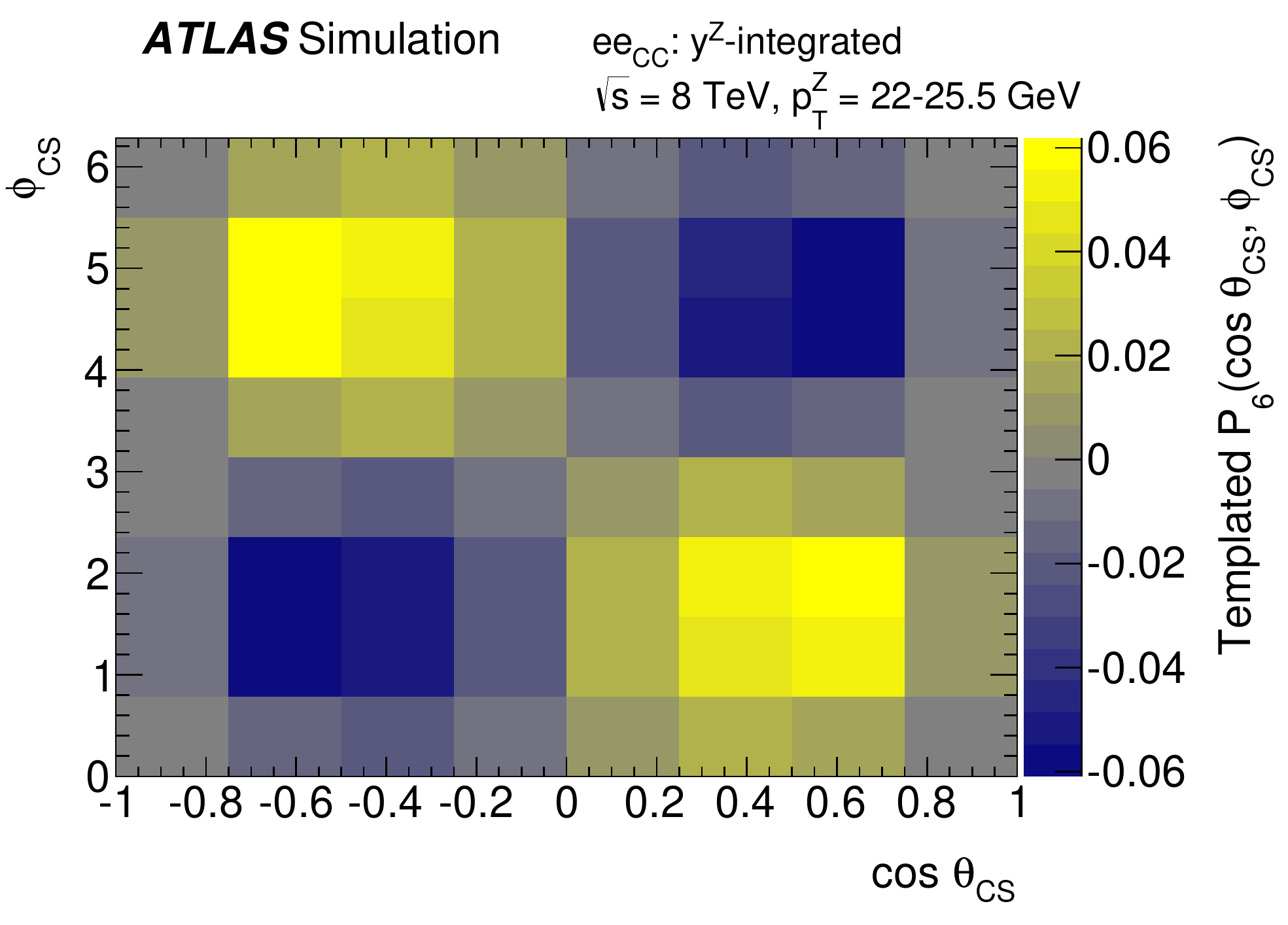}
    \includegraphics[width=5cm,angle=0]{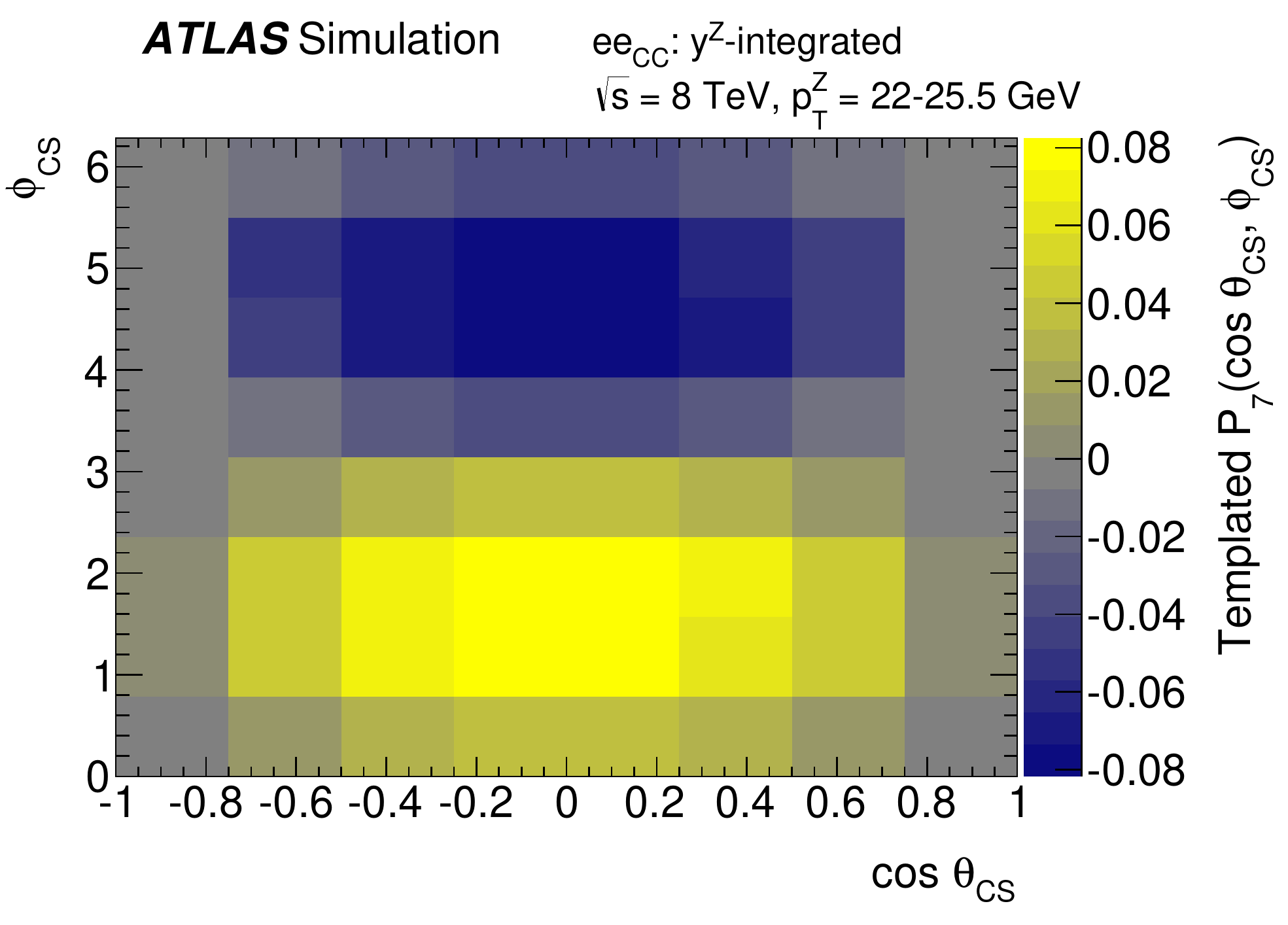}

    \includegraphics[width=5cm,angle=0]{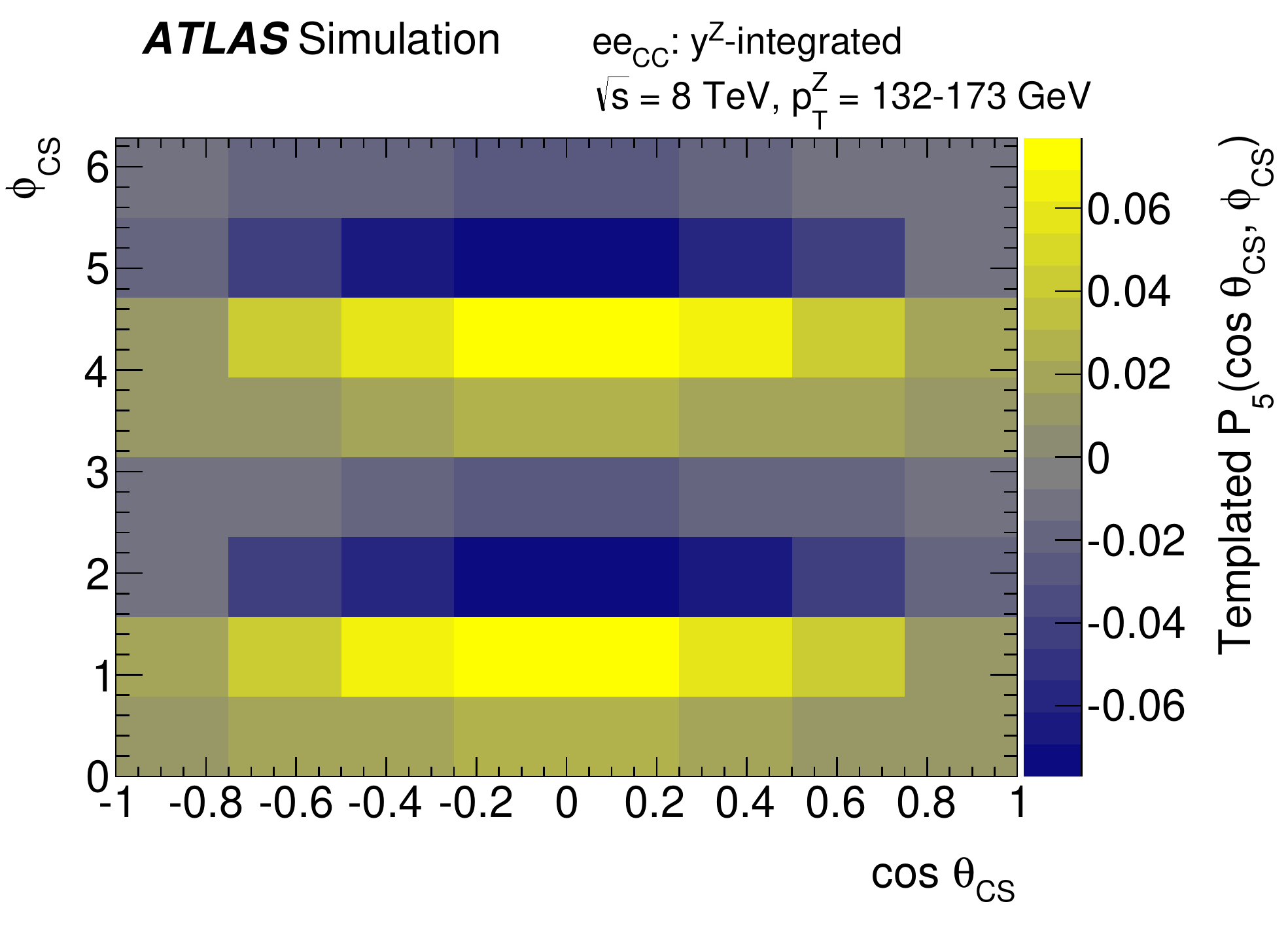}
    \includegraphics[width=5cm,angle=0]{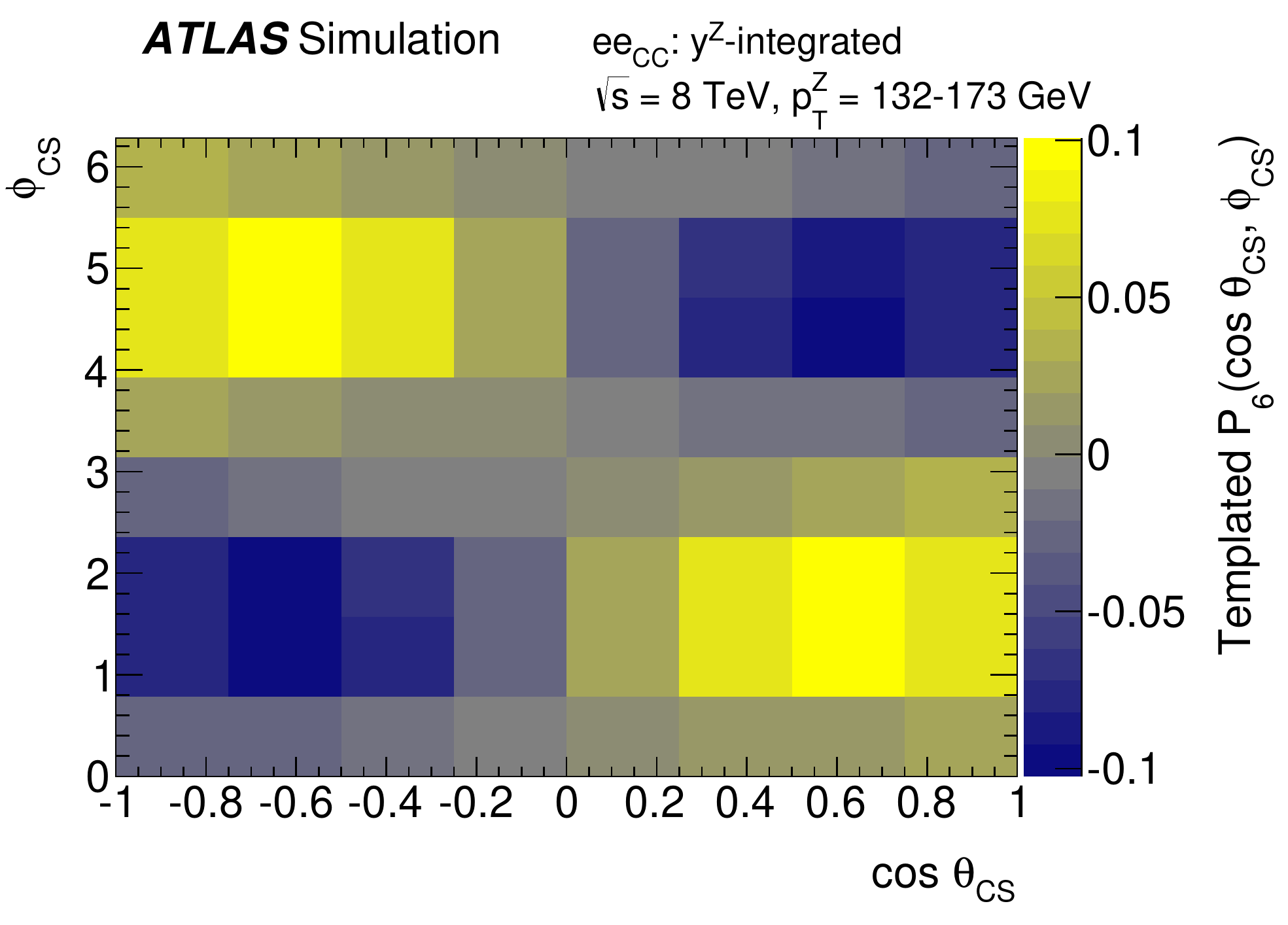}
    \includegraphics[width=5cm,angle=0]{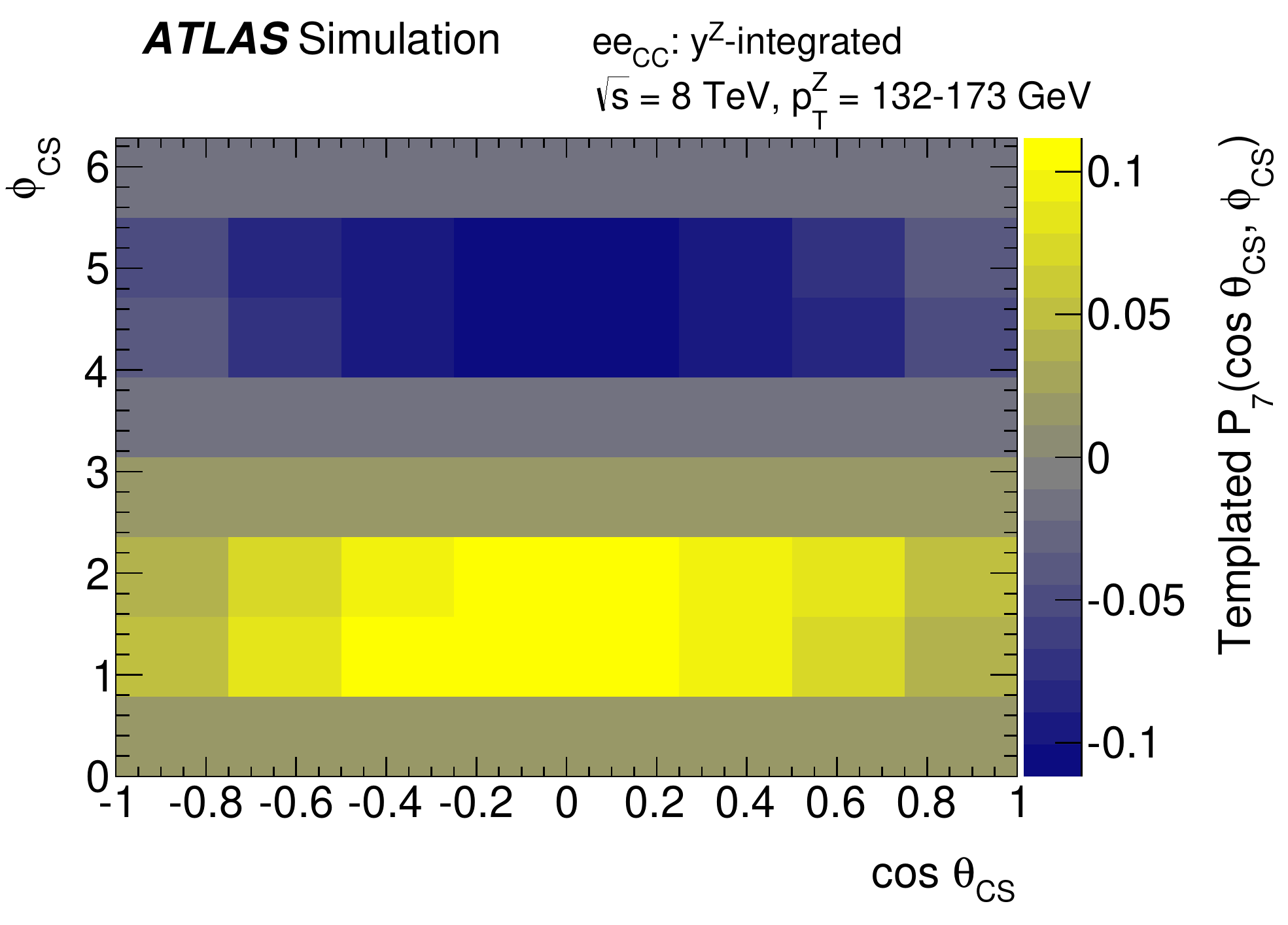}

}
\end{center}
\vspace{-5mm}
\caption{Shapes of the polynomials $P_{5,6,7}$ as a function of~$\costhetacs$ and $\phics$ (top). Below these are the templated polynomials for the $\yz$-integrated $ee_{CC}$~events at low~($5-8$~GeV), medium~($22-25.5$~GeV), and high~($132-173$~GeV) values of~$\ptz$.
\label{Fig:templates2D_Ai_3}}
\end{figure}

\section{Regularisation}
\label{sec:regularisation}

The migration of events between $\ptz$ bins leads to anti-correlations between the measured $\Ai$ in neighbouring $\ptz$ bins which enhance the effects of statistical fluctuations. To mitigate this effect and aid in resolving the underlying structure of the $\Ai$ spectra, the $\Ai$~coefficients are regularised by imposing a Gaussian penalty term on the significance of their higher-order derivatives with respect to $\ptz$. This penalty multiplies the likelihood in Eq.~(\ref{eq:likelihood}). 

The exact derivative order is chosen based on the expected reduction in statistical uncertainties of the measurement and the potential bias that the regularisation scheme may introduce. The smaller statistical uncertainties come with increased positive correlation between neighbouring coefficients. The $n^{th}$ derivative of $A_{ij}$ is defined as:

\begin{equation}
A^{(n)}_{ij} = \left\{
\begin{array}{l l}
 A^{(n-1)}_{i,j} - A^{(n-1)}_{i,j-1}, & \quad n\ \rm{odd} \\
 A^{(n-1)}_{i,j+1} - A^{(n-1)}_{i,j}, & \quad n\ \rm{even} , \\
\end{array}
\right . 
\end{equation}
\label{formula:reg_derivative}

where $A^{(0)}_{ij}\equiv A_{ij}$. The derivatives are staggered between even and odd orders in order to create a derivative definition more symmetric around each $\ptz$ bin. 

Since the measurement is determined from the exact likelihood expression for the coefficients, the covariance matrix ${\bf\Sigma}$ of the coefficients can be derived based on the second-order partial derivatives of the likelihood~\cite{Cowen2011}. Along with the uncertainties in $A_{i}$ and $A_{j}$, namely $\sigma(A_{i})$ and $\sigma(A_{j})$, their correlation can be computed as $\rho_{ij} = \Sigma_{ij}/\sigma(A_{i})\sigma(A_{j})$. This is done based on pseudo-data taken from \POWHEG+~\PYTHIA8 and is shown in Fig.~\ref{Fig:asimov_regularized_cov} before and after regularisation.

\begin{figure}
\begin{center}
\includegraphics[width=7.5cm,angle=0]{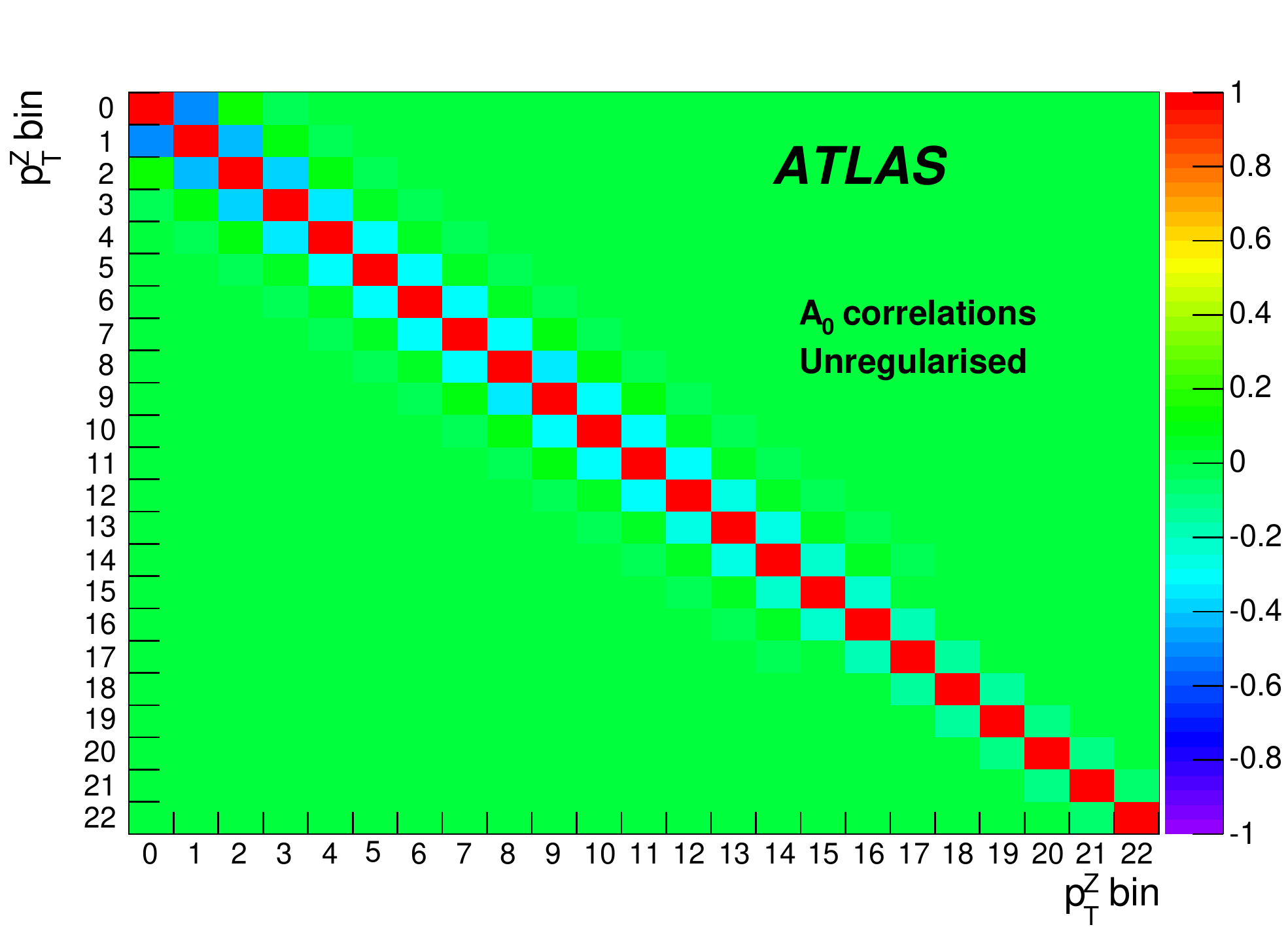}
\includegraphics[width=7.5cm,angle=0]{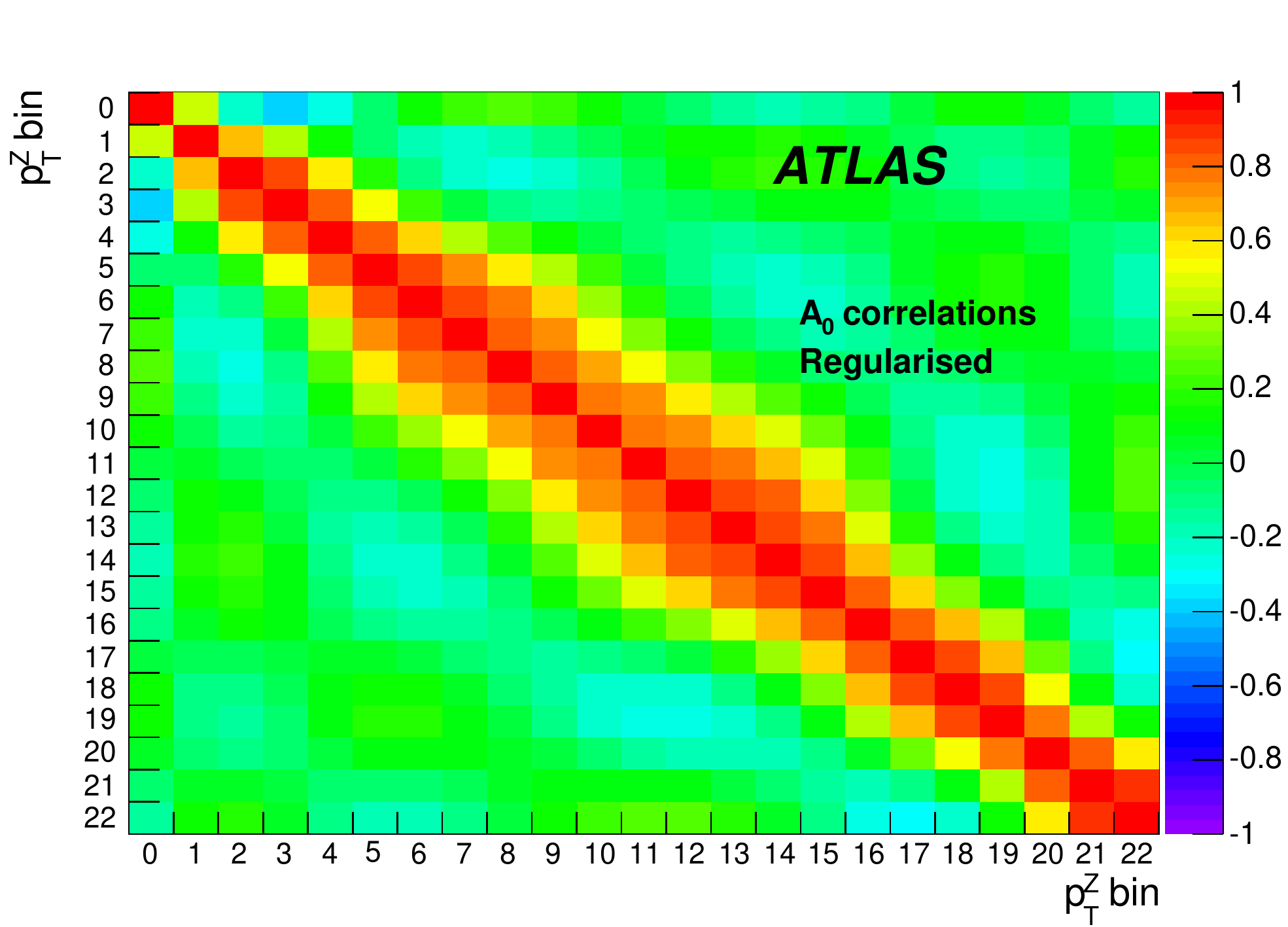}
\end{center}
\caption{Correlation matrix between the $\ptz$ bins of $A_{0}$ before (left) and after (right) regularisation.}
\label{Fig:asimov_regularized_cov}
\end{figure}

A Jacobian matrix ${\bf J}$ (and its transpose ${\bf J}^{T}$) is used to transform the covariance matrix of the coefficients to the covariance matrix of their derivatives. A regularisation strength $\gamma$ is introduced to control the amount by which the derivatives are penalised. The penalty term applied to the likelihood that controls the regularisation is therefore defined as

\begin{equation}
\mathcal{A}({\bf A}_{\rm{reg}}) = \exp\left\{-0.5\gamma \vec{A}^{(n)}({\bf J\Sigma J}^{T})^{-1}\vec{A}^{(n),T} \right\}.
\end{equation}
\label{formula:reg_penalty}

In all channels, a regularisation scheme using sixth-order derivatives is used.

In the limit that the regularisation procedure described above has infinite strength, an $n^{th}$-order derivative regularisation fixes the measured spectrum to be an $(n-1)^{th}$-order polynomial. This can be seen in Fig.~\ref{Fig:residual_meas_A0}, which shows the residual of a fifth-order polynomial fit to the $A_{0}$ spectrum regularised with sixth-order derivatives and strength $\gamma=100$; the fit is nearly perfect (the regularisation strength used in this case is large but not infinite and so there are some small non-zero residuals). Also shown is the residual of a fourth-order polynomial fit to this same spectrum; a fifth-order term can be clearly observed in the residual.
The regularisation bias $B[A_{ij}]$ in the coefficients is evaluated using pseudo-experiments based on the difference between the expectation value of the best-fit coefficient $ E[A_{ij}]$ and the value of the coefficient $y_{ij}$ used to randomise the data: $B[A_{ij}] = E[A_{ij}] - y_{ij}$. The choice of $y_{ij}$ is derived from a sixth-order polynomial fit to the \POWHEG+~\PYTHIA8 reference coefficients. 

The derived uncertainty due to the regularisation bias in the $\yz$-integrated $A_{0}$ coefficient in the $ee_{\text{CC}}+\mu\mu_{\text{CC}}$ channel is shown in~Fig.~\ref{Fig:reg_bias_comp} for four different regularisation strengths, along with the corresponding statistical uncertainty of the coefficient for each strength. As can be seen, the regularisation uncertainty increases with increasing regularisation strength, while the corresponding statistical uncertainty decreases, as expected. In the limit that the regularisation strength goes to zero, the statistical uncertainty approaches the unregularised one. Along with the decrease in statistical uncertainty comes an increase in correlation among the measured coefficients of neighbouring $\ptz$ bins. The regularisation bias uncertainty appears to plateau between $\gamma=10$ and $\gamma=100$, which corresponds to the limit that the spectrum is fixed to a sixth-order polynomial, as described above. Based on these studies, a strength of $\gamma=100$ is chosen for the $ee_{\text{CC}}$ and $\mu\mu_{\text{CC}}$ channels, while the scheme in the $ee_{\text{CF}}$ channel is based on a strength of $\gamma=5$.

\begin{figure}
\begin{center}
\includegraphics[width=7.5cm,angle=0]{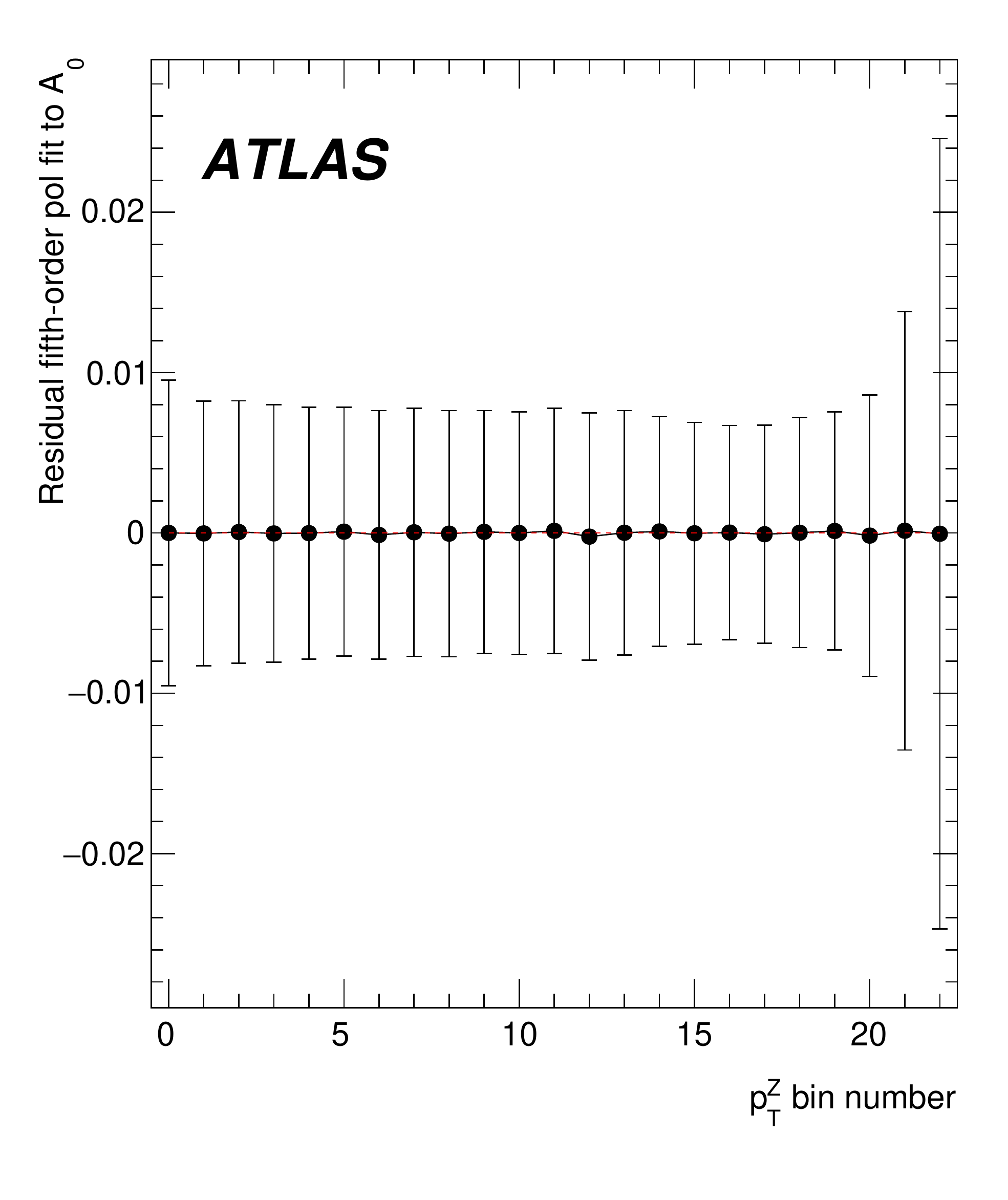}
\includegraphics[width=7.5cm,angle=0]{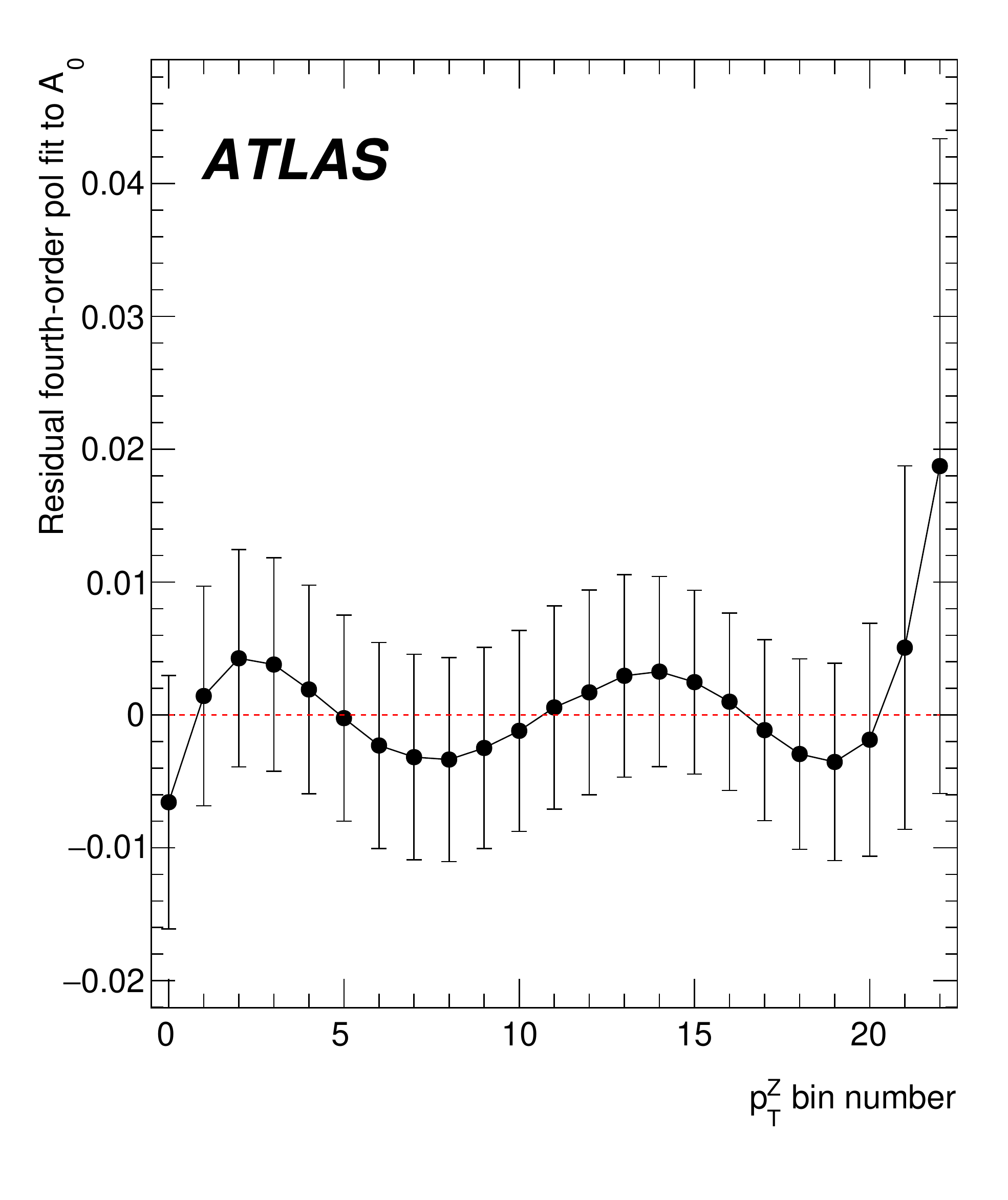}
\end{center}
\caption{Residuals of a fifth-order (left) and fourth-order (right) polynomial fit to the measured $A_{0}$ spectrum in the $ee_{\text{CC}}$ $\yz$-integrated channel, regularised with sixth-order derivatives.
\label{Fig:residual_meas_A0} }
\end{figure}

\begin{figure}
\begin{center}
\includegraphics[width=7.5cm,angle=0]{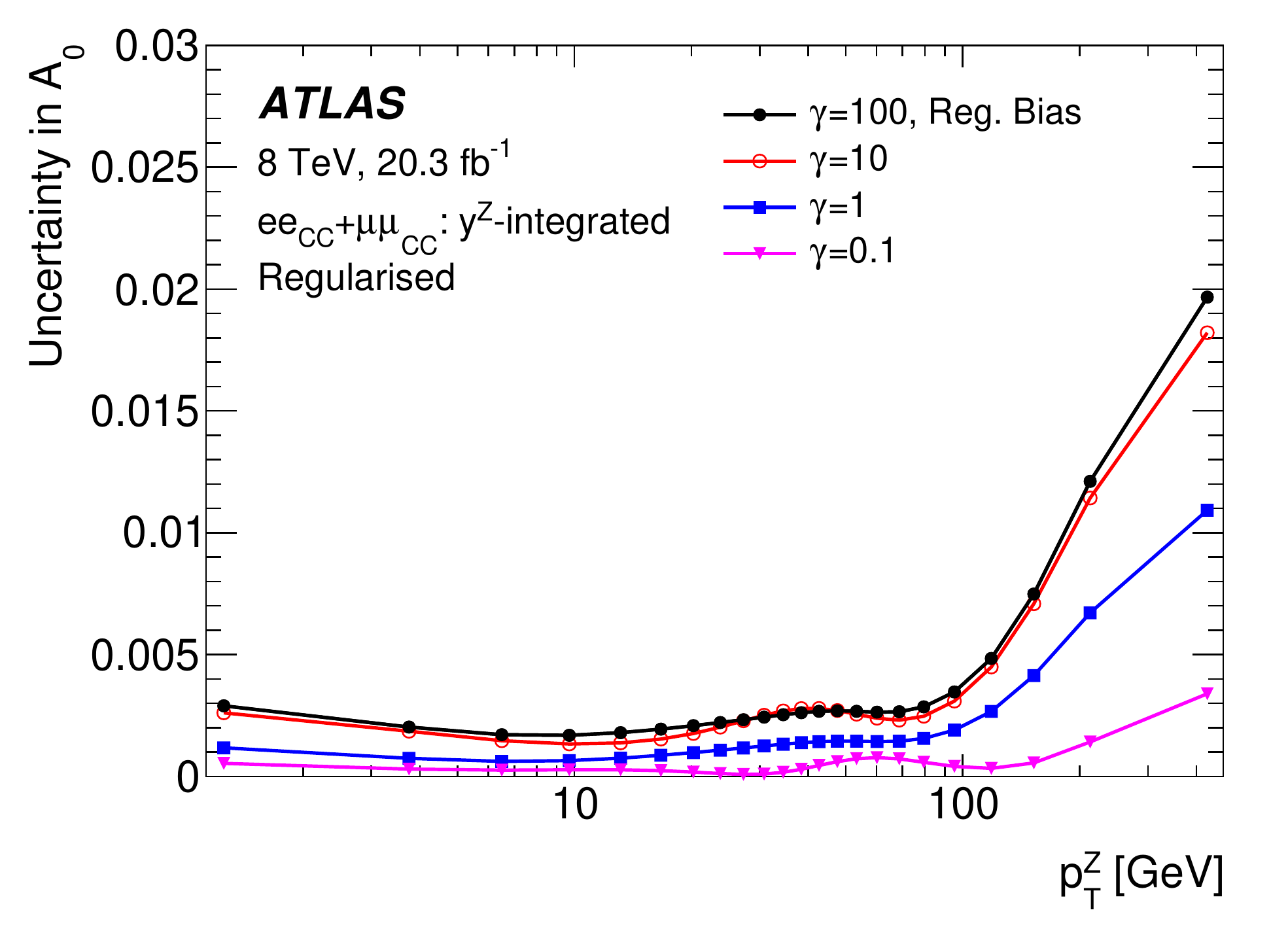}
\includegraphics[width=7.5cm,angle=0]{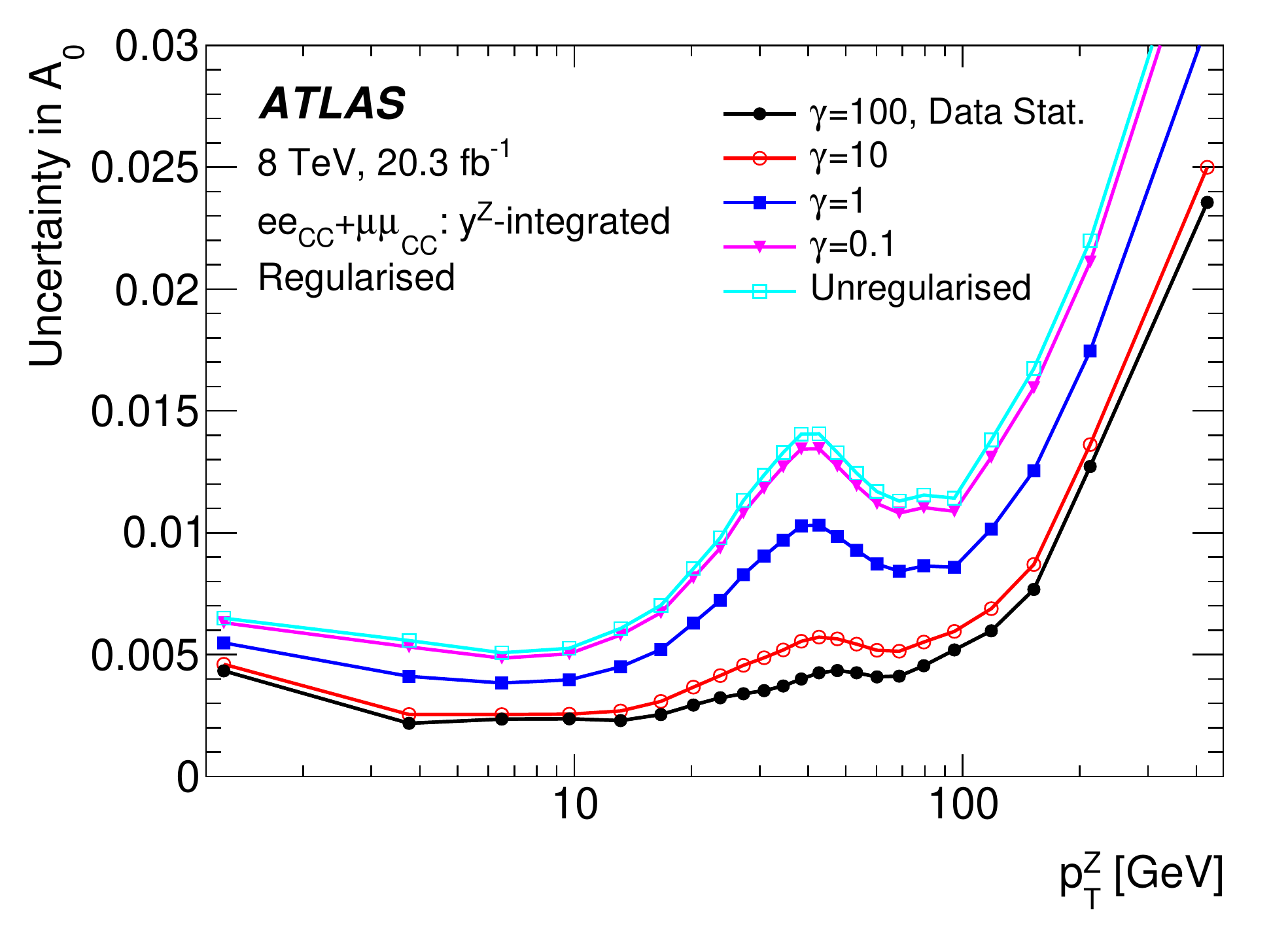}

\end{center}
\caption{For the $ee_{\text{CC}}+\mu\mu_{\text{CC}}$ channel, the derived regularisation bias uncertainty in the $\yz$-integrated $A_0$ coefficient for various regularisation strengths (left) along with the corresponding statistical uncertainty of the coefficient (right) versus $\ptz$. The unregularised statistical uncertainty is shown for comparison.
\label{Fig:reg_bias_comp} }
\end{figure}

Overlays of the regularised measurements for the $ee_{\text{CC}}+\mu\mu_{\text{CC}}$ channel in the $\yz$-integrated configuration are shown in~Fig.~\ref{Fig:reg_overlays_A07} for~$A_{0-7}$ and in~Fig.~\ref{Fig:reg_overlays_A02} for~$A_{0}-A_{2}$. In the unregularised results, there are many bin-to-bin fluctuations that enter primarily through anti-correlations between neighbouring measurements. In contrast, the regularised results are largely correlated from bin-to-bin and are much smoother.

\begin{figure}
\begin{center}
\includegraphics[width=7.5cm,angle=0]{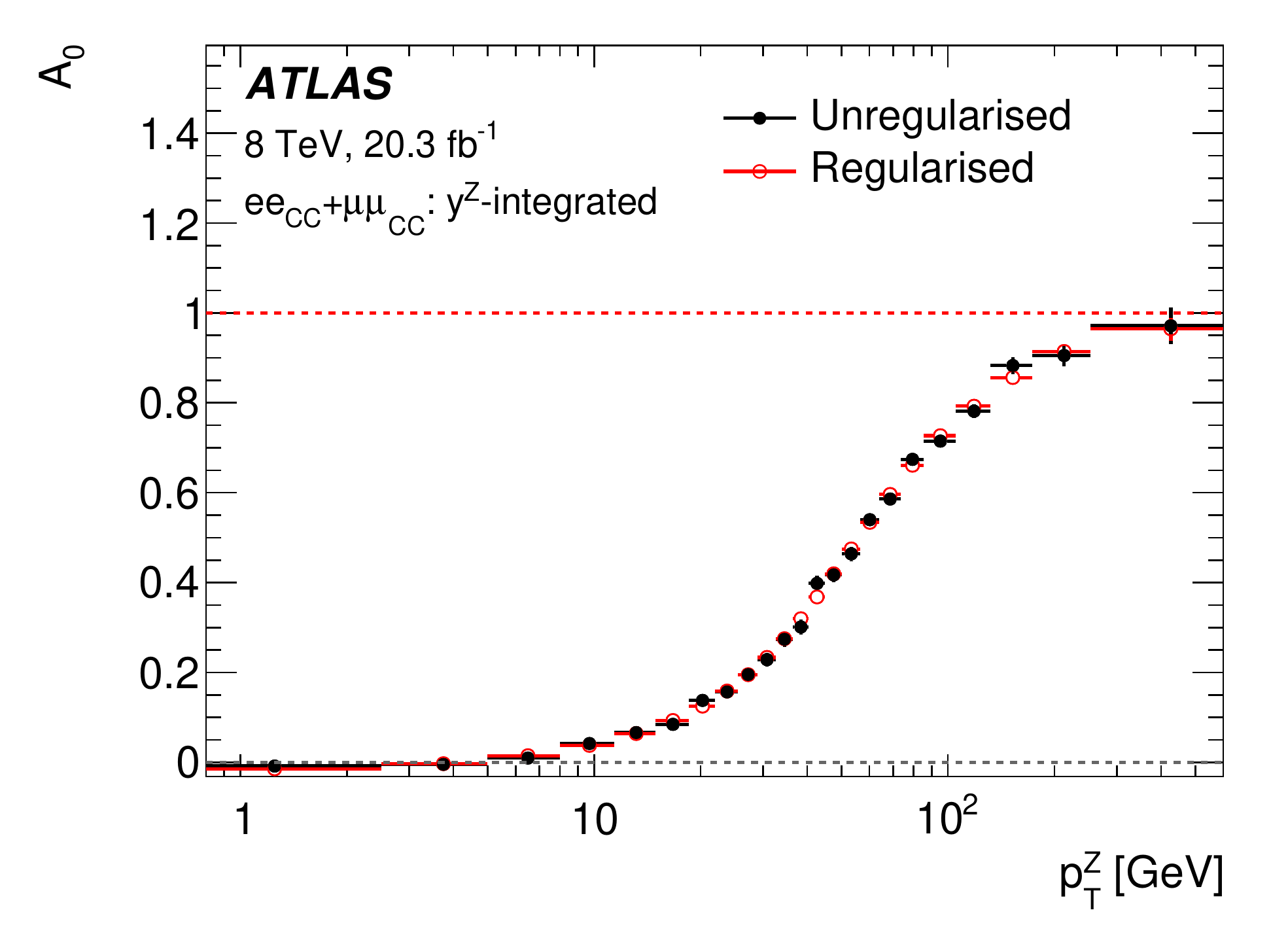}
\includegraphics[width=7.5cm,angle=0]{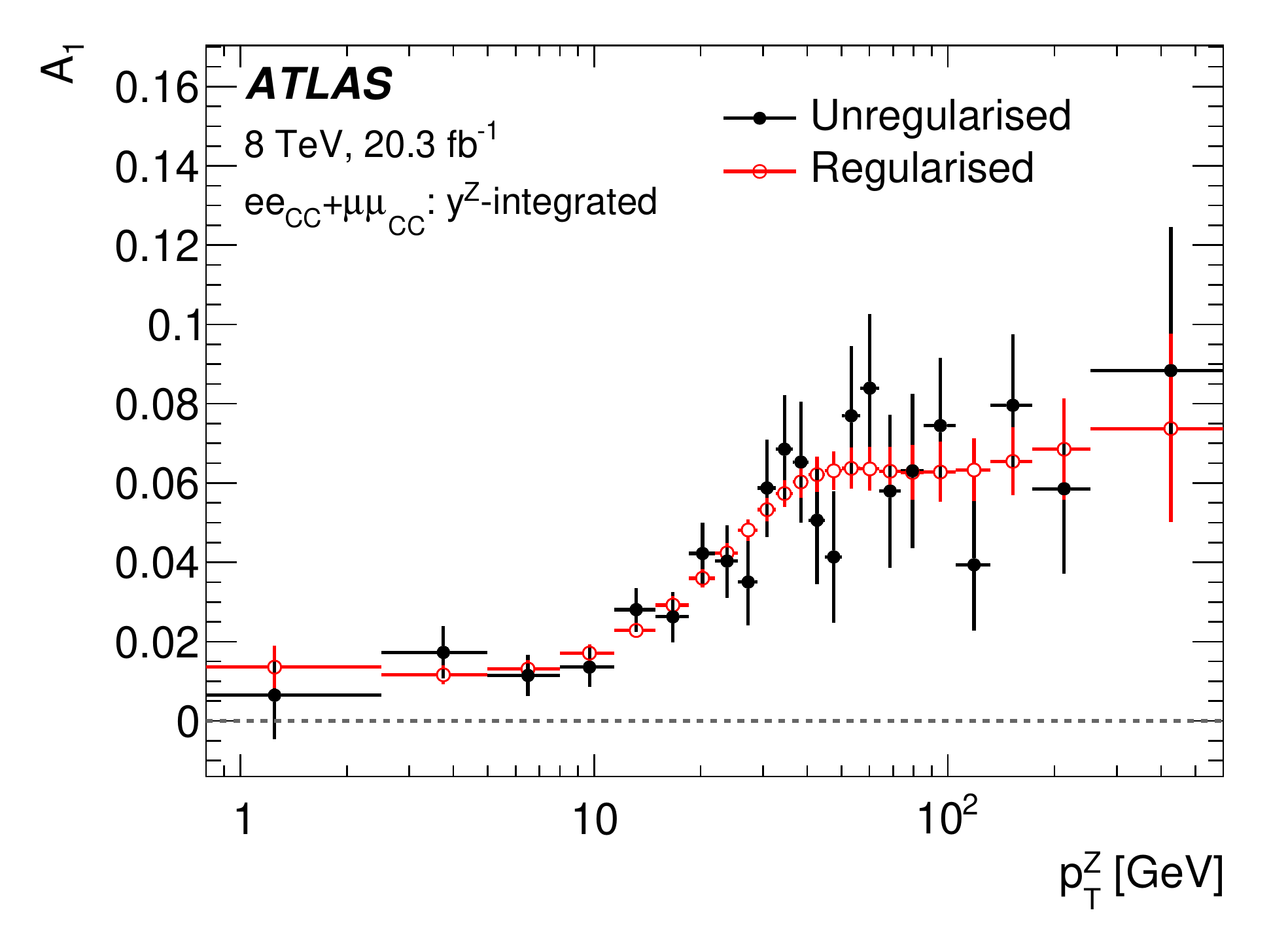}
\includegraphics[width=7.5cm,angle=0]{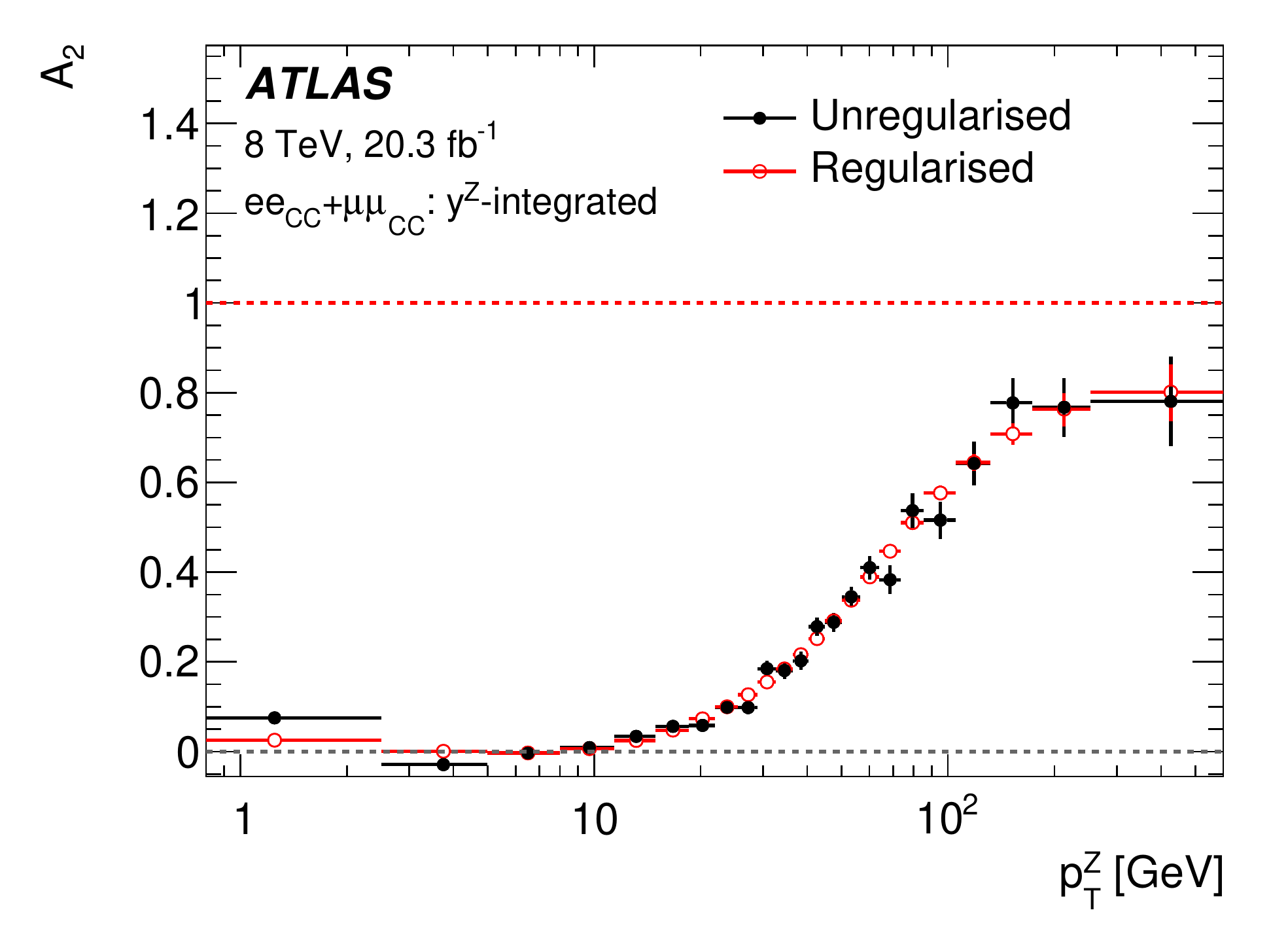}
\includegraphics[width=7.5cm,angle=0]{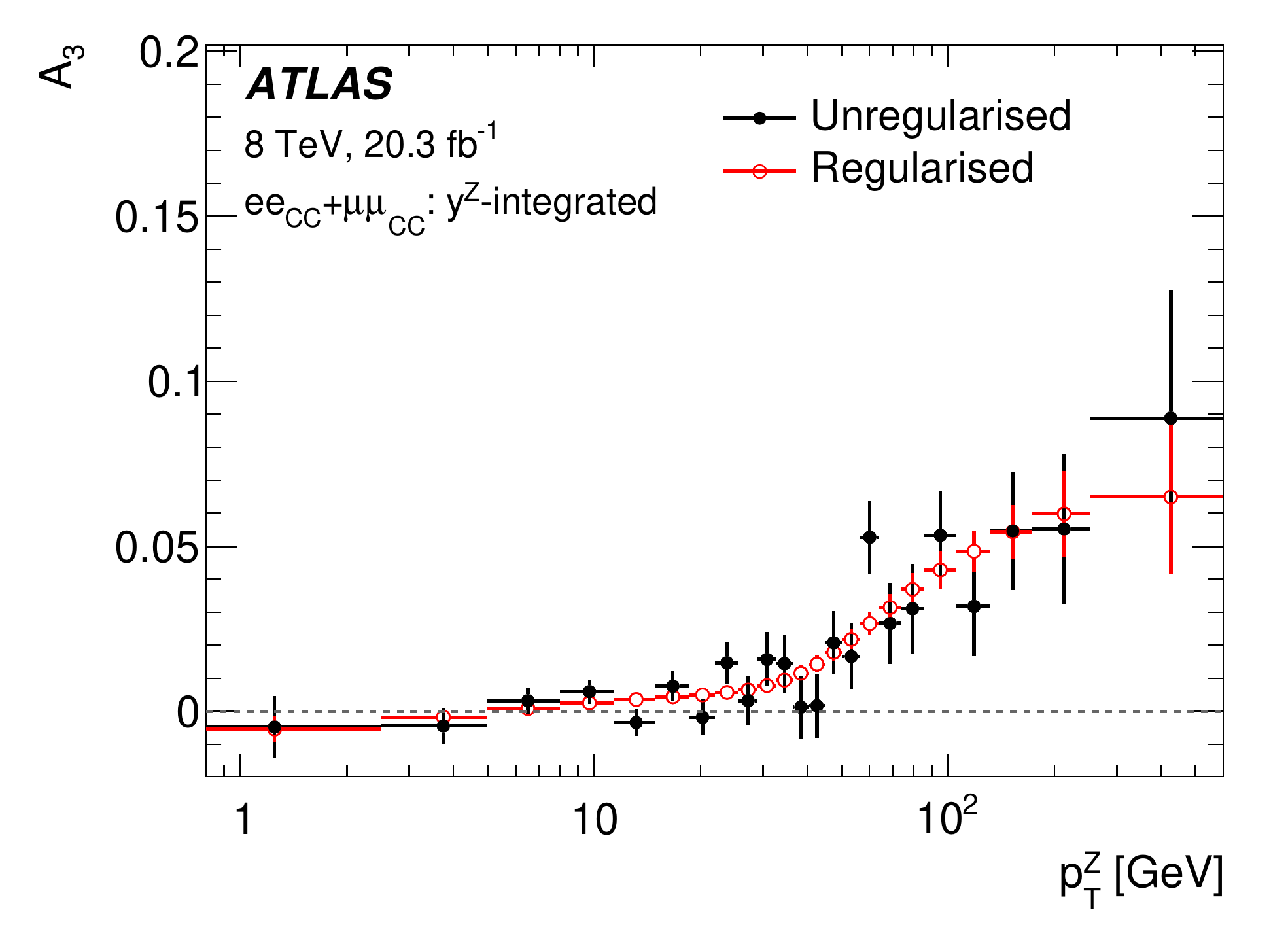}
\includegraphics[width=7.5cm,angle=0]{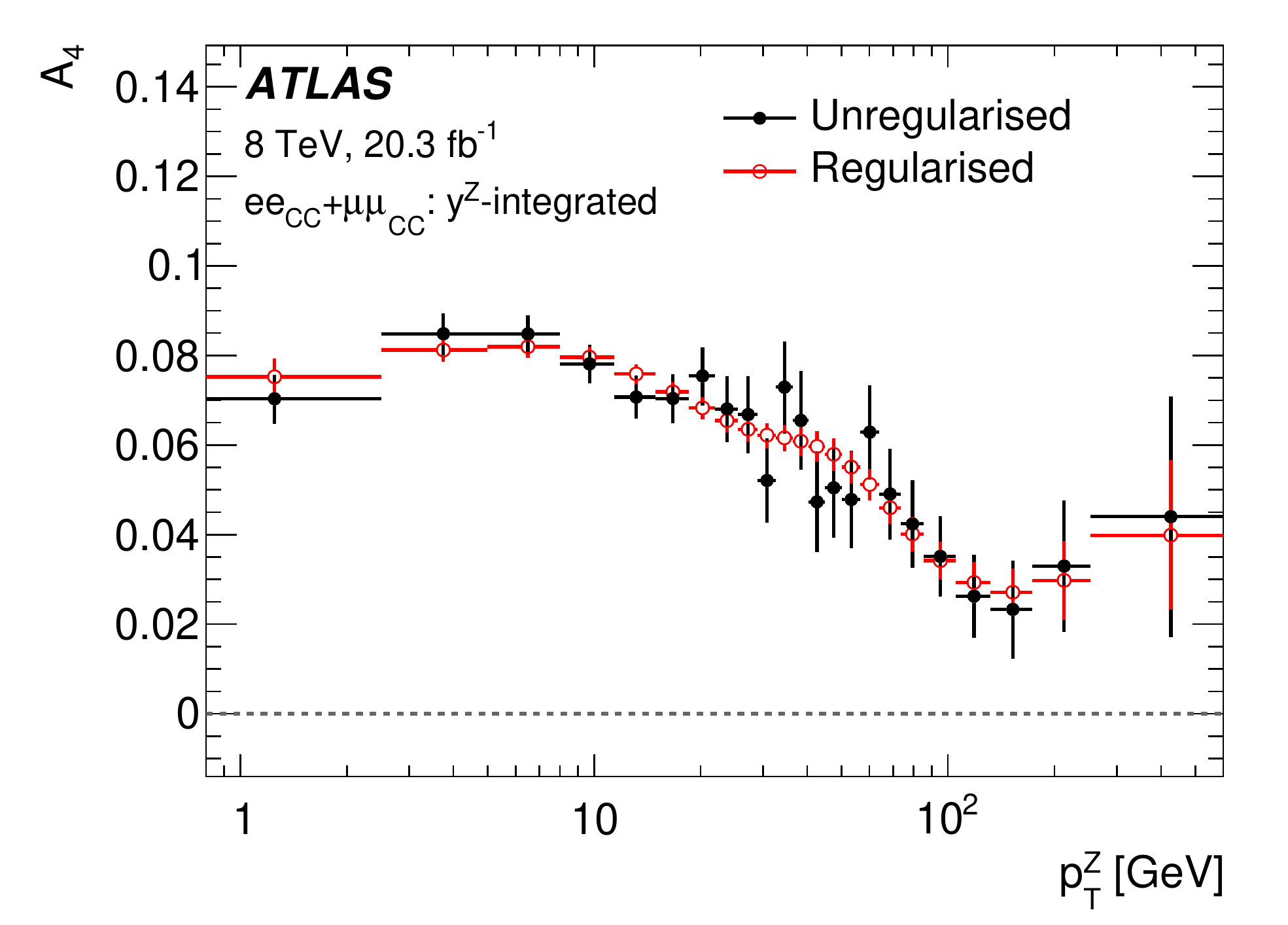}
\includegraphics[width=7.5cm,angle=0]{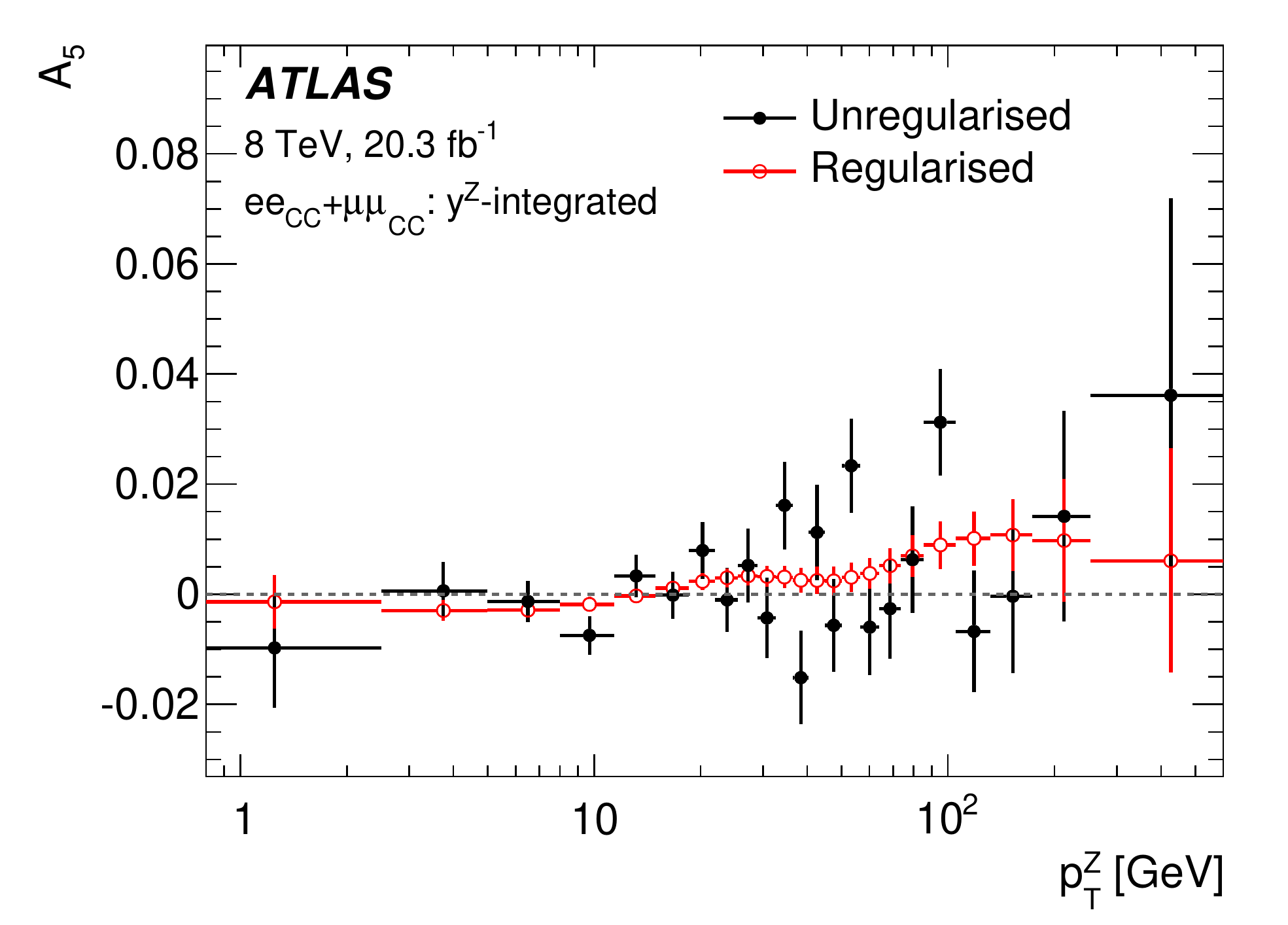}
\includegraphics[width=7.5cm,angle=0]{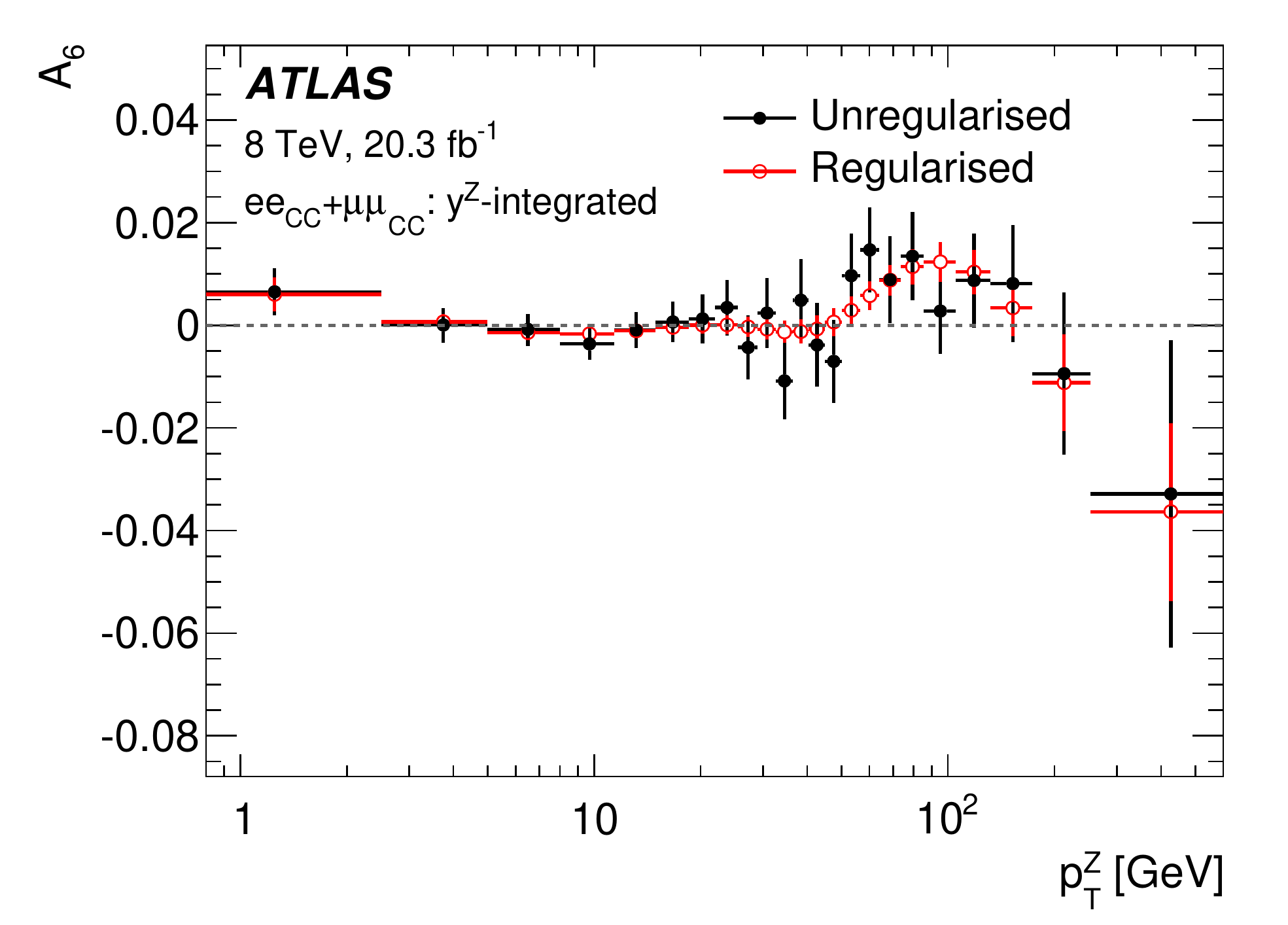}
\includegraphics[width=7.5cm,angle=0]{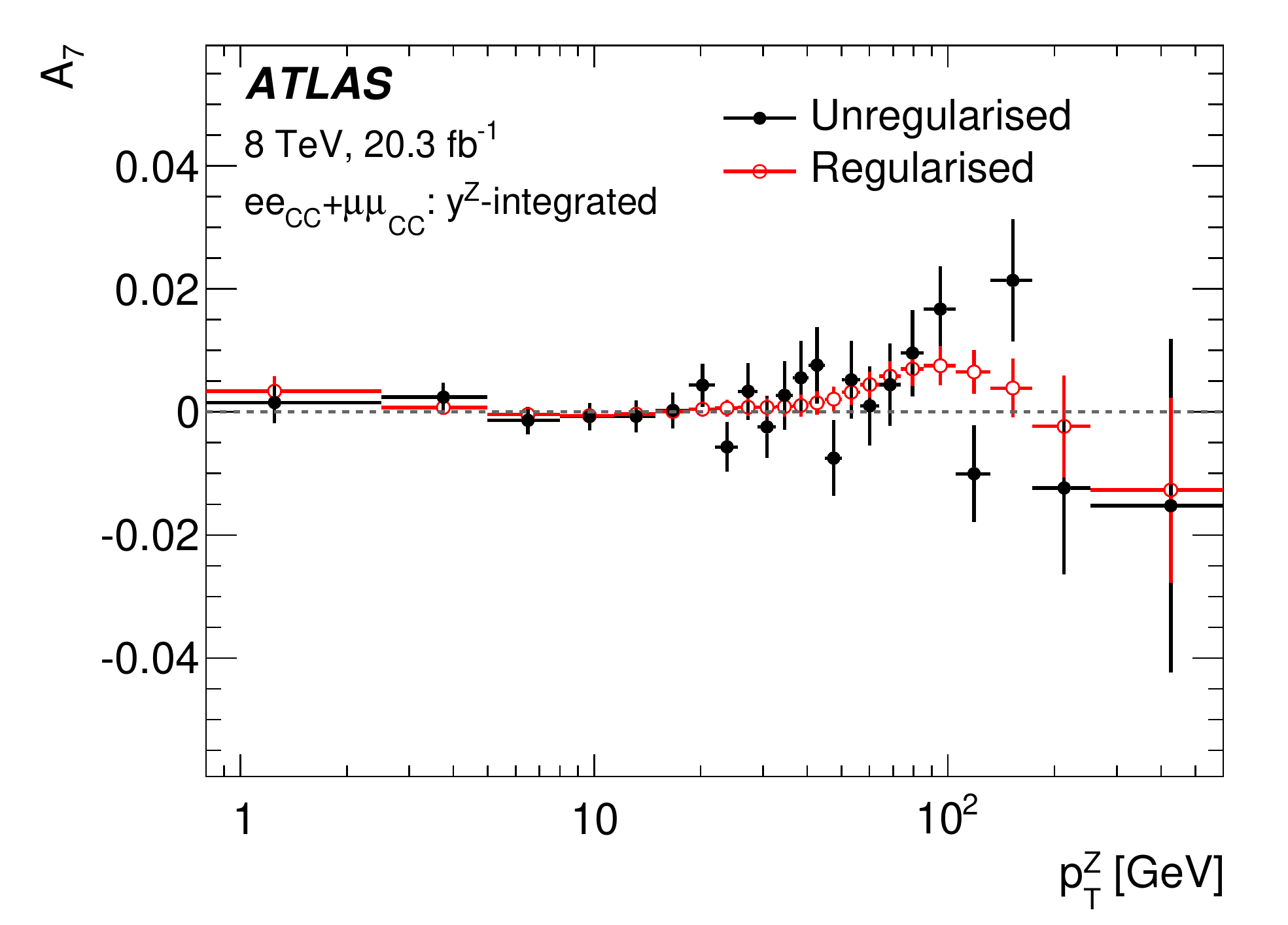}

\end{center}
\caption{For the $ee_{\text{CC}}+\mu\mu_{\text{CC}}$ channel in the $\yz$-integrated configuration, overlays of regularised with unregularised results are shown for $A_{0-7}$.
\label{Fig:reg_overlays_A07} }
\end{figure}

\begin{figure}
\begin{center}
\includegraphics[width=15cm,angle=0]{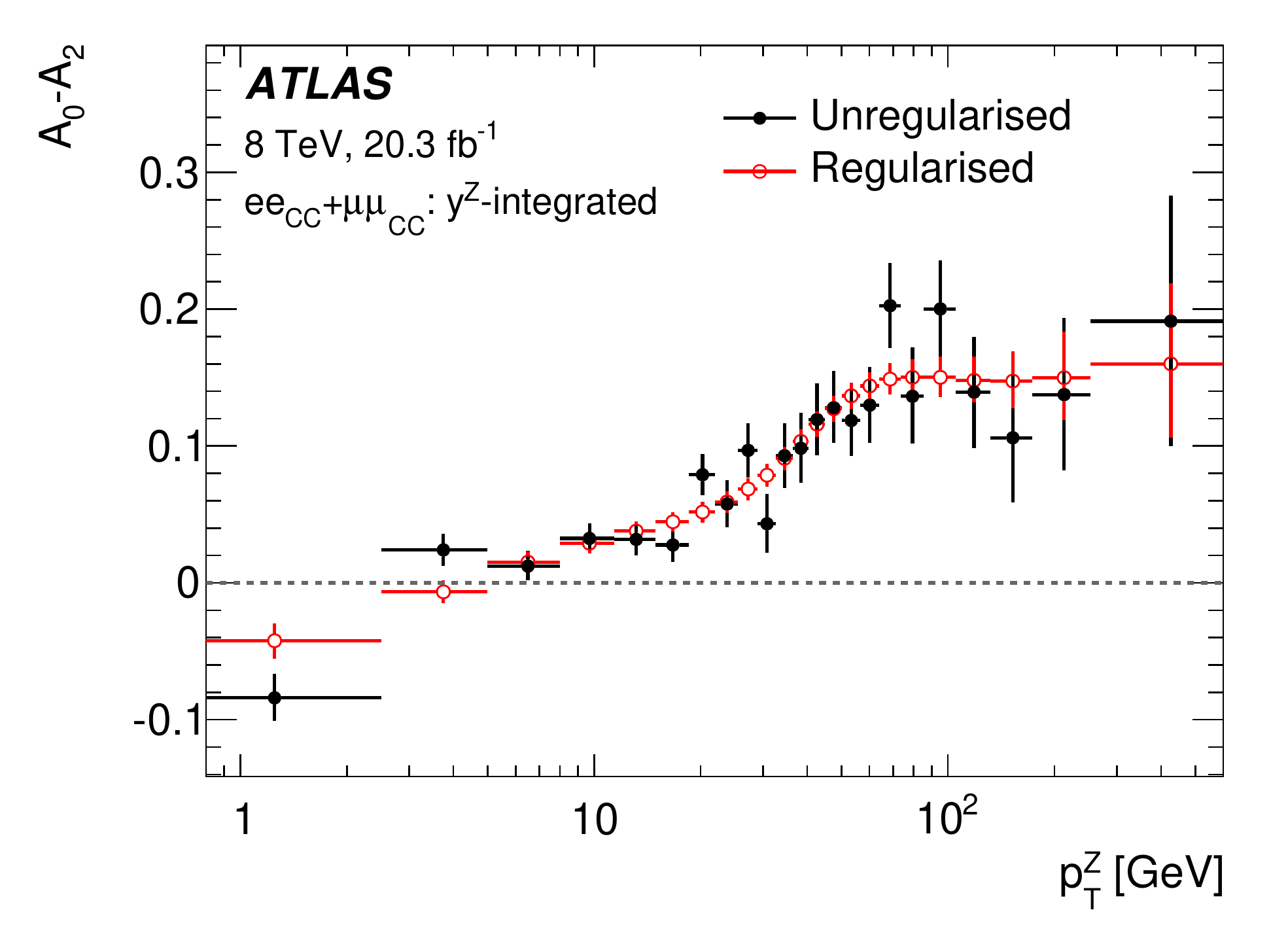}

\end{center}
\caption{For the $ee_{\text{CC}}+\mu\mu_{\text{CC}}$ channel in the $\yz$-integrated configuration, overlays of regularised with unregularised results are shown for $A_{0}-A_{2}$.
\label{Fig:reg_overlays_A02} }
\end{figure}
\clearpage

\section{Categorisation of statistical uncertainties}
\label{appendix:stat-cat}

The categorisation of statistical uncertainties is illustrated in Fig.~\ref{Fig:data-stat-cat} for~$A_{0}$ in $\ptz$~bin~0. Uncertainties due to the parameter of interest alone are labelled as ``Uncorr.-stat'' in the solid red box.
Boxes directly below the solid red box represent parameters common to the same $\ptz$ bin as the parameter of interest, and are therefore non-migration parameters. The other boxes represent parameters in different $\ptz$ bins, and are categorised as migration parameters. The categorisation can be broken down as follows:
\begin{itemize}
\item Parameters in the dashed green box are from {\it different} coefficient numbers but the {\it same} $\ptz$ bin and are labelled as ``Shape'' parameters.
\item Parameters in the dotted blue box are from the {\it same} coefficient number ($A_{0}$) but in a {\it different} $\ptz$ bin and are labelled as ``Self-migration'' parameters. 
\item The complements to these two categories are the parameters in the single-lined orange box and are labelled as ``Shape-migration''; they are outside of both the chosen $\ptz$ bin and coefficient number. 
\end{itemize}
These separations are done as well for the cross-section parameters, and are labelled as ``Norm'' and ``Norm-migration'' in the dot-dashed blue and double-lined purple boxes, respectively. 

An illustration of this categorisation of the various components of the statistical uncertainty is shown for the $ee_{\text{CC}}+\mu\mu_{\text{CC}}$, $\yz$-integrated measurement of the coefficient  $A_{0}$  in Fig.~\ref{Fig:incl_CC_A0_uncertainties_sta}.

\begin{figure}
\begin{center}
\includegraphics[width=9cm,angle=0]{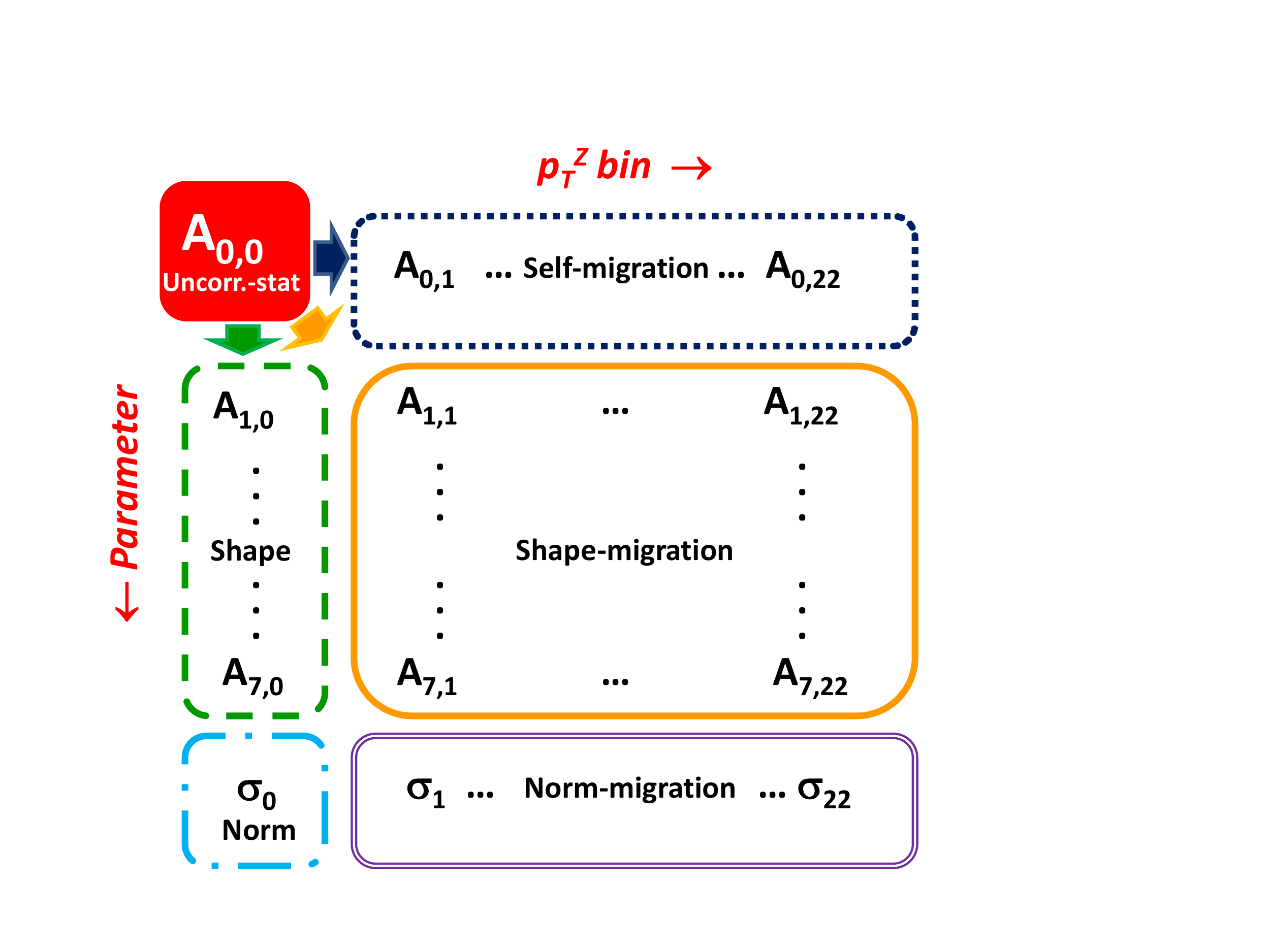}
\end{center}
\caption{Categorisation of parameters leading to the data statistical uncertainty in the measured coefficients illustrated by the uncertainty categorisation for $A_{0}$ in $\ptz$ bin 0.
\label{Fig:data-stat-cat} }
\end{figure}

\begin{figure}
\begin{center}
\includegraphics[width=7.5cm,angle=0]{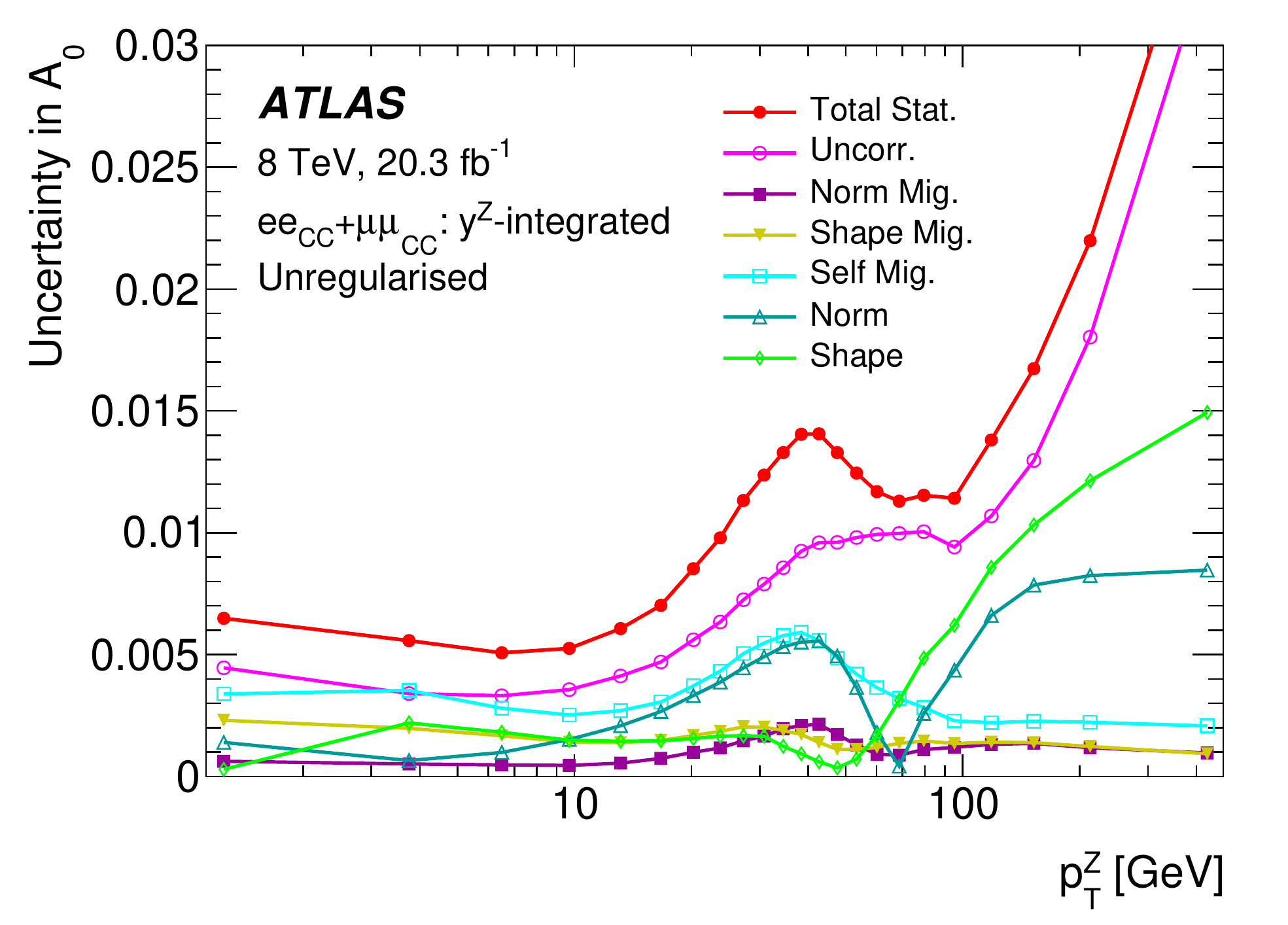}
\includegraphics[width=7.5cm,angle=0]{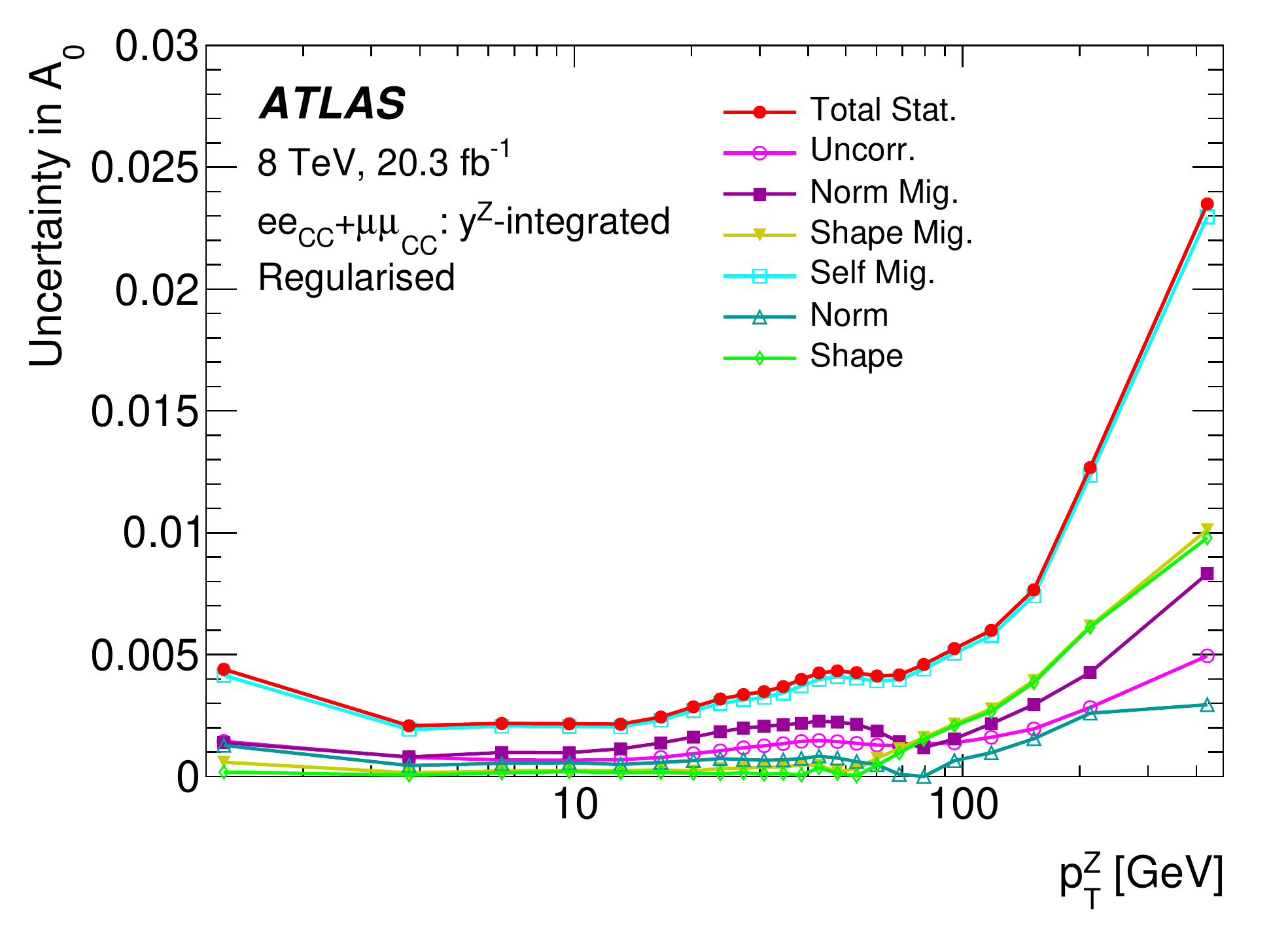}
\end{center}
\caption{Statistical uncertainty decomposition for the unregularised (left) and regularised (right) measurement of the $A_{0}$ coefficient in the $ee_{\text{CC}}+\mu\mu_{\text{CC}}$ channel for the integrated $\yz$ configuration. 
\label{Fig:incl_CC_A0_uncertainties_sta}}
\end{figure}

\clearpage
\section{Quantifying $A_{5,6,7}$}
\label{sec:app_a567}

The coefficients $A_{5,6,7}$ are expected to be zero at NLO, but are expected to receive NNLO contributions as large as 0.005 at high~$\ptz$. The data measurements appear to be consistent with this, although the level at which the data measurements are non-zero should be quantified. A simple method to quantify this would be a standard $\chi^2$ test of the measured spectra with respect to the null hypothesis of zero, but this has several disadvantages. First, the coefficients are expected to be non-zero only at high~$\ptz$, and therefore a $\chi^2$ test across the entire spectrum would be diluted by the low~$\ptz$ bins. Performing the test only for high $\ptz$ could improve this locally, although this introduces some model dependence due to the choice of $\ptz$ cutoff, as well as introducing a look-elsewhere-effect. Second, a $\chi^2$ test is insensitive to the sign of the measured coefficients in each bin. Finally, it does not optimally account for positive or negative trends in the observation.

A signed covariant test statistic $Q_{\rm{signed}}^{\rm{cov}}$ based on pseudo-experiments was developed for the purpose of quantifying the observed spectra. This takes into account pair-wise correlations between coefficients in neighbouring $\ptz$ bins, as well as correlations between the different coefficients in the same $\ptz$ bin. The contribution of each coefficient measurement to the test statistic is signed. Measurements below zero have a negative contribution, while measurements above zero have a positive contribution. $Q_{\rm{signed}}^{\rm{cov}}$ is computed both on observed data and simulated data based on DYNNLO~predictions at~NNLO. The distribution of $Q_{\rm{signed}}^{\rm{cov}}$ is obtained from ensemble tests under the null hypothesis. A p-value is obtained by integrating this distribution from the observed and simulated values to positive infinity, and converted to a one-sided statistical significance. 

To compute $Q_{\rm{signed}}^{\rm{cov}}$ (for any observed, simulated, or pseudo data), an initial set of pseudo-experiments are used to obtain the distribution of $\hat{A}_{5,6,7}^{\rm{pseudo}}$. A fit is first performed to the data under the null hypothesis $A_{5,6,7}=0$ to obtain $\hat{A}_{0-4}^{\rm{null}}$, $\hat{\sigma}^{\rm{null}}$, and $\hat{\theta}^{\rm{null}}$. Pseudo-data is then generated in each likelihood bin around the expected events $N^{n}_{\rm{exp}}(\hat{A}_{0-4}^{\rm{null}}, \hat{\sigma}^{\rm{null}}, \hat{\theta}^{\rm{null}})$ (see Eq.~(\ref{eq:Nexp})). A fit is performed to the pseudo-data to obtain $\hat{A}_{5,6,7}^{\rm{pseudo}}$.

$Q_{\rm{signed}}^{\rm{cov}}$ is computed based on $\hat{A}_{5,6,7}$ in any particular dataset in conjunction with the distribution of $\hat{A}_{5,6,7}^{\rm{pseudo}}$ from pseudo-data. It includes several components, which are described here. The significance $Z_{i}$ of the deviation from zero of each of the three $A_{5,6,7}$ coefficients in every $\ptz$ bin is computed as depicted in~Fig.~\ref{Fig:A567_Z}. A weight, $w_{ij} = (\rm{sign}(\mathit{Z_{i}})\mathit{Z_{i}^{2}}+\rm{sign}(\mathit{Z_{j}})\mathit{Z_{j}^{2}})/(Z_{i}^{2}+Z_{j}^{2})$, is computed based on the individual $Z_{i}$ values for every coefficient pair, both in coefficient number and bin in $\ptz$. A weight~$w_{ij}$ has the property that $w_{ij} = +1$ or~$-1$ if~$Z_{i}$ and~$Z_{j}$ are both above or below zero, respectively, while it is a weighted difference between them otherwise. The correlation coefficient~$\rho_{ij}$ between these pairs is extracted from their two-dimensional distributions. The pair-wise significance~$Z_{ij}$ is computed from the tail probability of the measurement being more outward in it's quadrant than is observed. An example of this for each quadrant is shown in~Fig.~\ref{Fig:A567_Z}. $Q_{\rm{signed}}^{\rm{cov}}$ is defined using these components as follows:

\begin{equation}
Q_{\rm{signed}}^{\rm{cov}} = \displaystyle\sum_{i}\rm{sign}(\mathit{Z_{i}})\mathit{Z_{i}^{2}} + \displaystyle\sum_{\mathit{i>j}}\mathit{w_{ij}Z_{ij}^2|\rho_{ij}|}\ .
\label{eq:qmu_A567}
\end{equation}

\begin{figure}
  \begin{center}
{
    \includegraphics[width=7.5cm,angle=0]{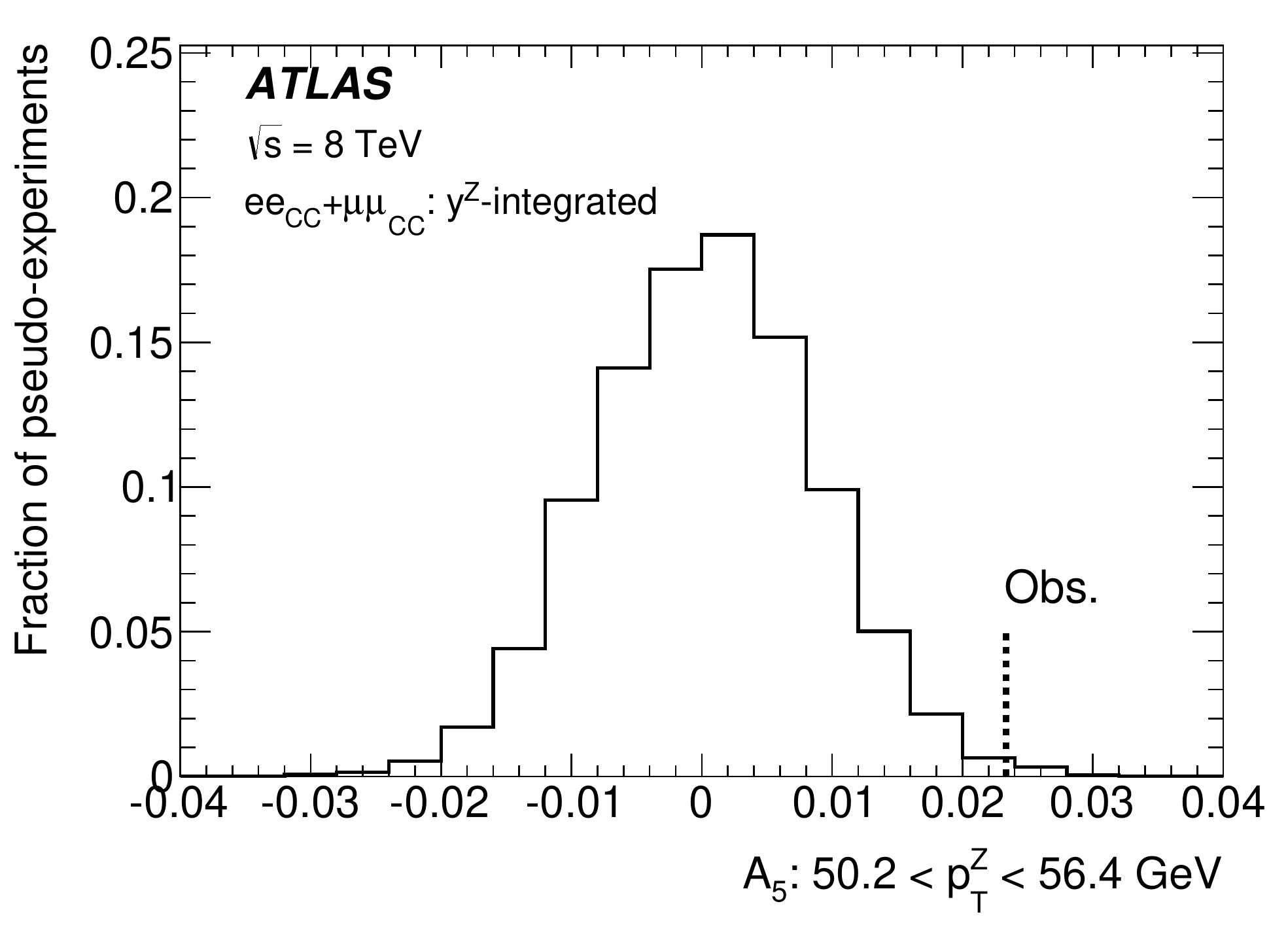}
    \includegraphics[width=7.5cm,angle=0]{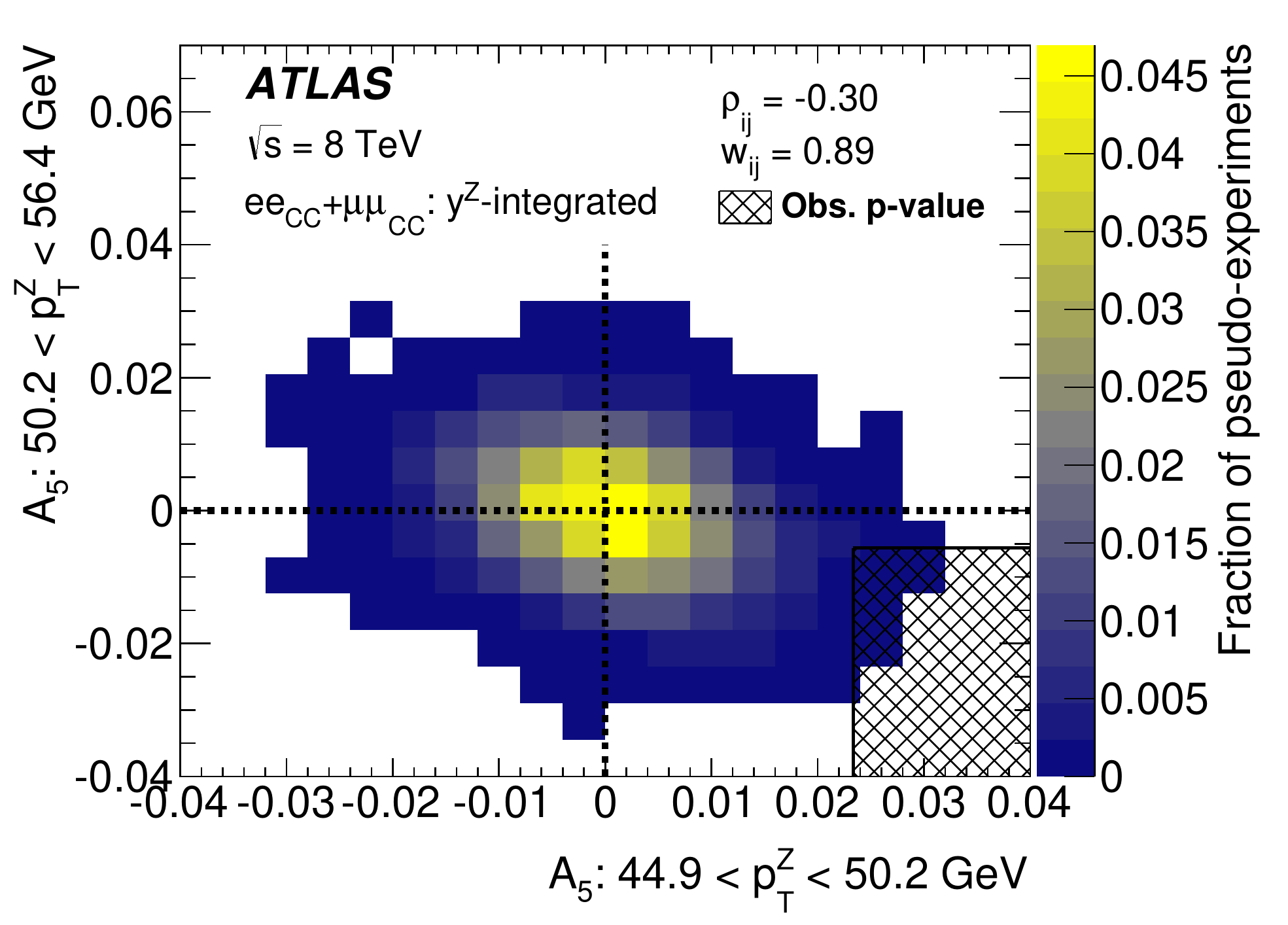}

}
\end{center}
\vspace{-5mm}
\caption{Left: Example of the distribution of the fitted value of $A_5$ in one $\ptz$ bin from pseudo-data along with the observed value represented by a dashed line. The $p$-value computed as the right-sided tail probability is used to calculate the individual values of $Z_i$. Right: Two-dimensional distribution of the fitted $A_5$ from pseudo-data for two neighbouring $\ptz$ bins. The upper left corner of the shaded area represents the value measured in the observed data, while the shaded area represents the $p$-value used to calculate $Z_{ij}$.
\label{Fig:A567_Z}}
\end{figure}

The distribution of $Q_{\rm{signed}}^{\rm{cov}}$ is finally obtained from a second set of pseudo-experiments. The observed value is computed along with the value from the DYNNLO expectation. The distribution is shown in~Fig.~\ref{Fig:A567_Q}, with vertical bars representing the observed and expected values. A total of 7800~pseudo-experiments were used in this computation. Integrating from the observed value to the right, the fraction of events in the tail is~0.14\%, corresponding to a significance of~$3.0\sigma$. Similarly, the expected significance is~$3.2\sigma$.

\begin{figure}
  \begin{center}
{
    \includegraphics[width=15cm,angle=0]{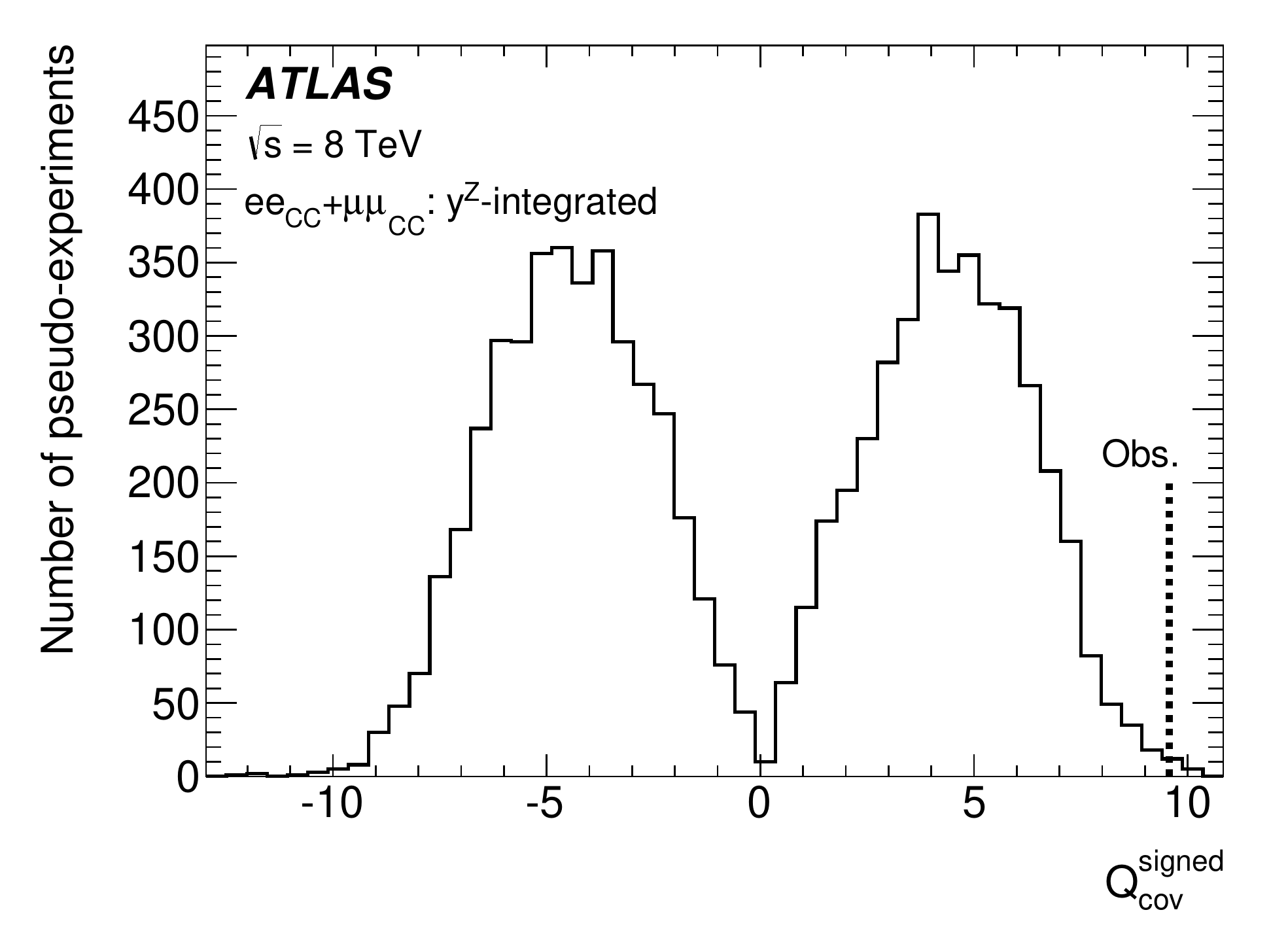}

}
\end{center}
\vspace{-5mm}
\caption{Distribution of the test statistic $Q_{\rm{signed}}^{\rm{cov}}$ from pseudo-experiments, along with the observed value represented by the vertical dashed line. The area to the right of the dashed line is used to compute the significance of non-zero positive values of the observed $A_{5,6,7}$. 
\label{Fig:A567_Q}}
\end{figure}

\section{Additional results}
\label{sec:additional_results}

Results are presented in Tables~\ref{Tab:measured_A0_A2}--\ref{Tab:measured_A5_A7} for the $\yz$-integrated measurements and in Tables~\ref{Tab:measured_A0_yz}--\ref{Tab:measured_A7_yz} in bins of $\yz$. Figure~\ref{Fig::AngCoefMeas_additional} shows the coefficients in bins of $\yz$.  Tables~\ref{Tab:SystSummary_ybin1_bin2_reg_A0}--\ref{Tab:SystSummary_ybin2_bin2_reg_A2} show uncertainty breakdowns for the coefficients in the first two $\yz$ bins.

\begin{table}
\caption{ Measured angular coefficients $A_0$, $A_2$ and difference $A_0-A_2$ with uncertainties $\pm \delta_{\rm{stat}} \pm \delta_{\rm{syst}}$ for the $\yz$-integrated measurement.
\label{Tab:measured_A0_A2} }
\begin{center}

\begin{tabular}{c||c|c|c}
\hline
\multicolumn{4}{c}{$\yz$-integrated}\\
\hline
 $\ptz$ range [GeV] & $A_{0}$ & $A_{2}$ & $A_{0}-A_{2}$ \\
\hline
 0.0  --  2.5 & --0.014 $\pm$ 0.004 $\pm$ 0.008 & 0.025 $\pm$ 0.009 $\pm$ 0.006 & --0.039 $\pm$ 0.010 $\pm$ 0.008 \\
 2.5  --  5.0 & --0.003 $\pm$ 0.002 $\pm$ 0.008 & 0.001 $\pm$ 0.004 $\pm$ 0.003 & --0.003 $\pm$ 0.004 $\pm$ 0.007 \\
 5.0  --  8.0 & 0.015 $\pm$ 0.002 $\pm$ 0.007 & --0.003 $\pm$ 0.003 $\pm$ 0.003 & 0.018 $\pm$ 0.003 $\pm$ 0.007 \\
 8.0  -- 11.4 & 0.038 $\pm$ 0.002 $\pm$ 0.007 & 0.007 $\pm$ 0.002 $\pm$ 0.002 & 0.031 $\pm$ 0.003 $\pm$ 0.007 \\
 11.4 -- 14.9 & 0.064 $\pm$ 0.002 $\pm$ 0.007 & 0.025 $\pm$ 0.002 $\pm$ 0.002 & 0.039 $\pm$ 0.003 $\pm$ 0.007 \\
 14.9 -- 18.5 & 0.093 $\pm$ 0.002 $\pm$ 0.007 & 0.048 $\pm$ 0.002 $\pm$ 0.002 & 0.045 $\pm$ 0.003 $\pm$ 0.006 \\
\hline
\end{tabular}
\end{center}
\end{table}

\begin{table}
\caption{ Measured angular coefficients $A_1$, $A_3$ and $A_4$ with uncertainties $\pm \delta_{\rm{stat}} \pm \delta_{\rm{syst}}$ for the $\yz$-integrated measurement.
\label{Tab:measured_A1_A4} }
\begin{center}

\begin{tabular}{c||c|c|c}
\hline
\multicolumn{4}{c}{$\yz$-integrated}\\
\hline
 $\ptz$ range [GeV] & $A_{1}$ & $A_{3}$ & $A_{4}$ \\
\hline
0.0 -- 2.5    & 0.014 $\pm$ 0.005 $\pm$ 0.003 & --0.005 $\pm$ 0.003 $\pm$ 0.002 & 0.075 $\pm$ 0.003 $\pm$ 0.003 \\
 2.5 -- 5.0   & 0.012 $\pm$ 0.002 $\pm$ 0.002 & --0.001 $\pm$ 0.001 $\pm$ 0.001 & 0.081 $\pm$ 0.001 $\pm$ 0.002 \\
 5.0 -- 8.0   & 0.013 $\pm$ 0.002 $\pm$ 0.002 & 0.001 $\pm$ 0.001 $\pm$ 0.001 & 0.082 $\pm$ 0.001 $\pm$ 0.002 \\
 8.0 -- 11.4  & 0.017 $\pm$ 0.002 $\pm$ 0.002 & 0.003 $\pm$ 0.001 $\pm$ 0.001 & 0.080 $\pm$ 0.001 $\pm$ 0.002 \\
 11.4 -- 14.9 & 0.023 $\pm$ 0.001 $\pm$ 0.001 & 0.004 $\pm$ 0.001 $\pm$ 0.001 & 0.076 $\pm$ 0.001 $\pm$ 0.002 \\
 14.9 -- 18.5 & 0.029 $\pm$ 0.002 $\pm$ 0.001 & 0.004 $\pm$ 0.001 $\pm$ 0.001 & 0.072 $\pm$ 0.002 $\pm$ 0.002 \\
\hline
\end{tabular}

\end{center}
\end{table}

\begin{table}
\caption{ Measured angular coefficients $A_5$, $A_6$ and $A_7$ with uncertainties $\pm \delta_{\rm{stat}} \pm \delta_{\rm{syst}}$ for the $\yz$-integrated measurement.
\label{Tab:measured_A5_A7} }
\begin{center}

\begin{tabular}{c||c|c|c}
\hline
\multicolumn{4}{c}{$\yz$-integrated}\\
\hline
 $\ptz$ range [GeV] & $A_{5}$ & $A_{6}$ & $A_{7}$ \\
\hline
0.0 -- 2.5   & --0.001 $\pm$ 0.004 $\pm$ 0.003 & 0.006 $\pm$ 0.003 $\pm$ 0.002 & 0.003 $\pm$ 0.002 $\pm$ 0.001 \\
 2.5 -- 5.0  & --0.002 $\pm$ 0.002 $\pm$ 0.001 & 0.001 $\pm$ 0.001 $\pm$ 0.001 & 0.001 $\pm$ 0.001 $\pm$ 0.001 \\
 5.0 -- 8.0  & --0.002 $\pm$ 0.001 $\pm$ 0.001 & --0.001 $\pm$ 0.001 $\pm$ 0.001 & 0.000 $\pm$ 0.001 $\pm$ 0.001 \\
 8.0 -- 11.4 & --0.001 $\pm$ 0.001 $\pm$ 0.001 & --0.001 $\pm$ 0.001 $\pm$ 0.001 & 0.000 $\pm$ 0.001 $\pm$ 0.001 \\
 11.4 -- 14.9 & 0.000 $\pm$ 0.001 $\pm$ 0.001 & --0.001 $\pm$ 0.001 $\pm$ 0.001 & 0.000 $\pm$ 0.001 $\pm$ 0.000 \\
 14.9 -- 18.5 & 0.001 $\pm$ 0.001 $\pm$ 0.001 & 0.000 $\pm$ 0.001 $\pm$ 0.001 & 0.000 $\pm$ 0.001 $\pm$ 0.001 \\
\hline
\end{tabular}

\end{center}
\end{table}

\begin{table}
\caption{ The angular coefficient $A_{0} \pm \delta_{\rm{stat}} \pm \delta_{\rm{syst}}$ in bins of $\yz$.   \label{Tab:measured_A0_yz}}
\begin{center}
\begin{tabular}{c||c|c|c}
\hline
\multicolumn{4}{c}{$\yz$-binned $A_{0}$}\\
\hline
 $\ptz$ range [GeV] & $0<|y^{Z}|<1$ & $1<|y^{Z}|<2$ & $2<|y^{Z}|<3.5$ \\
\hline
0.0 -- 2.5    & 0.004 $\pm$ 0.006 $\pm$ 0.005 & --0.016 $\pm$ 0.008 $\pm$ 0.008 &  0.093 $\pm$ 0.027 $\pm$ 0.044 \\
 2.5 -- 5.0   & 0.009 $\pm$ 0.003 $\pm$ 0.004 & --0.001 $\pm$ 0.004 $\pm$ 0.007 &  0.085 $\pm$ 0.019 $\pm$ 0.035 \\
 5.0 -- 8.0   & 0.022 $\pm$ 0.003 $\pm$ 0.004 & 0.017 $\pm$ 0.004 $\pm$ 0.007 &  0.070 $\pm$ 0.017 $\pm$ 0.033 \\
 8.0 -- 11.4  & 0.041 $\pm$ 0.003 $\pm$ 0.004 & 0.038 $\pm$ 0.004 $\pm$ 0.007 &  0.069 $\pm$ 0.016 $\pm$ 0.032 \\
 11.4 -- 14.9 & 0.064 $\pm$ 0.003 $\pm$ 0.004 & 0.061 $\pm$ 0.004 $\pm$ 0.006 &  0.070 $\pm$ 0.017 $\pm$ 0.032 \\
 14.9 -- 18.5 & 0.091 $\pm$ 0.003 $\pm$ 0.004 & 0.087 $\pm$ 0.004 $\pm$ 0.006 &  0.084 $\pm$ 0.019 $\pm$ 0.031 \\
\hline
\end{tabular}
\end{center}
\end{table}

\begin{table}
\caption{ The angular coefficient $A_{1} \pm \delta_{\rm{stat}} \pm \delta_{\rm{syst}}$ in bins of $\yz$. The $A_1$ measurements are missing from the third $\yz$ bin since they are inaccessible in the projections used in the $ee_{\text{CF}}$ channel.   \label{Tab:measured_A1_yz}}
\begin{center}
\begin{tabular}{c||c|c|c}
\hline
\multicolumn{4}{c}{$\yz$-binned $A_{1}$}\\
\hline
 $\ptz$ range [GeV] & $0<|y^{Z}|<1$ & $1<|y^{Z}|<2$ & $2<|y^{Z}|<3.5$ \\
\hline
0.0 -- 2.5    & 0.007 $\pm$ 0.006 $\pm$ 0.004 & 0.018 $\pm$ 0.008 $\pm$ 0.005 &   \\
 2.5 -- 5.0   & 0.004 $\pm$ 0.002 $\pm$ 0.002 & 0.016 $\pm$ 0.003 $\pm$ 0.002 &   \\
 5.0 -- 8.0   & 0.002 $\pm$ 0.002 $\pm$ 0.001 & 0.017 $\pm$ 0.002 $\pm$ 0.002 &   \\
 8.0 -- 11.4  & 0.002 $\pm$ 0.002 $\pm$ 0.001 & 0.021 $\pm$ 0.002 $\pm$ 0.002 &   \\
 11.4 -- 14.9 & 0.003 $\pm$ 0.002 $\pm$ 0.001 & 0.027 $\pm$ 0.002 $\pm$ 0.002 &   \\
 14.9 -- 18.5 & 0.004 $\pm$ 0.002 $\pm$ 0.001 & 0.032 $\pm$ 0.003 $\pm$ 0.002 &   \\
\hline
\end{tabular}
\end{center}
\end{table}

\begin{table}
\caption{ The angular coefficient $A_{2} \pm \delta_{\rm{stat}} \pm \delta_{\rm{syst}}$ in bins of $\yz$.   \label{Tab:measured_A2_yz}}
\begin{center}
\begin{tabular}{c||c|c|c}
\hline
\multicolumn{4}{c}{$\yz$-binned $A_{2}$}\\
\hline
 $\ptz$ range [GeV] & $0<|y^{Z}|<1$ & $1<|y^{Z}|<2$ & $2<|y^{Z}|<3.5$ \\
\hline
0.0 -- 2.5    & 0.032 $\pm$ 0.011 $\pm$ 0.007 & 0.039 $\pm$ 0.015 $\pm$ 0.011 &  0.198 $\pm$ 0.094 $\pm$ 0.063 \\
 2.5 -- 5.0   & 0.007 $\pm$ 0.005 $\pm$ 0.003 & 0.010 $\pm$ 0.006 $\pm$ 0.005 &  0.081 $\pm$ 0.071 $\pm$ 0.041 \\
 5.0 -- 8.0   & 0.003 $\pm$ 0.003 $\pm$ 0.002 & 0.003 $\pm$ 0.004 $\pm$ 0.003 &  0.071 $\pm$ 0.059 $\pm$ 0.035 \\
 8.0 -- 11.4  & 0.012 $\pm$ 0.003 $\pm$ 0.002 & 0.009 $\pm$ 0.004 $\pm$ 0.002 &  0.070 $\pm$ 0.048 $\pm$ 0.031 \\
 11.4 -- 14.9 & 0.028 $\pm$ 0.003 $\pm$ 0.002 & 0.025 $\pm$ 0.004 $\pm$ 0.003 &  0.091 $\pm$ 0.047 $\pm$ 0.026 \\
 14.9 -- 18.5 & 0.050 $\pm$ 0.003 $\pm$ 0.002 & 0.047 $\pm$ 0.004 $\pm$ 0.003 &  0.116 $\pm$ 0.050 $\pm$ 0.026 \\
\hline
\end{tabular}
\end{center}
\end{table}

\begin{table}
\caption{ The angular coefficient $A_{3} \pm \delta_{\rm{stat}} \pm \delta_{\rm{syst}}$ in bins of $\yz$.   \label{Tab:measured_A3_yz}}
\begin{center}
\begin{tabular}{c||c|c|c}
\hline
\multicolumn{4}{c}{$\yz$-binned $A_{3}$}\\
\hline
 $\ptz$ range [GeV] & $0<|y^{Z}|<1$ & $1<|y^{Z}|<2$ & $2<|y^{Z}|<3.5$ \\
\hline
0.0 -- 2.5    & --0.007 $\pm$ 0.004 $\pm$ 0.002 & 0.001 $\pm$ 0.006 $\pm$ 0.003 &  0.004 $\pm$ 0.064 $\pm$ 0.020 \\
 2.5 -- 5.0   & --0.006 $\pm$ 0.002 $\pm$ 0.001 & 0.004 $\pm$ 0.002 $\pm$ 0.001 &  0.020 $\pm$ 0.029 $\pm$ 0.009 \\
 5.0 -- 8.0   & --0.005 $\pm$ 0.001 $\pm$ 0.001 & 0.005 $\pm$ 0.002 $\pm$ 0.001 &  0.005 $\pm$ 0.019 $\pm$ 0.006 \\
 8.0 -- 11.4  & --0.004 $\pm$ 0.001 $\pm$ 0.001 & 0.005 $\pm$ 0.002 $\pm$ 0.001 &  0.003 $\pm$ 0.015 $\pm$ 0.005 \\
 11.4 -- 14.9 & --0.003 $\pm$ 0.001 $\pm$ 0.001 & 0.005 $\pm$ 0.002 $\pm$ 0.001 &  0.012 $\pm$ 0.015 $\pm$ 0.004 \\
 14.9 -- 18.5 & --0.003 $\pm$ 0.001 $\pm$ 0.001 & 0.005 $\pm$ 0.002 $\pm$ 0.001 &  0.009 $\pm$ 0.016 $\pm$ 0.005 \\
\hline
\end{tabular}
\end{center}
\end{table}

\begin{table}
\caption{ The angular coefficient $A_{4} \pm \delta_{\rm{stat}} \pm \delta_{\rm{syst}}$ in bins of $\yz$.   \label{Tab:measured_A4_yz}}
\begin{center}
\begin{tabular}{c||c|c|c}
\hline
\multicolumn{4}{c}{$\yz$-binned $A_{4}$}\\
\hline
 $\ptz$ range [GeV] & $0<|y^{Z}|<1$ & $1<|y^{Z}|<2$ & $2<|y^{Z}|<3.5$ \\
\hline
0.0 -- 2.5    & 0.019 $\pm$ 0.004 $\pm$ 0.002 & 0.048 $\pm$ 0.005 $\pm$ 0.003 &  0.095 $\pm$ 0.014 $\pm$ 0.007 \\
 2.5 -- 5.0   & 0.023 $\pm$ 0.002 $\pm$ 0.001 & 0.060 $\pm$ 0.002 $\pm$ 0.001 &  0.139 $\pm$ 0.010 $\pm$ 0.004 \\
 5.0 -- 8.0   & 0.023 $\pm$ 0.002 $\pm$ 0.001 & 0.065 $\pm$ 0.002 $\pm$ 0.001 &  0.142 $\pm$ 0.009 $\pm$ 0.004 \\
 8.0 -- 11.4  & 0.022 $\pm$ 0.002 $\pm$ 0.001 & 0.066 $\pm$ 0.002 $\pm$ 0.001 &  0.140 $\pm$ 0.008 $\pm$ 0.004 \\
 11.4 -- 14.9 & 0.020 $\pm$ 0.002 $\pm$ 0.001 & 0.064 $\pm$ 0.002 $\pm$ 0.001 &  0.127 $\pm$ 0.009 $\pm$ 0.004 \\
 14.9 -- 18.5 & 0.018 $\pm$ 0.002 $\pm$ 0.001 & 0.061 $\pm$ 0.003 $\pm$ 0.001 &  0.123 $\pm$ 0.010 $\pm$ 0.004 \\
\hline
\end{tabular}
\end{center}
\end{table}

\begin{table}
\caption{ The angular coefficient $A_{5} \pm \delta_{\rm{stat}} \pm \delta_{\rm{syst}}$ in bins of $\yz$.   \label{Tab:measured_A5_yz}}
\begin{center}
\begin{tabular}{c||c|c|c}
\hline
\multicolumn{4}{c}{$\yz$-binned $A_{5}$}\\
\hline
 $\ptz$ range [GeV] & $0<|y^{Z}|<1$ & $1<|y^{Z}|<2$ & $2<|y^{Z}|<3.5$ \\
\hline
0.0 -- 2.5    & --0.002 $\pm$ 0.005 $\pm$ 0.003 & 0.000 $\pm$ 0.007 $\pm$ 0.004 &  --0.030 $\pm$ 0.072 $\pm$ 0.025 \\
 2.5 -- 5.0   & --0.003 $\pm$ 0.002 $\pm$ 0.001 & --0.002 $\pm$ 0.003 $\pm$ 0.002 &  0.012 $\pm$ 0.026 $\pm$ 0.009 \\
 5.0 -- 8.0   & --0.003 $\pm$ 0.002 $\pm$ 0.001 & --0.001 $\pm$ 0.002 $\pm$ 0.001 &  0.013 $\pm$ 0.015 $\pm$ 0.005 \\
 8.0 -- 11.4  & --0.002 $\pm$ 0.002 $\pm$ 0.001 & 0.000 $\pm$ 0.002 $\pm$ 0.001 &  0.006 $\pm$ 0.013 $\pm$ 0.005 \\
 11.4 -- 14.9 & 0.000 $\pm$ 0.002 $\pm$ 0.001 & 0.003 $\pm$ 0.002 $\pm$ 0.001 &  0.004 $\pm$ 0.013 $\pm$ 0.005 \\
 14.9 -- 18.5 & 0.000 $\pm$ 0.002 $\pm$ 0.001 & 0.004 $\pm$ 0.002 $\pm$ 0.001 &  0.000 $\pm$ 0.014 $\pm$ 0.005 \\
\hline
\end{tabular}
\end{center}
\end{table}

\begin{table}
\caption{ The angular coefficient $A_{6} \pm \delta_{\rm{stat}} \pm \delta_{\rm{syst}}$ in bins of $\yz$.  The $A_6$ measurements are missing from the third $\yz$ bin since they are inaccessible in the projections used in the $ee_{\text{CF}}$ channel.   \label{Tab:measured_A6_yz}}
\begin{center}
\begin{tabular}{c||c|c|c}
\hline
\multicolumn{4}{c}{$\yz$-binned $A_{6}$}\\
\hline
 $\ptz$ range [GeV] & $0<|y^{Z}|<1$ & $1<|y^{Z}|<2$ & $2<|y^{Z}|<3.5$ \\
\hline
0.0 -- 2.5    & 0.008 $\pm$ 0.004 $\pm$ 0.003 & 0.003 $\pm$ 0.005 $\pm$ 0.003 &   \\
 2.5 -- 5.0   & 0.000 $\pm$ 0.002 $\pm$ 0.001 & 0.000 $\pm$ 0.002 $\pm$ 0.001 &   \\
 5.0 -- 8.0   & --0.003 $\pm$ 0.002 $\pm$ 0.001 & 0.000 $\pm$ 0.002 $\pm$ 0.001 &   \\
 8.0 -- 11.4  & --0.003 $\pm$ 0.002 $\pm$ 0.001 & 0.000 $\pm$ 0.002 $\pm$ 0.001 &   \\
 11.4 -- 14.9 & --0.002 $\pm$ 0.002 $\pm$ 0.001 & 0.001 $\pm$ 0.002 $\pm$ 0.001 &   \\
 14.9 -- 18.5 & --0.002 $\pm$ 0.002 $\pm$ 0.001 & 0.002 $\pm$ 0.002 $\pm$ 0.001 &   \\
\hline
\end{tabular}
\end{center}
\end{table}

\begin{table}
\caption{ The angular coefficient $A_{7} \pm \delta_{\rm{stat}} \pm \delta_{\rm{syst}}$ in bins of $\yz$.   \label{Tab:measured_A7_yz}}
\begin{center}
\begin{tabular}{c||c|c|c}
\hline
\multicolumn{4}{c}{$\yz$-binned $A_{7}$}\\
\hline
 $\ptz$ range [GeV] & $0<|y^{Z}|<1$ & $1<|y^{Z}|<2$ & $2<|y^{Z}|<3.5$ \\
\hline
0.0 -- 2.5    & 0.004 $\pm$ 0.003 $\pm$ 0.002 & 0.001 $\pm$ 0.003 $\pm$ 0.002 &  --0.023 $\pm$ 0.013 $\pm$ 0.007 \\
 2.5 -- 5.0   & 0.001 $\pm$ 0.001 $\pm$ 0.001 & 0.001 $\pm$ 0.002 $\pm$ 0.001 &  --0.005 $\pm$ 0.006 $\pm$ 0.003 \\
 5.0 -- 8.0   & 0.000 $\pm$ 0.001 $\pm$ 0.001 & 0.000 $\pm$ 0.001 $\pm$ 0.001 &  --0.004 $\pm$ 0.006 $\pm$ 0.003 \\
 8.0 -- 11.4  & 0.000 $\pm$ 0.001 $\pm$ 0.001 & 0.000 $\pm$ 0.001 $\pm$ 0.001 &  0.007 $\pm$ 0.006 $\pm$ 0.002 \\
 11.4 -- 14.9 & 0.001 $\pm$ 0.001 $\pm$ 0.001 & --0.001 $\pm$ 0.001 $\pm$ 0.001 &  0.006 $\pm$ 0.006 $\pm$ 0.002 \\
 14.9 -- 18.5 & 0.002 $\pm$ 0.001 $\pm$ 0.001 & --0.001 $\pm$ 0.001 $\pm$ 0.001 &  0.008 $\pm$ 0.007 $\pm$ 0.003 \\
\hline
\end{tabular}
\end{center}
\end{table}

\begin{figure}[p]
  \begin{center}                               
{
    \includegraphics[width=7.5cm,angle=0]{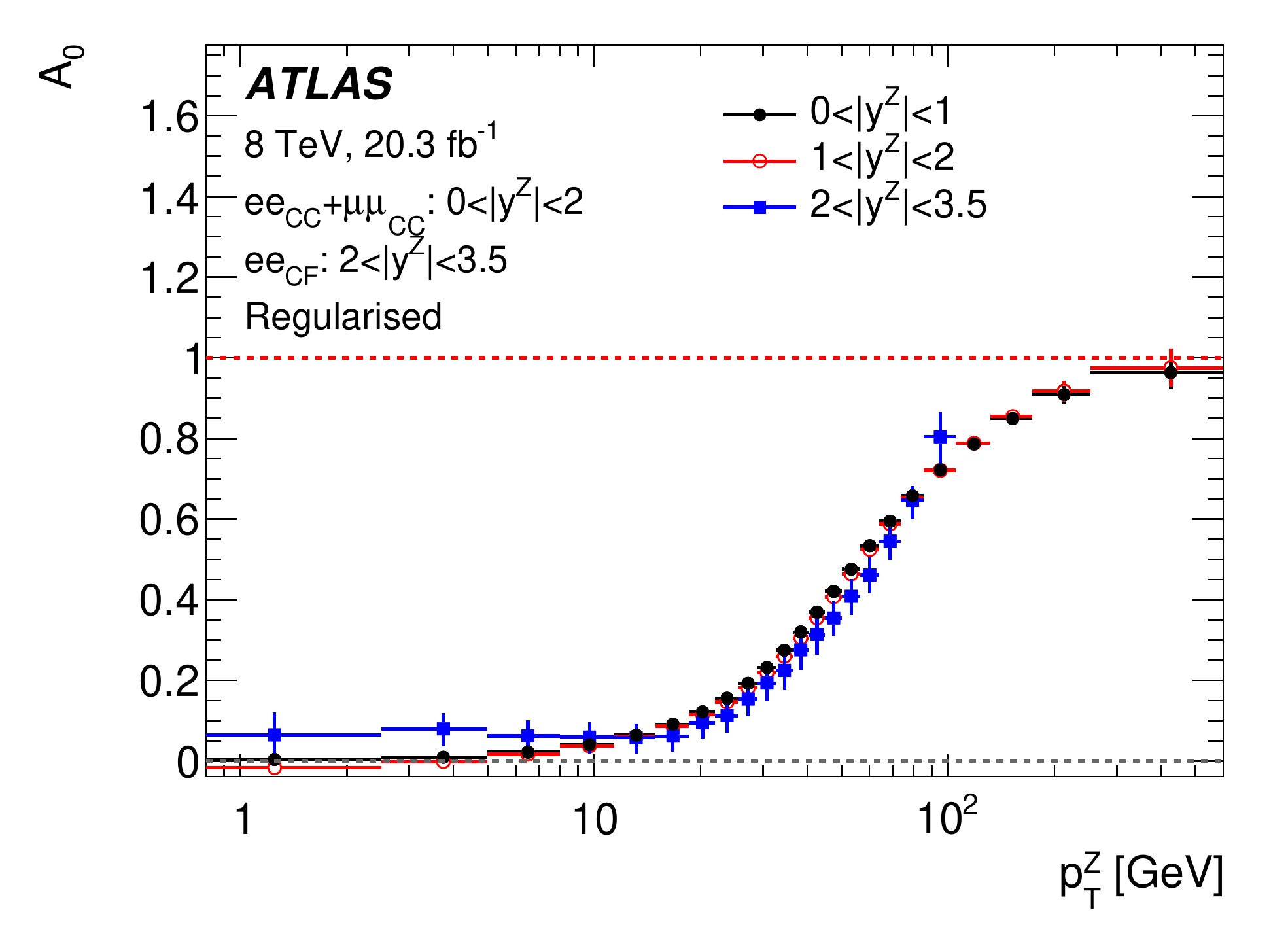}
    \includegraphics[width=7.5cm,angle=0]{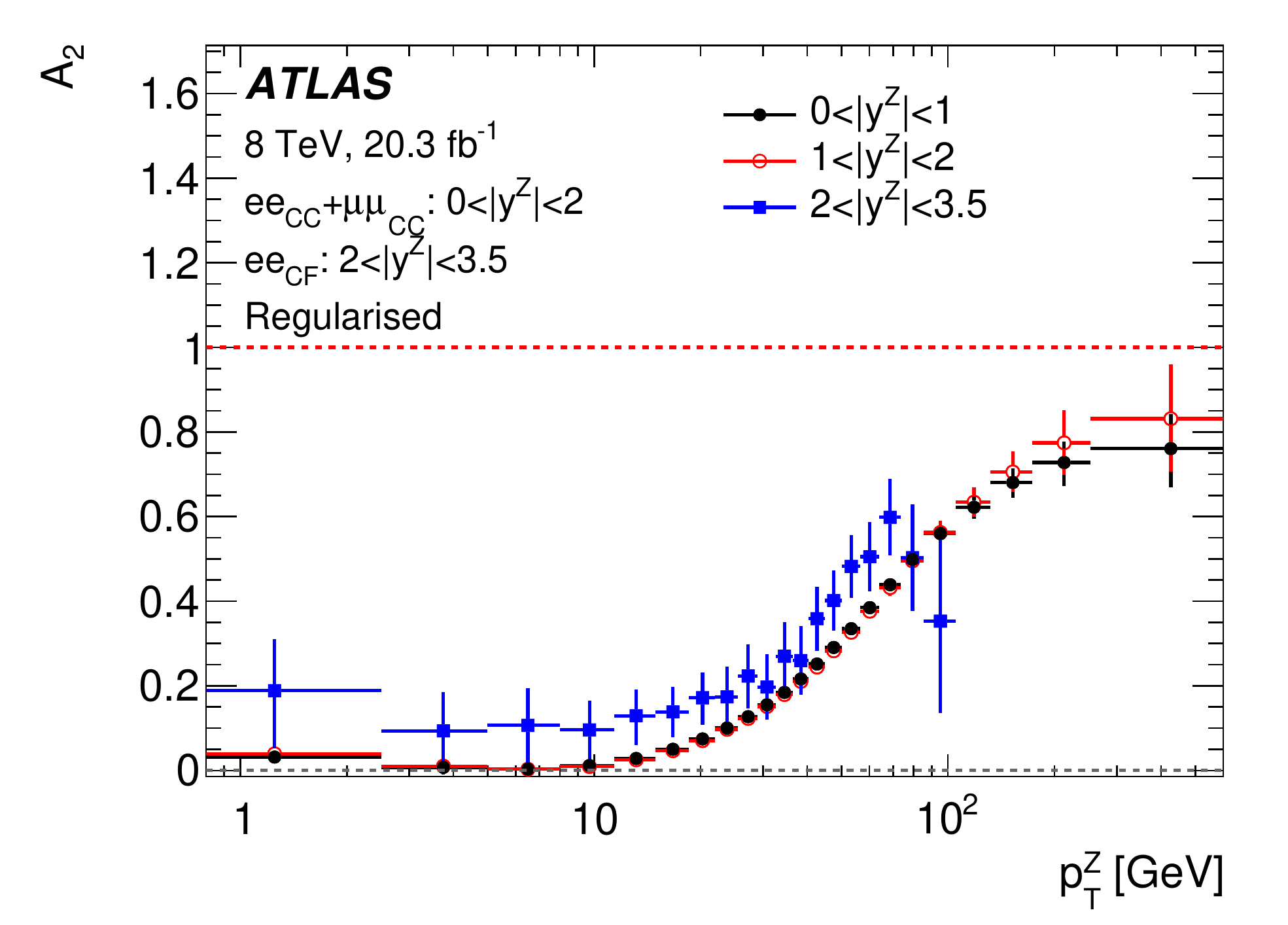}
    \includegraphics[width=7.5cm,angle=0]{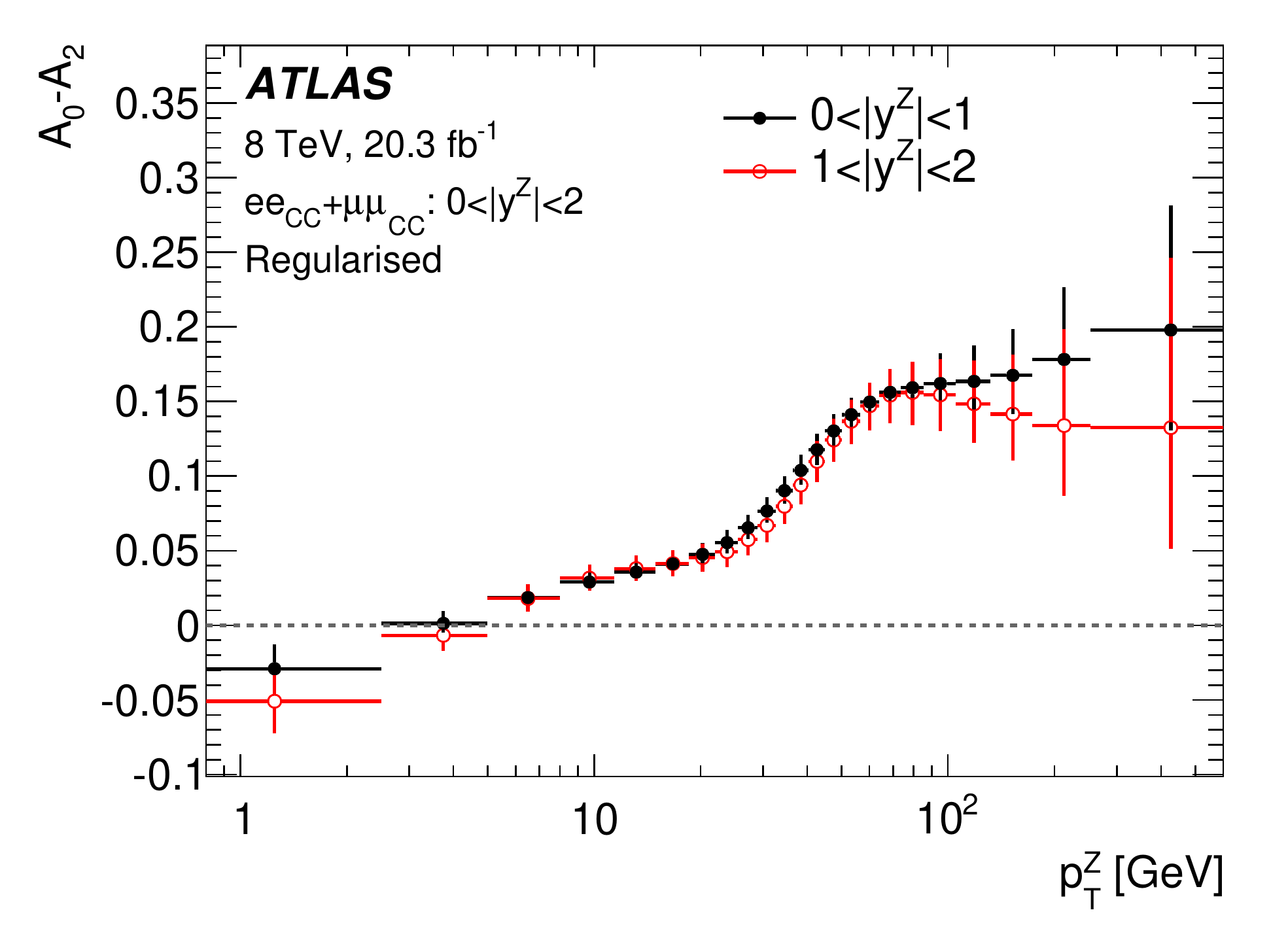}
    \includegraphics[width=7.5cm,angle=0]{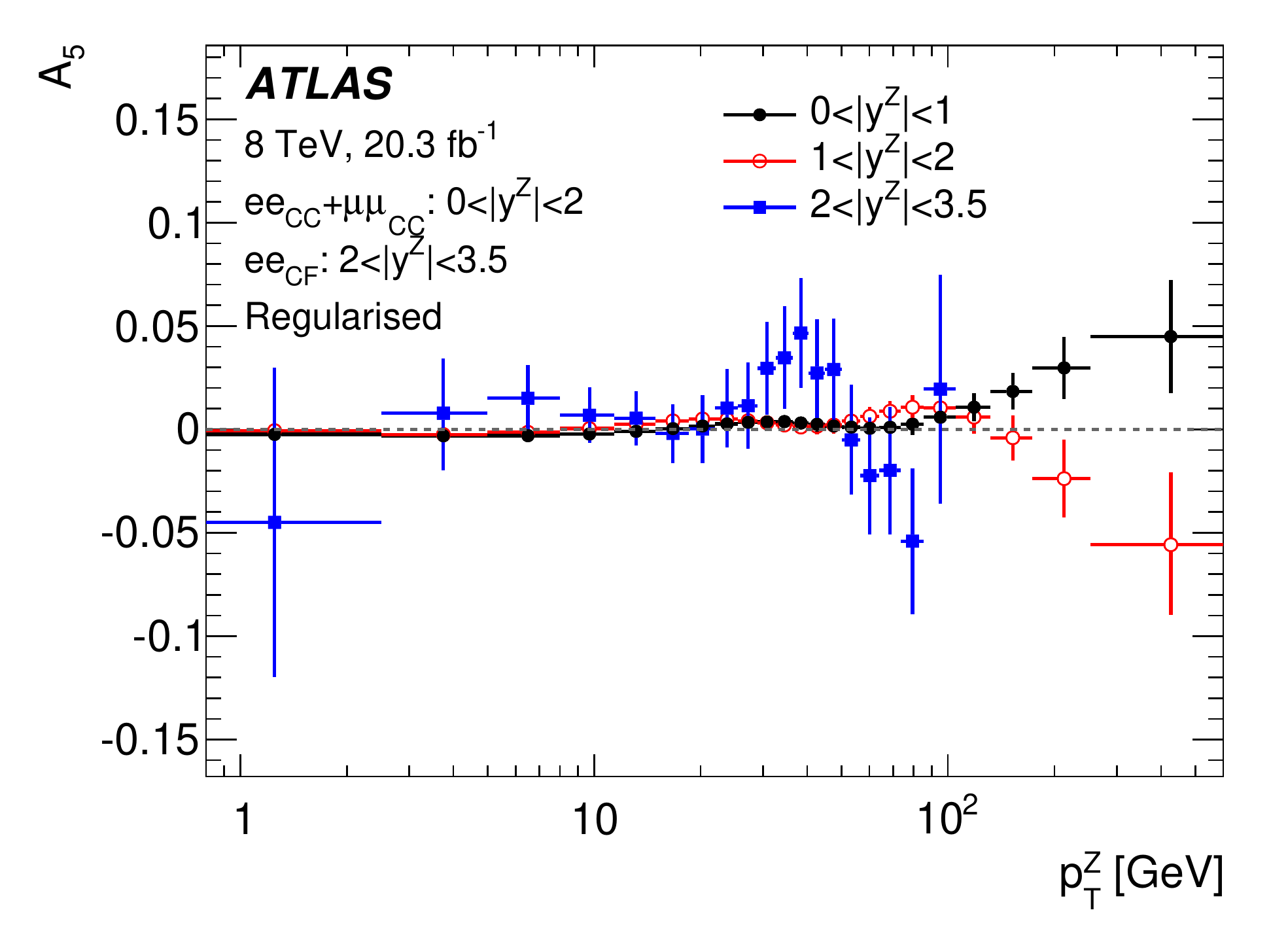}
    \includegraphics[width=7.5cm,angle=0]{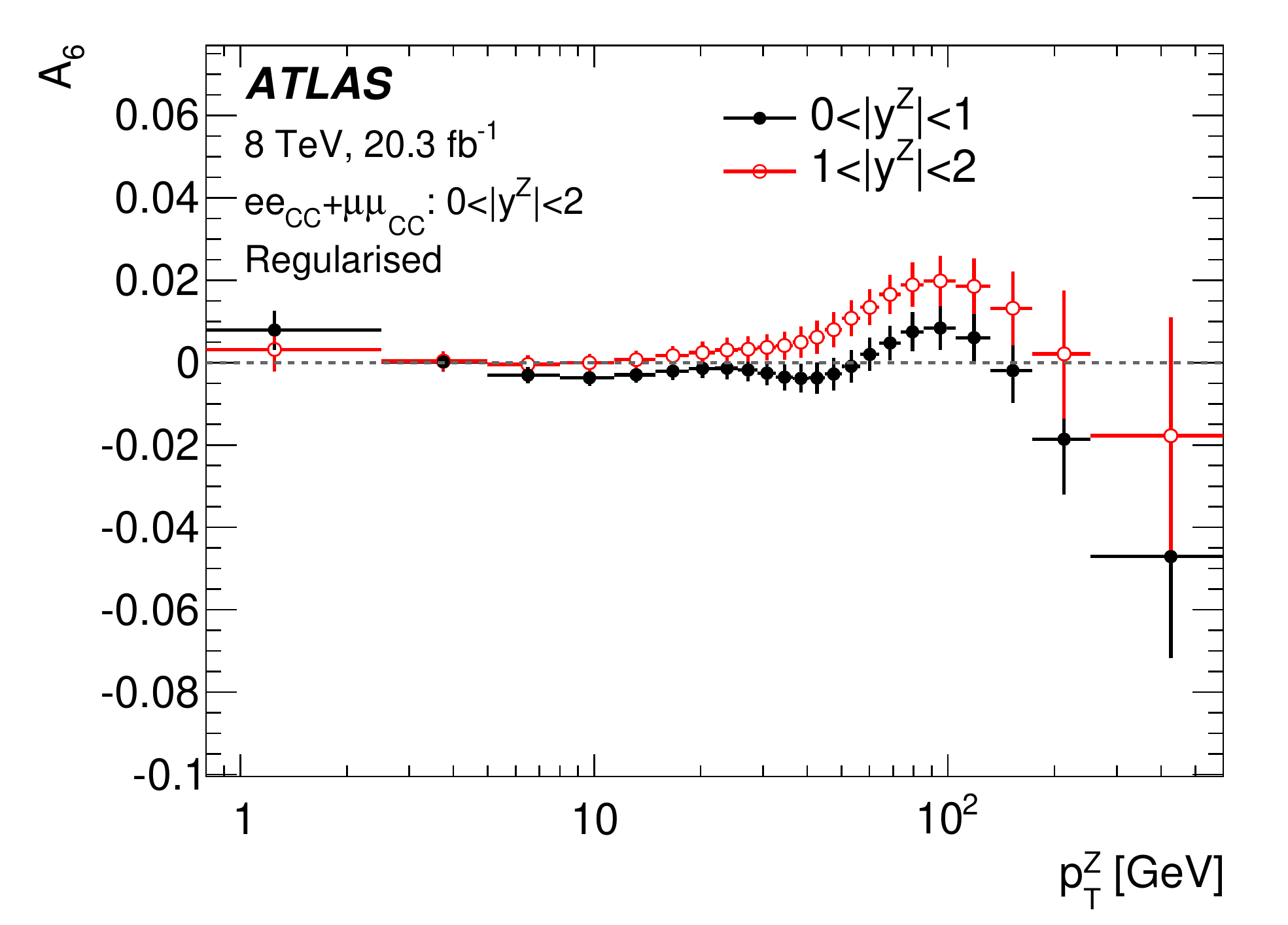}
    \includegraphics[width=7.5cm,angle=0]{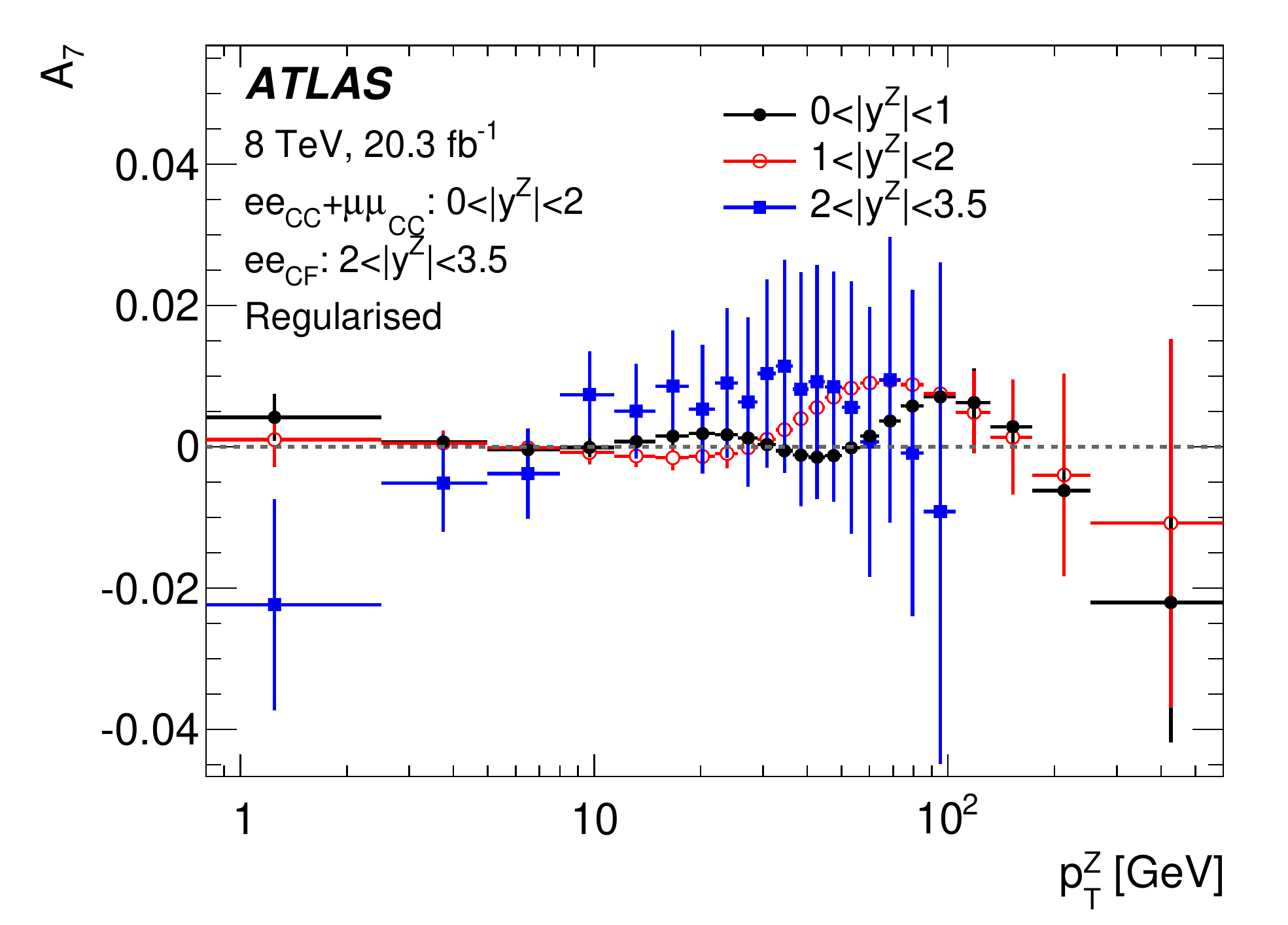}
}
\end{center}
\caption{ The measured angular coefficients $A_0$, $A_2$, $A_0-A_2$, $A_{5}$, $A_{6}$, and $A_{7}$ in bins of  $\yz$. 
\label{Fig::AngCoefMeas_additional} }
\end{figure}

\clearpage
\begin{table}
\caption{Summary of regularised uncertainties expected for $A_{0}$ and $A_{2}$ at low (5--8~GeV), mid (22--22.5~GeV), and high (132--173~GeV) \ptz\ for the $0<|\yz|<1$ configuration. The total systematic uncertainty is shown with the breakdown into its underlying components. Entries marked with ``-'' indicate that the uncertainty is below 0.0001.
\label{Tab:SystSummary_ybin1_bin2_reg_A0} }
\begin{center}
\begin{tabular}{l||c|c|c||c|c|c}
\hline
\hline
\multicolumn{7}{c}{$p_{T}^{Z} =$ 5--8 GeV} \\
\hline
\hline
Coefficient & \multicolumn{3}{c||}{$A_{0}$} & \multicolumn{3}{c}{$A_{2}$}\\
\hline
 Channel        & $ee$       & $\mu\mu$ & $ee$+$\mu\mu$    & $ee$    & $\mu\mu$ & $ee$+$\mu\mu$        \\
\hline
Total & 0.0098 & 0.0050 & 0.0053 & 0.0062 & 0.0055 & 0.0042\\
\hline
Data Stat. & 0.0043 & 0.0038 & 0.0028 & 0.0049 & 0.0048 & 0.0034\\
\hline
Syst. & 0.0088 & 0.0033 & 0.0045 & 0.0038 & 0.0029 & 0.0024\\
\: MC Stat. & 0.0023 & 0.0021 & 0.0015 & 0.0025 & 0.0026 & 0.0018\\
\: Lepton & 0.0072 & 0.0005 & 0.0017 & 0.0025 & 0.0015 & 0.0014\\
\hline
\hline
\end{tabular}
\end{center}
\end{table}

\begin{table}
\caption{Summary of regularised uncertainties expected for $A_{0}$ and $A_{2}$ at low (5--8~GeV), mid (22--22.5~GeV), and high (132--173~GeV) \ptz\ for the $1<|\yz|<2$ configuration. The total systematic uncertainty is shown with the breakdown into its underlying components. Entries marked with ``-'' indicate that the uncertainty is below 0.0001.
\label{Tab:SystSummary_ybin2_bin2_reg_A0} }
\begin{center}
\begin{tabular}{l||c|c|c||c|c|c}
\hline
\hline
\multicolumn{7}{c}{$p_{T}^{Z} =$ 5--8 GeV} \\
\hline
\hline
Coefficient & \multicolumn{3}{c||}{$A_{0}$} & \multicolumn{3}{c}{$A_{2}$}\\
\hline
 Channel        & $ee$       & $\mu\mu$ & $ee$+$\mu\mu$    & $ee$    & $\mu\mu$ & $ee$+$\mu\mu$        \\
\hline
Total & 0.0110 & 0.0081 & 0.0079 & 0.0094 & 0.0064 & 0.0058\\
\hline
Data Stat. & 0.0058 & 0.0048 & 0.0037 & 0.0070 & 0.0053 & 0.0044\\
\hline
Syst. & 0.0094 & 0.0065 & 0.0070 & 0.0063 & 0.0037 & 0.0039\\
\: MC Stat. & 0.0029 & 0.0025 & 0.0019 & 0.0034 & 0.0026 & 0.0020\\
\: Lepton & 0.0077 & 0.0009 & 0.0019 & 0.0055 & 0.0025 & 0.0023\\
\hline
\hline
\end{tabular}
\end{center}
\end{table}

\begin{table}
\caption{Summary of regularised uncertainties expected for $A_{1}$, $A_{3}$, and $A_{4}$ at low (5--8~GeV), mid (22--22.5~GeV), and high (132--173~GeV) \ptz\ for the $0<|\yz|<1$ configuration. The total systematic uncertainty is shown with the breakdown into its underlying components. Entries marked with ``-'' indicate that the uncertainty is below 0.0001.
\label{Tab:SystSummary_ybin1_bin2_reg_A1} }
\begin{center}
\begin{tabular}{l||c|c|c||c|c|c||c|c|c}
\hline
\hline
\multicolumn{10}{c}{$p_{T}^{Z} =$ 5--8 GeV} \\
\hline
\hline
Coefficient & \multicolumn{3}{c||}{$A_{1}$} & \multicolumn{3}{c||}{$A_{3}$} & \multicolumn{3}{c}{$A_{4}$} \\
\hline
 Channel        & $ee$       & $\mu\mu$ & $ee$+$\mu\mu$    & $ee$    & $\mu\mu$ & $ee$+$\mu\mu$         & $ee$       & $\mu\mu$ & $ee$+$\mu\mu$   \\
\hline
Total & 0.0042 & 0.0032 & 0.0026 & 0.0024 & 0.0023 & 0.0016 & 0.0033 & 0.0029 & 0.0022\\
\hline
Data Stat. & 0.0031 & 0.0027 & 0.0021 & 0.0021 & 0.0021 & 0.0015 & 0.0029 & 0.0026 & 0.0019\\
\hline
Syst. & 0.0028 & 0.0017 & 0.0015 & 0.0011 & 0.0011 & 0.0008 & 0.0015 & 0.0014 & 0.0010\\
\: MC Stat. & 0.0015 & 0.0014 & 0.0010 & 0.0010 & 0.0010 & 0.0007 & 0.0015 & 0.0014 & 0.0010\\
\: Lepton & 0.0019 & 0.0013 & 0.0011 & 0.0004 & 0.0002 & 0.0002 & 0.0003 & 0.0001 & 0.0002\\
\hline
\hline
\end{tabular}
\end{center}
\end{table}

\begin{table}
\caption{Summary of regularised uncertainties expected for $A_{1}$, $A_{3}$, and $A_{4}$ at low (5--8~GeV), mid (22--22.5~GeV), and high (132--173~GeV) \ptz\ for the $1<|\yz|<2$ configuration. The total systematic uncertainty is shown with the breakdown into its underlying components. Entries marked with ``-'' indicate that the uncertainty is below 0.0001.
\label{Tab:SystSummary_ybin2_bin2_reg_A1} }
\begin{center}
\begin{tabular}{l||c|c|c||c|c|c||c|c|c}
\hline
\hline
\multicolumn{10}{c}{$p_{T}^{Z} =$ 5--8 GeV} \\
\hline
\hline
Coefficient & \multicolumn{3}{c||}{$A_{1}$} & \multicolumn{3}{c||}{$A_{3}$} & \multicolumn{3}{c}{$A_{4}$} \\
\hline
 Channel        & $ee$       & $\mu\mu$ & $ee$+$\mu\mu$    & $ee$    & $\mu\mu$ & $ee$+$\mu\mu$         & $ee$       & $\mu\mu$ & $ee$+$\mu\mu$   \\
\hline
Total & 0.0046 & 0.0041 & 0.0030 & 0.0033 & 0.0025 & 0.0020 & 0.0041 & 0.0036 & 0.0027\\
\hline
Data Stat. & 0.0038 & 0.0032 & 0.0024 & 0.0029 & 0.0023 & 0.0018 & 0.0036 & 0.0031 & 0.0024\\
\hline
Syst. & 0.0026 & 0.0026 & 0.0019 & 0.0015 & 0.0011 & 0.0009 & 0.0019 & 0.0017 & 0.0013\\
\: MC Stat. & 0.0018 & 0.0016 & 0.0012 & 0.0014 & 0.0011 & 0.0008 & 0.0018 & 0.0016 & 0.0012\\
\: Lepton & 0.0017 & 0.0023 & 0.0015 & 0.0001 & 0.0001 & 0.0001 & 0.0004 & 0.0001 & 0.0003\\
\hline
\hline
\end{tabular}
\end{center}
\end{table}

\begin{table}
\caption{Summary of regularised uncertainties expected for $A_{5}$, $A_{6}$, and $A_{7}$ at low (5--8~GeV), mid (22--22.5~GeV), and high (132--173~GeV) \ptz\ for the $0<|\yz|<1$ configuration. The total systematic uncertainty is shown with the breakdown into its underlying components. Entries marked with ``-'' indicate that the uncertainty is below 0.0001.
\label{Tab:SystSummary_ybin1_bin2_reg_A2} }
\begin{center}
\begin{tabular}{l||c|c|c||c|c|c||c|c|c}
\hline
\hline
\multicolumn{10}{c}{$p_{T}^{Z} =$ 5--8 GeV} \\
\hline
\hline
Coefficient & \multicolumn{3}{c||}{$A_{5}$} & \multicolumn{3}{c||}{$A_{6}$} & \multicolumn{3}{c}{$A_{7}$} \\
\hline
 Channel        & $ee$       & $\mu\mu$ & $ee$+$\mu\mu$    & $ee$    & $\mu\mu$ & $ee$+$\mu\mu$         & $ee$       & $\mu\mu$ & $ee$+$\mu\mu$   \\
\hline
Total & 0.0027 & 0.0027 & 0.0019 & 0.0029 & 0.0027 & 0.0020 & 0.0020 & 0.0020 & 0.0014\\
\hline
Data Stat. & 0.0024 & 0.0024 & 0.0017 & 0.0025 & 0.0023 & 0.0017 & 0.0018 & 0.0018 & 0.0013\\
\hline
Syst. & 0.0013 & 0.0013 & 0.0010 & 0.0015 & 0.0014 & 0.0011 & 0.0010 & 0.0009 & 0.0007\\
\: MC Stat. & 0.0012 & 0.0013 & 0.0008 & 0.0013 & 0.0012 & 0.0009 & 0.0009 & 0.0009 & 0.0006\\
\: Lepton & 0.0002 & - & - & 0.0002 & 0.0001 & - & 0.0002 & 0.0001 & 0.0001\\
\hline
\hline
\end{tabular}
\end{center}
\end{table}

\begin{table}
\caption{Summary of regularised uncertainties expected for $A_{5}$, $A_{6}$, and $A_{7}$ at low (5--8~GeV), mid (22--22.5~GeV), and high (132--173~GeV) \ptz\ for the $1<|\yz|<2$ configuration. The total systematic uncertainty is shown with the breakdown into its underlying components. Entries marked with ``-'' indicate that the uncertainty is below 0.0001.
\label{Tab:SystSummary_ybin2_bin2_reg_A2} }
\begin{center}
\begin{tabular}{l||c|c|c||c|c|c||c|c|c}
\hline
\hline
\multicolumn{10}{c}{$p_{T}^{Z} =$ 5--8 GeV} \\
\hline
\hline
Coefficient & \multicolumn{3}{c||}{$A_{5}$} & \multicolumn{3}{c||}{$A_{6}$} & \multicolumn{3}{c}{$A_{7}$} \\
\hline
 Channel        & $ee$       & $\mu\mu$ & $ee$+$\mu\mu$    & $ee$    & $\mu\mu$ & $ee$+$\mu\mu$         & $ee$       & $\mu\mu$ & $ee$+$\mu\mu$   \\
\hline
Total & 0.0037 & 0.0028 & 0.0023 & 0.0034 & 0.0030 & 0.0022 & 0.0026 & 0.0021 & 0.0016\\
\hline
Data Stat. & 0.0033 & 0.0025 & 0.0020 & 0.0030 & 0.0026 & 0.0020 & 0.0023 & 0.0018 & 0.0014\\
\hline
Syst. & 0.0017 & 0.0013 & 0.0010 & 0.0016 & 0.0014 & 0.0010 & 0.0012 & 0.0010 & 0.0008\\
\: MC Stat. & 0.0016 & 0.0013 & 0.0010 & 0.0016 & 0.0014 & 0.0010 & 0.0012 & 0.0009 & 0.0007\\
\: Lepton & 0.0006 & 0.0002 & 0.0003 & 0.0003 & 0.0001 & 0.0001 & 0.0002 & 0.0001 & 0.0001\\
\hline
\hline
\end{tabular}
\end{center}
\end{table}

\clearpage
\printbibliography


\newpage 
\begin{flushleft}
{\Large The ATLAS Collaboration}

\bigskip

G.~Aad$^\textrm{\scriptsize 87}$,
B.~Abbott$^\textrm{\scriptsize 114}$,
J.~Abdallah$^\textrm{\scriptsize 65}$,
O.~Abdinov$^\textrm{\scriptsize 12}$,
B.~Abeloos$^\textrm{\scriptsize 118}$,
R.~Aben$^\textrm{\scriptsize 108}$,
O.S.~AbouZeid$^\textrm{\scriptsize 138}$,
N.L.~Abraham$^\textrm{\scriptsize 150}$,
H.~Abramowicz$^\textrm{\scriptsize 154}$,
H.~Abreu$^\textrm{\scriptsize 153}$
% ...
\bigskip
\\
% ...
$^{12}$ Institute of Physics, Azerbaijan Academy of Sciences, Baku, Azerbaijan\\
% ...
$^{65}$ University of Iowa, Iowa City IA, United States of America\\
% ...
$^{87}$ CPPM, Aix-Marseille Universit{\'e} and CNRS/IN2P3, Marseille, France\\
% ...
$^{108}$ Nikhef National Institute for Subatomic Physics and University of Amsterdam, Amsterdam, Netherlands\\
% ...
$^{114}$ Homer L. Dodge Department of Physics and Astronomy, University of Oklahoma, Norman OK, United States of America\\
% ...
$^{118}$ LAL, Univ. Paris-Sud, CNRS/IN2P3, Universit{\'e} Paris-Saclay, Orsay, France\\
% ...
$^{138}$ Santa Cruz Institute for Particle Physics, University of California Santa Cruz, Santa Cruz CA, United States of America\\
% ...
$^{150}$ Department of Physics and Astronomy, University of Sussex, Brighton, United Kingdom\\
% ...
$^{153}$ Department of Physics, Technion: Israel Institute of Technology, Haifa, Israel\\
$^{154}$ Raymond and Beverly Sackler School of Physics and Astronomy, Tel Aviv University, Tel Aviv, Israel\\
% ...
\end{flushleft}



\end{document}